\documentclass[phd,12pt,noupper,appendixpart,oneside]{ubcthesis}

\usepackage[numbers,sort&compress]{natbib}
\usepackage{graphicx}
\usepackage{amsmath}
\usepackage{amssymb}
\usepackage{youngtab}

\newcommand{\Tr}{\ensuremath{\mathop{\mathrm{Tr}}}}

\newcommand{\del}{\partial}
\def\ni{\noindent}
\def\be{\begin{equation}}
\def\ee{\end{equation}}
\def\bea{\begin{eqnarray}}
\def\eea{\end{eqnarray}}
\def\bsp{\be\begin{split}}
\def\la{\langle}
\def\ra{\rangle}
\def\dag{\dagger}
\def\wt{\widetilde}
\def\wh{\widehat}

\def\lr{\leftrightarrow}

\def\da{\dot{\alpha}}
\def\db{\dot{\beta}}
\def\dg{\dot{\gamma}}
\def\dd{\dot{\delta}}
\def\dl{\dot{\lambda}}
\def\dr{\dot{\rho}}
\def\ds{\dot{\sigma}}

\def\G{\Gamma}
\def\D{\Delta}
\def\L{\Lambda}
\def\S{\Sigma}
\def\a{\alpha}
\def\b{\beta}
\def\g{\gamma}
\def\k{\kappa}
\def\d{\delta}
\def\e{\epsilon}
\def\m{\mu}

\def\n{\nu}
\def\s{\sigma}
\def\r{\rho}
\def\l{\lambda}
\def\t{\tau}
\def\o{\omega}
\def\O{\Omega}

\def\mc{\mathcal}

\def\p{\partial}
\def\K{\widetilde{K}}

\def\lr{\leftrightarrow}
\def\bC {\mathbb{C}}
\def\bR {\mathbb{R}}
\def\bZ {\mathbb{Z}}
\def\w{\wedge}

\institution{The University Of British Columbia}
\institutionaddress{Vancouver, Canada}
\department{Department of Physics and Astronomy}
\numberofsignatures{0}          % Signature lines on title page

\previousdegree{B.Sc., McGill University, 2000}
\previousdegree{M.Sc., University of British Columbia, 2003}

\title{The AdS/CFT Correspondence: Classical, Quantum, and Thermodynamical Aspects}
\author{Donovan Young}
\copyrightyear{2006}
\submitdate{\today}
\renewcommand\thepart         {\Roman{part}}
\renewcommand\thechapter      {\arabic{chapter}}
\renewcommand\thesection      {\thechapter.\arabic{section}}
\renewcommand\thesubsection   {\thesection.\arabic{subsection}}
\renewcommand\thesubsubsection{\thesubsection.\arabic{subsubsection}}
\renewcommand\theparagraph    {\thesubsubsection.\arabic{paragraph}}

\begin{document}

% This starts numbering in Roman numerals as required for the thesis
% style.
\frontmatter

% The order of the following components is preserved.  The order
% listed here is the order currently required by the library.
\maketitle
%\authorizationform
\begin{abstract}
Certain aspects of the AdS/CFT correspondence are studied in detail.
We investigate the one-loop mass shift to certain two-impurity string
states in light-cone string field theory on a plane wave
background. We find that there exist logarithmic divergences in the
sums over intermediate mode numbers which cancel between the cubic
Hamiltonian and quartic ``contact term''. Analyzing the impurity
non-conserving channel we find that leading, non-perturbative terms
predicted in the literature are in fact an artifact of these
logarithmic divergences and vanish with them. We also argue that
generically, every order in intermediate state impurities contributes
to the mass shift at leading perturbative order.

The same mass shift is also computed using an improved 3-string vertex
proposed by Dobashi and Yoneya. The result is compared with the
prediction from non-planar corrections in the BMN limit of
$\mathcal{N}=4$ supersymmetric Yang-Mills theory. It is found to agree
at leading order -- one-loop in Yang-Mills theory -- and is close but
not quite in agreement at order two Yang-Mills loops. Furthermore, in
addition to the leading non-perturbative power in the 't Hooft
coupling, we find that two higher half-integer powers are also
miraculously absent. We extend the analysis to include discrete
light-cone quantization, considering states with up to three units of
$p^+$.

We study the weakly coupled plane-wave matrix model at finite
temperature. This theory has a density of states which grows
exponentially at high energy, implying that the model has a phase
transition. The transition appears to be of first order. However, its
exact nature is sensitive to interactions. We analyze the effect of
interactions by computing the relevant parts of the effective
potential for the Polyakov loop operator to three loop order. We show
that the phase transition is indeed of first order. We also compute
the correction to the Hagedorn temperature to two loop order.

Finally, correlation functions of 1/4 BPS Wilson loops with the
infinite family of 1/2 BPS chiral primary operators are computed in
$\mathcal{N}=4$ super Yang-Mills theory by summing planar ladder
diagrams. Leading loop corrections to the sum are shown to vanish. The
correlation functions are also computed in the strong-coupling limit
by examining the supergravity dual of the loop-loop correlator. The
strong coupling result is found to agree with the extrapolation of the
planar ladders. The result is related to known correlators of 1/2 BPS
Wilson loops and 1/2 BPS chiral primaries by a simple re-scaling of
the coupling constant, similar to an observation made in the
literature, for the case of the 1/4 BPS loop vacuum expectation
value.

\end{abstract}

\tableofcontents
%\listoftables
%\listoffigures

\acknowledgements 

I would like first to thank my supervisor, Gordon Semenoff, for more
than five years of tutelage and collaboration, and for introducing me
to the wonders of gauge theory and the gauge/string duality, which I
have grown very fond of. I would also like to thank Gordon, and my
other collaborators Gianluca Grignani, Marta Orselli, Bojan
Ramadanovic, and Shirin Hadizadeh, for sharing in long, complicated,
and exciting computations. I would like to thank the members of the
string group over the years, Greg Van Anders, Henry Ling, Hsien-Hang
(Brian) Shieh, Karene Chu, Dominic Brecher, Mark Laidlaw, Phil DeBoer,
Kazuyuki Furuuchi, and Kazumi Okuyama, for creating a rich environment
and for sharing in the learning process. I would also like to give
special thanks to Moshe Rozali and Mark Van Raamsdonk for pedagogy and
lively group meetings.  I'd also like to acknowledge the many summer
schools and conferences hosted by the department, PIMS, and PITP which
were fundamental in my education as a string theorist. I'd like to
thank Matt Hasselfield for entertaining my puzzlements and helping me
bounce them off the blackboard.

There are also teachers and people I consulted over the years, Kristin
Schleich, Don Witt, Douglas Scott, and Eric Zhitnitsky who I would
like to thank for that and for their general contribution to the
department and therefore to this doctoral work. I would like to thank
the staff of the physics department at UBC over the years, especially
Janet Johnson, Tony, Oliva Dela Cruz-Cordero, and Bridget Hamilton. I
would like to give a special thanks to Janis McKenna for much
administrative help and lively conversations.

Finally, on a personal note, I would like to thank my parents and
brother, and also friends Charles Boylan and Donna Petersen for being
supportive and interested. I'd also like to thank Tim and Dagmar
Sullivan for conversations and hospitality. During my Ph.D., I lost a
dear friend and mentor Ruth Taylor. She deserves a huge thank-you for
influencing the man who went on to do this work. I would also like to give
a very special thanks to my partner Tara for her confidence in me that
is as easy and sure as gravity, and for her bright, rebounding
optimism in which my fears about this thesis could never have hoped to
find their reflection.

% Force a new page.
\newpage

% Now regular page numbering begins.
\mainmatter

% ************************************************************************** %
% ************************************************************************** %
% ************************************************************************** %
\foreword

This thesis collects the work of four publications by the author
concerning quantum, classical, and thermodynamical aspects of the
AdS/CFT correspondence. The thesis begins with an introductory chapter
which provides the reader with the necessary background in non-abelian
gauge theory and the `t Hooft expansion, the non-renormalizable nature
of point-particle quantum gravity, supersymmetry, modern string
theory, and the AdS/CFT correspondence itself. The main matter of the
thesis begins with chapter \ref{chap:lcsft} where the reader is
introduced to the plane-wave version of the AdS/CFT correspondence and
to light-cone string field theory in that context. The original work
of the author published in \cite{Grignani:2005yv} is presented in
section \ref{sec:divcan}, while that of \cite{Grignani:2006en} is
presented in section \ref{sec:altvert}. These works concern divergence
cancellations in string loop corrections and the comparison of those
corrections with their gauge theory duals. Chapter
\ref{sec:matrixmodel} begins with an introduction to the matrix model
of M-theory, and specifically to the plane-wave matrix model. The
original work of the author \cite{Hadizadeh:2004bf} concerning the
deconfinement phase transition found in this model is presented in
section \ref{sec:FEPTPW}. Chapter \ref{sec:wilsonloop} begins with a
brief introduction to the Wilson loop in the AdS/CFT
correspondence. In section \ref{sec:wilsonquartersection}, original
work of the author \cite{Semenoff:2006am} concerning the two point
functions of chiral primary operators with a certain 1/4 BPS circular
Wilson loop is presented.

% ************************************************************************** %
% ************************************************************************** %
% ************************************************************************** %
\chapter{Introduction}
\label{sec:introchap}

{\small
\begin{quote}
His tongue, continuous before and apt\\
For utterance, severs; and the other's fork\\
Closing unites. That done, the smoke was laid.\\
The soul, transform'd into the brute, glides off,\\
Hissing along the vale, and after him\\
The other talking sputters;\\
\rightline{--- Dante's {\it Inferno}, Canto 25 $\;\;\;\;\;\;\;\;\;\;\;\;\;\;\;\;\;\;$}
\end{quote}}

It is a well known observation that in any endeavour, the tension
of apparent contradictions leads to a higher understanding - one
which naturally fuses those into a whole greater than the sum of
its parts. Such a tension exists in theoretical physics, between
the description of the strongest force in nature, and the weakest.
String theory in general and the AdS/CFT correspondence in
particular, are emerging as a fusion of the understanding of these
two forces; a symbiosis with the potential to answer questions
beyond the scope of either and to probe the very structure of
space and time themselves. In order to understand this
correspondence, we must know something about these two forces, and
their individual descriptions.

% ************************************************************************** %
\section{Strong nuclear force}

The strong nuclear force is responsible for the cohesion of matter
at the smallest known scales - inside the particles which compose
the nuclei of atoms - a scale of $10^{-15}$ m. The modern
description of this force is known as Quantum Chromodynamics or
QCD. QCD is a non-abelian gauge theory described by the Yang-Mills
action

\be\label{qcdaction}
S = -\frac{1}{4} \,\int d^4x
\,\Tr\, F_{\m \n} \,F^{\m \n}+S_m, \qquad F_{\m\n} = \del_\m
A_\n - \del_\n A_\m - i\,g_{YM}\, [A_\m, A_\n]
\ee

\ni where the $A_\m(x)$ are matrix-valued four-vectors in the adjoint
representation of $SU(3)$, and the coupling constant in the theory is
$g_{YM}$. The action for the matter content of the theory - the quarks
- has been indicated by $S_m$. As is usual in quantum field theories,
this bare coupling is renormalized, and the physical strength of the
force described by the theory is given by the renormalized coupling
$\tilde g_{YM}(k)$ which is a function of the energy scale $k$ of the
process being described.  A remarkable feature of this renormalization
earned Politzer, Gross, and Wilczek the Nobel prize in physics
2004. The coupling $g_{YM}$ of QCD, unlike other quantum field
theories, decreases with increasing energy scale $k$, so that at high
energies, the theory becomes free. This property therefore earned the
name {\it asymptotic freedom}. We see here that the strongest force in
nature, is actually weak if the relevant energy scales are high
enough.

Often in quantum field theory our only analytic tool is perturbation
theory. The same is true for QCD. If we would like to calculate the
expectation value of an observable ${\cal O}$, we need to evaluate the
path-integral\footnote{We are being schematic here, in an attempt to
  maintain clarity. A more precise statement is that $\la \O |\,{\cal T}
  \,{\cal O}\, | \O \ra = \lim_{T\rightarrow \infty(1-i\e)}\frac{ \int [d
    A_\m] [d\, \text{matter}] {\cal O} \exp \left[i \int_{-T}^T d^4x
    {\cal L} \right]}{ \int [d A_\m] [d \,\text{matter}] \exp \left[i
    \int_{-T}^T d^4x {\cal L} \right]}$, where $S=\int d^4x\, {\cal L}$,
$|\O\ra$ is the ground state of the interacting theory, and ${\cal T}$
  indicates time-ordering, c.f. \cite{Peskin:1995ev}.}

\be
\la {\cal O} \ra = \frac{ \int [d A_\m] \,{\cal O}\,e^{iS} }{ \int [d
    A_\m] e^{iS} }.
\ee

\ni This is accomplished by Taylor-expanding the exponential
$e^{iS}$, out to the desired order of accuracy. This procedure is
only sensible when $g_{YM}$ is small. For QCD, this procedure then
only works for very high-energy processes. Indeed this is the
regime where QCD has been tested in particle accelerators, and has
successfully described the dynamics witnessed there. But at
terrestrial energy scales, like those found roughly anywhere
cooler than inside the sun, $g_{YM} \sim 1$, and perturbation
theory is useless. Now we see that there are two issues, one is
that the strong force is only sometimes strong, and the second is
that we can only use our quantum field theory to (analytically)
describe its nature when it is weak. Lattice field theory is a
numerical technique which allows strong-coupling answers to be
squeezed out of (\ref{qcdaction}) and has been successful in
describing some aspects of those dynamics. However, an analytical
technique remains out of reach, and greatly desired.

% ************************************************************************** %
\section{Gravity and renormalization}

Gravity is the force responsible for structure at the largest
known scales - those of the known universe - some $10^{25}$ m in
size. The gravitational force is 40 orders of magnitude weaker
than the strong force. The modern description of gravity is
entirely classical, it says nothing about $\hbar$, the scale at
which quantum fluctuations become important. In this respect it is
radically different from QCD, for which {\it only} a quantum
description is sensible, due to its fantastically short range.
Gravity's modern description was given birth to by Einstein, who
successfully unified the force with the precepts of special
relativity - that is Lorentz invariance. It is captured by the
Einstein-Hilbert action

\be
S = \frac{1}{16\pi\,G}\,\int d^4x \, \sqrt{-g} \, R \qquad
R = g^{\m\n}\,R_{\m\n} \qquad R_{\m\n} = \del_\m \G_{\r \n}^\r -
\del_\r \G_{\m\n}^\r + \G_{\m\l}^\r \G_{\r\n}^\l - \G_{\r\l}^\r \G_{\m\n}^\l
\ee

\ni where $R$ is the Ricci scalar built out of $g_{\m\n}$, the metric
of space-time, and $G$ is the Newton constant, or universal constant
of gravitation. Already at this level we note similarities between
the descriptions of these vastly divergent forces. The Christoffel
connection $\G^\r_{\m\n}$ is analogous to the gauge field $A_\m$ of
(\ref{qcdaction}) and the Ricci tensor $R_{\m\n}$ is a sort of
``field-strength'' of $\G^\r_{\m\n}$ in the same sense that $F_{\m\n}$
is of $A_\m$. We see immediately that the two theories are non-linear
(non-abelian), and so share the characteristic that their fields are
sources for themselves. However, early attempts to push this analogy
further by quantizing gravity met with failure.

It is simple to see that there is a scale at which one expects gravity
to be modified by quantum mechanics. Take for example a black hole
formed by a very heavy particle. When the Compton wavelength of the
particle is comparable to the Schwarzschild radius

\be\label{QG}
\frac{\hbar}{m\,c} \sim \frac{2\,G\,m}{c^2}
\ee

\ni we expect that classical gravity ought to be invalid. This occurs
for $m = M_{pl} \simeq 10^{19}$ GeV, or for length scales $l_{pl}
\simeq 10^{-35}$ m. At these unimaginably high energies, an accurate
description of gravity would na\"{i}vely be given by a quantization of the
classical theory into a quantum field theory of gravity. It is,
however, precisely that characteristic of such theories which is
responsible for asymptotic freedom in QCD, which cripples such an
attempt at the first step.

The renormalization of the coupling constant in a quantum field
theory, such as QCD, arises in the treatment of integrals over the
momenta of intermediate virtual particles. These integrals
formally diverge, but may be made finite by placing an upper-bound
on the momenta being integrated over. This procedure is very sound
physically, because one expects the quantum field theory at hand
to be an {\it effective} theory, valid at the scale in question,
but eventually superseded at some higher energy, where new physics
is expected to be active. In condensed matter physics, this idea
was understood early on, because the cut-off is the very physical
scale of the atomic size. Once cut off, the integrals in QCD
produce pieces proportional to the cut-off but independent of the
energy scale of the process being described, and other pieces
independent of the cut-off, but dependent on the relevant energy
scale.  The cut-off dependent pieces are interpreted in much the
same way that an absolute potential energy is - it is irrelevant -
only potential differences are physical. It is then the cut-off
independent, energy scale dependent or {\it
  running} quantities which correspond to physical attributes of the
theory.

If the coupling constant in a quantum field theory is
dimensionless, then probability amplitudes may be expressed as
polynomials in it. This is the case for QCD. Should the coupling
constant $g$ have negative mass-dimension $-p$, then probability amplitudes can
only be described as polynomials in the dimensionless combination
$\L^p\,g$, where $\L$ is the momentum cut-off. This is a
non-renormalizable quantum field theory, whose
cut off momentum integrals do not contain pieces independent of the
cut-off scale. Because we cannot - regardless of the energy scale of
the process being described - arrive at a prediction independent of
the cut-off scale, and since we don't know with any precision what
this scale is, we cannot make any definite predictions with such a theory.
As can be seen from (\ref{QG}), the coupling constant in gravity $G$
is proportional to $m^{-2}$ if we set $\hbar = c =1$. Thus gravity
has a coupling constant with negative mass dimension and so is
a non-renormalizable quantum field theory.

It is this fact that set the strongest and weakest forces in the
universe at loggerheads. Indeed, it set gravity apart from all
three of the other fundamental forces, which were successfully
described by an aggregate, renormalizable quantum field theory
called the {\it standard model} by the 1970's.

% ************************************************************************** %
\section{Early string theory and large N}
\label{sec:largeN}

\begin{figure}[ht]
\begin{center}
\includegraphics*[bb= 0 0 190 60,width=2.5in]{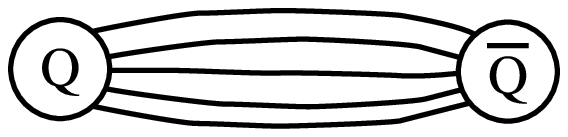}
\end{center}
\caption{The field lines between two quarks in the low energy regime of QCD.}
\label{fig:fluxtube}
\end{figure}

String theory was born in an attempt to describe the strong
nuclear force before the days of QCD. One of the early
observations was that there was a zoo of mesons, whose masses $m$
were related to their spins via $J = \a'\,m^2$, where $\a'$ is a
constant known as the Regge slope. It was soon realized that a
quantum string gave rise to such a relation. With the benefit of
hindsight, we can see how this stringy-ness is manifested in
mesons. We now know that a meson is a quark-antiquark bound state,
whose colour field lines are confined into a {\it flux tube} as
shown in figure \ref{fig:fluxtube}. It is this flux tube which behaves
as a string of a given tension.
An empirical formula for meson scattering was put forward by Veneziano
\cite{Veneziano:1968yb}, which was later shown to be derivable from
string theory. However in the late 60's experimental data began to
show that the Veneziano amplitude gave an incorrect large energy
behaviour, and soon after QCD was adopted as the correct description
of the strong force.

\begin{figure}[ht]
\begin{center}
\includegraphics*[width=4in, height=1.6in]{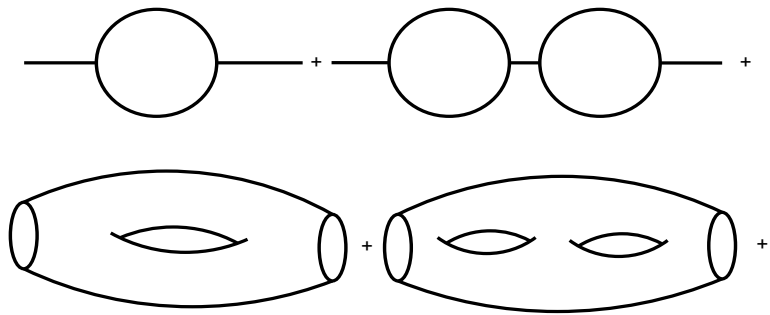}
\end{center}
\caption{Comparison of quantum field theory perturbative expansion to
  that of string theory. A closed string sweeps-out a two dimensional
  {\it worldsheet} whose genus represents the number of powers of the
  string coupling constant associated with the process.}
\label{fig:stringpert}
\end{figure}

This turn of events still left the question of how the action
(\ref{qcdaction}) could possibly encode string-like dynamics in the
strong coupling regime. In 1974 `t Hooft \cite{'tHooft:1973jz} made a
remarkable leap forward in this direction. String perturbation theory
naturally organizes itself into a {\it genus expansion}, see figure
\ref{fig:stringpert}. 
Whereas in a regular quantum field theory each vertex would
contribute a power (or two) of the coupling constant, the same
r\^{o}le in string theory is played by the genus of the worldsheet. `t
Hooft discovered that such a genus expansion lay hidden in the theory
described by (\ref{qcdaction}). In order to see this we take the gauge
group of the theory to be $SU(N)$, where we will eventually want to
consider $N$ large. We write the gauge fields as $A_\m = g_{YM}\,A^a_\m\,t^a$,
where $t^a$ are the generators of $SU(N)$ with $a=1,\ldots,N$, and we
have rescaled the fields by the coupling constant. This allows us to
write the action in the following form\footnote{We have left the
  matter action $S_m$ out of this discussion, for simplicity.}

\begin{equation}
S = -\frac{1}{2\,g_{YM}^2} \int d^4x \Tr F_{\mu\nu}\,F^{\m\n} 
\label{normaction}
\end{equation} 

\noindent where now $F_{\mu\nu}$ contains no factors of $g_{YM}$. 
We may express the gauge degrees of freedom as

\begin{equation}
A^a_\mu t^a_{{\bar i}j} \rightarrow \left( A_\mu \right)_{{\bar i}j}
\end{equation}  

\begin{figure}[ht]
\begin{center}
\includegraphics*[bb= 0 0 465 250, width=2.75in]{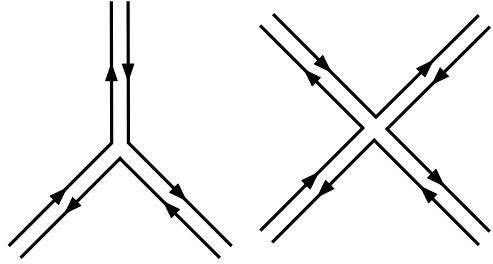}
%\includegraphics*[bb= 30 140 160 258, width=1.5in, height=1.5in]{thooft.ps}
%\hspace{0.75cm}
%\includegraphics*[bb= 22 284 154 372, width=1.5in, height=1.5in]{thooft.ps}
%\hspace{0.75cm}
%\includegraphics*[bb= 178 167 274 244, width=1.5in, height=1.5in]{thooft.ps}
\end{center}
\caption{Vertices in the `t Hooft model.}
\label{fig:verts}
\end{figure}

\ni where $\bar i, j$ is an (anti) fundamental index running from $1,\ldots,N$.
In this language the propagators have the following index structure

\be
\Bigr< \left( A_\mu (x) \right)_{{\bar i}j}
       \left( A_\nu (y) \right)_{{\bar k}l}  \Bigl> \,\sim
\bigl( \delta_{{\bar i}l} \delta_{{\bar k}j} 
	- \frac{1}{N} \delta_{{\bar i}j} \delta_{{\bar k}l} \bigr)
\label{aprop}
\ee

\noindent and therefore to leading order in the large $N$ limit, the
second term in the gauge field propagator may be ignored. In fact,
this second term disappears entirely if one considers the gauge group
$U(N)$ instead of $SU(N)$, and then everything that follows here is
exactly (instead of approximately) true.  In non-abelian gauge theory,
the gauge field $A^a_\mu(x)$ always transforms in the adjoint
representation of the gauge group. The interpretation of the picture
which emerges here is that an adjoint field $\phi^a(x)$ may be
represented as a direct product of fundamental and anti-fundamental
fields, $\phi_{\bar i}(x)\phi_j(x) = \phi_{{\bar i}j}(x)$.  In group
theory language this is the statement

\begin{equation}
{\bar N} \otimes N = \hbox{adjoint} \oplus \hbox{singlet}
\end{equation}

\noindent but for large $N$ the singlet contribution is suppressed, as
per (\ref{aprop}).  Thus the adjoint gauge fields are in a sense
quark/anti-quark composites, stressing the {\it flux tube}
interpretation. `t Hooft developed a diagram notation based on this
fact, called the {\it fat graph} notation,
\[
\begin{minipage}[bottom]{2in}
$\Bigr< \left( A_\mu (x) \right)_{{\bar i}j} \left( A_\nu (y)
\right)_{{\bar k}l} \Bigl>$
\end{minipage} \sim \hspace{0.85cm} 
\begin{minipage}[bottom]{2in}
\includegraphics*[bb= 0 0 155 60, width = 1.5in]{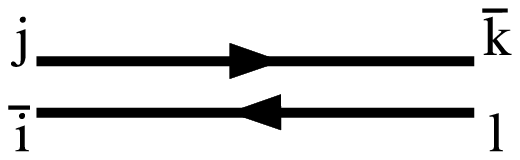}
\end{minipage}
\]
in this notation the three and four point vertices are given
by the diagrams shown in figure \ref{fig:verts}.

\begin{figure}[ht]
\begin{center}
\includegraphics*[bb= 0 0 230 210, width=1.5in]{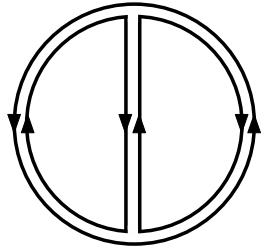}
\end{center}
\caption{A vacuum diagram of the 't Hooft model.}
\label{fig:vac}
\end{figure}

We now introduce the quantity $\lambda$ which is known as the `t Hooft
coupling, $\lambda = g_{YM}^2N$. Glancing back at (\ref{normaction}),
it can be seen that the three and four point vertices come with a
factor of $1/g_{YM}^2 = N/\lambda$ whereas the gauge field propagator
is proportional to $g_{YM}^2 = \lambda/N$. Also, in a given diagram,
when an arrowed line closes on itself (forms a loop) it supplies a
factor of $\delta_{ii} = N$. Consider now the set of all vacuum
diagrams, an example of which is shown in figure \ref{fig:vac}. A
diagram with $V$ vertices, $E$ propagators, and $F$ loops will
therefore be proportional to

\begin{equation}
N^{V-E+F} \lambda^{E-V}
\end{equation}

\begin{figure}[ht]
\begin{center}
\includegraphics*[bb= 0 0 230 210, width=1.5in]{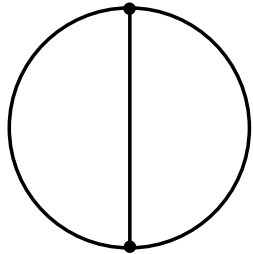}
\end{center}
\caption{The vertices and edges of the diagram in figure \ref{fig:vac}.}
\label{fig:vac2}
\end{figure}

\begin{figure}[ht]
\begin{center}
\includegraphics*[bb=0 0 230 210, width=1.5in]{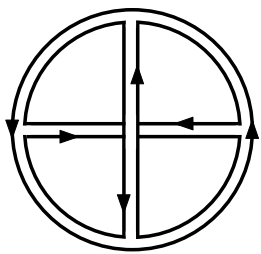}
\end{center}
\caption{A diagram associated with a surface of genus $g=1$.}
\label{fig:vac3}
\end{figure}

\noindent We have chosen the letters $V$ and $E$ because if we collapse
the double lines to single ones, then the diagram has $V$ vertices, and $E$ edges,
as in figure \ref{fig:vac2}. The letter $F$ is chosen because each loop forms
a {\it face} in the double-line diagram. 
The combination $\chi = V-E+F$ is recognized
as the Euler character, which implies a connection between the diagrams and surfaces
of a given genus $g$, since $\chi = 2-2g$. Thus we have that a given diagram is 
proportional to

\begin{equation}
N^{2-2g} \lambda^{E-V}
\end{equation}

\noindent and so diagrams corresponding to surfaces of higher genus are suppressed 
by successive powers of $1/N^2$. An example of a higher genus diagram is shown in
figure \ref{fig:vac3}.

Referring to figure \ref{fig:vac}, we see that $V=2$, $E=3$, and
$F=3$, and therefore the power of $N$ associated with this diagram is
$N^{2-3+3}=N^2$ or equivalently the genus $g$ is $0$. The genus $0$
graphs are given a special name, they are called {\it planar}
graphs. This is because they can be drawn on a plane. In contrast
figure \ref{fig:vac3} shows a graph with $V=4$, $E=6$, and $F=2$, and
thus is proportional to $N^0$ or is genus $1$ and obviously can not be
drawn on a plane due to the crossing central propagators; a handle
would need to be added to the plane in order to draw this graph. The
sum of all the vacuum diagrams takes the following form

\begin{eqnarray}
\label{oneoverN}
\sum_g N^{2-2g} {\cal F}_g(\lambda)\\
{\cal F}_g(\lambda) = \sum_n C^g_n \lambda^n
\end{eqnarray} 

\noindent where the function ${\cal F}_g(\lambda)$ is the sum of all diagrams
of genus $g$, which is naturally a power series in the `t Hooft coupling. For
example ${\cal F}_0(\lambda)$ would be the sum of all planar diagrams. The point
of interest here is that in the large $N$ limit, we have a perturbative genus 
expansion. In string theory, the very same type of expansion arises in the 
calculation of amplitudes. It is this connection to string theory which makes
the `t Hooft large $N$ expansion an important observation; it connects QCD-type
quantum field theory to string theory.

% ************************************************************************** %
\section{Modern string theory}
After losing the bid to describe the strong nuclear force, string
theory was revived in the 1980's when it was realized that it could be
used as a candidate for a quantum theory of gravity and perhaps even a
grand unified theory of physics. The divergences which plagued the
quantum field theory approach to quantizing gravity disappear with
string theory. This can be traced to the delocalization of the
interaction vertices enjoyed by stringy Feynman diagrams, as shown in
figure \ref{fig:stringpert}. We will review string theory from the
perspective of the Green-Schwarz superstring, as this will be most
relevant for the work on light-cone string field theory on the
plane-wave presented in chapter \ref{chap:lcsft}. We will not
reference very widely in this section; the relevant references are to
be found in the standard textbooks such as
\cite{Green:1987sp}\cite{Green:1987mn}.

% ========================================================================== %
\subsection{Preliminaries}
\label{sec:prel}

\begin{figure}[ht]
\begin{center}
\includegraphics*[bb= 0 0 580 340, height=2.5in]{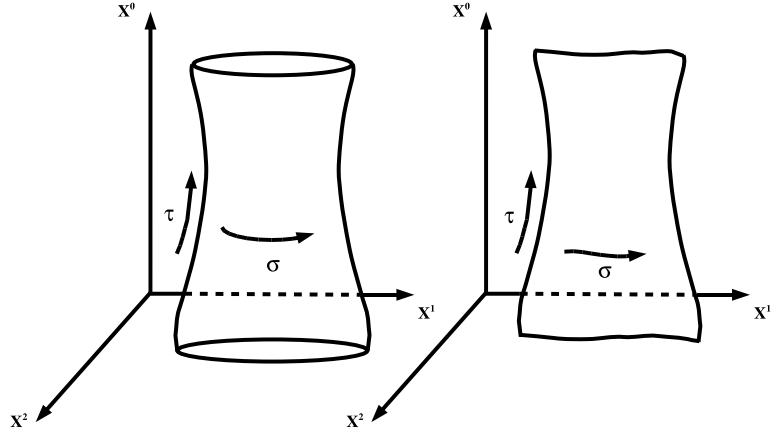}
\end{center}
\caption{A closed (left) or open (right) string worldsheet is described by
  embedding functions $X^\m(\s,\t)$.}
\label{fig:string}
\end{figure}

The dynamics of a string are defined by the requirement that at the
classical level, they lead to a minimization of the proper area
swept out by the string's worldsheet as it
propagates through a target spacetime $G_{\m\n}$, see figure \ref{fig:string}. 
The string worldsheet is
embedded into the target spacetime by $d$ embedding functions
$X^\m(\s,\t)$, where $(\s,\t)$ are the coordinates on the worldsheet
and $d$ is the dimension of the target spacetime. The Polyakov action
for the string is given by

\be\label{polyakov}
S_P = -\frac{T}{2}\,\int d\s \int d\t\sqrt{|\det h|} \, h^{a b}\, \del_a X^\m\,
G^{\m\n}(X)\, \del_b X^\n
\ee

\ni where $a$ and $b$ take on values 0 or 1 corresponding to the
coordinates $\t$ and $\s$ respectively,
and $h_{ab}(\s,\t)$ is known as the worldsheet metric. The energy per
unit length of the string, or {\it string tension}, is given by $T$.
The Polyakov action respects the symmetries of the target spacetime,
but also respects two further symmetries: Weyl (or conformal) invariance, and
reparametrization invariance. Weyl invariance is simple to
see. Consider rescaling the worldsheet metric as follows $h_{ab}' =
\o(\s,\t)\,h_{ab}$. Only for two-dimensional metrics will the
combination $\sqrt{|\det h|}\,h^{ab}$ be invariant. This is a very
powerful symmetry in string theory. It tells us that the worldsheet
theory is conformally invariant. Reparametrization invariance is the
statement that we may paint on to the worldsheet any coordinates we
see fit; the dynamics can not depend on the coordinate system
chosen. This symmetry may be expressed as follows

\be
\s \rightarrow \s'(\s,\t) \qquad 
\t \rightarrow \t'(\s,\t).
\ee

\ni We thus have three free functions with which to gauge-fix the
worldsheet metric $h_{ab}$ : two reparametrization and one Weyl
re-scaling. However, being a symmetric $2\times 2$ matrix, $h_{ab}$
has only three degrees of freedom. We are therefore free to set it to
the Minkowski metric $\text{diag}(-1,1)$.

In order to analyze the equations of motion for the fields $h_{ab}$
and $X^\m$, we will temporarily set the target space to flat
$d$-dimensional Minkowski space. The equations of motion are then

\be\label{streom}
\left( \del_\s^2 - \del_\t^2 \right) X^\m (\s,\t) = 0 \qquad
\frac{\d S_P}{\d h_{ab}} = 0
\ee

\ni and so we have the free two-dimensional wave equation governing
the embedding functions, while the equation of motion for $h_{ab}$ may
be restated using the energy-momentum tensor for the worldsheet
theory

\be\label{Tab}
T_{ab} \equiv -\frac{2}{T}\frac{1}{\sqrt{|\det h}} \frac{\d S_P}{\d
  h_{ab}} = \del_a X^\m \del_b X_\m -\frac{1}{2}\,h_{ab}\,h^{cd}
\del_c X^\m \del_d X_\m = 0.
\ee

\ni Once the Minkowski gauge has been chosen for the worldsheet
metric, (\ref{Tab}) must be imposed as a constraint - this is known as
the {\it Virasoro constraint}. The solutions to the wave equation for
the embedding functions come in two topologies, which are
differentiated by a choice of boundary condition. We may impose one
of

\be
X^\m(\s,\t) = X^\m(\s+\pi,\t) 
\qquad \text{or} \qquad
\del_\s X^\m(0,\t) = \del_\s X^\m(\pi,\t) = 0
\ee

\ni where we have taken the range of $\s$ to be $[0,\pi]$. The first
of these describes closed strings, and the second open strings. 

% ========================================================================== %
\subsection{Mode expansions and light-cone gauge}
\label{sec:modeslcg}

The solution of (\ref{streom}) for closed strings is as follows,

\bsp\label{modexp}
X^\m(\s,\t) &= X^\m_L + X^\m_R \qquad \text{where}\\
X^\m_L &= \frac{1}{2} x_L^\m + \a' p_L^\m (\t + \s) + \frac{i}{\sqrt{2\a'}}
\sum_{n\neq 0} \frac{1}{n} \wt \a^\m_n \, e^{-2in(\t+\s)}\\
X^\m_R &= \frac{1}{2} x_R^\m + \a' p_R^\m (\t - \s) + \frac{i}{\sqrt{2\a'}}
\sum_{n\neq 0} \frac{1}{n} \a^\m_n \, e^{-2in(\t-\s)}
\end{split}
\ee

\ni where we have introduced $\a' \equiv (2\pi T)^{-1}$, which is the
Regge slope. The quantities $\tiny{\frac{1}{2}}(x_L^\m + x_R^\m)$ and
$\tiny{\frac{1}{2}}(p_L^\m + p_R^\m)$ are the center of mass
coordinates and momenta\footnote{Note that momentum is defined as
$\int_0^\pi d\s \d L / \d \dot X^\m = T\int_0^\pi d\s \dot
X^\m(\s,0)$.}, respectively. Since the string is closed, we must take
$p_L^\m = p_R^\m$ in a topologically trivial target space\footnote{An
open string must obey the Neumann boundary condition $\p_\s
X^\m(\s)|_{\s=0,\pi} = 0$. This not only sets $p_L^\m = p_R^\m$ but
also enforces $\wt \a^\m_n = \a^\m_n$.}. The $\a_n^\m$ and $\wt
\a_n^\m$ are the amplitudes of the n-th left-moving and right-moving
vibration modes, respectively. Reality of $X^\m$ enforces $(\a^\m_n)^*
= \a^\m_{-n}$, and $(\wt\a^\m_n)^* = \wt\a^\m_{-n}$. The Virasoro
constraint (\ref{Tab}) is

\be\label{vira}
\dot X \cdot X' = 0 = \frac{1}{2} \left( \dot X^2 + X'^2 \right)
\ee

\ni where we use the prime to denote differentiation by $\s$ and the
dot for differentiation by $\t$. Further the dot product refers to
contraction of Lorentz indices.

There are different methods of quantizing the string, but we will
concentrate on light-cone quantization. This method is attractive
because it eliminates unphysical degrees of freedom at the outset, so
that every quantum state is physical, and there is no need to worry
about ghosts. The drawback of the method is that Lorentz invariance
becomes obscured and is no longer manifest. To begin, we note that
fixing $h_{ab}$ to the Minkowski metric has not completely used up the
gauge freedom. Indeed, under a reparametrization $\xi^0 = \del \t /
\del \t'$, $\xi^1 = \del \s /\del \s'$, $h_{ab}$ transforms as follows

\be
\d h^{ab} = \xi^c \del_c h^{ab} - \del_c \xi^a h^{cb} - \del_c \xi^b h^{ac}.
\ee

\ni If we accompany this by a Weyl rescaling $(1 + \o(\s,\t))$ such that

\be\label{resid}
\del^a \xi^b + \del^b \xi^a = \o \eta^{ab}
\ee

\ni where $\eta^{ab}$ is the Minkowski metric, then this combination
leaves the choice $h^{ab} = \eta^{ab}$ invariant. Consider the
following coordinates $\s^{\pm} = \t \pm \s$. In these coordinates
(\ref{resid}) for $a\neq b$ becomes

\be
\del_+ \xi^- = 0 \qquad \del_- \xi^+ =0.
\ee

\ni This implies that we are free to change $\s^+$ by any function of
$\s^+$: $\s^+ \rightarrow \tilde \s^+(\s^+)$, and similarly $\s^-
\rightarrow \tilde\s^-(\s^-)$. This is a powerful residual gauge
symmetry which allows for light-cone gauge quantization.

The manifest Lorentz invariance of the target space is broken in the
light-cone gauge, by singling-out two directions to be the so-called
light-cone directions 

\be
X^\pm = \frac{1}{\sqrt{2}}\left(X^0 \pm X^{d-1}\right) \qquad X^\m = (X^-,X^+,X^i)
\ee 

\ni where $i=1,\ldots,d-2$. In the previous paragraph, we saw that we
are free to reparametrize the worldsheet coordinates. To this end, and
in light of (\ref{modexp}) we choose 

\be
\s^+ = \frac{1}{\a'\,p^+}\left( X^+_L - \frac{1}{2} x^+ \right) \qquad
\s^- = \frac{1}{\a'\,p^+}\left( X^+_R - \frac{1}{2} x^+ \right).
\ee

\ni Now we have a rather ``natural'' embedding where the worldsheet time
$\t$ is simply given by the light-cone $+$ direction, since,

\be\label{x+}
X^+(\s,\t) = X^+_L+X^+_R = x^+ + 2\a'p^+\t.
\ee

\ni We still need to impose the Virasoro constraints
(\ref{vira}), which we may write equivalently as $(\dot X \pm
X')^2=0$. In light-cone coordinates, this is

\be
(\dot X^- \pm {X^-}') = \frac{1}{4\a' p^+}(\dot X^i \pm {X^i}')^2
\ee 

\ni where we have used (\ref{x+}). From this expression $X^-(\s,\t)$
is completely fixed in terms of the $X^i(\s,\t)$. What we see here is
that, in fact, there are only $d-2$ physical vibratory degrees of
freedom; these are the $X^i(\s,\t)$. The $X^\pm$ are non-dynamical,
and fixed by gauge freedom and the imposition of constraints. The
physical idea here is that longitudinal modes are non-physical and do
not correspond to string dynamics. The transverse oscillations
captured by the mode expansions of the $X^i(\s,\t)$, along with any
center of mass motion, completely capture the string's dynamics.

The Virasoro constraints are enforced through the Fourier components
of the stress-energy tensor (\ref{Tab}). In light-cone gauge, we have

\bsp
L_m = \frac{T}{2} \int_0^\pi d\s\, e^{-2im\s}\, T_{--}  = 
\a' \frac{p_R^2}{8} + \frac{1}{2}
\sum_{n \neq 0} \a_{m-n} \cdot \a_n\\
\wt L_m = \frac{T}{2} \int_0^\pi d\s\, e^{-2im\s}\, T_{++}  = 
\a' \frac{p_L^2}{8} + \frac{1}{2}
\sum_{n \neq 0} \wt \a_{m-n} \cdot \wt \a_n
\end{split}
\ee

\ni and must have that $L_m = \wt L_m = 0$ for all $m$. The condition
$L_0 + \wt L_0 = 0$ then gives us\footnote{When the string is
  quantized, the oscillators are promoted to operators and a normal
  ordering constant modifies this relation so that $L_0 + \wt L_0 - 2
  a$, instead, annihilates a physical state.} 

\be\label{masssquared}
(\text{mass})^2 = - p^2 = \frac{2}{\a'} \sum_{n=1}^\infty \left(
\a_{-n} \cdot \a_n + \wt \a_{-n} \cdot \wt \a_n \right)
\ee

\ni where we have used the fact that $p^\m_L = p^\m_R = p^\m$ for
closed strings. This is a truly beautiful result, for it tells us
that the internal excitations of the string worldsheet are reflected
as spacetime mass in the target space; an excited string is
heavy. Another important constraint arising from $L_0$ and $\wt L_0$
is the {\it level matching condition}. This stems from the condition
$L_0 - \wt L_0=0$. This tells us that

\be\label{levelmatch}
\sum_{n \neq 0} \a_{-n} \cdot \a_n = 
\sum_{n \neq 0} \wt \a_{-n} \cdot \wt \a_n
\ee

\ni or, in other words, the degree of excitation of left moving modes
must be matched by that of the right-moving modes. As we will see in a
later section, the level-matching condition is modified when the
target space contains topologically non-trivial cycles. 

% ========================================================================== %
\subsection{Supersymmetry: Why?}

\begin{figure}[ht]
\begin{center}
\includegraphics*[bb=30 45 550 290,height=2.5in]{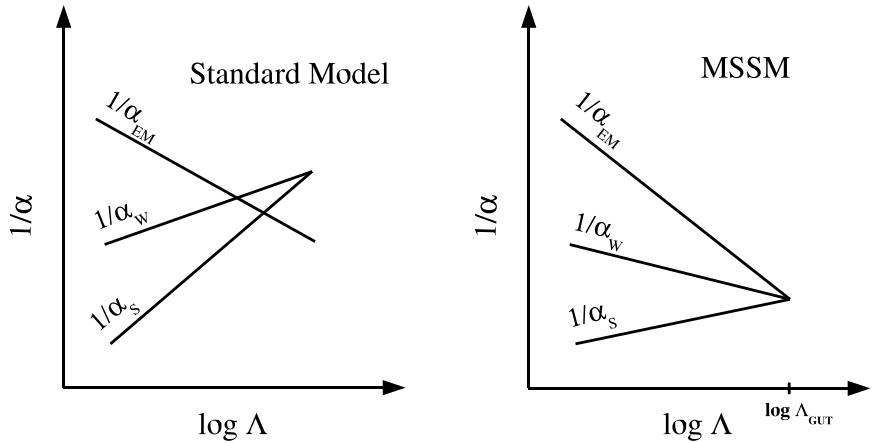}
\end{center}
\caption{Under the minimal supersymmetric extension of the standard
  model (MSSM), the extrapolated couplings for the strong, weak, and
  electromagnetic fields are equal at an energy scale $\L_{\text{GUT}}$, the
  {\it grand unified theory} scale.}
\label{fig:coupling}
\end{figure}

Supersymmetry is an enlargement of the symmetry group of spacetime
obtained by a grading of that algebra. In terms of particles and
fields propagating in spacetime, it is more simply understood as the
statement that there is a symmetry relating fermionic and bosonic
physical degrees of freedom such that, for every bosonic state of mass
m, there exists a fermionic {\it superpartner} of the same mass, and the
same quantum numbers generally, with the obvious exception of
spin. Supersymmetry has grown to be an attractive concept in
theoretical physics. Pessimistically, one might say this is because an enlargement of
symmetry allows for an enlargement of calculational techniques, or at
least an enlargement of ease in developing calculational
techniques. Optimistically, supersymmetry does go a certain distance
towards solving the cosmological constant problem, the hierarchy
problem, and when applied to
the standard model, predicts a unification of the strong, weak, and
electromagnetic coupling constants, see figure \ref{fig:coupling} for
a cartoon of this result. This last point peaks interest in
so far as the possible indication that at some high energy scale, a
grand unified supersymmetric theory may exist, which flows down to our
standard model at low energies.

The cosmological constant problem is an apparent mismatch between the
expected vacuum energy of the standard model, and the observed value
in nature inferred via cosmology. The discrepancy is an embarrassing
120 orders of magnitude. The vacuum energy in the standard model is
formally infinite as it corresponds to the sum of the zero-point
energies of all the modes of all the fields. This infinity is cut off
conservatively at the Planck scale $10^{19}$ GeV, since we expect the
standard model to lose its validity at least by this energy. A
standard model with unbroken supersymmetry would actually give zero
for the vacuum energy. This is a general statement about
supersymmetric theories: the ground state energy is always identically zero.
Of course the cosmological constant observed in nature is not zero,
but there are methods
available to softly break supersymmetry leading to a greatly reduced,
non-zero, vacuum energy.  

The hierarchy problem concerns the mass $m$ of the Higgs boson. The
renormalization of this mass is 
controlled by the quadratic divergence encountered in the quantum
corrections to its propagator
\[
\begin{minipage}[bottom]{2in}
\includegraphics*[bb= 0 0 110 50]{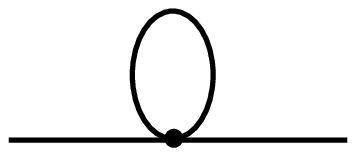}
\end{minipage} \rightarrow \hspace{0.3cm} 
\begin{minipage}[bottom]{2in}
\be\nonumber \d m^2 \propto M^2 \ee
\end{minipage}
\]
where $M$ is the mass of the particle in the loop. Therefore if
there are heavy particles in a theory, they will cause the
renormalized Higgs mass to be too large. For example, string theory
will give Planck mass particles, thus causing the Higgs mass and hence
the electroweak scale to be Planck scale. In fact the electroweak scale is
about $100$ GeV. In a supersymmetric theory, the corrections would also include a
fermion loop, which would cancel out the mass shift. This cancellation
would persist at higher order loops in perturbation theory effectively
protecting the Higgs mass against quantum corrections. Without such a
mechanism, the only recourse is to ``fine-tune'' the bare
(unrenormalized) Higgs mass so as to end-up with the observed value
after renormalization. This fine-tuning is viewed as an extremely
unnatural procedure in a fundamental theory of physics, and is
generally unpalatable to most researchers. Supersymmetry offers a more
universal resolution of this issue.

Obviously, supersymmetry is not an exact symmetry of nature at
currently probed energy scales; we do not see superpartners of the
elementary particles. This is generally taken to mean that
supersymmetry is a broken symmetry, and that there is a scale which
sets when this breaking occurs, and which determines the masses of the
superpartners. Particle experimentalists are ever pushing up the
energy of their accelerators in hopes that, amongst other things, the
superpartner masses will cross into view.

% ========================================================================== %
\subsection{Supersymmetry: Details and implementation on the string}
\label{sec:susydet}

The Poincar\'{e} group is a realization of the symmetries manifest in
flat Minkowski spacetime - translations, rotations, and boosts. The
generators of these transformations are given by $P_\m$, $J_i$, and
$K_i$ respectively and obey an algebra given by

\bsp
[J_i, J_j] = i \e_{ijk} J_k \qquad [K_i, K_j] = -i \e_{ijk} J_k
\qquad [J_i, K_j] = i \e_{ijk} K_k\\
[J_i,P_j] = i\e_{ijk}P_k \qquad [J_i, P_0]=0 \qquad 
[K_i,P_j]=-i\d_{ij}P_0 \qquad [K_i,P_0]=-iP_i
\end{split}
\ee

\ni where $\m=0,\ldots,d-1$ is a spacetime index, while
$i,j,k=1,\ldots,d-1$ are spacial indices. This is more compactly
expressed in terms of the Lorentz generators $M_{\m\n} = - M_{\n\m}$
defined as $M_{0i} = K_i$ and $M_{ij} = \e_{ijk}J_k$

\bsp\label{poin}
[P_\m,P_\n] = 0 \qquad [M_{\m\n},P_\r] = -i\eta_{\r\m}P_\n +i\eta_{\r\n}P_\m\\
[M_{\m\n}, M_{\r\s}] = i\eta_{\n\r}M_{\m\s} -i\eta_{\m\r}M_{\n\s} -i\eta_{\n\s}
M_{\m\r} +i\eta_{\m\s}M_{\n\r}.
\end{split}
\ee

\ni where $\eta_{\m\n} = \text{diag}(+,-,\ldots,-)$. Supersymmetry
enlarges this group via the introduction of spinorial generators
$Q_\a^I$ and $\bar Q_{\dot \a}^I$, where $\a$ and $\dot \a$ are spinor
indices, while $I=1,\ldots,{\cal N}$ labels the number of
supersymmetries. The enlargement of (\ref{poin}) is as follows

\bsp
[P_\m,Q_\a^I] = 0 \qquad [P_\m, {\bar Q}_{\dot \a}^I] = 0
\qquad [M_{\m\n}, Q_\a^I] = i \left(\G_{\m\n} \right)_\a^\b Q_\b^I
\qquad [M_{\m\n}, {\bar Q}_{\dot \a}^I] 
= i \left(\bar \G_{\m\n} \right)_{\dot \a}^{\dot \b} 
\bar Q_{\dot \b}^I\\
\{ Q_\a^I, \bar Q_{\dot \b}^J\} = 2 \G^\m_{\a \dot \b} P_\m \d^{IJ}
\qquad \{ Q_\a^I, Q_\b^J \} = \e_{\a \b} Z^{IJ} \qquad
\{ \bar Q_{\dot \a}^I, \bar Q_{\dot \b}^J \} = \e_{\dot \a \dot \b} 
\left( Z^{IJ} \right)^*
\end{split}
\ee

\ni where some definitions are in order. The bar is defined as
follows: $\bar Q = Q^\dag \G^0$. The $\G^\m$ are a
representation of the $d$-dimensional Clifford algebra

\be
\{\G^\m , \G^\n\} = 2 \eta^{\m\n}
\ee

\ni and $\G^{\m\n} \equiv - \frac{i}{4} [\G^\m,\G^\n]$. The $Z^{IJ}$
are an ${\cal N}\times {\cal N}$ matrix of {\it central charges} which
is necessarily antisymmetric in $I,J$ and therefore exists only for
${\cal N} > 1$. This is called {\it extended} supersymmetry.

There are two main methods of implementing supersymmetry on the string
worldsheet. They are referred to as the Neveu-Schwarz-Ramond or NSR
string, and the Green-Schwarz string. The latter makes the resulting
supersymmetry in the target spacetime more explicit, and will also be
more relevant for describing superstrings in the plane-wave background
of chapter \ref{chap:lcsft}. For these reasons, the Green-Schwarz
formalism is developed here. It is most instructive to consider the
supersymmetrized action of a point particle, rather than a string,
first. A massless point particle in Minkowski spacetime has the
following action

\be\label{bospoint}
S = \int d\t \frac{1}{h(\t)}\, \dot x^\mu(\t) \dot x_\mu(\t)
\ee 

\ni where $h$ is a worldline metric relating an interval in $\t$ to a
physical time interval. The embedding function $x^\m(\t)$ describes the
worldline of the particle through spacetime. We can render this action
supersymmetric through the introduction of some fermionic partners for
the bosonic fields $x^\m$. These we denote $\theta^A_a(\t)$ where
$A=1,\ldots,{\cal N}$ and $a$ is a spacetime spinor index which will
be suppressed in what follows. It can be verified that the following
generalization of (\ref{bospoint})

\be\label{susypoint}
S = \int d\t \frac{1}{h(\t)}\, \left(\dot x^\m - i \bar \theta^A \G^\m
\dot \theta^A \right)^2
\ee

\ni is invariant under the supersymmetry variations

\be
\d \theta^A = \e^A \qquad \d x^\m = i \bar \e^A \G^\m \theta^A \qquad
\d \bar \theta^A = \bar \e^A \qquad \d h = 0
\ee

\ni where $\e,\bar \e$ are spinors independent of $\t$. Thus there are
${\cal N}$ supersymmetries obeyed by this action. The equations of
motion of the fields in (\ref{susypoint}) are given by

\be
p^2 = 0 \qquad \dot p^\m = 0 \qquad \G \cdot p \, \dot \theta^A = 0 
\ee

\ni where $p^\m \equiv \dot x^\m - i \bar \theta^A \G^\m \dot
\theta^A$. In fact, this shows that half of the components of each
$\theta^A$ are left entirely unfixed by these equations. This is
because the matrix $\G\cdot p$ is nilpotent by the equations of
motion, i.e. it squares to zero: $(\G\cdot p)^2 = p^2 = 0$. This
indicates that its rank is half of of its dimension ${\cal N}$. Since
$\dot \theta^A$ appears in the action only in the combination $(\G
\cdot p) \dot \theta^A$, half of $\theta^A$'s components have no
dynamics; they are are not physical propagating degrees of freedom,
and therefore we have over estimated the fermionic content of our
theory. The reason for this is something called $\k$ symmetry, which
the action (\ref{susypoint}) is invariant under. It may be expressed
as

\be
\d \theta^A = i \G \cdot p \, \k^A \qquad \d x^\m = i\bar \theta^A
\G^\m \d \theta^A \qquad \d h = 4 h \dot {\bar \theta}^A \, \k^A 
\ee

\ni where $\k^A(\t)$ is a set of $A$ {\it local} spinors; $\k$
symmetry, unlike supersymmetry, is not global. This symmetry will be a
necessary ingredient in the superstring action in order to ensure its
supersymmetry.

The superstring action may be constructed for flat target spacetime in
much the same way that the superparticle action was
found. Generalizing to a non-flat target is an extremely non-trivial
exercise which we won't discuss here. The flat space action may be
expressed as

\be\label{sustr}
S  = -\frac{T}{2} \int d^2\s \sqrt{h} \,h^{ab} \,\Pi_a \cdot \Pi_b + S_\k
\ee

\ni where $\Pi_a^\m \equiv \p_a X^\m - i\bar \theta^A \G^\m \p_a
\theta^A$, and $S_\k$ is a term which must be added in order to enforce
the local $\k$ symmetry. In fact, it turns out that this symmetry
cannot be realized for arbitrary ${\cal N}$, we must take the number
of supersymmetries to be $\leq 2$. We will present $S_\k$ for ${\cal
  N}=2$; the other cases may be obtained by setting one or both the
$\theta^A$'s to zero. The form of $S_\k$ involves coupling of the
bosonic and fermionic degrees freedom, as well as a four-Fermi term

\be
S_\k = T \int d^2\s \biggl( -i \e^{ab} \p_a X^\m \left( \bar \theta^1
\G_\m \p_b \theta^1 - \bar \theta^2 \G_\m \p_b \theta^2 \right) +
\e^{ab} \bar \theta^1 \G^\m \p_a\theta^1 \, \bar \theta^2 \G_\m \p_b
\theta^2 \biggr).
\ee

\ni This addition must also obey the global ${\cal N}=2$
supersymmetry. This ends up setting constraints on the type of spinor
$\theta$ may be, and on the spacetime dimension $d$. There are four
choices involving $d=3,4,6$, and 10. We will see in the next section
that only the $d=10$ choice will lead to a consistent quantum theory. 

% ========================================================================== %
\subsection{Light-cone gauge quantization and critical dimension}
\label{sec:lcgq}

Our acquaintance with the superstring thus far has shown us some
remarkable features. First the classical superstring can only have ${\cal
  N} \leq 2$. Second, in order for this to be true the dimension of
the target spacetime must be 3, 4, 6, or 10. This power of the string to set
parameters was one of the early attractions of the theory - it was
hoped that the superstring would give a theory of everything where the
number of parameters, or true ``constants of nature'', would be
minimal. We will now see that the discretion of quantum mechanics goes
further in this direction and requires the spacetime dimension to be 10. 

The recipe for quantization is to replace Poisson brackets of fields
and their conjugate momenta with commutators. In this way mode
amplitudes (c-numbers) are promoted to operators which act
upon a vacuum to create and annihilate states. If all of the gauge
freedom is used-up prior to quantization, then one is guaranteed to
have every state (made by acting the creation operators on the vacuum)
be physical. The light-cone gauge quantization presented here is such
a regime. We begin by fixing the gauge freedom afforded us by our
superstring action (\ref{sustr}). The $\k$ symmetry allows us to set

\be\label{slcgcond}
\G^+\theta^1 = \G^+\theta^2 = 0
\ee

\ni where, as in section \ref{sec:modeslcg}, $\G^{\pm} = (\G^0 \pm
\G^9)/\sqrt{2}$. As the matrices $\G^\pm$ are nilpotent, this gauge
choice fixes-out exactly half of the components of the spinors - as we
saw in section \ref{sec:susydet}, this is exactly the r\^{o}le of $\k$
symmetry. Because, for $d=10$, the spinors $\theta^A$ are Majorana-Weyl, the 32
complex degrees of freedom associated to a generic spinor in $d=10$ are
reduced to 16 real degrees of freedom.  The additional constraint
(\ref{slcgcond}) reduces this further to 8 real components per
spinor. Thus, in this gauge, the $\theta^A$ constitute an eight
dimensional spinor representation of the group $SO(8)$. We showed in
section \ref{sec:modeslcg} that a similar reduction occurs for the
embedding coordinates $X^\m$. There the $X^\pm$ were fixed by gauge
symmetry, leaving only the eight fields $X^i$ as physical degrees of
freedom. Taking these gauge choices for the $\theta^A$ and $X^\pm$, we
find that the equations of motion resulting from (\ref{sustr}) are
immensely simplified

\be
\left( \p_\s^2 - \p_\t^2 \right) X^i = 0 \qquad
\left( \p_t + \p_\s \right) \theta^1 = 0 \qquad
\left( \p_t - \p_\s \right) \theta^2 = 0 .
\ee

\ni Notice that $\theta^1$ is exclusively left-moving while
$\theta^2$ is exclusively right-moving. We have seen the mode
expansions for the $X^i$ in (\ref{modexp}). For our fermionic
partners, for closed superstrings, we have

\be\label{flatspacespinors}
\theta^{1\a}(\s,\t) = \sum_{n=0}^\infty \b^\a_n e^{-2in(\t-\s)}\qquad
\theta^{2\a}(\s,\t) = \sum_{n=0}^\infty \wt \b^\a_n e^{-2in(\t+\s)}
\ee

\ni where $\a$, the $SO(8)$ spinor index, has been restored to
emphasize that the fermionic oscillators $\b$ and $\wt \b$ are
spinors. The Poisson brackets between the fields and their
conjugate momenta are

\bsp
[ \dot X^i(\s,\t), X^j(\s',\t) ]_{\text{P.B.}} &= \pi \d(\s-\s')
\d^{ij}\\
\{ \theta^{A\a}(\s,\t), \theta^{B\b}(\s',\t) \}_{\text{P.B.}} &= i\pi \d^{\a \b}
\d^{AB} \d(\s-\s'),
\end{split}
\ee

\ni this implies

\bsp\label{comrel}
[ \a^i_m, \a^j_n ]_{\text{P.B.}} = i m \d_{m+n} \d^{ij} \qquad
[\wt \a^i_m, \wt \a^j_n ]_{\text{P.B.}} &= i m \d_{m+n} \d^{ij} \qquad
[\a^i_m, \wt \a^j_n]_{\text{P.B.}}=0 \\
\{ \b^\a_m, \b^\b_n \}_{\text{P.B.}} = i m \d_{m+n} \d^{\a\b} \qquad
\{\wt \b^\a_m, \wt \b^\b_n \}_{\text{P.B.}} &= i m \d_{m+n} \d^{\a\b} \qquad
\{\b^\a_m, \wt \b^\b_n\}_{\text{P.B.}}=0
\end{split}
\ee

\ni Quantization amounts to the replacement
$[A,B]_{\text{P.B.}}\rightarrow i[A,B]$, and
$\{A,B\}_{\text{P.B.}}\rightarrow i\{A,B\}$, which implies that the
oscillators are promoted to creation and annihilation operators. 

We have alluded to the fact that quantization selects for us a target
spacetime dimension $d$. In fact this selection can be seen in an
anomaly arising in the Lorentz group algebra (\ref{poin}). The gauge
choice (\ref{x+}) explicitly breaks Lorentz invariance by choosing a
preferred direction. Under quantization, those elements of $M_{\m\n}$
which mix the + direction with the others can and do develop an
anomaly. Specifically, the commutator $[M^{i-},M^{j-}]$, which must be
zero classically, is no longer so after quantization. The proof of
this proceeds as follows: first $M^{\m\n}$ is constructed using the
standard method

\be
M^{\m\n} = T \int_0^\pi d\s \left( X^\m \dot X^\n - X^\n \dot X^\m +
\bar \theta^A \G^{\m\n} \theta^A \right).
\ee

\ni The mode expansions are then inserted, giving an expression in
terms of creation and annihilation operators. The algebra (\ref{poin})
is then evaluated using the (quantum versions of the) commutation
relations (\ref{comrel}). The end result is that $[M^{i-},M^{j-}]=0$
if and only if the target spacetime dimension is $d=10$, while the
rest of the algebra (including the supersymmetric extension) is
anomaly free.

% ========================================================================== %
\subsection{Closed string spectrum, background fields, and low energy
  effective actions}
\label{sec:clos}

The superstring has selected for us the amount of supersymmetry and
the spacetime dimension $d$. The next natural question to ask is what
the particle content of the theory is, and what the interactions are
between those particles. Again the notion of a quantum anomaly is
important here. We saw in section \ref{sec:prel} that the string
worldsheet possessed conformal or Weyl invariance. Should we place our
superstring in a general target space background (a given metric, and
possibly other gauge fields and superpartners), we will generically
develop a quantum conformal anomaly on the worldsheet. The requirement
that this anomaly vanish gives us equations of motion for the
background fields which are exactly those obeyed by the string modes
themselves - i.e. a string will develop a conformal anomaly unless it
is placed in a background of strings. Thus, not only does the
superstring choose its supersymmetry and spacetime dimension, but also
tells us that everything is made of string.  This is a further
manifestation of self-reference and internal consistency. It therefore
seemed to early researchers that superstring theory could indeed be a
theory of everything.

The closed string spectrum is generated by tensoring the left and
right moving modes, which are representations of the $SO(8)$ symmetry
enjoyed by (\ref{sustr}). The bosonic modes of the $X^i$ are obviously
in the vector representation ${\bf 8_v}$, whereas the fermionic modes
of the $\theta^A$ are $SO(8)$ spinors which come in two chiralities on
account of them being Weyl; these are labelled as ${\bf 8_c}$ and ${\bf
8_s}$. The lowest energy (massless) states of the string theory
correspond to two-mode excitations (one right and one left moving,
since for closed strings the number of left and right-movers must be
equal, see (\ref{levelmatch})). We are free to take the left-moving
and right-moving modes to have same or opposite spinor chirality; the
choice will lead to different string theories. If we take them to be
opposite, the following massless spectrum is generated

\bsp\label{iiaspec}
({\bf 8_v} + {\bf 8_c})_L \otimes ({\bf 8_v} + {\bf 8_s})_R = 
({\bf 1} + {\bf 28} + {\bf 35_v} + {\bf 8_v} + {\bf 56_v})_B\\
+ ({\bf 8_s} + {\bf 8_c} + {\bf 56_s} + {\bf 56_c})_F 
\end{split}
\ee

\ni where the subscript $B$ stands for bosons and $F$ for
fermions. This is the spectrum of type IIA supergravity. If we take
the same chirality for left and right-movers, we obtain

\bsp
({\bf 8_v} + {\bf 8_c})_L \otimes ({\bf 8_v} + {\bf 8_c})_R = 
({\bf 1} + {\bf 28} + {\bf 35_v} + {\bf 1} + {\bf 28} + {\bf
  35_c})_B\\
+ ({\bf 8_s} + {\bf 8_s} + {\bf 56_s} + {\bf 56_s})_F 
\end{split}
\ee

\ni which is the spectrum of type IIB supergravity. The lesson is that
if we restrict ourselves to the least excited strings, those with the
smallest energy (which happen to be massless), we obtain the particle
content of something called supergravity, a theory we will explain
below. Indeed, the interactions between these string modes are also
identical to the supergravity interactions, leading us to the
conclusion that the low-energy effective dynamics of closed
superstring theory is supergravity.

Supergravity is a supersymmetrization of Einstein gravity. Unlike the
global supersymmetry of field theories, in supergravity the
supersymmetry is promoted to a local symmetry whose gauge connection
is an object of spin 3/2. It is believed that massless particles of
spin $> 2$ cannot be coupled consistently in any field theory. This fact
places an upper limit on the spacetime dimension a supergravity theory
may live in. It turns out that if $d > 11$, local supersymmetry
requires the presence of massless particles whose spin is greater than
two. Therefore, 11-dimensional supergravity plays a privileged r\^{o}le,
and some lower dimensional supergravities may be realized through
toroidal compactification of this theory. We will present only the
bosonic content of the supergravities, as the fermionic content can be
obtained from supersymmetry. The 11-dimensional supergravity contains
a spacetime metric $G_{MN}$ and a 3-form gauge potential $A_3$. Its
action is as follows

\be\label{S11}
S^{\text{bos.}}_{11} = \frac{1}{2\k_{11}^2} \int d^{11}x \sqrt{-G} 
\left( R - \frac{1}{2} |F_4|^2 \right) - \frac{1}{6}\int A_3 \w F_4 \w F_4
\ee

\ni where $R$ is the Ricci scalar built from $G_{MN}$ and $F_4 \equiv
dA_3$. Compactifying one direction of this theory with period $2\pi R$
causes the 11-dimensional metric to be mapped to a 10-dimensional
scalar $\Phi$, vector $A_1$, and traceless symmetric tensor (metric)
$G_{mn}$. The 3-form is mapped to another 3-form (we'll keep the label
$A_3$) and a 2-form $B_2$. A glance at (\ref{iiaspec}) reveals
precisely this pattern - the ${\bf 8_v}$ are the physical
propagating degrees of freedom corresponding to a massless vector in
ten dimensions, the ${\bf 28}$ is that corresponding to an
antisymmetric rank-2 tensor (i.e. $B_2$), etc. This is type IIA
supergravity, whose action may be written as

\bsp\label{IIA}
S_{\text{IIA}}^{\text{bos.}} = \frac{1}{2 \k_{10}^2} &\int d^{10}x 
\sqrt{-G} e^{-2\Phi} \left(R + 4\p_\m\Phi \p^\m\Phi -
\frac{1}{2}|H_3|^2 \right)\\
-\frac{1}{4\k_{10}^2} &\int d^{10}x \sqrt{-G} \left(
|F_2|^2 + |\wt F_4|^2 \right)
-\frac{1}{4\k_{10}^2} \int  B_2 \w F_4 \w F_4
\end{split}
\ee
 
\ni where $F_{M+1} \equiv dA_M$, $H_3 = dB_2$, $\wt F_4 = dA_3 - A_1
\w H_3$, $\k_{10}^2 = \k_{11}^2/2\pi R$, and we have rescaled the
metric by the factor $\exp(-2\Phi/3)$. The field $\Phi$, called the
dilaton, plays a very important r\^{o}le here. This is because the
effective coupling of the theory, i.e. the 10-dimensional universal
constant of gravitation is given by $8\pi G_{10} = (\k_{10}
e^\Phi)^2$. Therefore the dilaton sets the coupling strength - the
coupling constant in string theory is {\it dynamical}\footnote{In
general the closed string coupling is denoted by $g_s \equiv
\exp(\la\Phi\ra)$. This is the quantity which weights the vertices
where strings interact, analogous to $\a$ in Quantum
Electrodynamics. The value of of $\k_{10}$ may be determined via a
closed string exchange, it is equal to $[(2\pi)^7
{\a'}^4/2]^{1/2}$.}. This means that it does not need to be set as a
parameter, the theory determines it for us self-consistently. The
corresponding action for type IIB supergravity contains, instead of 1
and 3-form fields, 0, 2, and 4-form fields labelled $C_0$, $C_2$, and
$C_4$. Its action may be written as

\bsp\label{IIB}
S_{\text{IIB}}^{\text{bos.}} = \frac{1}{2 \k_{10}^2} &\int d^{10}x 
\sqrt{-G} e^{-2\Phi} \left(R + 4\p_\m\Phi \p^\m\Phi -
\frac{1}{2}|H_3|^2 \right)\\
-\frac{1}{4\k_{10}^2} &\int d^{10}x \sqrt{-G}  \left(
|F_1|^2 + |\wt F_3|^2 + \frac{1}{2}|\wt F_5|^2  \right)
-\frac{1}{4\k_{10}^2} \int C_4 \w H_3 \w F_3
\end{split}
\ee

\ni where $F_{M+1} \equiv d C_M$, $\wt F_3 = F_3 - C_0 \w H_3$, $H_3=d
B_2$, and $\wt F_5 = F_5 -\tiny{\frac{1}{2}} C_2 \w H_3 +
\tiny{\frac{1}{2}} B_2 \w F_3$.

We have now seen the effective dynamics of closed superstrings when
the energy scale is low enough not to excite massive string modes. The
equations of motion following from the actions (\ref{IIA}) and
(\ref{IIB}), and therefore the interactions between the particles of
their fields, are recovered by superstring interactions. Further, we
remind the reader that only when the superstring is placed in a
background obeying these equations of motion will the worldsheet
theory be free of the conformal anomaly. To a very large extent
``string theory'' is concerned with type IIA and type IIB
supergravity; genuine perturbative stringy effects are difficult to
calculate and therefore do not appear in any considerable volume in
the literature. However, we shall see in the next section that open
strings do give rise to tractable and immensely powerful
non-perturbative objects which have no analogue in point-particle
theories, the D-branes. It should be noted that there are two very
important closed string theories, and one closed + open string theory
that have not been presented here - the heterotic string theories and
the type I superstring respectively. These theories may be obtained
via various dualities which act upon the type II theories. Details of
these string theories may be found in the standard textbooks
\cite{Polchinski:1998rq,Polchinski:1998rr,Green:1987sp,Green:1987mn}.

% ========================================================================== %
\subsection{Open strings, T-duality, and D-branes}
\label{sec:dbranes}

It was realized in the late 1980's \cite{Buscher:1987qj} that strings
see the geometry of their target spacetime in a radically different
way than point particles. It is easiest to see this from the point of
view of closed strings. Consider the zero modes of the closed string
(\ref{modexp}) (i.e. set $\a^\m_n = \wt \a^\m_n = 0$) propagating on a
flat target space which contains an $S^1$. Let the the radius of the
$S^1$ be $R$ and consider the zero mode of the embedding function in
this direction. We have

\be
X(\s,\t) = \frac{1}{2}(x_L + x_R) +  \a'(p_L + p_R)\,\t + \a' (p_L -
p_R) \,\s.
\ee

\ni Imposing the closed string boundary conditions is different in
this space because $X \sim X + 2\pi R$. Therefore we must only have
that $X(\s,\t) = X(\s+\pi,\t) + 2\pi R w$, where $w \in \bZ$ is called
the winding (or wrapping) number as it counts the number of times the
string winds the $S^1$ before closing back on itself. We can also note
that quantum mechanical momentum on a circle is quantized in units of
the inverse radius. Given these facts we see that (see (\ref{modexp}))

\be
p_L + p_R = \frac{2}{R}n, \qquad p_L - p_R = \frac{2R}{\a'} w
\ee

\ni where $n \in \bZ$ is the momentum quantum number\footnote{Note
  that the level matching condition (\ref{levelmatch}) is now modified
  since $p_L \neq p_R$. The result is that $nw$ is added to the
  RHS. Also note that the tension of the wound string and its momentum
  in the compact direction will now contribute to its mass
  (\ref{masssquared}).}. Now consider the following target space
  transformation, $R \rightarrow \a'/R$. Under this operation the
  closed string zero mode simply sees an exchange of $n$ and $w$, and
  nothing else. More precisely a closed string zero mode cannot tell
  whether it is propagating on a circle of radius $R$ or $\a'/R$. This
  effect, dubbed {\it Target space duality} or {\it T-duality} was
  shown to extend beyond the level of the zero modes
  \cite{Buscher:1987qj} and is a symmetry of the full string theory.
  The T-dual operation, extended to all moments of the momenta,
  amounts to

\be
\a_n \rightarrow -\a_n, \qquad \wt\a_n \rightarrow \wt\a_n
\ee

\ni and so may be realized via the replacement $X(\s,\t) \rightarrow
X_L(\s,\t) - X_R(\s,\t)$, which is referred to as the {\it T-dual
  coordinate}. 

It is instructive to consider what effect this operation has on open
strings. The open string cannot wrap a compact direction because it
does not close. Thus the periodicity of the target space does not
effect the mode expansion, which is

\be
X(\s,\t) = x + 2 \a' p \t + i\sqrt{\frac{2}{\a'}}
\sum_{n\neq 0} \frac{1}{n} \a_n \, e^{-in\t} \cos(n\s)
\ee

\ni where $\s\in[0,\pi]$. We are free to write this in terms of a
sum of a function of $(\t + \s)$ and one of $(\t - \s)$, i.e. as a sum
of left and right-moving pieces

\bsp
X(\s,\t) &= X_L + X_R \qquad \text{where}\\
X_L &= \frac{1}{2} x_L + \a' \frac{n}{R} (\t + \s) + \frac{i}{\sqrt{2\a'}}
\sum_{n\neq 0} \frac{1}{n} \a_n \, e^{-in(\t+\s)}\\
X_R &= \frac{1}{2} x_R + \a' \frac{n}{R} (\t - \s) + \frac{i}{\sqrt{2\a'}}
\sum_{n\neq 0} \frac{1}{n} \a_n \, e^{-in(\t-\s)}.
\end{split}
\ee

\ni What does this open string embedding look like in the T-dual
theory? The T-dual coordinate is

\be
X_L - X_R = x_0 + 2\a' \frac{n}{R} \s + \sqrt{\frac{2}{\a'}}
\sum_{n\neq 0} \frac{1}{n} \a_n \, e^{-in\t} \sin(n\s).
\ee

\begin{figure}[ht]
\begin{center}
\includegraphics*[bb=0 0 360 420,width = 3.0in,height=2.5in]{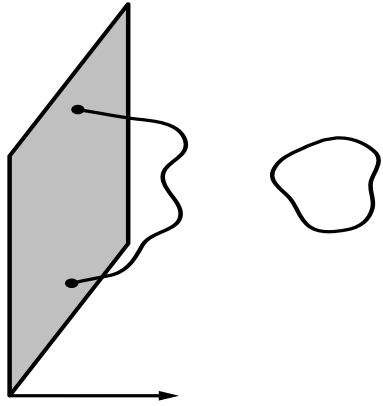}
\end{center}
\caption{A D-brane supports the endpoints of open strings, which are
  free to move in the brane's worldvolume. The bulk of the open
  string is not confined, nor are closed strings propagating in the spacetime.
}
\label{fig:dbrane}
\end{figure}

\ni There are two things to realize about this embedding. First the
endpoints of the open string ($\s=0,\pi$) do not oscillate (the $\sin$
function is zero there) and second, they are fixed at $x_0$ and $x_0 +
2\pi n \wt R$, where $\wt R$ is the T-dual radius $\a'/R$. In the
T-dual theory these points are identified in the target space. The
embedding functions in the other directions are unaffected by this of
course, so we have an open string whose bulk fluctuates in the full
spacetime but whose endpoints are confined to a $d-1$ dimensional
hyperplane. If we take in addition $R \rightarrow 0$, the bulk of the
string sees a (T-dualized) compact direction of infinite radius. What
we have found is a {\it Dirichlet (d-2)-brane} or {\it
D(d-2)-brane}. This is displayed in figure \ref{fig:dbrane}.
We did not discuss open superstrings in section \ref{sec:clos},
however there is an open supersymmetric string theory called type I,
which includes both open and closed strings and is unoriented. If we
T-dualize any odd number $p$ of dimensions we will end up with a
D$(10-p-1)$-brane and type IIA closed strings far away from the
brane. If we T-dualize any even number of dimensions $q$, we will end
up with a D$(10-q-1)$-brane and type IIB closed strings far away from
the brane\footnote{This is not precisely true, as things are
  complicated by the fact that type I theory has an $SO(32)$ gauge
  group. In fact, multiple D-branes are produced upon T-dualizing type
  I, as well as objects called {\it orientifold planes}, a consequence
  of the unoriented nature of type I strings. See \cite{Polchinski:1998rr},
  pg. 138 for details.}. What is the theory governing the strings on
these branes?  The bosonic degrees of freedom of the Dp brane are
described by a $(p+1)$-dimensional gauge field $A_a$, corresponding to
string excitations in the worldvolume of the brane and $10-(p+1)$
scalars $\Phi_I$ describing transverse string excitations. The vacuum
expectation values (VEV's) or zero modes of these fields describe the
embedding of the brane into the target space, $X^\m(\xi^a)$, where the
$\xi_a$ are the worldvolume coordinates of the brane. Therefore the
$X^I$ describe the shape of the brane while the $X^a$ describe any
constant gauge field backgrounds turned-on on the worldvolume. The
bosonic action is given by the Dirac-Born-Infeld \cite{Born:1934gh} or
DBI action\footnote{The second term involving the coupling to the form
  potentials $C_M$ is called the {\it Chern-Simons}
  \cite{Chern:1974ft} term.}

\bsp\label{DBI}
S_{Dp} = -T_p \int d^{p+1} \xi\, e^{-\Phi} &\left[ 
-\det \left( G_{ab} + B_{ab} + 2\pi \a' F_{ab} \right) \right]^{1/2}\\
&+ i\m_p \int_{p+1} \exp \left( 2\pi \a' F_2 + B_2 \right) \w \sum_q
C_q 
\end{split}
\ee

\ni where 

\be
G_{ab} = G_{\m\n}(X) \p_a X^\m \p_b X^\n \qquad B_{ab} = B_{\m\n}(X)
\p_a X^\m \p_b X^\n = P[B_2]
\ee

\ni i.e. the metric and antisymmetric B-field from the target space
are pulled back onto the worldvolume of the D-brane. Here $F_{ab}$ is
the field strength built on $A_a$, i.e. $F_{ab} = \p_a A_b - \p_b
A_a$. Note that $\Phi$ above is the dilaton (from (\ref{IIA}) for
example) and not the transverse scalars. Note also the appearance of
the space-time form potentials $C_q$; the D-branes carry charge $\m_p$
under these potentials. Note that the expansion of the exponential
will give forms of various rank, but the integral will only be
non-zero for those combinations which amount to a $(p+1)$-form. The
D-brane charge and tension are calculable via a closed string exchange
amplitude. The result is that $T_p = \m_p = (2\pi)^{-p}
(\a')^{-(p+1)/2}$.  However, since we have a factor of $e^{-\Phi}$ in
front of (\ref{DBI}), the effective D-brane tension is inversely
proportional to the string coupling (see discussion beneath
(\ref{IIA})). Therefore the D-brane is a non-perturbative object,
infinitely heavy at zero coupling. We could never have hoped to
discover it through perturbative techniques.

% ************************************************************************** %
\section{AdS/CFT correspondence}
\label{sec:adscftcorr}

{\small
\begin{quote}
{\bf Glaucon}: You have shown me a strange image, and they are strange
prisoners.\\ {\bf Socrates}: Like ourselves, I replied; and they see
only their own shadows, or the shadows of one another, which the fire
throws on the opposite wall of the cave?\footnote{It was Polyakov
\cite{Polyakov:1998ju} who originally noted the appropriateness of the
classic allegory to holography.}\\ \rightline{--- Plato's {\it
The Republic}, Book VII$\;\;\;\;\;\;\;\;\;\;\;\;\;\;\;\;\;\;$}
\end{quote}}

The AdS/CFT correspondence, in its most celebrated form, is a
conjectured duality between type IIB string theory on the background
space $AdS_5 \times S^5$ (with a background 4-form potential), and
${\cal N}=4$ supersymmetric Yang-Mills theory in four spacetime
dimensions. There are many other manifestations of this duality, which
is really a much deeper statement about the connection between gauge
theories and gravity. It is also an instance of {\it holography},
where a higher dimensional gravity is entirely captured by a lower
dimensional quantum field theory. That gravity has this suspicious
scaling of its physical degrees of freedom was hinted at in the 1970's
by Beckenstein \cite{Bekenstein:1972tm}, who associated the area of a
black hole's horizon with an entropy. This picture was later
strengthened by Hawking's discovery \cite{Hawking:1974sw} that
semi-classically, black holes produce a thermal spectrum of radiation,
whose characteristic temperature $T$ is related to the surface gravity
$\k$ of the black hole via $T=\k/2\pi$. Indeed the four laws of
thermodynamics may be applied to the black hole with these
identifications \cite{Bardeen:1973gs}. The situation was much improved
upon the successful microscopic computation of the black hole entropy
using string theory \cite{Strominger:1996sh}. This calculation
depended upon the concept that fluctuation modes of D-branes which
served as the central ``mass'' of the black holes, embodied the
microstates responsible for the macroscopic, spacetime entropy. The
emerging duality between the brane dynamics and those of the curved
spacetime which they source, including the realization that absorption
cross sections could be calculated from either perspective
\cite{Klebanov:1997kc, Gubser:1997yh, Gubser:1997se}, led Maldacena
\cite{Maldacena:1997re} to the AdS/CFT correspondence in 1997. The
significance of this discovery is twofold. On the one hand it offers
insights into gravity via quantum field theory. On the other it
affords the long sought-after string description of (at least some)
gauge theories, and their strong coupling dynamics. In fact, Polyakov
\cite{Polyakov:1998ju} had already realized that the string
description of gauge theories required the string to propagate in a
higher dimensional spacetime, before the Maldacena conjecture
appeared. We will give below a general introduction to the AdS/CFT
correspondence, its main features, and a cross section of results
pertinent to this thesis.

% ========================================================================== %

\subsection{Supergravity p-branes and string theory D-branes}
\label{sec:sugrapbranes}

Before the discovery of D-branes, solutions of supergravity were
discovered which were solotonic hyperplanes \cite{Horowitz:1991cd,
Duff:1991pe, Duff:1994an}. These solutions, called {\it p-branes},
exist in both type IIA and type IIB supergravities, their general form
given by

\bsp ds^2 = H^{-1/2}(r) &\left[ -f(r) dt^2 + (dx^i)^2 \right] +
H^{1/2}(r) \left[f^{-1}(r) dr^2 + r^2 d\O_{8-p}^2 \right],\\ 
&e^{\Phi} = H^{\frac{(3-p)}{4}}(r), \qquad F_{t i_1 \ldots i_p r} =
\e_{i_1\ldots i_p} \frac{1}{H^2(r)}\frac{Q}{r^{8-p}},\\
&H(r) = 1 + \left(\frac{R}{r}\right)^{7-p}, \qquad f(r) = 1-
\left(\frac{r_0}{r}\right)^{7-p}
\end{split}
\ee

\ni where $F_{t i_1 \ldots i_p r}$ is the $(p+2)$-form field strength
from (\ref{IIA}) or (\ref{IIB}), so that $p$ must be odd for type IIB
solutions and even for type IIA. $Q$ is the charge of the solution
under this form field. In general these solutions have horizons at $r
= r_0$, and hence are extended black hole solutions. Unlike the
standard Schwarzschild black hole \cite{Schwarzschild}, whose
singularity is point-like, here the singularity is extended in a
p-spatial-dimensional hyperplane, covered by the coordinates
$x_i$. These solutions are also charged, and so may be viewed as
generalizations of the Reissner-Nordstr\"{o}m solution
\cite{Reissner}. Like that solution, there is a bound relating the
mass and charge $M\geq Q$, both of which are functions of $R$ and
$r_0$; when this bound is saturated $r_0=0$, and the solution is
called {\it extremal}. In equation (\ref{DBI}), and the discussion
beneath it, we saw that D-branes carry a charge $\m_p$ which is equal
to their tension or ``mass'' $T_p$. It should not be surprising that
the flat Dp-branes are extremal p-brane solutions, viewed in the low
energy limit where the supergravity description is appropriate. In
fact the equality of mass and charge is a reflection of supersymmetry;
the D-brane (or extremal p-brane) preserves 1/2 of the 32
supersymmetries of the original closed string theory (or
supergravity). This is often referred to as 1/2 BPS, where BPS is
named after Bogomol'nyi, Prasad, and Sommerfield
\cite{Bogomolny:1975de,Prasad:1975kr}. The case of the extremal
3-brane is the most important for the AdS/CFT correspondence. In fact,
in the realm of 10-dimensional theories, only the 3-brane will give
Anti-deSitter or $AdS$ space in a given limit; conversely only $AdS$
space will have a conformal theory on its boundary\footnote{We will discuss
11-dimensional versions of AdS/CFT in chapter
\ref{sec:matrixmodel}. It should also be mentioned that there is a
version of AdS/CFT dealing with the space $AdS_3\times S^3 \times M^4$
where $M^4$ is a compact manifold. This theory will not be discussed
in this thesis.}.  

Setting $p=3$ and $r_0=0$, we arrive at the following solution for the
extremal 3-brane

\be\label{extrem3}
ds^2 = \left(1+\frac{R^4}{r^4}\right)^{-1/2} \left(-dt^2 +
dx_i^2\right) + \left(1+\frac{R^4}{r^4}\right)^{1/2}\left(dr^2 + r^2
d\O_5^2\right).
\ee

\ni We note that the dilaton is constant, and so the string
coupling is the same everywhere. There is also a self-dual five-form
given by 

\be\label{extrem3F}
F_5 = (1+ *) dt \w dx_1 \w dx_2 \w dx_3 \w
d\left(\left[1+\frac{R^4}{r^4}\right]^{-1}\right).  
\ee

\ni The first order of business is to relate the parameters of the
D3-brane to this solution. In fact, we will be interested in a stack
of $N$ parallel D3-branes. We are free to do this because parallel
D-branes do not interact with each other - a consequence of
supersymmetry which ensures that their gravitational attraction is
balanced exactly by their ``electro-magnetic'' (in the sense of the
form potentials) repulsion. The first thing to do is to equate the
tension of the $N$ D3-branes to the ADM \cite{Arnowitt:1962hi} mass
of the spacetime (\ref{extrem3}). The ADM mass is the general
relativistic measure of the stress energy responsible for the
curvature of spacetime; it is the gravitational ``charge''. The ADM
mass has been calculated in \cite{Lu:1993vt}, the result is

\be\label{madm}
M_{ADM} = \frac{2\pi^3}{8\pi G_{10}} R^4 = \frac{R^4}{32 \pi^4 g_s^2 \a'^4} 
\ee

\ni where we have used $8\pi G_{10} = (\k_{10}\, g_s)^2$ and
$\k_{10}^2=(2\pi)^7 {\a'}^4/2$ in the second equality, as per section
\ref{sec:clos}. The D-brane tension was given in section
\ref{sec:dbranes}, multiplying this by $N$ we have

\be\label{Nd3tension}
\t_{ND3} = g_s^{-1}T_{ND3} = \frac{N}{8\pi^3 g_s \a'^2}.
\ee

\ni where we have noted the dilaton factor in (\ref{DBI}). Equating
(\ref{madm}) and (\ref{Nd3tension}), we arrive at a special relation

\be\label{radius}
\boxed{
R^4 = 4\pi g_s N \a'^2} ~.
\ee

So far we have analyzed the N D3-branes in terms of the low energy
supergravity description. Recall that this is the picture seen by
closed strings propagating in the bulk, see figure
\ref{fig:dbrane}. We should also ask ourselves what the open strings
attached to the D-brane are doing. To answer this we take the low
energy limit ($\a' \rightarrow 0$) of the DBI action (\ref{DBI})
describing our D3-brane. We have no $B_2$ field in the background, and
the dilaton $\Phi$ is a constant defining $g_s$. Further, the D-brane
is in flat 10-dimensional space, and so $G_{\m\n} =
\eta_{\m\n}$. We take the embedding to be as follows

\be
X^a (\xi^a) = \xi^a, \qquad X^I(\xi^a) = \sqrt{2} \pi \a' \Phi^I(\xi^a)
\ee

\ni where $a=0,\ldots,3$ are the worldvolume coordinates of the
D3-brane and $I=4,\ldots,9$ are the transverse directions. Ignoring
the coupling to the space-time form potentials, we have

\be
S_{D3} = -\frac{1}{(2\pi)^3 g_s \a'^2} \int d^4\xi \sqrt{ -\det \left(\eta_{ab} +
  2\pi\a' F_{ab} + (\sqrt{2}\pi \a')^2 \p_a \Phi^I \p_b \Phi^I \right) } 
\ee

\ni where we have explicitly indicated the D3-brane tension. Expanding
to leading order in $\a'$, we have

\be\label{lowen}
S_{D3} = -\frac{V_4}{(2\pi)^3 g_s \a'^2} - \frac{1}{4\pi g_s}\int
d^4\xi \left(\frac{1}{4} (F_{ab})^2  + \frac{1}{2} (\p_a \Phi^I)^2
\right) + \dots
\ee

\begin{figure}[ht]
\begin{center}
\includegraphics*[bb=0 0 185 300, height = 2.5in]{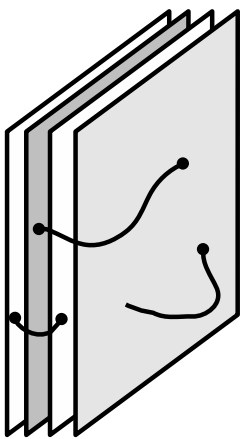}
\end{center}
\caption{A stack of $N$ D-branes, intended to be coincident as well as
  parallel, but shown separated for clarity. Open strings may begin
  and end on any two (or the same) branes, without penalty in
  energy. The effect is that worldvolume fields $X^\mu(\xi^a)$ are
  promoted to $N\times N$ matrices, leading to a non-abelian
  supersymmetric Yang-Mills theory.  
}
\label{fig:stack}
\end{figure}

\ni where $V_4$ is the infinite volume of the brane. Apart from this
constant, we have a free gauge theory with six scalars.  Our task is
not quite this however, since we would like to describe the
worldvolume theory of open strings on $N$ coincident, parallel
D3-branes. The generalization required is not difficult to understand,
see figure \ref{fig:stack}.
Open strings are free to begin and end on any of the $N$ branes,
without penalty in energy since the branes are coincident. In the
$N=1$ case, the modes of a string parallel to the brane described a
gauge field $A^a(\xi^a)$. Now this field has a factor of $N^2$ times
the number of components in order to allow for the specification of
the string end-points' branes-of-residence. Thus $A^a(\xi^a)$ is
promoted to an $N \times N$ unitary matrix, and similarly for the
scalars describing the transverse position of the stack. The full
generalization of the action (\ref{DBI}) to this case is known
\cite{Myers:1999ps}, but rather than indicating it explicitly here, we
give the $\a' \rightarrow 0$ limit of it for the D3-brane, i.e. the
generalization of (\ref{lowen})

\be\label{SYM}
S_{ND3} \simeq -\frac{1}{4\pi g_s} \Tr \int d^4x \left( \frac{1}{2}
(F_{\m\n})^2 + (D_\m \Phi^I)^2 - \frac{1}{2} [\Phi^I, \Phi^J]^2 +
\text{fermions} \right) + \ldots
\ee

\ni where the trace is over the $U(N)$ matrix indices, we have changed
the worldvolume coordinates to $x_\m$, and the leading constant
proportional to volume has been dropped. Here $F_{\m\n} = \p_\m A_\n -
\p_\n A_\m - i [A_\m,A_\n]$ and $D_\m\Phi^I = \p_\mu\Phi^I -
i[A_\m,\Phi^I ]$. We have indicated ``+ fermions'' to remind the
reader that even equation (\ref{DBI}) is only the bosonic portion of
the action. All of these objects are supersymmetric and so fermions
must be added in the appropriate manner. The action (\ref{SYM}) is
${\cal N}= 4$ supersymmetric Yang-Mills theory in four space-time
dimensions. It is a non-abelian $U(N)$ gauge theory, as we encountered
in section \ref{sec:largeN}, see equation
(\ref{normaction}). Comparing the forms of the action, we arrive at
the second fundamental relation of the AdS/CFT correspondence

\be\label{couplings}
\boxed{
4\pi g_s = g_{YM}^2 }
\ee   

\ni i.e. the Yang-Mills coupling constant is related to the
square-root of the closed string coupling.

% ========================================================================== %

\subsection{Absorption cross-sections}
\label{sec:absxs}

\begin{figure}[ht]
\begin{center}
\includegraphics*[height=2.5in]{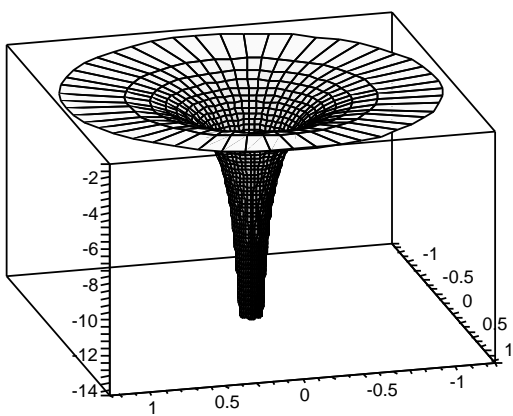}
\end{center}
\caption{The ratio of proper radial distance to coordinate radial
  distance $d r_P/dr = (1+R^4/r^4)^{1/4}$ is plotted on the vertical
  axis for the extremal 3-brane geometry (\ref{extrem3}). The angular
  variable may be thought of as representing the five-sphere. The
  worldvolume coordinates $x_i$ are suppressed. Here $R$ is set to one
  and $dr_P/dr$ is multiplied by $-1$ for visual effect.
}
\label{fig:extremal}
\end{figure}

We have seen that a stack of $N$ D3-branes, seen from the low-energy
supergravity limit, i.e. away from the branes (see figure
\ref{fig:dbrane}), looks like a curved spacetime (\ref{extrem3}) with
a five-form field strength turned on. We have also seen that the
low-energy limit of the theory on the stack of D-branes is ${\cal
N}=4$ supersymmetric Yang-Mills with gauge group $U(N)$. One of the
indications that these two pictures might be equivalent came from the
consideration of the absorption of closed string modes by either the
geometry (\ref{extrem3}) or by the stack of D-branes
\cite{Klebanov:1997kc, Gubser:1997yh, Gubser:1997se}. The geometry may
be envisioned as having a central throat from which it is difficult
for particles to escape, see figure \ref{fig:extremal}.
We begin by considering the absorption of the fluctuations of the
dilaton $\Phi$, which we will call $\phi$. The equation of motion for
this field can be obtained from (\ref{IIB}), using the solution
(\ref{extrem3}), (\ref{extrem3F}). It turns out that it is simply
$\Box \phi = 0$, where the D'Alembertian $\Box$ is defined by the
metric (\ref{extrem3}). We take the following form for $\phi$

\be
\phi(X) = R(r) \Theta(\O_5) e^{-i\o t}  
\ee

\ni where $X$ indicates the full 10-dimensional coordinates, $r$ is
the radial coordinate from (\ref{extrem3}) and $\O_5$ is shorthand
for the coordinates on the five-sphere. The energy of the fluctuation
is given\footnote{As is customary, the units chosen in this
  thesis are such that $\hbar = c = 1$.} by $\o$. Notice that we have
suppressed dependence on the brane coordinates $x_i$; we will not be
interested in these fluctuations as they will not contribute to the
absorption cross-section. It is straightforward to
calculate the D'Alembertian, which then gives the following equation
of motion

\be
\Box \phi = \left[ -\left(1+\frac{R^4}{r^4}\right) \p_t^2 + \p_r^2 +
  \frac{5}{r} \p_r + r^2 D_{\O_5}^2 \right] \phi(X) = 0
\ee

\ni where $D_{\O_5}^2$ indicates the Laplacian on the five-sphere. We
will consider only the s-wave or $D_{\O_5}^2 \Theta = 0$ modes, and
hence calculate the s-wave absorption cross-section. The resulting
radial equation is

\be
\left[ \o^2\left(1+ \frac{R^4}{r^4} \right) + r^{-5} \p_r (r^5 \p_r)
  \right] R(r) = 0.
\ee

\ni It is simpler to solve this equation after the following change of
variables $r = R e^{-z}$, $R(r) = e^{2z} \psi$, then

\be
\left[ \p_z^2 + 2\o^2 R^2 \cosh 2z \right] \psi(z) = 0
\ee

\ni and our problem reduces to a Schr\"{o}dinger equation with
potential $-2\o^2 R^2 \cosh 2z$. This is a barrier problem where the
incoming wave (from $z=-\infty$) has zero ``energy'' and the top of
the potential is also at zero energy. Thus $\psi$ is on the border
between tunnelling or conventionally transmitting from $z=-\infty$ to
$z=\infty$, i.e. from asymptotic flat space at $r=\infty$ to the
center of the throat.  The equation may be solved easily in the $z
\rightarrow \infty$ and in the $z \rightarrow -\infty$ limits, the
solutions are as follows

\be\label{solutions}
\psi_{-\infty} (z) \simeq a J_2(\o R e^{-z}), \qquad 
\psi_\infty (z) \simeq iH^{(1)}_2(\o R e^z)
\ee

\begin{figure}[ht]
\begin{center}
\includegraphics*[bb=82 270 475 700,height=2.5in]{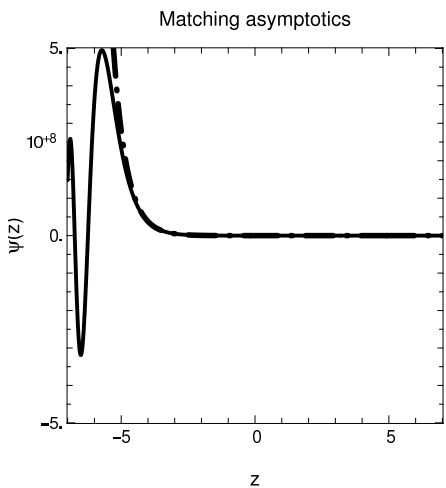}
\end{center}
\caption{The solutions (\ref{solutions}) are plotted for $\o R =
  0.01$, where the solid line is $\psi_{-\infty} (z)$ and the dot-dash
  the real part of $\psi_{\infty} (z)$ (the imaginary part is
  negligible). It is seen that they overlap in the region $\ln \o R \ll
  z \ll -\ln \o R$.  }
\label{fig:asympt}
\end{figure}

\ni where $H^{(1)}_m$ is the $m^{\text{th}}$ Hankel function of the
first kind, and $J_2$ is the second Bessel function. In the region
$\ln \o R \ll z \ll -\ln \o R$, i.e. for low energies $\o$, the two
solutions are simultaneously valid, see figure \ref{fig:asympt}. This
allows for the determination of the in-going ``amplitude'' $a$, by
matching the two solutions in their overlapping region.
Investigating the asymptotics for $z=0$ and $\o R \ll 1$, we find

\be
 J_2(\o R e^{-z}) \simeq \frac{(\o R)^2}{8}, \qquad 
iH^{(1)}_2(\o R e^z) \simeq \frac{4}{\pi (\o R)^2}
\ee

\ni and so 

\be
a = \frac{32}{\pi} (\o R)^{-4}. 
\ee

\ni The opposite asymptotics give

\bsp
&aJ_2(\o R e^{-z}) \simeq -\sqrt{\frac{2}{\pi \o R e^{-z}}}
\frac{a}{2}\left( e^{i(\o R e^{-z} - \pi/4)} +  e^{-i(\o R e^{-z} -
  \pi/4)}\right), \qquad z \rightarrow -\infty \\
&iH^{(1)}_2(\o R e^z)  \simeq -i\sqrt{\frac{2}{\pi \o R e^z}}
e^{i(\o R e^z - \pi/4)}, \qquad z \rightarrow \infty
\end{split}
\ee

\ni from which we can read-off the incident, reflection, and
transmission coefficients. The absorption probability is the squared norm of
the ratio of the transmission coefficient to the incident coefficient,
or

\be
{\cal P} = \left| \frac{1}{a/2} \right|^2 = \frac{\pi^2}{16} ( \o R )^8
\ee  

\ni The task of translating this into a cross-section is rather
involved in the general case \cite{Gubser:1997qr}, the result is the
following prescription

\be\label{grx}
\s = \frac{(2\pi)^n}{\o^n \O_n} {\cal P} = \frac{\pi^4}{8} \o^3 R^8
\ee

\ni where $n$ is the dimension of the sphere in the geometry, in our
case $n=5$, while $\O_n = 2\pi^{(n+1)/2}/\G((n+1)/2)$ is the volume
of the n-sphere. We have therefore found the absorption cross-section
for the dilaton s-waves in the extremal 3-brane geometry.

How do we envision this process from the point of view of the stack of
D3-branes? Although we made no mention of it, the action (\ref{DBI})
clearly contains a coupling to the dilaton, i.e. the $e^\Phi$
factor. We may therefore ask the question, what is the total
cross-section for dilaton absorption by a stack of $N$ D3-branes? In
order to answer this question we should begin by analyzing the
low-energy limit of our D3-branes and the subsequent dilaton
coupling. This is most easily accomplished by placing the action
(\ref{IIB}) into canonical form by rescaling the metric 

\be
G_{\m\n} \rightarrow \wt G_{\m\n} =  e^{\frac{1}{2}(\Phi_0 - \Phi)} G_{\m\n}
\ee

\ni where the dilaton has been shifted by a constant $\Phi_0$. We will be
interested only in the action for the canonically normalized dilaton
$\wt \Phi = \Phi - \Phi_0$. Applying the rescaling to (\ref{IIB}), we
have\footnote{Note that $\wt R = e^{\wt \Phi /2} \left[ R - \frac{9}{2}
    \nabla^2 \wt \Phi + \frac{9}{2} \p_\m \wt \Phi \p^\m \wt \Phi
    \right]$, see for example Appendix E of \cite{Ortin:2004ms}.} 

\be\label{daction}
S_{10} = \frac{1}{2(\k_{10} e^{\Phi_0})^2} \int d^{10}x \sqrt{- \wt G}
\left( \wt R - \frac{1}{2} \p_\m \wt \Phi \p^\m \wt \Phi + \ldots\right).
\ee

\ni This frame is referred to as the {\it Einstein frame}, while the
pre-scaled version is dubbed the {\it string frame}. Notice that the
gravitational coupling is determined by the constant dilaton shift,
$\wt \k_{10} = \k_{10} e^{\Phi_0} = \k_{10} g_s$. Shifting to the
Einstein frame in our D-brane action (\ref{DBI}), we see that only the
$G_{ab}$ term is affected. The result for the low-energy action of our
stack of D3-branes is

\be\label{coupling}
S_{4} = -\frac{1}{2 g^2_{YM}} \Tr \int d^4x \left( e^{-\wt \Phi}
(F_{\a\b})^2 + \ldots \right) 
\ee

\ni where the ``$\ldots$'' refers to terms not coupled to the dilaton
and fermion terms\footnote{The in-coming dilaton s-wave cannot be
converted into a pair of fermions on the brane because the coupling,
involving the kinetic term $\bar \psi \displaystyle{\not} \p \psi$,
gives an odd power of the momentum.}. Also, we have omitted couplings
to the dilaton involving the transverse scalars on the D-brane; these
correspond to higher-than-s-wave dilaton couplings. We now take $\wt
G_{\m\n} = \eta_{\m\n}$ since we take the D3-branes to be sitting in
flat ten-dimensional space. Further, we put the dilaton action
(\ref{daction}) into standard canonical form by defining $\phi = \wt
\Phi/\sqrt{2}{\wt \k_{10}}$; we do the same for the coupling in
(\ref{coupling}) by rescaling $A_\a$ by $g_{YM}$. We then obtain

\be
S_{10}+S_4=
-\frac{1}{2}\int d^{10}x \, \p_\m \phi\, \p^\m \phi - \int d^4
\xi \left( \frac{1}{4} (F_{\a\b}^a)^2 + \frac{\wt \k_{10}}{\sqrt{2}}
\,\phi \, \p_\a A^a_\b\, \p^\a A^{a \b} \right)   
\ee

\ni where we have used the fact that the generators $T^a$ of $U(N)$
are normalized by $\Tr(T^aT^b) = \tiny{\frac{1}{2}}\d^{ab}$ where the
index $a$ runs from 1 to $N^2$. The problem of calculating the
cross-section is now straightforward, and can be found in any textbook
on quantum field theory, see for example \cite{Peskin:1995ev},
pg. 107. The field $A^a_\a$ will have two physical polarizations for
each $a$. At leading order in $\l = g_{YM}^2N$, each will couple to an
in-coming dilaton $\phi$ (solid line below) via the following Feynman
diagram

\begin{center}
\includegraphics*[bb= 200 600 430 710]{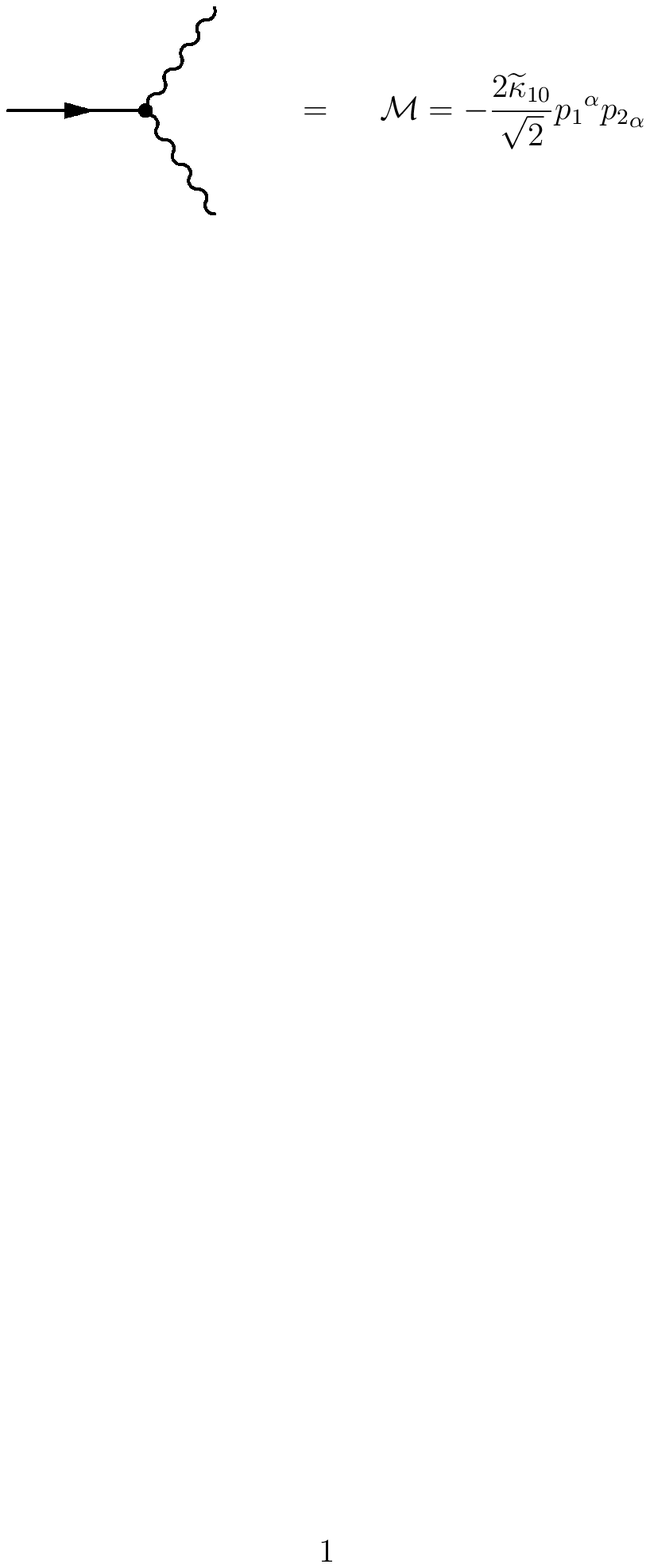}
\end{center}

\ni where the ${p_i}^\a$ are the four-momenta of the final state
photons (wiggly lines above). The cross-section is then given by

\be \s = \frac{1}{2} \frac{1}{2 \o} \int \frac{d^3 p_1}{(2 \pi)^3}
\frac{1}{2 E_1} \int \frac{d^3 p_2}{(2 \pi)^3} \frac{1}{2 E_2}
(2\pi)^4 \d(E_1 + E_2 - \o) \d^3(\vec p_1 + \vec p_2) \,| {\cal M}|^2
= \frac{\wt \k_{10}^2 \o^3}{64 \pi}
\ee

\ni where $\o$ is the energy of the in-coming dilaton and the leading
factor of $1/2$ accounts for the fact that the final-state photons are
identical. Since we have $2N^2$ species of photons, where the $2$
counts the number of physically distinct polarizations, we have

\be\label{dbx}
\s_{tot} = \frac{N^2\wt \k_{10}^2 \o^3}{32 \pi}.
\ee

\ni Using the fact that $\wt \k_{10}^2 = (2\pi)^7 g_s^2 \a'^4/2$, and
(\ref{radius}) we see that this is identical to
(\ref{grx}). Therefore, for the dilaton s-wave, the stack of D3-branes
in flat 10-dimensional space absorbs exactly the same as the throat of
the geometry (\ref{extrem3}). In fact the agreement is suspicious. The
supergravity result (\ref{grx}) is valid in the supergravity
approximation, i.e. when $R^4/\a'^2 = g_{YM}^2 N \gg 1$, but we
performed our D-brane calculation only to leading order in $\l =
g_{YM}^2N$, i.e. in the $R^4/\a'^2 \ll 1$ limit. In fact the higher
order corrections to (\ref{dbx}) vanish as a result of a
non-renormalization theorem \cite{Gubser:1997se}. The cross-sections
for other closed string modes, namely fermions and gravitons, were
found to agree similarly in \cite{Gubser:1997yh}.

This result was an indication that there was a duality emerging
between strongly coupled super-Yang-Mills and IIB strings on a weakly
curved background. In fact there was already an indication from a
study of entropy \cite{Gubser:1996de}, that a correspondence may be at
play. It was found that the entropy of the weakly coupled
super-Yang-Mills theory at small temperature agreed with that of the
corresponding near-extremal black hole entropy, up to a factor of
$4/3$. This near match for quantities at opposite ends of the coupling
spectrum was notable. Maldacena \cite{Maldacena:1997re} eventually
codified these observations into a conjecture which has largely shaped
string theory research in the interim.

% ========================================================================== %

\subsection{The Maldacena conjecture}
\label{sec:maldacon}

The central theme in the Maldacena conjecture is a {\it decoupling
limit}. Specifically, Maldacena considered taking $R,\a' \rightarrow
0$ while keeping $R^4/\a'^2 = \l = 4\pi g_sN$ fixed. This limit sends
the cross-section (\ref{dbx}) to zero; the strings in the throat of
(\ref{extrem3}) (on the stack of D-branes) are decoupled from those in
the asymptotic region (away from the stack of D-branes). This results
in two pictures, each containing two decoupled theories. In the
throat-geometry picture we have closed, type IIB strings propagating
in the $r \ll R$ region of (\ref{extrem3})

\be\label{nearhorizon}
ds^2 = \frac{r^2}{R^2} \left(-dt^2 +
dx_i^2\right) + \frac{R^2}{r^2}\left(dr^2 + r^2
d\O_5^2\right)
\ee

\ni which are decoupled from closed IIB strings propagating out at $r \gg
R$, which is just 10-dimensional flat space. In the D-brane picture we
have ${\cal N}=4$ super-Yang-Mills in four spacetime dimensions with
gauge group $U(N)$ on the stack of D-branes, while decoupled closed,
type IIB strings propagate in the bulk spacetime which is
10-dimensional flat space. The two pictures share a decoupled theory:
closed type IIB strings in 10-dimensional flat space. The Maldacena
conjecture \cite{Maldacena:1997re} posits that the other two decoupled
theories are also equivalent, that is\\

\fbox{\begin{minipage}[bottom]{5.85in}
\begin{center}
{\it ${\cal N}=4$ super-Yang-Mills in four spacetime dimensions with
gauge group $U(N)$ is dual to type IIB string theory on the background
$AdS_5 \times S^5$}
\end{center}
\end{minipage}
}\\

\begin{figure}[ht]
\begin{center}
\includegraphics*[bb=0 0 300 275,height=2.5in]{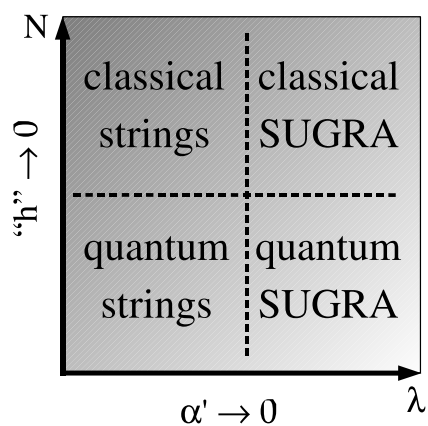}
\end{center}
\caption{The parametric limits of the AdS/CFT correspondence. The rank
  of the super-Yang-Mills gauge group $N$ controls quantum corrections
  in the closed string theory; these correspond to non-planar gauge
  theory processes. The 't Hooft coupling controls the ratio of the
  closed string background curvature to the string length; large 't
  Hooft coupling corresponds to point-particle supergravity
  (``SUGRA''). Here ``h'' refers to the effective Planck's constant -
  i.e. the parameter controlling ``quantumness''.}
\label{fig:Nvslambda}
\end{figure}

\ni where we have identified (\ref{nearhorizon}) as the metric of five
dimensional anti de-Sitter space times a five-sphere. We should also
not forget that this background includes the form-field strength
(\ref{extrem3F}). What does this duality imply about the two theories?
The super-Yang-Mills has two parameters; $N$ the rank of the gauge
group, and $\l = g^2_{YM} N$ the 't Hooft coupling. The decoupling
limit keeps $\l$ fixed. This means that when $N$ is varied, $g^2_{YM}
= 4\pi g_s$ varies inversely. Thus the closed string coupling varies
inversely with $N$, so that when $N$ is large the closed string
dynamics are captured by their tree-level or classical limit, see
figure \ref{fig:Nvslambda}. The 't Hooft coupling itself may also be
varied. From (\ref{radius}) we see that $\l$ measures the ratio of the
AdS (and five-sphere) curvature to the string length
($\sqrt{\a'}$). Thus when $\l$ is large, i.e. the gauge theory is
strongly coupled, the ``stringiness'' of the closed strings is
suppressed and they are well approximated by their low-energy
point-particle limit: type IIB supergravity. 
The beauty of this correspondence is that (at least for large
$N$) it is precisely where the analytical techniques fail in the gauge
theory, i.e. at large $\l$, that the dual string theory simplifies to
classical supergravity on a weakly curved background. We have
analytical control over the dual theory in this regime, and so we have
finally realized a string description of a strongly coupled gauge
theory that will allow analytical calculations.   

% ========================================================================== %

\subsection{Preliminary evidence for AdS/CFT: symmetries}
\label{sec:prelim}

We have presented a conjecture in section \ref{sec:maldacon} with no
evidence or proof. What is the main motivation, above and beyond the
cross-section calculations of section \ref{sec:absxs}, for suggesting
this duality? The answer is that the two theories share the same
symmetry groups. ${\cal N}=4$, $d=4$ supersymmetric Yang-Mills theory is
believed to be a {\it conformal field theory} or CFT. The meaning of
this is that the full interacting, quantum mechanical theory is
invariant under conformal transformations. These are angle-preserving
transformations which include global rescalings

\be\label{scale}
x^\m \rightarrow \L x^\m
\ee

\ni and the so-called {\it special conformal transformations}

\be\label{spconf}
x^\m \rightarrow \frac{x^\m + a^\m x^2}{1 + 2 a_\m x^\m + a^2 x^2}
\ee

\ni where $a^\m$ is some real d-vector. A natural consequence of
invariance under (\ref{scale}) is that the $\b$-function for the
coupling $g_{YM}$ is identically zero. This means that the coupling in
a CFT does not run at all - it is a free parameter in the theory. This
is important for the duality of AdS/CFT since we would like to be able
to vary the 't Hooft coupling $\l$ freely (see figure
\ref{fig:Nvslambda}). The conformal group extends the symmetry of flat
space (\ref{poin}), i.e. the Poincar\'{e} group, via the inclusion of
the generators of (\ref{scale}) $D$, and those of (\ref{spconf})
$K_\m$. The extra algebraic relations are

\bsp
[D,P_\m] = -i P_\m, \qquad [P_\m,K_\n] &= 2i M_{\m\n} - 2i \eta_{\m\n} D,
\qquad [D, K_\m] = i K_\m, \\ [M_{\m\n},D]=0, \qquad
&[M_{\m\n},K_\r] = -i ( \eta_{\m\r}K_\n - \eta_{\n\r}K_\m).
\end{split}
\ee

\ni In fact the conformal group is isomorphic to $SO(2,d)$. This can
be seen via the following assignments

\be
J_{\m\n} = M_{\m\n}, \qquad J_{\m d} = \frac{1}{2}(K_\m - P_\m), \qquad
J_{\m \, (d+1)} = \frac{1}{2}(K_\m + P_\m), \qquad J_{(d+1)\,d} =D,
\ee

\ni for which $[J_{MN}, J_{IJ}] = i\eta_{N I}J_{M J}
-i\eta_{M I}J_{N J} -i\eta_{N J} J_{M I} +i\eta_{M J}J_{N I}$ simply
gives the conformal group relations, where $I,J,M,N = 0, \ldots, d+1$
with signature $(-,+,\ldots,+,-)$. ${\cal N}=4$ super-Yang-Mills (\ref{SYM})
contains another important symmetry group, called R-symmetry. This
symmetry rotates the six scalar fields $\Phi^I$ into one another;
therefore this group is $SO(6)$. On the level of bosonic symmetries,
this is it. Thus the bosonic symmetries of ${\cal N}=4$
super-Yang-Mills is $SO(2,4) \times SO(6)$.

Anti de-Sitter space may be defined as the embedding of a hyperboloid
in a two-time signature space

\be
ds^2 = -dx_0^2 -dx_{d+1}^2 + (dx_i)^2, \qquad  i = 1,\ldots,d.
\ee

\ni The hyperboloid is embedded as

\be\label{hyper}
x_0^2 + x_{d+1}^2 - (x_i)^2 = R^2
\ee

\ni which is manifestly $SO(2,d)$ invariant. It is then plain to see
that the isometry group of the space $AdS_5 \times S^5$ is $SO(2,4)
\times SO(6)$. We have therefore an exact matching of the bosonic
symmetries between ${\cal N}=4$ super-Yang-Mills and fields on $AdS_5
\times S^5$. In fact it can be shown that the full supergroup of type
IIB strings on $AdS_5 \times S^5$ and ${\cal N}=4$ super-Yang-Mills is
$SU(2,2|4)$, of which $SO(2,4) \times SO(6)$ is the bosonic
subgroup. 

It will be useful to explore anti de-Sitter space more
thoroughly. The hyperboloid (\ref{hyper}) may be coordinatized using
the following relations

\be
x_0 = R \cosh \r \cos \t, \qquad x_{d+1} =  R \cosh \r \sin \t,
\qquad x_i = R \sinh \r\, \O_i, \qquad (\O_i)^2 = 1
\ee

\ni where $\O_i$ are an embedding of the $(d-1)$-sphere. This leads to
so-called {\it global AdS}

\be\label{globads}
ds^2 = R^2 ( -\cosh^2 \r \,d\t^2 + d\r^2 + \sinh^2 \r \,d \O_{d-1}^2 )  
\ee

\ni where the coordinate $\t$ is unwrapped from its fundamental domain
$[-\pi,\pi]$ to $[-\infty,\infty]$ in order to avoid closed timelike
curves, while $\r \in [0,\infty]$. Another important coordinatization
is

\bsp x_0 = \frac{1}{2r} &\left( 1 + r^2(R^2 + \vec z^2 - t^2 ) \right),
\qquad x_{d} = \frac{1}{2r} \left( 1 - r^2(R^2
- \vec z^2 + t^2 ) \right), \\
&~~~x_{a} = R r z_a, \qquad  x_{d+1} = R r t, \qquad a=1,\ldots,d-1 
\end{split}
\ee

\ni where $r \geq 0$ and $t$ and $\vec z$ are unconstrained. This
gives the {\it Poincar\'{e} patch}

\be\label{poinads}
ds^2 = \frac{r^2}{R^2} \left(-dt^2 +
d\vec z^2\right) + \frac{R^2}{r^2} dr^2
\ee

\ni which we saw in (\ref{nearhorizon}). The global coordinates
(\ref{globads}) cover the entire hyperboloid (\ref{hyper}) once, the
Poincar\'{e} patch, as we will see, covers only half of it. The
relationship between the global and Poincar\'{e} coordinates systems
is most easily seen when $d=2$, i.e. for $AdS_3$. In this case the
$\O_i$ parametrizes a circle whose parameter we will take as $\phi$,
and we have

\bsp r = \frac{1}{R} \left( \cosh \r \cos \t - \sinh \r \cos \phi
\right),~~~~~~~~~~~~~~\\ \qquad t = \frac{R \cosh \r \sin \t}{\cosh \r
\cos \t - \sinh \r \cos \phi }, \qquad z = \frac{R \sinh\r \sin
\phi}{\cosh \r \cos \t - \sinh \r \cos \phi }.
\end{split}
\ee

\begin{figure}[ht]
\begin{center}
\includegraphics*[bb=0 0 215 330,height=2.5in]{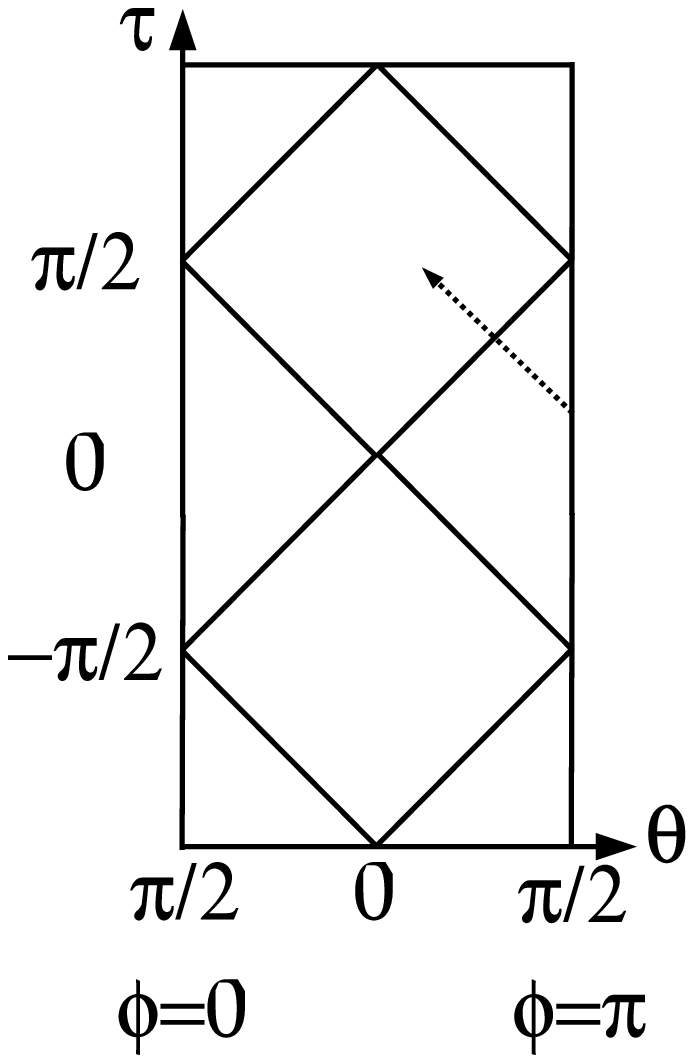}
\includegraphics*[bb=70 260 490 710,height=2.5in]{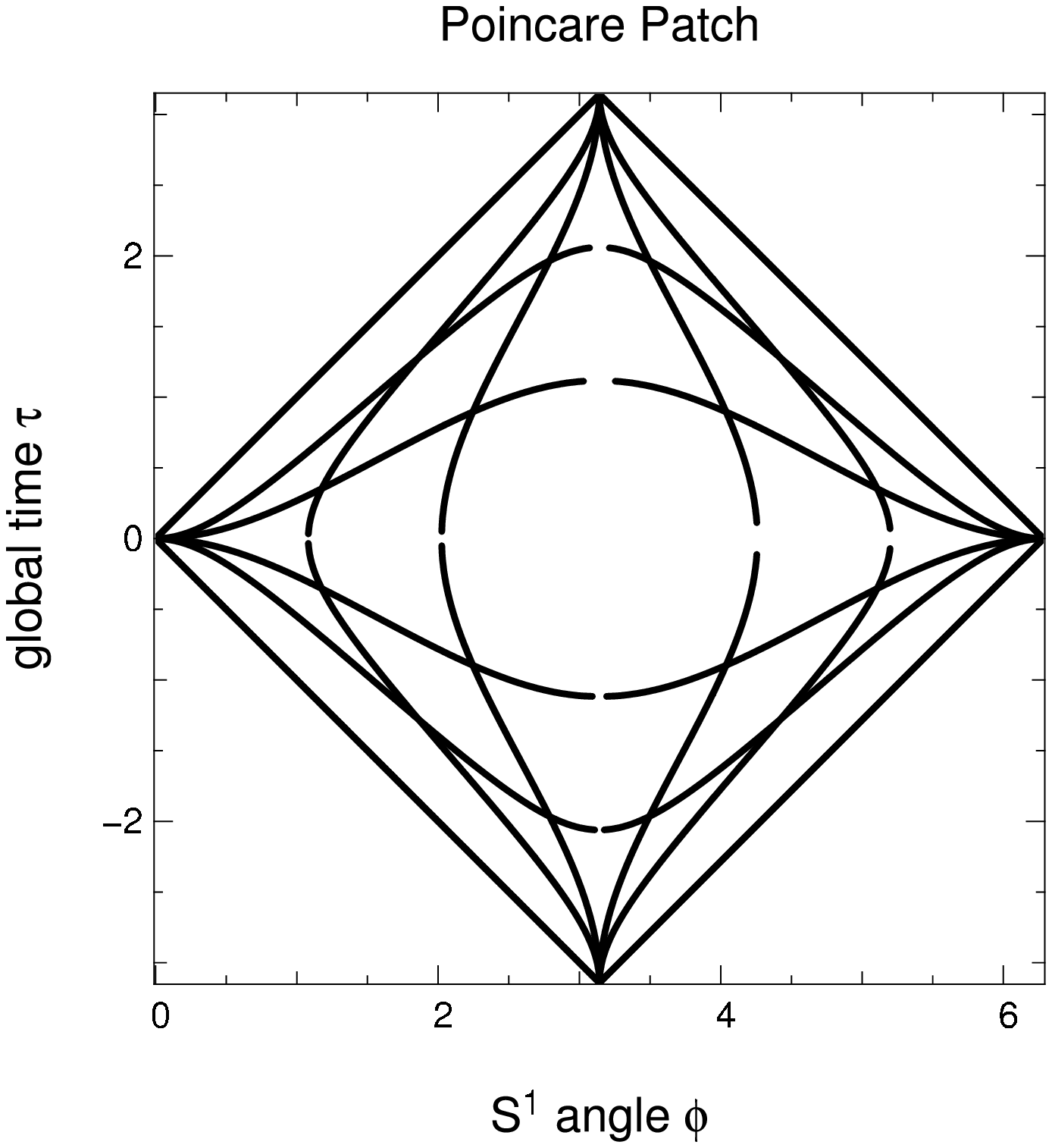}
\end{center}
\caption{A Penrose diagram of anti de-Sitter space (\ref{penrads}) is
shown on the left. The diamonds are the paths of light-rays beginning
at $\theta=0$, $\t=-\pi$ and reflecting off the boundary at
$\theta=\pi/2$. A signal from the boundary (which is at spatial
infinity) may propagate into the spacetime in finite coordinate
time. Note that the diagram should be understood to be a fundamental
domain which is periodically continued to $\t=[-\infty,\infty]$ The
angle $\phi$ is understood as the azimuthal angle in the $\O_i$
portion of the metric. On the right the boundary ($r=\infty$) of the
Poincar\'{e} patch (\ref{poinads}) is displayed for $d=2$. The
boundary of global $AdS_3$ may be envisioned as a cylinder with a
coordinate $\t$ running along the length of the cylinder and the angle
$\phi$ going around it. The Poincar\'{e} patch fits into half a
fundamental domain, bounded by null surfaces at spatial infinity
($z=\pm \infty$) forming a diamond shape which wraps around the
cylinder. The horizontal curves are lines of constant $z$, while the
vertical curves are lines of constant $t$.}
\label{fig:poinpatch}
\end{figure}

\ni Since $r \geq 0$, we must have $\cos \t \geq \tanh \r \cos
\phi$. The boundary of the patch is therefore given by the following
curves (see figure \ref{fig:poinpatch}, right panel, where $\r$ is set
to infinity)

\be
\t = \pm \arccos \left( \tanh \r \cos \phi \right).
\ee

\ni One may verify that the area (in the $\t$-$\phi$ plane) of the
patch is therefore given by

\be
4 \int_0^{\pi} d\phi  \arccos \left( \tanh \r \cos \phi \right) = 2\pi^2
\ee

\ni independently of $\r$. This is one half of the total area,
$4\pi^2$ (since $\t$ and $\phi$ are $\in[0,2\pi]$). 

Anti de-Sitter space also has a time-like boundary which is most
easily seen through the change of coordinates $\tan \theta = \sinh
\r$, where $\theta \in [0, \pi/2]$. The global $AdS$ metric
(\ref{globads}) then takes the following form

\be\label{penrads}
ds^2 = \frac{R^2}{\cos^2 \theta} ( -d\t^2 + d\theta^2 +
\sin^2\theta d\O_i^2). 
\ee
 
\ni The boundary is found at spatial infinity ($\theta=\pi/2$ or $\r =
\infty$) and is of the form $\bR \times S^{d-1}$. A Penrose diagram of
the space is shown in figure (\ref{fig:poinpatch}).
The boundary is time-like and it takes a finite amount of coordinate
time for a signal to propagate from the boundary at spatial infinity
to any point in the space. This implies that information may be gained
from or lost to the boundary, and in this respect anti de-Sitter space
is very similar to Minkowski space in a box. This will be important
for us in the next section. The boundary in the Poincar\'{e} patch is
at $r=\infty$ and is given by $ds^2 = r^2 (-dt^2 +d \vec z^2)/R^2$;
i.e. it is a conformal rescaling of flat d-dimensional space. In this
sense the Poincar\'{e} coordinates ``lose the point at infinity''
required to restore the $\bR \times S^{d-1}$ topology of the boundary.

% ========================================================================== %

\subsection{The field-operator dictionary and the GKP-W relation}
\label{sec:GKPW}

The AdS/CFT correspondence alleges an equivalence between a conformal
field theory and a string theory on $AdS_5 \times S^5$, but how is the
equivalence seen? In order to specify what is equivalent to what, a
dictionary is required which translates a problem posed in one setting
into the language of the other, and vice-versa. Such a dictionary is
given by the GKP-W relation \cite{Gubser:1998bc,Witten:1998qj}, named
for Gubser, Klebanov, Polyakov, and Witten. The first question we
should ask is what are the meaningful (physical) quantities in each
of the theories that are eligible for comparison. A conformal field
theory or CFT has no scale, it is therefore meaningless to discuss
asymptotically free wave-packets, and this precludes an S-matrix and
the concept of particles with definite mass. An important class of
invariants that a CFT does possess is the scaling dimensions of
operators. These must relate to some other invariant in the gravity
(string) theory. Supergravity in $AdS_5 \times S^5$ does have a scale
and asymptotic mass eigenstates. We mentioned in section
\ref{sec:prelim} that in some respects $AdS$ space is similar to
Minkowski space in a box. In fact, a unique solution to the Laplace
equation for a (for example scalar) field $\phi(r;\vec z,t)$ on $AdS$
requires the specification of boundary data $\phi_0(\vec z,t) =
\phi(\infty; \vec z,t)$. Keeping these facts in mind, the absorption
cross-section calculations of section \ref{sec:absxs} hint at what the
relation between the CFT and the supergravity should be. Recall that
the coupling of the dilaton to the D-brane worldvolume theory (at low
energy) was (\ref{coupling}) $\sim \Tr\int d^4x \phi_0 F^2$. That is,
the bulk closed string mode described by the field $\phi$, interacts
with the worldvolume theory via a local (i.e. proportional to the
value of $\phi$ on the space where the worldvolume theory lives, that
is, $\phi_0$) coupling to an operator of the CFT. In the throat of the
geometry, $\phi$ represents a minimization of the supergravity action
on $AdS_5$, subject to the boundary condition $\phi_0$. This minimized
action therefore appears equivalent to the addition of the operator
${\cal O}=\Tr F^2$ to the worldvolume theory's action, $S_{WV}
\rightarrow S_{WV} + \int d^4x \phi_0 {\cal O}$. To be more precise,
the GKP-W relation is

\be\label{GKPW}\boxed{
\left< \exp \left( \int d^4x \,\phi_0\,{\cal O}\right) \right>_{CFT}
= \exp \left( -S_{\text{SUGRA}} (\phi) \right)}.
\ee

It may seem confusing that $\phi_0$ is at once the value of the field
$\phi$ on the stack of D-branes and the value on the boundary of
$AdS_5$. What must be remembered is that in the decoupling limit (see
section \ref{sec:maldacon}), the throat region, which is identified
with the position of the D-branes, is blown-up to the entire space
$AdS_5 \times S^5$. In this {\it near horizon} geometry, where has the
corresponding position of the stack of D-branes gone? The answer is to
the boundary of $AdS_5$. Indeed we saw in section \ref{sec:prelim}
that this boundary is conformally equivalent to four-dimensional flat
space. In fact, the $SO(2,4)$ isometry group of $AdS_5$ acts upon this
boundary as the four-dimensional conformal group. It should be
emphasized that it is a mistake to think of the CFT as living on the
boundary of $AdS_5$ {\it simultaneously} with the supergravity in the
bulk. They are conjectured as equivalent descriptions of the same
physics; we can either work with the full $AdS_5$ supergravity or we
can throw that away and answer the same questions with the holographic
CFT. To see that this is so we can work out a simple example in which
it will be revealed that the relation (\ref{GKPW}) actually implies an
equivalence between the scaling dimension of the operator ${\cal O}$
and the mass of the associated field $\phi$. It is simplest to employ
the Poincar\'{e} metric (\ref{poinads}), in Euclidean signature, and
with the coordinate redefinition $r \rightarrow R^2/y$,

\be\label{stanADS}
ds^2 = \frac{R^2}{y^2}(dy^2 + dx_i^2)
\ee

\ni where the Euclideanized boundary space at $y=0$ is now covered by the
coordinates $x_i$. The Green's function corresponding to a field
$\phi(y,x_i)$ specified by boundary data $\phi(0,x_i)$ is most easily
obtained from an $SO(1,d+1)$ transformed version depending only on $y$
\cite{Witten:1998qj}

\be\label{kx0}
K(y) = C y^d
\ee

\ni where $C$ is some constant. This ans\"{a}tz obeys the equation of
motion $\Box \phi = 0$ and the boundary condition $K(\infty) =
\infty$. Under the $SO(1,d+1)$ inversion

\be
y \rightarrow \frac{y}{y^2 + x_i^2}, \qquad x_i \rightarrow \frac{x_i}{y^2 + x_i^2}
\ee

\ni the Green's function (\ref{kx0}) becomes

\be
K(y,x_i) = C \frac{y^d}{(y^2 + x_i^2)^d}
\ee

\ni which gives (for properly chosen $C$) the behaviour $K(0,x_i) =
\d^d(x_i)$ at the boundary point $y=0$. This is the desired boundary
condition for a {\it bulk-to-boundary propagator}, as it defines the
bulk field $\phi(y,x_i)$ in terms of its boundary value $\phi(0,x_i)$
in the following manner

\be\label{phisol}
\phi(y,x_i) = \int d^d \wt x_i \frac{C y^d}{ (y^2 + (x_i-\wt x_i)^2)^d }
\, \phi(0,\wt x_i).
\ee

\ni We should now plug this solution into the RHS of (\ref{GKPW}). The
action for a massless scalar field is

\be S[\phi] = \frac{1}{2} \int d^d x_i \, \int dy \, \sqrt{g} \,\p_\m \phi
\,\p^\m \phi \ee

\ni where $\m$ runs over all $d+1$ coordinates of (\ref{stanADS})
which is also the metric $g_{\m\n}$ refers to. Plugging the solution
(\ref{phisol}) into this action gives only a surface term at $y=0$,
since the main part of the action vanishes by the equation of
motion. It is straightforward to show that

\be\label{twopoint}
S[\phi(y,x_i)] = \frac{Cd}{2} \int d^d{x_i} \int d^d{\wt x_i}  \frac{
  \phi(0,x_i) \,\phi(0,\wt x_i) }{ (x_i - \wt x_i)^{2d} }.
\ee

\ni The two-point function of an operator ${\cal O}(x)$ in the CFT of
conformal weight $d$ has the following behaviour

\be 
\la {\cal O} (x_i) \, {\cal O}(\wt x_i) \ra =
\frac{{\cal C}}{(x_i-\wt x_i)^{2d}}
\ee

\ni where ${\cal C}$ is a constant. It is immediately seen that for
the appropriate choice of the constant $C$, the LHS of (\ref{GKPW}) is
exactly (\ref{twopoint}), where we have taken only the quadratic term
in the expansion of the exponentials, i.e. we are comparing two-point
functions. This whole story is repeated for the case of a massive
field ($(\Box-m^2) \phi=0$) with the replacement\footnote{The number
of CFT spacetime dimensions remain $d$. The other difference is in the
relation of the CFT scalar $\phi_0(x_i)$ to the boundary behaviour of
the bulk field $\phi(y,x_i)$; the general relationship is
$\lim_{y\rightarrow0} \phi(y,x_i) = y^{d-\D} \phi_0(x_i)$.}

\be\label{massdelta}
d \rightarrow \D =\frac{1}{2} \left( d + \sqrt{d^2 + 4m^2} \right) .
\ee

\ni We therefore have that the dual of the scaling dimension $\D$ of an
operator ${\cal O}$ in the CFT is related to the mass $m$ of the dual
AdS field by (\ref{massdelta}).

% ========================================================================== %

\subsection{Beyond two-point functions}

\begin{figure}[ht]
\begin{center}
\includegraphics*[bb=70 265 480 675, height=3.0in]{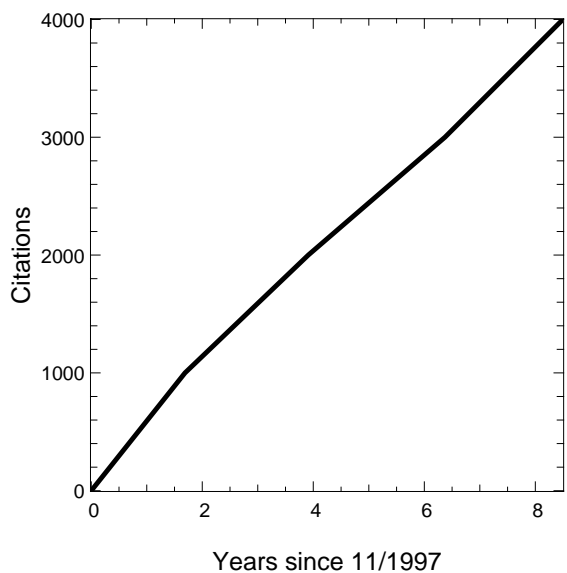}
\end{center}
\caption{Citations of Maldacena's original paper
  \cite{Maldacena:1997re} as a function of time, loosely interpreted 
  from the citebase website \cite{citebase}. The current SPIRES count 
  is more than 4500.}
\label{fig:malda}
\end{figure}

The AdS/CFT correspondence has passed many tests beyond the two-point
function presented in section \ref{sec:GKPW}. We will not give a
detailed account of the various successes of the correspondence
here, as that would fill several review papers. Three point functions
are well understood \cite{Lee:1998bx}, and are protected by conformal
invariance in the CFT. Four point functions do not share this
protection, but have been studied extensively (see references
\cite{Cornalba:2006xm} through \cite{Liu:1998ty}). There are also
large bodies of work concerning Wilson loops (see chapter
\ref{sec:wilsonloop}), M-theory (see chapter \ref{sec:matrixmodel}),
thermodynamics, D-brane states, macroscopic strings, viscosity,
black-hole entropy and information loss, and more. Indeed Maldacena's
original paper \cite{Maldacena:1997re} has been cited over 4500 times
at the time of writing, and appears to be increasing roughly linearly
with time, see figure \ref{fig:malda}.
The AdS/CFT correspondence has been the main focus of string theory
research for the past decade. Although a proof of the correspondence
is still lacking, no test of AdS/CFT has returned a negative
result. Much of the work on the correspondence has naturally focused
on the classical regime where large-$N$ super-Yang-Mills at strong
coupling is compared with supergravity. True quantum tests, involving
$1/N$ corrections, are less prevalent and less definitive, as we will
see in chapter \ref{chap:lcsft}. The quantum and stringy limits of
AdS/CFT are the most important avenues to study, in order to further
the evidence that the correspondence is indeed correct.

% ************************************************************************** %
\section{Summary}

We began this chapter with the clash of two titans, strongly coupled
gauge theory and gravity. String theory, devised to study the former,
was retired and then refitted to study the latter. In the end it led
to a remarkable observation, that in a certain context, the two titans
are reflections of each other, descriptions of the same physics
encoded in a different language. Further, the questions that are most
difficult to answer in one description are the easiest to answer in
the other. The stage is thus set for an exploration of this
gauge/string duality. In the following three chapters of this thesis,
we will present work concerning the quantum (chapter
\ref{chap:lcsft}), thermodynamic (chapter \ref{sec:matrixmodel}),
and classical (chapter \ref{sec:wilsonloop}) aspects of the AdS/CFT
correspondence. This research represents some important steps in the
very long journey to fully understanding the breadth and mechanisms of
this remarkable duality.

% ************************************************************************** %
% ************************************************************************** %
% ************************************************************************** %

\chapter{Light-cone string field theory on the plane-wave}
\label{chap:lcsft}

{\small
\begin{quote}
I call the loose snow to fly\\ from your rooftops, the timorous
doves\\ from your rafters' sanctum.\\
\rightline{--- Ruth Taylor {\it The Dragon Papers}, Emanations
$\;\;\;\;\;\;\;\;\;\;\;\;\;\;\;\;\;\;\;\;\;\;\;\;\;\;\;\;\;\;\;\;\;\;\;\;\;\;\;\;\;\;\;
$}
\end{quote}}

\begin{figure}[ht]
\begin{center}
\includegraphics*[bb=0 0 395 110,height=1.5in]{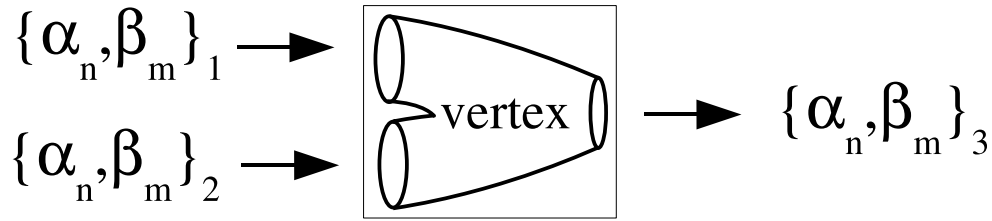}
\end{center}
\caption{In light-cone string field theory the cubic interaction
  vertex may be considered as a map between the collections of
  oscillators $\a_n$ and $\b_m$ which completely specify each string.}
\label{fig:vertex}
\end{figure}

String theory, despite all its complexity and scope, remains a
first-quantized theory. It is fundamentally the quantum dynamics of a
single relativistic string. The theory lacks a Lagrangian which would
dictate the full interacting theory of a field of strings. From the
point particle analogue point of view, we have only quantum mechanics
- not quantum field theory. This lack of completeness has not
prevented string theorists from considering on-shell string
interactions in some detail (c.f. \cite{Polchinski:1998rq,
Polchinski:1998rr, Green:1987sp, Green:1987mn}), guided simply by the
symmetries inherent in the first quantized theory, and by the manner
in which strings interact with background fields. To be more precise,
we have some control over interactions in the supergravity limit (as
discussed in section \ref{sec:clos}), where we know the full
point-particle Lagrangian is given by supergravity. On-shell
scattering amplitudes for low-lying string modes (i.e. few
excitations), with small numbers of loops ($< 3$), were calculated
very early on. Continuing string interactions off the mass-shell has
enjoyed a certain amount of success using Witten's cubic string field
theory \cite{Witten:1986qs}, see \cite{Taylor:2004rh} for a recent
review. From the point of view of light-cone gauge quantization (see
section \ref{sec:lcgq}), the description of the most basic interaction
- the three string vertex - is simply a map between the collections of
oscillators (\ref{comrel}) which completely specify each string, see
figure \ref{fig:vertex}.

\ni Such a map was developed by Green, Schwarz, and Brink
\cite{Green:1983hw, Green:1984fu} in the early 1980's, for type II
superstrings in a flat target space background. In 2002, Spradlin and
Volovich \cite{Spradlin:2002ar, Spradlin:2002rv} generalized this work
to the plane-wave background. In that context, the light-cone string
field theory may be used to calculate one-loop shifts to the masses of
string states. These are dual to anomalous dimensions of certain
operators in ${\cal N}=4$ super-Yang-Mills (in a given limit) and may
be compared as a check on the AdS/CFT correspondence\footnote{For
  reviews of the plane-wave string / gauge theory duality see
  \cite{Plefka:2003nb} and \cite{Sadri:2003pr}.}. This chapter
will introduce this program, and present original work by the author
which clarified some oblique issues concerning the importance of
intermediate string states in the one-loop mass shifts, proved the
finiteness of those shifts, and culminated in the best match so far
discovered between the gauge and string theory results using a version
of the string interaction vertex proposed by Dobashi and Yoneya
\cite{Dobashi:2004nm}.

% ************************************************************************** %
\section{The plane-wave background and the BMN limit of ${\cal N}=4$ SYM}
\label{sec:BMN}

In section \ref{sec:adscftcorr} the AdS/CFT correspondence was
introduced as a conjectured equivalency between type IIB superstrings
on $AdS_5 \times S^5$ and ${\cal N}=4$ SYM. An early stumbling block
encountered by investigators of the correspondence was solving the
string sigma model (curved space analogue of \ref{sustr}) on this
background. This remains an obstacle which has yet to be
overcome. Without the rudimentary information provided by the free
string spectrum on $AdS_5 \times S^5$, understanding or even testing
the AdS/CFT correspondence beyond the supergravity limit is extremely
limited. In 2001, a new type IIB background space was found
\cite{Blau:2001ne} which, like $AdS_5 \times S^5$, is maximally
supersymmetric. This is the so-called {\it plane-wave} background and
is given by

\be\label{ppmet}
ds^2 = -2 \,dx^- dx^+ - \m^2 (x_i)^2 (dx^+)^2 + dx_i^2, \qquad
F_{+1234} = F_{+5678} = \m \times \text{const.}
\ee

\ni where $+$ and $-$ denote light-cone directions, $i=1,\ldots,8$,
and $\m$ is a real, positive constant. This space may be
obtained via a Penrose limit \cite{Penrose:1976} of $AdS_5 \times S^5$
where the neighbourhood of a null geodesic (an equator of the
five-sphere) is ``zoomed-in'' upon. To see this consider global $AdS_5
\times S^5$, given by the following metric

\be
ds^2 = R^2 \left( -\cosh^2 \r\,dt^2 + d\r^2 + \sinh^2 \r \,d\O_3^2 + d\theta^2 +
 \cos^2\theta\,d\psi^2 + \sin^2 \theta \,d{\O'}^2_3 \right).
\ee

\ni The geodesic whose neighbourhood will be magnified is given by
$\r=\theta=0$, i.e. the equator parametrized by the angle $\psi$. To
realize this magnification we define $\hat x^+ = \tiny{\frac{1}{2}} (t +
\psi)$, $\hat x^- = (t - \psi)$, then re-scale the coordinates as follows

\be
\hat x^+ \rightarrow x^+ = \frac{1}{\m} \hat x^+, \qquad \hat x^-
\rightarrow x^- = \m R^2 \hat x^-,
\qquad \r = \frac{r}{R}, \qquad \theta = \frac{y}{R}, \qquad
R\rightarrow \infty
\ee

\ni it is then easy to see that the metric (\ref{ppmet}) is
obtained. It was not long before the $\k$-symmetric Green-Schwarz
superstring action was found and the string equations of motion were
solved on this simplified background \cite{Metsaev:2001bj,
Metsaev:2002re}. The free string spectrum, given in terms of the
light-cone Hamiltonian, is as follows

\be\label{ppstrspec}
H_{\text{l.c.}} = p^- = \frac{1}{\a' p^+} \sum_n N_n \sqrt{n^2 + (\m
  \a' p^+ )^2}
\ee

\ni where $N_n$ denotes the occupation number (number operator for
fermionic and bosonic oscillators), and positive $n$ denotes
left-moving modes while $n<0$ denotes right-moving modes. 

Because the plane-wave was obtained from $AdS_5 \times S^5$ through a
continuous scaling procedure, we may use the AdS/CFT correspondence to
provide a translation of the quantities found here to analogous ones in
${\cal N}=4$ SYM. We saw in section \ref{sec:prelim} that the
R-symmetry of SYM is the analogue of the $SO(6)$ symmetry group of the
five-sphere in $AdS_5 \times S^5$. We thus expect the (appropriate)
R-charge $J$ of operators in the SYM to be dual to the angular
momentum $-i\p_\psi$ of string states about the five-sphere. The
energy of a string state $i\p_t$ should be dual to the conformal
dimension $\D$ of those operators. We therefore have

\bsp\label{p+p-}
&p^- = i\p_{x^+} = i\m \,\p_{\hat x^+} = i \m (\p_t + \p_\psi) =
\m(\D-J) \\
&p^+ = i \p_{x^-} = \frac{i}{\m R^2}\, \p_{\hat x^-} = \frac{i}{\m R^2}
\frac{1}{2} (\p_t - \p_\psi) = \frac{\D+J}{2\m\sqrt{g^2_{YM} N} \,\a'}
\end{split}
\ee

\ni where we have used (\ref{radius}) and (\ref{couplings}). Since we
are taking the limit $R\rightarrow \infty$, states with finite $p^+$
must have $J \sim R^2 \sim \sqrt{g_{YM}^2N}$. Further, finite $p^-$
then implies $\D \sim J$. This allows us to rewrite the string
spectrum (\ref{ppstrspec}) in terms of gauge theory quantities, since
according to these scalings $\m \a' p^+ = J/\sqrt{\l}$ where $\l =
g_{YM}^2N$ is the 't Hooft coupling. We therefore have

\be\label{ppstrspecgauge}
\frac{1}{\m} p^-  = \sum_n N_n \sqrt{1 + \frac{n^2\l}{J^2}}. 
\ee

The specific operators which are
dual to free strings were identified in a milestone paper by
Berenstein, Maldacena, and Nastase (BMN) in 2002
\cite{Berenstein:2002jq}. The single string vacuum is labelled by its
light-cone momentum $p^+$, which is a free parameter in the
theory. This corresponds to a state with $p^-=0$, i.e. an operator
with $\D = J$, specifically the BMN operator in this case is

\be\label{vacop}
{\cal O}_J = \frac{1}{\sqrt{JN^J}} \Tr Z^J \lr |0;p^+ \ra
\ee 

\ni where $Z = \Phi^5 + i \Phi^6$, corresponds to the plane in $\bR^6$
(i.e. the 5-6 plane) which the five-sphere equator parametrized by
$\psi$ sits in. It is clear that the R-charge (corresponding to the
5-6 plane) of this operator is $J$, as it contains $J$ factors of the
field $Z$. Further, one may verify the conformal dimension at zero
Yang-Mills coupling (equivalently zero string coupling) is precisely
$J$ in the large $N$ (planar) limit, i.e.  

\be \la {\cal O}_J(x)\, {\cal O}_{J'}(0) \ra =
\left(\frac{g_{YM}^2}{4\pi}\right)^J \frac{\d_{JJ'}}{(x^2)^J}. 
\ee

\ni The BMN operators corresponding to excited string states were also
identified in \cite{Berenstein:2002jq}. The first excited states of
the string are those obtained by acting on the vacuum with two
oscillators, such that the total worldsheet momentum vanishes (as
required by level matching (\ref{levelmatch})). The reflection of
string oscillators on the gauge theory side of the correspondence are
the insertion into the operator ${\cal O}_J$ fields dubbed {\it
impurities}. We will encounter the full treatment of the string theory
in the next section; for now it suffices to say that there are 8
transverse bosonic oscillators $\a_n^i$ (labelled by the spacetime
index $i=1,\ldots,8$) and 8 fermionic super-counterparts $\b_n^a$
(labelled by an $SO(8)$ spinor index $a$). These oscillators are in
one-to-one correspondence with the fields of the gauge theory in the
following way

\bsp\label{impmap}
&\a^{\dag j}_n \rightarrow D_j Z \qquad j=1,\ldots,4\\
&\a^{\dag k}_n \rightarrow \Phi^{k-4} \qquad k=5,\ldots,8  \\
&\b^{\dag a}_n \rightarrow \chi^a
\end{split}
\ee

\ni where $D_j$ is the gauge-covariant derivative and $\chi^a$ are the
fermionic fields of ${\cal N}=4$ SYM. These impurities are interleaved
into the trace of (\ref{vacop}) by adding a position dependent phase
to each. Even though a single impurity state is unphysical, we present
it here as an instruction

\be
\frac{1}{\sqrt{J N^{J+1}}}
\sum_{l=0}^{J-1} e^{\frac{2\pi i n l}{J }}\Tr Z^l \Phi^1 Z^{J-l} \lr
\a_n^{\dag 5} | 0; p^+ \ra. 
\ee

\ni Note that by cyclicity all of the traces in the sum are equivalent,
leading to an overall factor of $\sum_{l=0}^{J-1} \exp(2\pi i n l /J) =
0$, since $n \in \bZ$. The unphysical nature of the state thus takes
care of itself by being identically zero. Note that this operator has
$J+1$ fields, while its 5-6 plane R-charge remains $J$. Therefore
$\D-J=1$, corresponding to one unit of light-cone energy. A
two-impurity state is built in the same way, by simply adding a second
impurity with its own phase factor and summing over positions of
insertion in the original chain of $J$ $Z$'s,

\be
\sum_{0 \leq k \leq l \leq J-1}\left[ 
\Tr Z^k \Phi^1 Z^{l-k} \Phi^2 Z^{J-l} 
e^{\frac{2\pi i n k}{J}} e^{-\frac{2\pi i n l}{J}} + 
\Tr Z^k \Phi^2 Z^{l-k} \Phi^1 Z^{J-l} 
e^{\frac{2\pi i n l}{J}} e^{-\frac{2\pi i n k}{J}} \right]
%\lr \a_n^{\dag 5} \a_{-n}^{\dag 6} |0; p^+\ra
\ee

\ni where we have dropped the normalization. Using the cyclicity of
the trace, this expression is simplified to the compact form

\be\label{bmn2imp}
\a_n^{\dag 5} \a_{-n}^{\dag 6} |0; p^+\ra \lr \frac{1}{\sqrt{J N^{J+2}}}
\sum_{l=0}^J e^{\frac{2\pi i n l}{J}} \Tr \Phi^1 Z^l
\Phi^2 Z^{J-l} = {\cal O}_J^{12}
\ee

\ni where the normalization has been restored and the dual string
state indicated. This procedure may be generalized to include any
number of impurities, c.f. \cite{Constable:2002hw}.

We now have a picture of perturbative strings as operators which
consist of a very long string of $J \sim R^2$ fields, with a few
impurities sprinkled along it. We also know that the 't Hooft coupling
$\l \sim R^4$ is taken to infinity. It would not be surprising to find
that a new coupling $\l' \equiv \l/J^2$ might arise in the
interactions between the BMN operators, since it is tunably small and
serves as the perturbative parameter in the expansion of the free
string energy (\ref{ppstrspecgauge}). We have found that in the zero
't Hooft coupling limit, $\D-J$ for the BMN operators are simply
integers, corresponding to (\ref{ppstrspecgauge}) with $\l=0$. If we
turn this coupling on, we expect to reproduce the entire Taylor
expansion of the square root. Indeed this has been verified to a few
orders in perturbation theory \cite{Kristjansen:2002bb,
Constable:2002hw, Bianchi:2002rw, Beisert:2002bb, Constable:2002vq,
Beisert:2002ff, Beisert:2003tq}, and to all orders via a superspace
formalism proof \cite{Santambrogio:2002sb}. The leading contribution
is derived from the quartic scalar interaction in the action for
${\cal N}=4$ SYM (\ref{SYM})

\be\label{V}
V = -4 g_{YM}^2 \biggl( \Tr \left| [Z,\Phi^1] \right|^2 
+ \Tr \left|[Z,\Phi^2]\right|^2 \biggr)
\ee

\begin{figure}[ht]
\begin{center}
\includegraphics*[bb=0 0 200 195, height=2.0in]{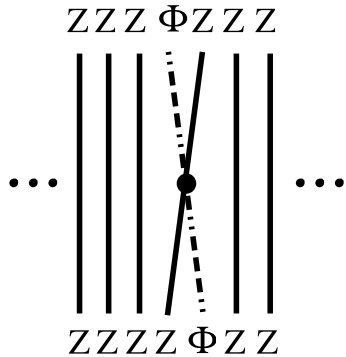}
\end{center}
\caption{The leading correction to the scaling dimension of the
  operator (\ref{bmn2imp}) is given by the quartic vertex which
  connects the terms of ${\cal O}(x)$ and ${\cal O}^\dag(0)$ in which
  one of the impurities is shifted by one unit.}
\label{fig:bmnint}
\end{figure}

\ni where we have shown the terms which will be important for the
operator (\ref{bmn2imp}). The leading correction to the two-point
function $\la {\cal O}^{12}_J(x)\, {\cal O}^{12\, \dag}_J(0) \ra$ is
depicted in figure \ref{fig:bmnint}. The interaction (\ref{V})
connects the term in ${\cal O}(x)$ with that of ${\cal O}^\dag(0)$ in
which the impurity $\Phi$ is moved along by one $Z$ field.
This gives the leading contribution to the anomalous dimension as follows

\be\label{anomdim}
\la {\cal O}^{12}_J(x)\, {\cal O}^{12\, \dag}_J(0) \ra
\sim \frac{1}{(x^2)^{J + 2 + \g}}, \qquad \g = n^2 \l' + \ldots 
\ee

\ni thus reproducing the leading term in the expansion of
(\ref{ppstrspecgauge}) for $N_n = 2$, i.e. $\D-J = 2 + n^2 \l'$.

\begin{figure}[ht]
\begin{center}
\includegraphics*[bb=0 0 175 130, height=1.85in]{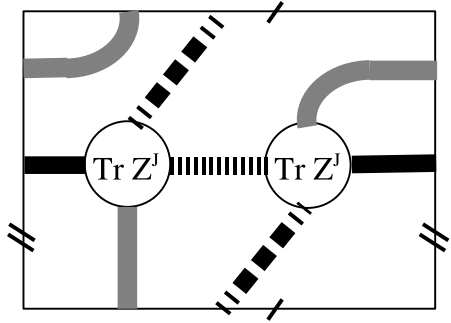}
\end{center}
\caption{The torus diagram for the two-point function of the operator
  (\ref{vacop}) in the free-field limit is given by separating the $J$
  propagators into four groups as shown above. The group depicted by
  the grey line may be removed, leaving a contribution from separation
  into three groups. These diagrams may not be drawn on the sphere
  without crossing lines. The $J$ fields of each operator have been
  arranged in a circle to reflect the cyclicity of the trace.}
\label{fig:bmntorus}
\end{figure}

\begin{figure}[ht]
\begin{center}
\includegraphics*[bb=0 0 175 130, height=1.85in]{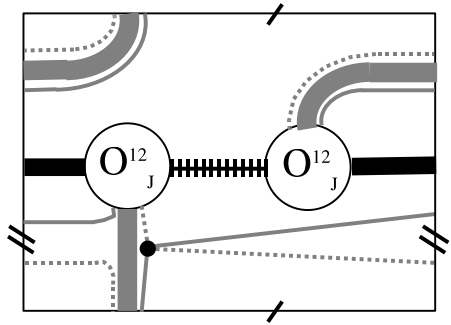}
\end{center}
\caption{A leading $g_2$ contribution to the anomalous dimension of
  the operator (\ref{bmn2imp}) is given by the diagram pictured above.
  The dotted grey line represents an impurity field which interacts
  via the quartic vertex with a $Z$ field. The second impurity sits in
  the central group (indicated by the dashed black line).}
\label{fig:bmntorusint}
\end{figure}

As was discussed in section \ref{sec:largeN}, string loop diagrams are
reflected in the dual gauge theory by non-planar diagrams, which are
suppressed by powers of $1/N$ compared to the planar diagrams. In
fact in the BMN limit, the string-loop counting parameter turns out to
be $J^2/N \equiv g_2$. In order to see this one must consider the
non-planar contributions to the two-point function of BMN
operators. Already in the free-field limit, when $\l' = 0$, the
coupling $g_2$ emerges readily. Consider the genus-1 contribution to
the two-point function of ${\cal O}_J$ given in (\ref{vacop}). There
are two classes of diagrams which can be drawn on a torus without
crossing lines, which cannot be drawn on a sphere. They correspond to
splitting the $J$ propagators into either 3 or 4 groups as shown in
figure \ref{fig:bmntorus}.
\ni There are therefore ${\tiny \begin{pmatrix} J \cr 3 \end{pmatrix}}
+ {\tiny \begin{pmatrix} J \cr 4 \end{pmatrix}} \simeq J^4/4!$ ways of
contracting the fields, and each is suppressed by $1/N^2$ compared to
its planar counterpart. Therefore the quantity $J^4/N^2 = g_2^2$
emerges naturally. When the coupling $\l'$ is turned on, torus (and
higher genus) diagrams will contribute to the process shown in figure
\ref{fig:bmnint}, for example. This leads to $g_2$ terms in the
anomalous dimension of the BMN operators. Figure \ref{fig:bmntorusint}
shows one of the torus diagrams which contributes to the leading $g_2$
contribution to $\g$ in (\ref{anomdim}).
We will not delve any further into the details of these gauge theory
calculations, and refer the reader to the references \cite{Kristjansen:2002bb,
Constable:2002hw, Bianchi:2002rw, Beisert:2002bb, Constable:2002vq,
Beisert:2002ff, Beisert:2003tq} for details. Suffice it to say that
the most important BMN operators concerning this thesis are the
two-impurity operators used as examples in this section. The state of
the art concerning the full $\l'$ and $g_2$ expansion of the anomalous
dimension may be summarized as

\be\label{gaugeresult}
\Delta-J=2+n^2\lambda'-\frac{1}{4}n^4\lambda'^2+\frac{1}{8}n^6
{\lambda'}^3...+\frac{g_2^2}{4\pi^2} \left(
\frac{1}{12}+\frac{35}{32n^2\pi^2}\right)\left(\lambda'-
\frac{1}{2}{\lambda'}^2 n^2\right) +\ldots 
\ee

\ni where the first set of dots should be understood to mean that the
entire square root in (\ref{ppstrspecgauge}) has in principle been
proven in the gauge theory \cite{Santambrogio:2002sb}.

Reproducing (\ref{gaugeresult}) from string theory involves
reproducing the $g_2$ terms via the consideration of string
interactions on the plane-wave geometry (\ref{ppmet}), in the limit
where both $\l'$ and $g_2$ are small. From our dictionary stemming
from  (\ref{p+p-}), we have that

\be\label{ppcoups}
\l' = \frac{g_{YM}^2 N}{J^2} = \frac{1}{(\m \a' p^+)^2}, 
\qquad g_2 = \frac{J^2}{N} = 4 \pi g_s (\m \a' p^+)^2
\ee

\ni we therefore would like to take the large $\m$, small $g_s$ limit
of the string theory, and calculate the one-loop shift to the energy
of a string excited by two oscillators. In order to achieve this, we
must have the machinery to calculate general string interactions at
our disposal. Such a machinery has been developed in the literature,
and the author has made some important contributions to it. In the
next section an overview of the state of the art prior to the author's
work will be given.

Before we do this it is important to note that the anomalous dimension
in (\ref{gaugeresult}) also receives non-perturbative corrections
known as {\it instanton}\footnote{For a review of instantons, see
  \cite{Teper:1980zs}.}  corrections. These corrections correspond to
similar non-perturbative processes in the dual string theory, arising
as contributions of D-branes with point-like worldvolume, the
so-called {\it D-instantons} or D(-1)-branes, see
\cite{Gaberdiel:2002hh}. In a series of important papers
\cite{Green:2005rh,Green:2005pg,Green:2005md,Green:2005ib} Green,
Sinha, and Kovacs calculated these contributions in both the gauge and
string theory finding remarkable agreement between the two.

% ************************************************************************** %
\section{Light-cone string field theory on the plane-wave: Introduction}

At the start of this chapter (chapter \ref{chap:lcsft}), we introduced
the basic idea of light-cone string field theory. Indeed, the reader
should have in mind figure \ref{fig:vertex}. The specific map for the
plane-wave superstring was developed originally by Spradlin and
Volovich \cite{Spradlin:2002ar, Spradlin:2002rv} but was revised and
elaborated in a long and technically complicated literature
\cite{Chu:2002eu, Pankiewicz:2002gs, Dobashi:2002ar, Gomis:2002wi,
  Pankiewicz:2002tg, Chu:2002wj, He:2002zu, Roiban:2002xr,
  Gomis:2003kj, DiVecchia:2003yp, Pankiewicz:2003kj, Spradlin:2003xc,
  Gutjahr:2004dv, Dobashi:2004ka, Dobashi:2006fu}. This work took its
cue directly from the work of the flat-space light-cone string field
theory for type IIB superstrings developed in the 1980's by Green,
Schwarz, and Brink \cite{Green:1983hw, Green:1984fu}, and elaborated
in a series of subsequent papers \cite{Greensite:1986gv,
  Greensite:1987sm, Green:1987qu, Greensite:1987hm}. Rather than trace
through the historical development, we will strive to give a
self-contained presentation of the state of the art, in order to lay
the ground to introduce the author's contributions in subsequent
sections. We begin with the free string, and then introduce
interactions.

% ========================================================================== %
\subsection{The free string on the plane-wave background}
\label{sec:freeppstr}

\begin{figure}[ht]
\begin{center}
\includegraphics*[bb=0 0 445 160, height=1.85in]{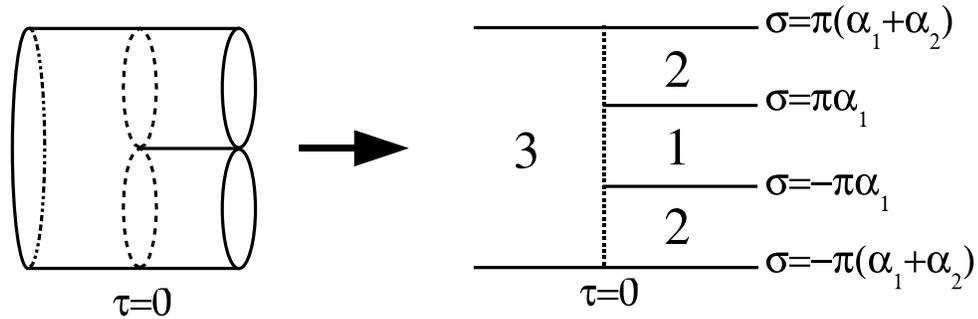}
\end{center}
\caption{The interaction between three closed superstrings is given by
the ``pair of pants'' diagram on the left. Cutting this diagram and
unfolding it (as shown on the right) reveals a convenient
parametrization of the worldsheet spatial coordinate $\s$. The lines
at $\s = \pm \pi (\a_1+\a_2)$ are identified, as are those at $\s =
\pm \pi \a_1$. See text for further explanation.}
\label{fig:strcoords}
\end{figure}

We begin by adopting an unusual convention for the coordinate length
around a closed string. Figure \ref{fig:strcoords} depicts the
three-string interaction which we are interested in. On the right, this
``pair of pants'' diagram has been cut and un-folded to reveal the
desired parametrization of the three strings; here $\a_r \equiv \a'
p^+_r$, where $r=1,2,3$ labels the string in question. The convention
chosen here is that $\s \in [-\pi|\a_r|, \pi |\a_r|]$ for each string. The
convention is to further set $\a_3 <0$, while $\a_1,\a_2>0$ such that
$\a_1+\a_2+\a_3=0$ as required by conservation of $p^+$. This
convention obviously ensures a constant $p^+$-density over the string
worldsheet, which turns out to be convenient.
\ni The light-cone Green-Schwarz action for the type IIB strings in
the pp-wave background is then given by \cite{Metsaev:2001bj, Metsaev:2002re}

\begin{eqnarray}\label{ppstrac}
S = \frac{e(\alpha)}{4\pi\alpha'}\int d\tau
\int_0^{2\pi\vert\alpha\vert} d\sigma\left( {\partial_\tau
X}^I{\partial_\tau X}^I-{\partial_\sigma X^I}{\partial_\sigma
X^I}-\mu^2X^IX^I\right)+\nonumber \\ +\frac{1}{8\pi}\int
d\tau\int_0^{2\pi|\alpha|}d\sigma\left(i\bar\vartheta\partial_\tau\vartheta+
i\vartheta\partial_\tau\bar\vartheta-\vartheta\partial_\sigma\vartheta
+\bar\vartheta\partial_\sigma\bar\vartheta-2\mu\bar\vartheta\Pi\vartheta\right)
\end{eqnarray}

\ni where $I=1,...,8$, $e(\alpha)={\rm sign}(\alpha)$, $\alpha =
\alpha'p^+$, $\theta$ is an 8-component, complex, positive chirality
spinor of SO(8), and $\Pi=\g^1\g^2\g^3\g^4$ is a symmetric, traceless
projection operator, $\Pi^2=1$. Here $\g^I$ are the $SO(8)$ Weyl
matrices.\footnote{ The $SO(8)$ gamma-matrices are
$\Gamma^I=\left(\begin{pmatrix} 0 & \gamma^I \cr \bar \gamma^I & 0
\cr\end{pmatrix}\right)$.} Notice that the two real spinors analogous
to $\theta^1$ and $\theta^2$ in the flat-space case
(\ref{flatspacespinors}) have been combined into a complex spinor in
which $\vartheta = (\theta^1 + i \theta^2)$ while $\bar\vartheta
= i(\theta^1 - i \theta^2)$. The equations of motion resulting
from the action (\ref{ppstrac}) are as follows

\bsp\label{ppwaveeom}
\left( \p_\t^2 - \p_\s^2 \right) X^I &+ \m^2 X^I = 0\\
i\p_\t \bar\vartheta  - \p_\s \vartheta + \m \Pi \bar\vartheta = 0,
&\qquad 
i\p_\t \vartheta  + \p_\s \bar\vartheta - \m \Pi \vartheta = 0.
\end{split}
\ee

\begin{figure}[ht]
\begin{center}
\includegraphics*[bb=0 0 485 305, height=1.85in]{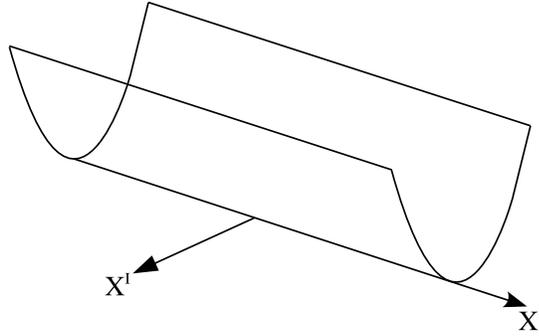}
\end{center}
\caption{The pp-wave geometry (\ref{ppmet}) as viewed by a particle or
  string. In the transverse directions the string sees a potential
  $\sim \m^2 x^{I^2}$. In this way only massive (i.e. excited) strings
  venture appreciably off the light-cone direction $x^-$, which is a
  flat direction in the geometry.  }
\label{fig:trough}
\end{figure}

\ni The equations (\ref{ppwaveeom}) show that the eight transverse
directions form a parabolic ``trough'' lending a mass $\sim \m^2$ to
the fields $X^I, \vartheta$, see figure \ref{fig:trough}. In this
sense massless particles (and strings) race down the light-cone
direction $x^-$ (at the bottom of the trough) at the speed of
light. In order for a string to (substantially) visit the transverse
directions, it requires an excitation on the order of at least $\m$
(in energy)\footnote{The existence of asymptotically free states in
  such a potential is by no means guaranteed. Indeed, since only the
  $x^+$ and $x^-$ directions are flat, we can only hope to separate
  wave-packets in this plane. In an important paper by Bak and
  Sheikh-Jabbari \cite{Bak:2002ku}, convincing evidence was given for
  the existence of a 1+1-dimensional S-matrix for massive modes on the
  plane-wave background. The massless (i.e. supergravity) modes
  present a curious riddle because they all propagate at the same
  speed down the trough - they can never catch-up to each other to
  scatter. This fact appears related to the gauge theory observation
  that multi-trace operators corresponding to states containing
  multiple supergravity modes are severly degenerate; for a recent
  discussion see \cite{Jevicki:2005ms,Jevicki:2006tr}. The issue of
  existence of an S-matrix is important for attempts to relate three
  and four-point functions of BMN operators to their string theory
  counterparts as in \cite{Dobashi:2004ka} and
  \cite{Dobashi:2006fu}. The author thanks Mohammad M. Sheikh-Jabbari
  for pointing this out to him.}. Thus massive strings extend into the
transverse directions $x^I$. The free, massive, 2-d Klein-Gordon
equation (\ref{ppwaveeom}), supplemented by the closed string boundary
conditions $X^I(\t,\s+2\pi\a) = X^I(\t,\s)$ may be easily solved via
the ans\"{a}tz
\be
X^I = \sum_n \left( x_n e^{i\left(\frac{n \s}{\a} + \e_n \t\right)} +
\wt x_n e^{i\left(\frac{n \s}{\a} - \e_n \t\right)} \right)
\ee 

\ni which yields, upon application of (\ref{ppwaveeom})

\be
- \e_n^2 + \frac{n^2}{\a^2} + \m^2 =0.
\ee

\ni To simplify things we define $\o_n \equiv \sqrt{n^2 + (\m \a)^2}$
so that $\e_n = \o_n/|\a|$. The solution for the fermionic field
proceeds in a similar way; at the end of the day we may express the
mode expansions for the fields (and their conjugate momenta $P^I$ and
$\l$) at $\t = 0$ (i.e. where the interaction will be taking place) in
the following convenient form

\begin{eqnarray}
&&X^I(\sigma) = x_0^I+\sqrt{2}\sum_{n=1}^\infty \left(
x_n^I\cos\frac{n\sigma}{|\alpha|} + x_{-n}^I
\sin\frac{n\sigma}{|\alpha|}\right) \\
&&P^I(\sigma) =
\frac{1}{2\pi|\alpha|}\left[p_0^I+\sqrt{2}\sum_{n=1}^\infty \left(
p_n^I\cos\frac{n\sigma}{|\alpha|} + p_{-n}^I\label{ppmodexp}
\sin\frac{n\sigma}{|\alpha|}\right)\right]\\
&&\vartheta^a(\sigma) = \vartheta_0^a+\sqrt{2}\sum_{n=1}^\infty \left(
\vartheta_n^a\cos\frac{n\sigma}{|\alpha|} + \vartheta_{-n}^a
\sin\frac{n\sigma}{|\alpha|}\right) \\
&&\lambda^a(\sigma) =\frac
{1}{2\pi|\alpha|}\left[ \lambda^a_0+\sqrt{2}\sum_{n=1}^\infty \left(
\lambda_n^a\cos\frac{n\sigma}{|\alpha|} + \lambda_{-n}^a
\sin\frac{n\sigma}{|\alpha|}\right)\right]
\end{eqnarray}

\ni where $2 \lambda_n^a = |\alpha|\bar\vartheta_n^a$ and $a$ is an
$SO(8)$ spinor index. Quantization then proceeds in a straightforward manner;
the non-vanishing (anti-)commutators of the Fourier modes are

\begin{equation}
\left[x_m^I,p_n^J\right]=i\delta^{IJ}\delta_{mn} ~~,~~
\left\{\vartheta^a_m,
\lambda^b_n\right\}=\delta^{ab}\delta_{mn}\end{equation}  and ensure
that \begin{equation} \left[
X^I(\sigma),P^J(\sigma')\right]=i\delta^{IJ}\delta(\sigma-\sigma')
~~,~~ \left\{
\vartheta^a(\sigma),\lambda^b(\sigma')\right\}=\delta^{ab}\delta(\sigma
- \sigma').
\end{equation}

\ni The modes can then be written in terms of oscillators

\begin{equation} x_n^I = i\sqrt{\frac{\alpha'}{2\omega_n}}\left(
a^I_n - a^{I\dagger}_n\right) ~~,~~ p_n^I =
\sqrt{\frac{\o_n}{2\a'}}\left(a^I_n +
a^{I\dagger}_n\right) ~~,~~ \left[a^I_m,
a^{J\dagger}_n \right] = \delta^{IJ}\delta_{mn}
\end{equation}
\begin{eqnarray}
&&\vartheta^a_n = \frac{c_n}{\sqrt{\vert\alpha\vert}}\left[
\left(1+\rho_n\Pi\right)b^a_n+e(n\alpha)\left(
1-\rho_n\Pi\right)b_{-n}^{a\dagger}\right]\label{fnm}\\
&&\lambda^a_n =
\frac{\sqrt{\vert\alpha\vert}c_n}{2}\left[\left(
1+\rho_n\Pi\right)b_{n}^{a\dagger}+e(n\alpha)
\left(1-\rho_n\Pi\right)b^a_{-n}\right]\label{fnmm}\\
&&\left\{ b_m^a,b_n^{b\dagger}\right\}=\delta^{ab}\delta_{mn}\label{bs}
\end{eqnarray}

\ni where 

\be
\rho_n=\frac{\omega_n - |n|}{\mu\alpha},\qquad
c_n=\frac{1}{\sqrt{1+\rho_n^2}}.
\ee

\ni This rather bizarre transformation of variables for the fermionic
oscillators was introduced in \cite{Spradlin:2002ar} in order that the
Hamiltonian appear in the canonical form (\ref{h2ab}). The free string
Hamiltonian for the $r$-th string is

\begin{eqnarray}
 &&\hspace{-1cm}H_2^{(r)}=
\frac{e(\a)}{2}\int_0^{2\pi|\a_r|}d\sigma\left[
2\pi\alpha' P^{(r)2}+\frac{1}{2\pi\alpha'}(\partial_\sigma X^{(r)})^2
+\frac{1}{2\pi\alpha'}\mu^2X^{(r)2})\right]\\
&&+\frac{1}{2}\int_{0}^{2\pi|\a_r|}d\sigma\left[
-2\pi \a' \lambda^{(r)}\partial_\sigma \lambda^{(r)}+
\frac{1}{2\pi\a'}\theta^{(r)}\partial_\sigma\theta^{(r)}
+2\mu\lambda^{(r)}\Pi\theta^{(r)}\right]\nonumber
\end{eqnarray}

\ni and in this Fock space basis reduces to

\begin{eqnarray}
 H^{(r)}_2=\sum_{n=-\infty}^{\infty}
\frac{\omega_n^{(r)}}{\a_r}
\left(
a_n^{(r)I\dagger}a_n^{(r)I}+b_n^{(r)a\dagger}b_{n}^{(r)a}
\right).
\label{h2ab}
\end{eqnarray}

\ni The states of the (single) string are then built upon its vacuum
$|0;\a\ra$ (recall that $\a = \a'p^+$) which is annihilated by
the lowering operators

\be\label{PANKvacc}
a_n |0;\a\ra = b_n |0;\a\ra = 0, \qquad \forall n.
\ee

\ni Physical string states $|\Psi\ra$ are built by acting the
$a^\dag_n$ and $b^{\dag}_n$ upon this vacuum, subject to the
level-matching condition (\ref{levelmatch})

\be
\sum_n n \left( \d_{IJ} a^{I\dag}_{-n} a_n^J + \d_{ab} b^{a\dag}_{-n}
b^b_n \right) |\Psi \ra = 0.
\ee 

Because the basis (\ref{fnm}) breaks the $SO(8)$ symmetry of the
plane-wave to $SO(4)\times SO(4) \times \bZ_2$ , it will be easier to
introduce a new basis for the $\g$-matrices in which \cite{Pankiewicz:2003kj}

\be
\Pi = \begin{pmatrix}
\d^{\b_1}_{\a_1} \d^{\b_2}_{\a_2} & 0 \cr
0 & -\d_{\db_1}^{\da_1} \d_{\db_2}^{\da_2}
\end{pmatrix}
\ee

\ni where we label representations of $SO(4)_1\times SO(4)_2$ through
$(SU(2)\times SU(2))_1\times (SU(2)\times SU(2))_2$ spinor indices.
With this decomposition of the R-charge index, the fermionic fields
$\vartheta^a$ and $\lambda^a$, are expressed in terms of creation
operators $b^{\dagger}_{\a_1\a_2}$ and
$b^{\dagger}_{\dot\a_1\dot\a_2}$ which transform in the
$(1/2,0,1/2,0)$ and $(0,1/2,0,1/2)$ representations of 
$(SU(2)\times SU(2))_1\times (SU(2)\times SU(2))_2$, respectively;
$\alpha_k$,$\da_k$ being two-component Weyl indices of $SO(4)_k$, see
appendix \ref{app:fermrep} for details. The Hamiltonian (\ref{h2ab})
in this basis is then

\be\label{BMNH2ab}
 H^{(r)}_2=\sum_{n=-\infty}^{\infty}
\frac{\omega_n^{(r)}}{\a_r}
\left(
a_n^{(r)I\dagger}a_n^{(r)I}+b_{n(r)}^{\a_1\a_2\dagger}b_{n(r)\,\a_1\a_2}
+b_{n(r)}^{\da_1\da_2\dagger}b_{n(r)\,\da_1\da_2}
\right)
\ee

\ni while the commutation relations are given by

\be\label{dotbasis}
\left\{ b_{n(r)\,\a_1\a_2}, b_{m(s)}^{\b_1\b_2\dagger} \right\}
= \d^{\b_1}_{\a_1} \d^{\b_2}_{\a_2} \d_{mn} \d_{rs}, \qquad
\left\{ b_{n(r)\,\da_1\da_2}, b_{m(s)}^{\db_1\db_2\dagger} \right\}
= \d^{\db_1}_{\da_1} \d^{\db_2}_{\da_2} \d_{mn} \d_{rs}.
\ee

% ========================================================================== %
\subsection{Local and non-local isometries}
\label{sec:dynkingen}

The super-Poicar\'{e} group of isometries respected by the plane-wave
involve the following generators

\bsp\label{ppgens}
P^+, P^I, &J^{+I}, J^{ij}, J^{i' j'}, Q^{+}, \bar Q^{+}\\
&H, Q^-, \bar Q^-
\end{split}
\ee

\ni where $i,j=1,\ldots,4$ and $i',j'=5,\ldots,8$. These are the same
objects we saw in section \ref{sec:susydet} where $J^{IJ}$ here
corresponds to $M^{IJ}$ used there. The commutation relations which
are different from those of flat space are given by

\bsp\label{ppalg}
[H,P^I] = &i\m^2 J^{+I}, \qquad [P^I, Q^-] = \m \Pi \g^I Q^+, \qquad
[H,Q^+] = \m \Pi Q^+,\\
&\left\{ Q^-, \bar Q^- \right\} = 2 H + i \mu \g_{ij} \Pi J^{ij} + i\m
  \g_{i',j'} \Pi J^{i' j'}.
\end{split}
\ee

\ni The generators (\ref{ppgens}) fall into two fundamentally
different categories; those on the top line and those on the bottom
line. The generators on the top line are {\it local}, or {\it
  kinematical} generators. This means that they act at a point on the
string world sheet. Consequently, they are not capable of splitting or
joining strings as this is a clearly non-local operation (see figure
\ref{fig:strcoords}). Conversely, the Hamiltonian $H$ and the
supercharges $Q^-$ and $\bar Q^-$ are non-local. The Hamiltonian
involves naturally a derivative of the light-cone time $x^+$, while
the supercharges involve $\s$ derivatives. These non-local or {\it
  dynamical} generators do induce string interactions and therefore
receive corrections from those interactions. The kinematical
generators remain unchanged by string interactions, and this allows us
to largely forget about them. 

The free string Hamiltonian was given in (\ref{h2ab}). The free string
supercharges are given formally by

\bea
\label{Q+}
&&Q^+_{(r)} = \sqrt{\frac{2}{\a'}}\int_0^{2\pi|\a_r|}d\s_r\,\sqrt{2}\l_r\,,\\
\label{qfield}
&&Q^-_{(r)}
=\sqrt{\frac{2}{\a'}}\int_0^{2\pi|\a_r|}d\s_r\,
\left[2\pi\a'e(\a_r)p_r\g\l_r-ix'_r\g\bar{\l}_r-i\m
x_r\g\Pi\l_r\right]\, 
\eea

\ni where
$\bar{Q}^{\pm}_{(r)}=e(\a_r)\bigl[Q_{(r)}^{\pm}\bigr]^{\dag}$.
Plugging the mode expansions of the fields into the expression for
$Q^-$, we obtain 

\bea\label{q-mode} Q^-_{(r)}&
=&\frac{e(\a_r)}{\sqrt{|\a_r|}}\g
\Bigl(\sqrt{\m}\left[a_{0(r)}(1+e(\a_r)\Pi)+a_{0(r)}^{\dag}(1-e(\a_r)\Pi)\right]\l_{0(r)}\cr
&+&\sum_{n\neq 0}\sqrt{|n|}\left[a_{n(r)}P_{n(r)}^{-1}b_{n(r)}^{\dag}
+e(\a_r)e(n)a_{n(r)}^{\dag}P_{n(r)}b_{-n(r)}\right]\Bigr)\,, \eea

\ni where

\begin{equation}
P_{n(r)}\equiv\frac{1-\r_{n(r)}\Pi}{\sqrt{1-\r_{n(r)}^2}}
=\frac{1+\Pi}{2}U_{|n|(r)}^{1/2}+\frac{1-\Pi}{2}U_{|n|(r)}^{-1/2}\,,\qquad
U_{n(r)}\equiv\frac{\o_{n(r)}-\m\a_r}{n}\,.
\end{equation}

\ni As advertised, $Q^-$ and $H$ will be corrected beyond the free
string result by string interactions. In the next section we will
trace the construction of these corrections. 

% ========================================================================== %
\subsection{The string field and determination of the interaction vertices}
\label{sec:detintvert}

Light-cone string field theory, as its name implies, deals with a {\it
  string field} - a field which strings are elementary excitations
  of. Consider a bosonic string, it has an infinite number of modes,
  here labelled by $k$, and each is a harmonic oscillator with
  creation operator $a^\dag_k$. In order to specify a completely
  general string state, one must specify the occupation numbers
  $\{n_k\}$, giving the level of excitation (i.e. the number of
  $a^\dag_k$'s) for each mode. There is also the further multiplicity
  of the spacetime dimensions, which we will here label by $i$, and so
  we should in fact say that a string will be entirely specified by
  the set $\{{n_k}_i\}$. Instead of relying on this {\it number basis}
  to define the string field, we would rather like to describe it as a
  momentum distribution. Indeed, the mode expansion of the conjugate
  momentum $P(\s)$ involves the set $\{{p_k}_i\}$ of Fourier
  coefficients, c.f. (\ref{ppmodexp}). We can then define a string
  field $\Phi[P(\s)]$, which acting on the vacuum creates a string
  (at $\t=0$) defined by it's conjugate momentum $P(\s)$. The
  expression may be given as

\be\label{strfield}
\Phi[P(\s)] = \sum_{\{{n_k}_i\}} \varphi_{\{{n_k}_i\}} 
\prod_{k=1}^\infty \prod_{i} \psi_{{n_k}_i} ({p_k}_i)
\ee

\ni where $\varphi_{\{{n_k}_i\}}$ creates the number basis state given
by $\{{n_k}_i\}$

\be
\varphi_{\{{n_k}_i\}}  = \prod_{k=1}^\infty \prod_{i} \bigl( a^{i
  \dag}_k \bigr)^{{n_k}_i} 
\ee

\ni while $\psi_{{n_k}_i} ({p_k}_i)$ is the momentum space
wavefunction of the $k_i^{\text{th}}$ simple harmonic oscillator in
the ${n_k}_i^{\text{th}}$ excited state

\be\label{oscwf}
\psi_n(p) = \la n | p \ra, \qquad |p\ra \sim \exp \left( -\frac{1}{4}
p^2 + p \,a^\dag - \frac{1}{2} a^\dag a^\dag \right) |0 \ra.
\ee

\ni Finally the sum in (\ref{strfield}) is over all possible
combinations of occupation numbers - i.e. all physical states of the
string. The string field operator $\Phi[P(\s)]$ has the property that

\be
\wh P(\s) \Phi[P(\s)] |0\ra = P(\s) \Phi[P(\s)] |0\ra
\ee

\ni that is, $\Phi[P(\s)] |0\ra$ is an eigenstate of the total
momentum operator $\wh P(\s)$. This construction may be easily
generalized to include fermionic modes created by $b^{a \dag}_k$.
Rather than explicitly tracing the fermionic construction here, we
will continue with the bosonic modes and construct the first
correction to the Hamiltonian. 

As was mentioned in section \ref{sec:dynkingen}, the Hamiltonian is a
dynamical generator which we expect to be corrected by string
interactions. Let $\k$ be the coupling constant controlling string
interactions, we expect that

\be
H = H_2 + \k H_3 + \k^2 H_4 + \ldots
\ee

\ni where $H_n$ is an operator which maps one string to $n-1$ strings;
i.e. an operator involving a product of $n$ string fields $\Phi$. The
quadratic Hamiltonian $H_2$ is simply the free string one given in
(\ref{h2ab}). In constructing $H_3$, we will be guided by
symmetry. The plane-wave background is not translationally invariant
in the eight transverse directions, as is plain from figure
\ref{fig:trough}, or from the first commutation relation in
(\ref{ppalg}). However, somewhat miraculously, the commutator of $H$
and $P^I$ is proportional to a {\it kinematical} generator. This means
that although $[H_2,P^I] \neq 0$, we have that $[H_{n>2}, P^I]=0$ as
it is in the flat space case. From the point of view of the
interaction Hamiltonian, the transverse momentum is conserved. A
natural ans\"{a}tz for the cubic Hamiltonian is then

\bsp\label{h3form}
H_3 &= \int d{\cal M}\,h_3\,
\Phi[P_1(\s)] \,\Phi[P_2(\s)] \,\Phi[P_3(\s)],\\
d{\cal M} &= \left( \prod_{r=1}^3 d \a_r \, D P_r(\s)\right)\, 
\d\left(\sum \a_r\right) \d\left( \sum P_r(\s) \right) 
\end{split}
\ee

\ni where $h_3$ is called the ``prefactor'' and is an as yet
undetermined function. The object is now to reduce (\ref{h3form}) into
an expression involving only the oscillators of the strings. The first
step in this direction is accomplished by representing the delta
function enforcing conservation of transverse momentum $P^I$ in the
Fourier basis of the third string

\be
\D[f(\s)] = \prod_m \d \left( \int_0^{2\pi |\a_3|} e^{im\s/|\a_3|}
f(\s) \, d\s \right)
\ee

\ni so that, for example

\bsp \D[P_2(\s)] = \prod_{m=-\infty}^\infty &\d \Biggl( \int_0^{2\pi
|\a_3|} d\s\,e^{im\s/|\a_3|} \frac{1}{2\pi|\alpha_2|}\\
&\times\left[p_0^{(2)I}+\sqrt{2}\sum_{n=1}^\infty \left(
p_n^{(2)I}\cos\frac{n\sigma}{|\alpha_2|} + p_{-n}^{(2)I}
\sin\frac{n\sigma}{|\alpha_2|}\right)\right] \Theta\left( |\s| - \pi
\a_1\right) \Biggr)
\end{split}
\ee

\ni where the Heaviside function $\Theta\left( |\s| - \pi \a_1\right)$
enforces the limits of the second string's worldsheet (see figure
\ref{fig:strcoords}). Performing the integral over $\s$, and
similarly for strings 1 and 3, we arrive at

\be\label{deltafour3}
\d\left( \sum P_r(\s) \right) =
\D \left[ \sum_{r=1}^3 P_r \right] = \prod_{I=1}^8 \prod_{m=-\infty}^\infty 
\d \left( p^{(3)I}_m + \sum_{n=-\infty}^\infty \left( X_{mn}^{(1)}\,
p^{(1)I}_n + X_{mn}^{(2)}\,p^{(2)I}_n \right) \right)
\ee

\ni where the matrices $X^{(r)}_{mn}$ perform the transformation
  between the Fourier basis of the third string and those of strings 1
  and 2, so that $X^{(3)}_{mn} = {\bf 1}$. Next we turn to the product
  of three string fields in (\ref{h3form}). In fact the definition of
  the string field (\ref{strfield}) is slightly redundant. Consider
  the simple one-dimensional simple harmonic oscillator. The following
  sum has a simple form 

\be
\sum_n |n\ra \psi_n(p) = \sum_n |n\ra\la n|p\ra = |p\ra
\ee

\ni in fact (\ref{strfield}) is nothing but a generalization of this
same form. This allows us to write the product of string fields as

\bsp\label{pposcwf}
&\prod_r \Phi[P_r] = {\cal N} \exp\left[ \sum_{n,r,I}\left(
-\frac{1}{\o_{n(r)}} \left( p_n^{(r)I} \right)^2 +
  \frac{2}{\sqrt{\o_{n(r)}}} p_n^{(r)I} a_n^{(r)I\dag} - \frac{1}{2}
  a_n^{(r)I\dag} a_n^{(r)I\dag} \right) \right], \\
&\text{with} \qquad {\cal N} = \prod_{n,r,I} \left(\frac{2}{\pi \o_{n(r)}} \right)^{1/4}
\end{split}
\ee

\ni where we have used the correctly normalized wavefunctions
appropriate to the string on the plane-wave, i.e. correct version of
(\ref{oscwf}). Postponing a discussion of the prefactor $h_3$ until
later, we can now proceed with the Gaussian integration resulting from
plugging (\ref{pposcwf}) and (\ref{deltafour3}) into
(\ref{h3form}). The integration proceeds over the transverse momentum
modes, taking (for now) $h_3=1$, and leaving the integration over the
light-cone momentum $\a_r$ until later. The result may be summarized
as follows

\be\label{bosvert}
\int \left( \prod_{n,r,I} d p_n^{(r)I} \right) \prod_r \Phi[P_r] \,
\D \left[ \sum_{r=1}^3 P_r \right] = {\cal C} \exp \left(
\frac{1}{2} \sum_{r,s=1}^3 \sum_{m,n=-\infty}^\infty  a_m^{(r)I\dag}
 \bar N^{rs}_{mn} a_n^{(s)I\dag} \right)
\ee

\ni where ${\cal C}$ is an overall constant which won't be important
for us, while $N^{rs}_{mn}$ are known as the ``Neumann matrices'' and may
be expressed in terms of the $X^{(r)}_{mn}$ as \cite{Spradlin:2002ar}

\be
\bar N^{rs}_{mn} = \d^{rs} \d_{mn} - 2 \sqrt{\o_{m(r)} \o_{n(s)}} \left(
{X^{(r)}}^T \G^{-1} X^{(s)} \right)_{mn}, ~~
\G_{mn} = \sum_r \sum_{p=-\infty}^\infty \o_{p(r)} X^{(r)}_{mp} X^{(r)}_{np}. 
\ee

At this point it is useful to step back and summarize what we have
found. The expression (\ref{bosvert}) is the oscillator manifestation
of transverse momentum conservation. Interpreted from a spatial point
of view, the oscillator map (\ref{bosvert}) ensures that the three
strings touch at $\t=0$, the moment of the interaction. There are
however, other symmetries which the interaction Hamiltonian should
satisfy. In the same way that $P^I$ commuted with the interacting
piece of the Hamiltonian, a look at (\ref{ppalg}) reveals that
$[H_{n>2},Q^+]=0$ since the full commutator is proportional to a
kinematical generator ($Q^+$ itself). Of course the string field
(\ref{strfield}) also needs to be amended to reflect the fermionic
modes of the superstring. Including the fermionic modes and enforcing
the $[H_{n>2},Q^+]=0$ symmetry is quite literally the supersymmetric
reflection of the bosonic construction which was detailed in the
previous paragraph. The details of the construction may be gleaned
from \cite{Spradlin:2002ar, Chu:2002wj, Pankiewicz:2003kj}, the result
is that the fermionic equivalent of (\ref{bosvert}) is

\be
\exp\left( \sum_{r,s=1}^3 \sum_{m,n \geq0} \left(
b_{-m(r)}^{\a_1\a_2\,\dag} b_{n(s)\,\a_1\a_2} +
b_{-m(r)}^{\da_1\da_2\,\dag} b_{n(s)\,\da_1\da_2}\right)
\bar Q^{rs}_{mn} \right)
\ee

\ni where $Q^{rs}_{mn}$ is a fermionic Neumann matrix which will be
given explicitly later. The progenitor of an interaction vertex
$|V\ra$ may then be built as follows

\bsp\label{fudge}
|V\ra = &\d\left(\sum \a_r\right) \exp \left(
\frac{1}{2} \sum_{r,s=1}^3 \sum_{m,n=-\infty}^\infty  a_m^{(r)I\dag}
 \bar N^{rs}_{mn} a_n^{(s)I\dag} \right)\\
&\times \exp\left( \sum_{r,s=1}^3 \sum_{m,n \geq0} \left(
b_{-m(r)}^{\a_1\a_2\,\dag} b_{n(s)\,\a_1\a_2} +
b_{-m(r)}^{\da_1\da_2\,\dag} b_{n(s)\,\da_1\da_2}\right)
\bar Q^{rs}_{mn} \right) |0;\a_1\ra \otimes|0;\a_2\ra
\otimes|0;\a_3\ra .
\end{split}
\ee

There is a subtlety here concerning the fact that $\a_3$ is
negative. It concerns the definition of the adjoint for string
\#3. Already it should have seemed suspicious that the free Hamiltonian
(\ref{h2ab}) is not strictly positive, since $\a_3 < 0$. In fact the
adjoint on the full Hilbert space ${\cal H} = \oplus_m {\cal H}_m$,
where ${\cal H}_m$ is the $m$-string Hilbert space, is not the same as
the single string Hilbert space adjoint
\cite{Green:1982tk}\cite{Spradlin:2002ar}. Objects such as $V$ in
(\ref{fudge}) may be viewed as operators from ${\cal H}_1 \rightarrow
     {\cal H}_2$, or as states in ${\cal H}_3$

\be
\la 3| V |2\ra |1 \ra \equiv \la 1| \la 2| \la 3' | V \ra.
\ee

\ni The prime denotes the fact that the adjoint on string \#3 is
modified by a sign, so that, for example, if $| \phi^{(3)} \ra$ is
some state built on $|0;\a_3\ra$ so that $| \l^{(3)}\ra = A^{(3)} |
\phi^{(3)} \ra$ where $A^{(3)}$ is some one-string operator
(i.e. from  ${\cal H}_1 \rightarrow {\cal H}_1$), then

\be\label{subtelty}
\la \l^{(3)} | |2 \ra| 1\ra \equiv \la \phi^{(3)}| (-A^{(3)\dag}) |
2 \ra | 1 \ra 
\ee

\ni whereas this sign is absent for strings \#1 and \#2, as $\a_1$
and $\a_2$ are taken positive. This ensures the positivity of the free
string energy

\be
\la H_2 \ra = \la 3 | H_2 |2 \ra |1\ra = \sum_{r=1}^3 e(\a_r)
H_2^{(r)} > 0.
\ee

\ni Below we will construct the cubic Hamiltonian and supercharges as
states in ${\cal H}_3$. The subtlety (\ref{subtelty}) will only arise
when considering operators in string \#3's Hilbert space ${\cal H}_1^{(3)}$.

The vertex (\ref{fudge}) respects the super-locality symmetry, but
there is one last symmetry which we have yet to enforce. That is the
commutation relations between the $Q$'s given on the second line of
(\ref{ppalg}). We see from (\ref{ppalg}) that $Q^-$ and $\bar Q^-$,
like $H$, commute with $P^I$ and $Q^+$ to give kinematical
generators. Therefore we can build $Q^-_3$ and $\bar Q^-_3$ using the
progenitor vertex $|V\ra$ as well. Recall that we included a prefactor
$h_3$ in our definition (\ref{h3form}), which we then set to 1 and
forgot about. The idea is to now restore this prefactor (and similar
$q_3^-$ and $\bar q_3^-$ for the supercharges) via an operator acting
on $|V\ra$

\be\label{stat}
|H_3\ra = h_3 |V\ra, \qquad |Q_3^-\ra = q_3^- |V\ra, \qquad 
|\bar Q_3^-\ra = \bar q_3^- |V\ra
\ee  

\ni and then to determine the specific form of the prefactors by
ensuring the closure of the supersymmetry algebra (\ref{ppalg}). This
process is simplified in the (\ref{dotbasis}) basis for the
fermions. There, as shown in appendix \ref{app:fermrep}, linear
combinations of $Q_3^-$ and $\bar Q_3^-$ may be taken so that 

\be\label{commutHQ}
\left\{Q_{\a_1\da_2},Q_{\b_1\db_2}\right\}=
-2\epsilon_{\a_1\b_1}\epsilon_{\da_2\db_2}H
\ee

\ni i.e. we can factor away the dependence on $J^{ij}$ and  $J^{i'j'}$
in (\ref{ppalg}). At first order in $\k$, we will have schematically
$\left\{Q_2,Q_3\right\} \sim H_3$. Written using the state language, as in
(\ref{stat}), we have

\bea
\label{1}
&&\sum_{r=1}^3 Q_{(r)\,\a_1\da_2}|Q_{3\,\b_1\db_2}\ra+\sum_{r=1}^3
Q_{(r)\,\b_1\db_2}|Q_{3\,\a_1\da_2}\ra =
-2\e_{\a_1\b_1}\e_{\da_2\db_2}|H_3\ra\,,\\
\label{2}
&&\sum_{r=1}^3Q_{(r)\,\da_1\a_2}|Q_{3\,\db_1\b_2}\ra+\sum_{r=1}^3
Q_{(r)\,\db_1\b_2}|Q_{3\,\da_1\a_2}\ra =
-2\e_{\da_1\db_1}\e_{\a_2\b_2}|H_3\ra\,,\\
\label{3}
&&\sum_{r=1}^3Q_{(r)\,\a_1\da_2}|Q_{3\,\db_1\b_2}\ra+\sum_{r=1}^3
Q_{(r)\,\db_1\b_2}|Q_{3\,\a_1\da_2}\ra = 0\,
\eea

\ni where $Q_{(r)\,\b_1\db_2}$ and $Q_{(r)\,\db_1\b_2}$ are the
quadratic, free string supercharges $Q_2$. Finding a solution for the
prefactors which obeys these relations is a non-trivial (and
non-unique) undertaking. We will not step the reader through this
process, and instead refer to the literature \cite{Pankiewicz:2003kj},
where the following result is obtained

\bsp\label{H3andQ3full}
&|H_3\ra =
g_2\,f(\m\a_3\,,\,\frac{\a_1}{\a_3})\frac{\alpha'}{8\,\a_3^3}
\Bigl[\bigl(K_i\K_j-\frac{\m\k}{\a'}\d_{ij}\bigr)v^{ij}
-\bigl(K_{i'}\K_{j'}-\frac{\m\k}{\a'}\d_{i'j'}\bigr)v^{i'j'}\\
&\qquad-K^{\da_1\a_1}\K^{\da_2\a_2}s_{\a_1\a_2}(Y)s^*_{\da_1\da_2}(Z)
-\K^{\da_1\a_1}K^{\da_2\a_2}s^*_{\a_1\a_2}(Y)s_{\da_1\da_2}(Z)\Bigr]|V\ra\,,\\
&|Q_{3\,\b_1\db_2}\ra =
 g_2\,\eta\,f(\m\a_3\,,\,
\frac{\a_1}{\a_3})\frac{1}{4\, \a_3^3}\,\sqrt{-\frac{\a'\k}{2}}
\Bigl(s_{\dg_1\db_2}(Z)t_{\b_1\g_1}(Y)\K^{\dg_1\g_1}\\
&\qquad\qquad\qquad\qquad\qquad\qquad\qquad\qquad\qquad\qquad
+ is_{\b_1\g_2}(Y)t^*_{\db_2\dg_2}(Z)\K^{\dg_2\g_2}\Bigr)|V\ra\,,\\
&|Q_{3\,\db_1\b_2}\ra = g_2\,{\bar \eta}\,f(\m\a_3\,,\,
\frac{\a_1}{\a_3})\frac{1}{4\, \a_3^3}\,\sqrt{-\frac{\a'\k}{2}}
\Bigl(s^*_{\g_1\b_2}(Y)t^*_{\db_1\dg_1}(Z)\K^{\dg_1\g_1}\\
&\qquad\qquad\qquad\qquad\qquad\qquad\qquad\qquad\qquad\qquad
+is^*_{\db_1\dg_2}(Z)t_{\b_2\g_2}(Y)\K^{\dg_2\g_2}\Bigr)|V\ra\,.
\end{split}
\ee

\ni where $\k\equiv\a_1\a_2\a_3$, $K^I$, $\wt{K}^I$ are expressions
linear in bosonic oscillators and are defined in (\ref{k}), while $Y$
and $Z$ are their fermionic counter-parts and are given in (\ref{yz}).
Also note that $\k$, the coupling constant, has been replaced by $g_2$
(\ref{ppcoups}). The string coupling must be this value according to
the AdS/CFT correspondence; it cannot be fixed by first principles,
hence it is a matter of choice to set $\k=g_2$. Further

\begin{equation}
K^{\dg_1\g_1} \equiv K^i{\s^i}^{\dg_1\g_1}\,,~~
K^{\dg_2\g_2} \equiv K^{i'}{\s^{i'}}^{\dg_2\g_2}\,,~~
\K^{\dg_1\g_1} \equiv \K^i{\s^i}^{\dg_1\g_1}\,,~~
\K^{\dg_2\g_2} \equiv \K^{i'}{\s^{i'}}^{\dg_2\g_2}\,
\end{equation} 

\ni where the $\s$-matrices are defined in appendix
\ref{app:fermrep}. We also have

\bsp
&v^{ij}  =
\d^{ij}\Bigl[1+\frac{1}{12}\bigl(Y^4+Z^4\bigr)+\frac{1}{144}Y^4Z^4\Bigr]\nonumber\\
&\qquad-\frac{i}{2}\Bigl[{Y^2}^{ij}\bigl(1+\frac{1}{12}Z^4\bigr)
-{Z^2}^{ij}\bigl(1+\frac{1}{12}Y^4\bigr)\Bigr]
+\frac{1}{4}\bigl[Y^2Z^2\bigr]^{ij}\,,\\
&v^{i'j'}  =
\d^{i'j'}\Bigl[1-\frac{1}{12}\bigl(Y^4+Z^4\bigr)+\frac{1}{144}Y^4Z^4\Bigr]\nonumber\\
&\qquad-\frac{i}{2}\Bigl[{Y^2}^{i'j'}\bigl(1-\frac{1}{12}Z^4\bigr)
-{Z^2}^{i'j'}\bigl(1-\frac{1}{12}Y^4\bigr)\Bigr]
+\frac{1}{4}\bigl[Y^2Z^2\bigr]^{i'j'}\,.
\end{split}
\ee

\ni Here we defined

\begin{equation}
{Y^2}^{ij} \equiv \s^{ij}_{\a_1\b_1}{Y^2}^{\a_1\b_1}\,,\quad
{Z^2}^{ij} \equiv \s^{ij}_{\da_1\db_1}{Z^2}^{\da_1\db_1}\,,\quad
\bigl(Y^2Z^2\bigr)^{ij} \equiv {Y^2}^{k(i}{Z^2}^{j)k}
\end{equation}

\ni and analogously for the primed indices.  We have also introduced
the following quantities quadratic and cubic in $Y$ and symmetric in
spinor indices

\begin{equation}\label{y2}
Y^2_{\a_1\b_1} \equiv Y_{\a_1\a_2}Y^{\a_2}_{\b_1}\,,\qquad
Y^2_{\a_2\b_2} \equiv Y_{\a_1\a_2}Y^{\a_1}_{\b_2}\,
\end{equation}

\begin{equation}\label{y3}
Y^3_{\a_1\b_2} \equiv Y^2_{\a_1\b_1}Y^{\b_1}_{\b_2}=-Y^2_{\b_2\a_2}Y^{\a_2}_{\a_1}\,,
\end{equation}

\ni and quartic in $Y$ and antisymmetric in spinor indices

\begin{equation}
Y^4_{\a_1\b_1} \equiv
Y^2_{\a_1\g_1}{Y^2}^{\g_1}_{\b_1}=-\frac{1}{2}\e_{\a_1\b_1}Y^4\,,\qquad
Y^4_{\a_2\b_2} \equiv
Y^2_{\a_2\g_2}{Y^2}^{\g_2}_{\b_2}=\frac{1}{2}\e_{\a_2\b_2}Y^4\,
\end{equation}

\ni where

\begin{equation}\label{y4}
Y^4 \equiv Y^2_{\a_1\b_1}{Y^2}^{\a_1\b_1}=-Y^2_{\a_2\b_2}{Y^2}^{\a_2\b_2}\,.
\end{equation}

\ni The spinorial quantities $s$ and $t$ are defined as

\begin{equation}
s(Y) \equiv Y+\frac{i}{3}Y^3\,\ ,~~~~~t(Y) \equiv \e+iY^2-\frac{1}{6}Y^4\,.
\end{equation}

\ni Analogous definitions can be given for $Z$. The normalization of
the dynamical generators is not fixed by the superalgebra at ${\cal
O}(g_2)$ and can be an arbitrary (dimensionless) function
$f(\m\a_3\,,\,\frac{\a_1}{\a_3})$ of the light-cone momenta and $\m$
due to the fact that $P^+$ is a central element of the algebra.

The definitions of the quantities $Y$, $Z$, $K$, and $\wt K$, along
with the bosonic and fermionic Neumann matrices are most easily
expressed in the so-called ``BMN basis'' for the oscillators

\bea 
&&\sqrt{2}a_n^i \equiv \a_n^i+\a_{-n}^i\,,~~~i\sqrt{2}a_{-n}^i
\equiv \a_n^i-\a_{-n}^i\,\cr &&\sqrt{2}a_n^{i'}\equiv
\a_n^{i'}+\a_{-n}^{i'}\,,~~~i\sqrt{2}a_{-n}^{i'} \equiv
\a_n^{i'}-\a_{-n}^{i'}\, ,\cr &&\sqrt{2}b_n^{\a_1\a_2}\equiv
\b_n^{\a_1\a_2}+\b_{-n}^{\a_1\a_2}\,,~~~ i\sqrt{2}b_{-n}^{\a_1\a_2}
\equiv \b_n^{\a_1\a_2}-\b_{-n}^{\a_1\a_2}\,,\cr
&&i\sqrt{2}b_n^{\da_1\da_2}\equiv -
\b_n^{\da_1\da_2}+\b_{-n}^{\da_1\da_2}\,,~~~
\sqrt{2}b_{-n}^{\da_1\da_2} \equiv
\b_n^{\da_1\da_2}+\b_{-n}^{\da_1\da_2}\,~~~~
\label{BMNbasis}
\eea

\ni for $n>0$, and

\be 
a_0^i \equiv \a_0^i\,~~~a_0^{i'} \equiv
\a_0^{i'}\,~~~b_0^{\a_1\a_2} \equiv \b_0^{\a_1\a_2}\,
~~~b_{0}^{\da_1\da_2} \equiv \b_0^{\da_1\da_2}\, 
\ee 

\ni for $n=0$. The commutation relations for the oscillators are then

\begin{equation}
[\a_m^i,\a_n^{\dag\,j}] = \d_{mn}\d^{ij}\,, \quad
\{\bigl(\b_m\bigr)_{\a_1\a_2},\bigl(\b_n^{\dag}\bigr)^{\b_1\b_2}\} =
\d_{mn}\d^{\b_1}_{\a_1}\d^{\b_2}_{\a_2}\,.
\end{equation}

\ni In order to perform the string-field theory calculations we are
interested in comparing to gauge theory, we require the large-$\m$
limit of all quantities. These were worked out in \cite{He:2002zu} and
are given in appendix \ref{app:neumann}. We find simpler expressions
for them, which are summarized in the BMN basis as

\begin{equation}
|V \rangle = |E_{\alpha}\rangle |E_{\beta}\rangle\delta(\sum_{r=1}^3 \alpha_r)
\ee

\ni where $|E_{\alpha}\rangle$ and $|E_{\beta}\rangle$ are
exponentials of bosonic and fermionic oscillators respectively

\be
|E_{\alpha}\rangle=
\exp \left( \frac{1}{2} \sum_{r,s=1}^{3} \sum_{m,n = -\infty}^{\infty}
\alpha^{\dagger K}_{m\,(s)} {\widetilde N}^{st}_{mn} \alpha^{\dagger K}_{n\,(t)} \right )
| \alpha \rangle _{123}
\label{eb}
\end{equation}

\ni and

\begin{equation}
|E_{\beta}\rangle=\exp\left(
\sum_{r,s=1}^3\sum_{m,n= -\infty}^{\infty}
\bigl(\b^{\a_1\a_2\,\dag}_{m(r)}\b^{\dag}_{n(s)\,\a_1\a_2}-
\b^{\da_1\da_2\,\dag}_{m(r)}\b^{\dag}_{n(s)\,\da_1\da_2}\bigr)
\wt{Q}^{rs}_{mn}\right)| \alpha \rangle _{123}
\label{ef}
\end{equation}

\ni where $| \alpha \rangle _{123}=|0;\alpha_1 \rangle \otimes|0;\alpha_2
\rangle\otimes|0;\alpha_3 \rangle$. We further have that

\be
\label{k}
K^I = \sum_{s=1}^3\sum_{n\in{\bZ}}K_{n(s)}\a_{n(s)}^{I\,\dagger}\,,~~~
\wt{K}^I=
\sum_{s=1}^3\sum_{n\in{\bZ}}K_{n(s)}\a_{-n(s)}^{I\,\dagger}\, \ee

\be
\label{yz}
Y^{\a_1\a_2} =
\sum_{s=1}^3\sum_{n\in{\bZ}}G_{|n|(s)}\b^{\dag\,\a_1\a_2}_{n(s)}\,,\qquad
Z^{\da_1\da_2} =
\sum_{s=1}^3\sum_{n\in{\bZ}}G_{|n|(s)}\b^{\dag\,\da_1\da_2}_{n(s)}\,,
\ee

\ni where the large-$\m$ limits of these quantities are repackaged
from the expressions found in appendix \ref{app:neumann} and are
expressed as \footnote{These expressions (\ref{N3r}-\ref{Gq}) are also
valid for $q,p=0$, except in the case of $\wt N^{rs}_{00} = - \wt
N^{rs}_{qp}|_{q,p=0}$, and in the case of $\wh Q^{3r}_{00} =
-ir^{-1}\b_r \wh Q^{3r}_{np}|_{n,p=0}$.}

\be\label{N3r} 
\wt N^{3 \,r}_{n\,q} = -\frac{\sin(n\pi r) \sqrt{\b_r}
\left(\L^+_n \L^+_q+ \L^-_n \L^-_q \right) }{2\pi\sqrt{\o_n \o_q}
\left( q-\b_r n \right)}, \qquad \wt N^{r\,s}_{q\,p} = \frac{
\sqrt{\b_r \b_s} \left( \L^+_q \L^+_p + \L^-_q \L^-_p \right) } {4\pi
\sqrt{ \o_q \o_p } \left( \b_s \o_q + \b_r \o_p \right) }, 
\ee

\be
\wh Q^{3 \,r}_{n\,q} =  \frac{i\sin(|n|\pi r) 
  \left(\o_q+\b_r \o_n\right) }{2\pi\sqrt{\o_n \o_q} \left( q-\b_r n \right)},
\qquad
\wh Q^{r\,s}_{q\,p}  =
\frac{i\left( \b_s q - \b_r p \right) }
{4\pi \sqrt{ \o_q \o_p } \left( \b_s \o_q + \b_r \o_p \right) }
\ee

\noindent where $\wh Q^{s \,r}_{n\,q} = \wt Q^{s \,r}_{n\,q} -\wt Q^{r
\,s}_{q\,n}$, $\b_r \equiv -\a_r/\a_3$ for $r=1,2$, and where we remind the
reader that $\a_3<0$ while $\a_1,\a_2>0$. Also $r \equiv \b_1$ while
$\b_2 = 1-r$. The mode number $n$ is associated with string 3, while
$p$ and $q$ are used for either string 1 or 2. We also drop
the string label on $\o_q$, $K_q$, $G_q$ etc. as it is obvious 
from the quantity given. For example $\o_q$ in $\wt N^{3r}_{nq}$
should be understood as $\o_q^{(r)}$. Continuing, we also have

\be
K_n = +\a_3 \sin(n\pi r)  \sqrt{\frac{r(1-r)}{\pi\a'}} 
\frac{\L^-_n - \L^+_n}{\sqrt{\o_n}},
\ee

\be
K_q = -\a_3  \sqrt{\frac{r(1-r)}{\pi\a'\b_r}} 
\frac{\L^+_q - \L^-_q}{2\sqrt{\o_q}},
\ee

\be\label{Gq}
G_q = \frac{1}{\sqrt{4\pi\o_q}},
\qquad G_n = -\frac{\sin(|n|\pi r)}{\sqrt{\pi\o_n}}
\ee

\noindent where

\be \L^+_q = \sqrt{\o_q - \b_r \m \a_3}, \qquad \L^-_q = e(q)
\sqrt{\o_q + \b_r \m \a_3}, 
\ee

\be \L^+_n = \sqrt{\o_n - \m \a_3}, \qquad \L^-_n = e(n) \sqrt{\o_n +
\m \a_3}.  
\ee

% ========================================================================== %
\subsection{The contact interaction}
\label{sec:contact}

Consider the commutation relation (\ref{commutHQ}) at the next order
in the coupling constant - i.e. ${\cal O}(g_2^2)$, we have

\bsp\label{ppalgg2}
&\{Q_{2\,\a_1\da_2},Q_{4\,\b_1\db_2}\}+\{Q_{4\,\a_1\da_2},Q_{2\,\b_1\db_2}\}
+ \{Q_{3\,\a_1\da_2},Q_{3\,\b_1\db_2}\}  = -2\e_{\a_1\b_1}\e_{\da_2\db_2}H_4\,.\\
%&\{\wt{Q}_{2\,\a_1\da_2},\wt{Q}_{4\,\b_1\db_2}\}+\{\wt{Q}_{4\,\a_1\da_2},
%\wt{Q}_{2\,\b_1\db_2}\}
%+ \{\wt{Q}_{3\,\a_1\da_2},\wt{Q}_{3\,\b_1\db_2}\} =
%-2\e_{\a_1\b_1}\e_{\da_2\db_2}H_4\,,\\
%&\{Q_{2\,\a_1\da_2},\wt{Q}_{4\,\b_1\db_2}\} +
%\{Q_{4\,\a_1\da_2},\wt{Q}_{2\,\b_1\db_2}\} +
%\{Q_{3\,\a_1\da_2},\wt{Q}_{3\,\b_1\db_2}\}= 0
\end{split}
\ee

\ni Determining $Q_4$ has been a long sought-after but yet to be
realized undertaking since the early days of light-cone string field
theory on flat space \cite{Greensite:1986gv, Greensite:1987sm,
Green:1987qu, Greensite:1987hm}. Since it remains to be determined,
the solution in the plane-wave case has been to simply set it to
zero. This is a self-consistent choice which gives rise to the
so-called {\it contact interaction} (see appendix \ref{app:fermrep})

\be
H_4 = \frac{1}{4} Q_3^{\a_1 \da_2} Q_{3\,{\a_1 \da_2} }.
\ee

\begin{figure}[ht]
\begin{center}
\includegraphics*[bb=0 0 330 110, height=1.75in]{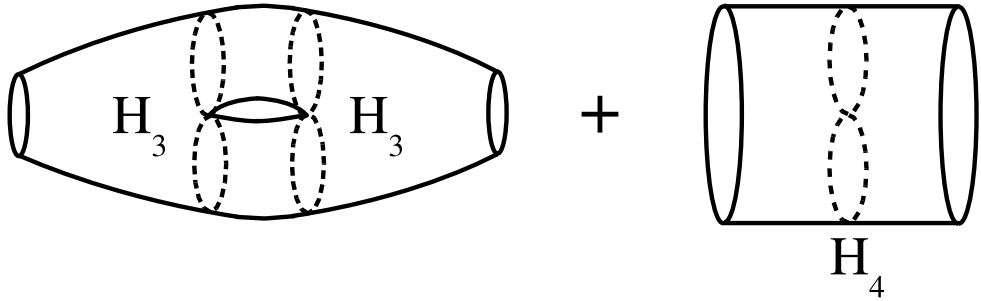}
\end{center}
\caption{The one-loop process contributing to the shift of the energy
  or mass of a string state. Two $H_3$ vertices may be combined to
  form a standard one-loop diagram. The contact interaction $H_4$
  (shown on the right) also contributes to this process. Unlike the
  first diagram, $H_4$ acts at a single time, while each $H_3$ acts at
  a different time; hence the name {\it contact}.}
\label{fig:oneloop}
\end{figure}

\ni In flat space \cite{Greensite:1986gv, Greensite:1987sm,
Green:1987qu, Greensite:1987hm}, the $2\rightarrow 2$ string process
requires a contribution from $Q_4$ to close the algebra
(\ref{ppalgg2}). Here, we will be concerned with the plane-wave
$1\rightarrow 1$ string process; specifically the one-loop mass shift
depicted in figure \ref{fig:oneloop}. Here two $H_3$ vertices
alternately split and then rejoin the strings at separated light-cone
times, while the contact interaction coalesces the splitting and
joining to a single event (the moment of {\it contact}).
It was argued in \cite{Roiban:2002xr}
that $Q_4$ cannot contribute to the $1\rightarrow 1$ string process on
account of it being quartic in string fields at tree level. Later, in
analyzing the $1\rightarrow 1$ string process on the plane-wave
\cite{Gutjahr:2004dv} argued less restrictively that although setting
$Q_4=0$ in this setting still allows the algebra to close, this is
only a necessary but not sufficient condition. It is the opinion of
the author that this issue has not been fully resolved; however the
work in this thesis follows the fashion of setting this quartic
supercharge to zero only because of the lack of another
option. Determining the full expression for $Q_4$ in the flat space or
in the plane-wave background remains a potentially crucial element in
the development of the light-cone string field theory. 

% ========================================================================== %
\subsection{One-loop mass shift: impurity conserving channel}
\label{sec:pankmassshift}

We are now in a position to attempt the string theory calculation of
the gauge theory result (\ref{gaugeresult}). The gauge theory result
is valid for a general two-$\Phi^i$-impurity operator (see
(\ref{impmap}), (\ref{bmn2imp})), independent of the $SO(4)\times
SO(4)$ representation (i.e. the spacetime index structure of the
impurities). This allows for a choice of string state to consider for
the calculation. Ideally, we would like to choose an $SO(4)\times
SO(4)$ representation which can only be constructed out of bosonic
oscillators. In this way, one circumvents having to worry about mixing
between different string states of the same uncorrected energy. For
example consider the representation which is scalar in both $SO(4)$'s

\be
|[{\bf 1},{\bf 1}] \ra = \begin{cases}
\a^{\dag\,k}_n \a^{\dag\,k}_{-n} |\a\ra \cr
\b_{n\,\a_1\a_2}^\dag \b_{-n}^{\dag\,\a_1\a_2} |\a\ra
\end{cases}
\ee

\ni i.e. it can either be constructed out of two fermionic, or two
bosonic oscillators (or to mirror the gauge theory discussion
``impurities''). These two states have the same energy at $g_2=0$, but
when interactions are turned on, generically they will mix. To avoid
this unpleasantness \cite{Gutjahr:2004dv} used the following
state\footnote{The normalization of this state is
$1+\frac{1}{2}\delta^{ij}$. One could have equally chosen $|[{\bf
      1},{\bf 9}]\ra$; the string field theory would not produce a
  different result for the one-loop mass shift.}

\be\label{91}
\left. |[{\bf 9}, {\bf
1}]\ra^{(ij)}\right. =
\frac{1}{\sqrt{2}}
\left(\a^{\dag\,i}_n\a^{\dag\,j}_{-n}+\a^{\dag\,j}_n\a^{\dag\,i}_{-n}
-\frac{1}{2}\d^{ij}\a^{\dag\,k}_n\a^{\dag\,k}_{-n}\right)|\a\ra
\ee

\ni whose $SO(4)\times SO(4)$ representation is unique. The one-loop
mass shift proceeds using standard quantum mechanical perturbation
theory

\be \label{shift} 
\delta E_n^{(2)} = \langle \phi_n|H_3\frac{{\cal
 P}}{E_n^{(0)}-H_{2}^{\text{int}}} H_3|\phi_n\rangle +\langle
 \phi_n|H_4|\phi_n\rangle 
\ee

\ni where $|\phi_n\ra$ represents the state whose shift we are
calculating (i.e. (\ref{91})), which we take to be string \#3 with
uncorrected energy $E_n^{(0)}$, and where ${\cal P}$ is a projection
operator on the space of two-string states. Finally $H_2^{\text{int}}$
is the free Hamiltonian (\ref{BMNH2ab}), restricted to the internal
strings 1 and 2. In practice it is not feasible to consider the full
range of intermediate two-string states; instead, a cue is taken from
the gauge theory computation where the total number of impurities
contained in intermediate states is equal to that of the external
state\footnote{There is no reason for this logic to be extended to the
string theory picture. Indeed, the results of the next section show
that it is an unjustified truncation.}. In the string theory
computation, this is the so-called impurity-conserving channel, which
is realized as follows

\be\label{expshift}
(1+\d^{ij}) \,\d E_n^{(2)} =  
{}^{(ij)}\la [{\bf 9}, {\bf 1}] | H_3 \frac{{\bf
    1}_B}{E_n^{(0)}-H_{2}^{\text{int}}} 
 H_3 | [{\bf 9}, {\bf 1}] \ra^{(ij)} +
\frac{1}{4} {}^{(ij)}\la [{\bf 9}, {\bf 1}] | Q_3^\dag \,{\bf 1}_F  \,
 Q_3 | [{\bf 9}, {\bf 1}] \ra^{(ij)}
\ee

\ni where\footnote{Oscillators act only on the vacuum closest to them.}

\bsp\label{projectors}
&{\bf 1}_B= \sum_{K,L=1}^8\int_0^1\frac{dr}{2\,r(1-r)} \biggl( \sum_{p}
\a_{p}^{\dag\,K}\,\a_{-p}^{\dag\,L}\,| {\a}_1\ra\, |
{\a}_2\ra\la {\a}_2|\,\la {\a}_1|\,\a_{-p}^L\,\a_{p}^K\\
&~~~~~~~~~~~~~~~~~~~~~~~~~~~~~~~~~~~~~~~~~ + \a_{0}^{\dag\,K}\,|
{\a}_1\ra\,\a_{0}^{\dag\,L}\, | {\a}_2\ra\la
{\a}_2|\,\a_{0}^L\la {\a}_1|\,\a_{0}^K \biggr)\\ &{\bf 1}_F=
\sum_{\S_1,\S_2}
\int_0^1\frac{dr}{r(1-r)} \biggl( \sum_{p}
\a_{p}^{\dag\,K}\,\b_{-p}^{\dag\,\Sigma_1\,\Sigma_2}\,| {\a}_1\ra\,
| {\a}_2\ra\la {\a}_2|\,\la
{\a}_1|\,\b_{-p}^{\Sigma_1\,\Sigma_2}\,\a_{p}^K\\
&~~~~~~~~~~~~~~~~~~~~~~~~~~~~~~~~~ + \a_{0}^{\dag\,K}\,|
{\a}_1\ra\, \b_{0}^{\dag\,\Sigma_1\,\Sigma_2}\,| {\a}_2\ra\la
{\a}_2|\,\b_{0}^{\Sigma_1\,\Sigma_2}\,\la {\a}_1|\,\a_{0}^K
\biggr).
\end{split}
\ee

\ni where $r \equiv -\a_1/\a_3$ so that $1-r = -\a_2/\a_3$ where we
remind the reader that $\a_3 <0$, while $\a_{1,2} >0$. The indices
$\S_1$ and $\S_2$ are shorthand for indicating a sum over both dotted
and un-dotted fermions. These projectors obey the condition ${\bf
1}_{B,F}^2 = {\bf 1}_{B,F}$, where we note further that the 
vacuua are normalized by

\be
\la \a_1 | \la \a_2 | \a_2 \ra | \a_1 \ra = r(1-r), \qquad \la \a_3 |
\a_3 \ra = 1.
\ee 

\ni Note that strictly we should have added a two fermion state to
${\bf 1}_B$. For the $|[{\bf 9},{\bf 1}]\ra$ state however, this
contributes nothing as it requires a trace of the $i,j$ indices.

In order to calculate $\d E_n^{(2)}$, we will require the following
matrix elements \cite{Gutjahr:2004dv}

\begin{equation}\label{H3elem}
\begin{split}
 ^{(ij)}\la{\bf[9,1]}|\,\la {\a}_2|\,\a_0^l\,\la {\a}_1|\,\a_0^k
\,|H_3\ra &= - 2\,r(1-r) \Big( \frac{\omega_{n(3)}}{\a_3} + \mu \Big)
\,\wt{N}^{31}_{n,0}\,\wt{N}^{32}_{n,0}\,\D^{ijkl} \\
^{(ij)}\la{\bf[9,1]}|\,\la {\a}_2|\,\la
{\a}_1|\,\a_{-p}^l\,\a_p^k \,|H_3\ra &= - 2\,r(1-r) \Big(
\frac{\omega_{n(3)}}{\a_3} - \frac{\omega_{p(1)}}{\a_3 r}\Big)
\,\wt{N}^{31}_{n,p}\,\wt{N}^{31}_{n,-p}\,\D^{ijkl} \\
\end{split}
\end{equation} 

\ni where we have used (\ref{L2N}) and (\ref{L2N3}),  and 

\begin{equation}\label{Q3elem}
\begin{split}
 ^{(ij)}\la{\bf[9,1]}|\,\la {\a}_2|\,(\b_0)^{\ds_1 \ds_2}\,\la
{\a}_1|\,\a_0^k\,|Q_{3\,\b_1\db_2}\ra &= \\ - 2i\, \bar{C}\,
G_{0(2)}\, \big(K_{n(3)} &+ K_{-n(3)}\big)\, \wt{N}^{31}_{n,0}\,
\D^{ijkl}(\s^l)^{\ds_1}_{\b_1}\,\d^{\ds_2}_{\db_2}\\
^{(ij)}\la{\bf[9,1]}|\,\la {\a}_2|\,\la
{\a}_1|\,(\b_{-p})^{\ds_1 \ds_2}\,\a_p^k\,|Q_{3\,\b_1\db_2}\ra &=\\
- 2i\, \bar{C}\, G_{|p|(1)}\, \big(K_{n(3)}\wt{N}^{31}_{n,p} &+
K_{-n(3)}\wt{N}^{31}_{n,-p}\big)\,
\D^{ijkl}(\s^l)^{\ds_1}_{\b_1}\,\d^{\ds_2}_{\db_2}
\end{split}
\end{equation}

\ni where $\D^{ijkl} \equiv
\frac{1}{\sqrt{2}}\big\{\d^{ik}\d^{jl}+\d^{il}\d^{jk}-\frac{1}{2}\d^{ij}\d^{kl}\big\}$
and $\bar{C} \equiv \frac{\bar{\eta}}{4} \,
\sqrt{-\frac{\a'}{2\,\a_3^3}}\, \sqrt{r(1-r)}$. The energy shift
(\ref{expshift}) is then calculated by taking the modulus squared of the
matrix elements (\ref{H3elem}) and dividing by the energy denominator

\be
\frac{-\a_3}{2\left(\o_n - r^{-1}\o_p\right)}
\ee

\ni then adding $\tiny\frac{1}{4}$ of the modulus squared of the
matrix elements (\ref{Q3elem}), and then summing over the intermediate
mode number $p$. Also note that the normalization $1+\frac{1}{2}\d^{ij}$ results
from

\be \sum_{k,l} \D^{ijkl} \D^{ijkl} = 1+\frac{1}{2}\d^{ij}, \qquad
\sum_{k,l}\sum_{\ds_1,\ds_2}
\left| \D^{ijkl} \left( \s^l \right)^{\ds_1}_{\b_1} \d^{\ds_2}_{\db_2}
\right|^2 = 4
\left( 1+\frac{1}{2}\d^{ij}\right). 
\ee

\ni The forms of the summand in the sum over $p$ may be massaged into
two classes

\be
F_1 = \sum_p \frac{P(p)}{Q(p) \sqrt{p^2 + (r \m \a_3)^2}},\qquad
F_2 = \sum_p \frac{P(p)}{Q(p)}
\ee

\ni where $Q(p)$ and $P(p)$ are polynomials in $p$. In order to
extract the large-$\m$ behaviour of the sums, the contour integral
method is employed

\begin{equation}\label{contmeth}
\sum_{p=-\infty}^{\infty} f(p) = -\frac{i}{2}\oint dz \, f(z) \, \cot(\pi z).
\end{equation}

\ni Rotating and scaling the integration variable through the
substitution $z \rightarrow i x r z$, where $x = -\mu\alpha_3$, turns
the cotangent into $\coth(\pi x r z)$ which can be set to one in the
large $x$ limit\footnote{The terms neglected by this approximation are
of order $\exp(-\m|\a_3|)$.}. If the summand $f(z)$ has no poles on
the real axis, the procedure simply replaces $p$ by $p' = r x\,p$ and
integrates

\begin{equation}\label{nopoles}
\sum_{p=-\infty}^{\infty} f(p) = \int_{-\infty}^\infty dp' f(p')
\end{equation}

\ni yielding the large $x$ behaviour. If there are poles on the real axis,
one must evaluate their residue using the integrand in (\ref{contmeth})
and then integrate along any cut which $f(z)$ may possess along the
imaginary axis. Thus we have

\bsp\label{howtodosums}
F_1 = -\pi \, \sum_i \text{Res} 
&\left( \cot(\pi p) \frac{P(p)}{Q(p)} , p_i \in \left\{p\, |\,
Q(p)=0\right\} \right)\\
&+ \int_1^\infty dz \frac{ \left[Q(ixrz)\right]^* P(ixrz) + \text{c.c.}}
{\left| Q(ixrz) \right|^2 \, \sqrt{z^2-1}}
\end{split}
\ee

\ni while for $F_2$ the second term is dropped as there is no
cut. This calculation was originally presented in
\cite{Gutjahr:2004dv}. The author of this thesis finds error in the
result reported there\footnote{The error is made in the evaluation of
the sum using the contour-integral method.}, as regards the
$\l'^{3/2}$ and $\l'^{5/2}$ powers. A careful recalculation reveals
the following result \cite{Grignani:2006en}\footnote{The undetermined
  function $f$ in (\ref{H3andQ3full}) is set to $r^{-1}(1-r)^{-1}$ here.}

\be\label{pankN4}\boxed{
\begin{split}
\d E_n^{(2)}/\m &=  \frac{g_2^2}{4\pi^2}\left[
\left(\frac{1}{24} + \frac{65}{64\pi^2 n^2}\right)\l'
+\frac{3}{16}\left(\frac{1}{\pi^2}+\frac{1}{2\pi}\right)\l'^{3/2}\right.\cr
&- \left. n^2\left(\frac{1}{48}+\frac{89}{128\pi^2 n^2}\right)\l'^2
-\frac{9\,n^2}{32}\left(\frac{1}{\pi^2}+\frac{1}{2\pi}\right)\l'^{5/2}
\right. \cr &+ \left.n^4\left(\frac{1}{64}+\frac{339}{512\pi^2
n^2}\right)\l'^{3} + n^4 \left( \frac{59}{160\pi^2} +
\frac{45}{256\pi}\right)\l'^{7/2} +{\cal O}(\l'^4) \right]
\end{split}}
\ee

\ni where we note that the final integration over $r$ in
(\ref{projectors}) is performed as the last step. The appearance of
half-integer powers of $\l'$ is disconcerting, as it is hard to see
how such terms could ever arise in the gauge theory; they are clearly
absent from (\ref{gaugeresult}). These terms appear to be generic to
light-cone string field theory on the plane-wave
\cite{Klebanov:2002mp} and so must find a way to cancel-out if a true
matching to gauge theory is to be realized.

% ************************************************************************** %
\section{Divergence cancellation and impurity non-conserving channel}
\label{sec:divcan}

This section is a presentation of the author's original work published
in \textsf{arXiv:hep-th/0508126} \cite{Grignani:2005yv}. In what
follows, some passages are taken directly from that publication.\\

\ni The result (\ref{pankN4}) does not match the gauge theory result
(\ref{gaugeresult}) even at leading order. In an earlier attempt at
this calculation \cite{Roiban:2002xr}, a reflection symmetry factor of
$\frac{1}{2}$ was added in front of the $H_3$ term in
(\ref{expshift}), which the authors of \cite{Gutjahr:2004dv} argued
was incorrect. This factor produced a leading order agreement with
gauge theory. The subleading orders were calculated by the author of
this thesis \cite{Grignani:2006en}

\bea \frac{1}{\m}
\d E^{(2)\text{ref. symm.}}_n &=& \frac{g_2^2}{4\pi^2} \left[
\left(\frac{1}{12}+\frac{35}{32\pi^2 n^2}\right)
\left(\l'-\frac{n^2}{2}\l'^2\right) +
\frac{n^2}{16\pi^2}\l'^{5/2}\right.\cr &+&
\left.n^4\left(\frac{1}{32}+\frac{117}{256\pi^2 n^2}\right) \l'^3
-\frac{7 n^4}{80\pi^2}\l'^{7/2} +{\cal O}(\l'^4) \right] \eea

\ni where it was found that the agreement persists up to ${\cal
  O}(\l'^2)$. Although this agreement is tantalizing, the factor of
  $\frac{1}{2}$ can not be justified, for reasons beyond the arguments
  of \cite{Gutjahr:2004dv}, who pointed out that (\ref{expshift}) is
  standard quantum mechanical perturbation theory, not a field theory
  Feynman diagram prescription. The author's work
  \cite{Grignani:2005yv} provided very convincing evidence in support
  of \cite{Gutjahr:2004dv}, which will be presented in this
  section. 

The disagreement of (\ref{pankN4}) with gauge theory led
  \cite{Gutjahr:2004dv} to conclude that (modulo the absence of $Q_4$)
  the truncation to the impurity conserving channel may be the source
  of the discrepancy. In fact, in the earlier work
  \cite{Roiban:2002xr}, a statement was made concerning the
  four-impurity channel. The claim was that the mode number sums
  diverge linearly if the large-$\m$ limit is taken pre-summation;
  this means that if the sum is evaluated first, and then the
  large-$\m$ limit taken (the method applied in section
  \ref{sec:pankmassshift}), this would result in a contribution to the
  mass shift which goes as $\sqrt{\l'}$. The idea is that (roughly)

\bsp
&\sum_p \frac{1}{p^2 + \m^2} \rightarrow 
\text{large-$\m$ limit} \rightarrow
\frac{1}{\m^2} \sum_p 1 \sim
\l' \cdot \infty\\
&\sum_p \frac{1}{p^2 + \m^2} \simeq \frac{1}{\m} \int dx
  \frac{1}{x^2+1} \sim \sqrt{\l'}.
\end{split}
\ee

\ni The prediction was therefore that the four-impurity channel should
give a contribution larger than the impurity-conserving channel in the
large-$\m$ limit. Further, the $\sqrt{\l'}$ indicated a
non-perturbative origin in the gauge theory. In the paper
\cite{Grignani:2005yv}, the author of this thesis and his
collaborators undertook a proper investigation of the four-impurity
channel (while making arguments concerning higher impurity channels)
to verify the claim of \cite{Roiban:2002xr}. What was discovered was
that $\sqrt{\l'}$ behaviour is a reflection of real (logarithmic)
divergences in the $H_3$ and contact amplitudes which cancel, taking
with them the $\sqrt{\l'}$ terms. Finiteness and the perturbative
nature of the mass-shift were thus established in concert. The
analysis revealed further that generically, every order in
intermediate state impurities contributes a leading $\l'$ contribution
to the mass shift; a discouraging result as regards matching to the
gauge theory.

% ========================================================================== %
\subsection{Invitation: trace state}
\label{sec:trace}

The logarithmic divergences found in the four impurity channel are at
play in a simpler setting. A careful calculation of the
impurity-conserving channel contribution to the mass shift of the
normalized bosonic trace state 

\be\left|[{\bf 1}, {\bf 1}]\ra\right. =
\frac{1}{{2}}\alpha^{i\dagger}_n \alpha^{i \dagger}_{-n} |\alpha
\rangle\label{singlet}\ee 

\ni reveals the same divergences and cancellation mechanism.
In~\cite{Gomis:2003kj}, this calculation was performed by taking the
large $\mu$ limit first, then summing over mode numbers. That
procedure found a finite result. However, if $\mu$ is kept finite,
there are logarithmically divergent summations which must be dealt
with before the large $\mu$ limit is taken. The $H_3$ matrix element
contributing to the mass shift is

\bsp
\langle \alpha_3 |\frac{1}{2} \alpha^i_{n} \alpha^i_{-n} \, \langle
{\a}_2 | \langle {\a}_1 | \alpha^{K}_{p} \alpha^{L}_{-p} |H_3
\rangle = -g_2\,\frac{r\,(1-r)}{8} \left[8\left(
\frac{\omega_{n}^{(3)}}{\alpha_3} + \frac{\omega_{p}^{(1)}}{\alpha_1}
\right) {\widetilde N}_{-n\,p}^{3\,1} {\widetilde N}_{n\,p}^{3\,1}
\,\delta^{kl}\right. \cr
\left.+16\,\frac{\omega_{n}^{(3)}}{\alpha_3} {\widetilde
N}_{n\,n}^{3\,3} {\widetilde N}_{p\,-p}^{1\,1} \,\delta^{KL} + 16
\,\frac{\omega_{p}^{(1)}}{\alpha_1} {\widetilde N}_{n\,-n}^{3\,3}
{\widetilde N}_{p\,p}^{1\,1} \,\Pi^{KL}\right] \label{trmatel} 
\end{split}
\ee

\ni where the index $i=1,\ldots,4$ is summed over. Note that $K,L =
1,\ldots,8$, while $\delta^{kl}$ is non-zero only for
$k=l=1,\ldots,4$. The matrix $\Pi^{KL}$ is given by

\be 
\Pi^{KL} = \mbox{diag}(1,1,1,1,-1,-1,-1,-1).
\ee 

\ni When calculating the $H_3$ contribution to the mass shift it is
only the very last term in (\ref{trmatel}) which is divergent.
Singling-out its contribution, one finds (using the projectors
(\ref{projectors}))

\begin{equation}\label{divH3}
\delta E^{\mbox{div}}_{H_3} = \int_0^1 dr\,\left(g_2 \frac{r(1-r)}{8}
\right)^2 \,\frac{-\alpha_3}{2\,r\,(1-r)} \sum_{KL}
\sum_{p=-\infty}^{\infty}\frac{ \left[
16\,\frac{\omega_{p}}{-r\,\alpha_3} {\widetilde N}_{n\,-n}^{3\,3}
\,{\widetilde N}_{p\,p}^{1\,1} \,\Pi^{KL} \right]^2} {2\,\omega_{n}
-2\,r^{-1}\omega_{p} }
\end{equation}

Inspection of the forms of the Neumann matrices (see appendix
\ref{app:relations}) reveal that the numerator in (\ref{divH3}) goes
like a constant for large $|p|$, and thus the sum as a whole goes like
$1/|p|$ for $|p| \gg |\mu\alpha_3|$. This is a logarithmically
diverging sum. In~\cite{Gomis:2003kj} the strict large $\mu$ limit was
taken for the energy denominator, leading to a convergent $1/p^2$
behaviour instead. Here we will stick with the finite $\mu$ expressions
and show that the divergence is removed by the contact term. Note that
a double fermionic impurity intermediate state also contributes to the
$H_3$ piece, however it does not display any divergent behaviour. In
appendix \ref{app:example}, section \ref{appsec:example}, the contribution
from this channel is calculated, as an example of how these
calculations are performed in general. Further, the $\alpha^\dagger_0
|\alpha_1\rangle \alpha^\dagger_0 |\alpha_2\rangle$ intermediate state
is unimportant as it does not contain a mode number sum. The
contribution from the contact term stems from the following matrix
element

\bsp
\label{contactme}
&\left( g_2 \frac{\eta}{4} \sqrt{\frac{r\,(1-r)\,
\alpha'}{-2\,\alpha_3^3}} \right)^{-1} \langle \alpha_3 |
\frac{1}{2}\alpha^i_{n} \alpha^i_{-n} \, \langle {\a}_2 | \langle
{\a}_1 | \alpha^{K}_{p} \beta^{\Sigma_1\,\Sigma_2}_{-p} |Q_{3\,
\beta_1 {\dot \beta}_2} \rangle\\ &\qquad\qquad= \biggl( G_{|p|}^{(1)}
\, K_{n}^{(3)} {\widetilde N}^{3\,1}_{n\,p} + G_{|p|}^{(1)} \,
K_{-n}^{(3)} {\widetilde N}^{3\,1}_{-n\,p} \biggr)(
\sigma^k)^{\dot{\sigma_1}}_{ \beta_1 } \delta^{{\dot \sigma_2}}_{{\dot
\beta_2}} +4\, G_{|p|}^{(1)} \, K_{-p}^{(1)} {\widetilde
N}^{3\,3}_{n\,-n} ( \sigma^K)^{\Sigma}_{ \beta }
\delta^{\Sigma}_{\beta}~~~~~
\end{split}
\ee 

\ni Here $K=1,\ldots,8$ while the $\Sigma$ and $\beta$ indices are
either dotted or undotted as required by the particular SO(4)
representation indicated by $K$.  The last term in (\ref{contactme})
gives rise to a log-divergent sum. For large positive $p$,
$(K_{-p}^{(1)})^2$ goes as a constant, and so the sum is controlled by
$(G_{|p|}^{(1)})^2$ which goes as $1/p$, and hence diverges
logarithmically. For $p$ negative, the sum converges. Thus, the
divergent contribution to $\delta E^{(2)}$ is found to be (again using
(\ref{projectors}))

\begin{equation}\label{divH4}
\delta E^{\mbox{div}}_{H_4} = 8 \int_0^1 dr\, \left( g_2 \frac{1}{4}
\sqrt{\frac{r\,(1-r)\, \alpha'}{-2\,\alpha_3^3}} \right)^2
\frac{1}{r(1-r)} \sum_{p=1}^{\infty} \biggl( 4\,G_{|p|}^{(1)}
\,K_{-p}^{(1)} {\widetilde N}^{3\,3}_{n\,-n} \biggr)^2
\end{equation}

\ni where the leading factor of 8 comes from the sum over $K$.  Again
the intermediate state $~\alpha^\dagger_0 |\alpha_1\rangle
\beta^\dagger_0 |\alpha_2\rangle$ is unimportant to convergence and is
ignored here. In taking the large $p$ limits of the summands in
(\ref{divH3}) and (\ref{divH4}), one finds,

\begin{equation}\label{svpsdivH3}
\delta E^{\mbox{div}}_{H_3} \sim -\frac{1}{2} \int_0^1 dr
\,\frac{g_2^2\,r(1-r)}{r\,|\alpha_3|\,\pi^2} \left( {\widetilde
N}^{3\,3}_{n\,-n} \right)^2 \, \frac{1}{|p|},
\end{equation}

\begin{equation}\label{svpsdivH4}
\delta E^{\mbox{div}}_{H_4} \sim +\int_0^1 dr
\,\frac{g_2^2\,r(1-r)}{r\,|\alpha_3|\,\pi^2} \left( {\widetilde
N}^{3\,3}_{n\,-n} \right)^2 \, \frac{1}{p}.
\end{equation}

\ni Noting that in the $H_3$ contribution the divergence is found for
both positive and negative $p$, while in the $H_4$ contribution the
divergence occurs only for positive $p$, and hence a relative factor
of 2 is induced in the $H_3$ term, one sees that the logarithmically
divergent sums cancel identically between the $H_3$ and contact terms,
leaving a convergent sum. As promised at the beginning of section
\ref{sec:divcan}, this cancellation fixes the relative weight of the
$H_3$ and contact terms to that employed in \cite{Gutjahr:2004dv}. It
contradicts the reflection symmetry factor of $1/2$ originally given
in \cite{Roiban:2002xr}; finiteness of the string theory amplitude
requires the absence of this factor.

% ========================================================================== %
\subsection{Four impurity channel}

We now consider the mass shift of the $\left|[{\bf 9}, {\bf
    1}]\ra^{(ij)}\right.$ string state (\ref{91}) due to intermediate
states which contain four impurities. In the explicit expression for
the matrix element to be quoted below, we shall see that the parameter
$\mu\alpha_3$ occurs only in combinations involving $\omega_p$ and
there is a duality between the large $p$ and the large $\mu\alpha_3$
limits.  Therefore, since a logarithmic divergence in the sums
indicates that the summands have as many (inverse) powers of the
summation variables as there are summation variables, this translates
into vanishing $\mu\alpha_3$ dependence for this contribution to
$\delta E^{(2)}$, leaving $\delta E^{(2)}/\mu \sim
\sqrt{\lambda'}$. It is thus seen that $\sqrt{\lambda'}$ behaviour is
simply the result of log divergences, which should, if pp-wave
light-cone string field theory is to make any sense, cancel out
entirely. We begin with the $H_3$ contribution to the mass shift. We
consider the following intermediate state\footnote{From now on, the
  sum over intermediate state spacetime indices is implied, rather
  than explicitly indicated.}

\begin{equation} \label{projector}
{\bf 1}_B= \int_0^1\frac{dr}{4!\,r(1-r)} \sum_{p_1\,p_2\,p_3\,p_4}
\a_{p_1}^{\dag\,K}\,\a_{p_2}^{\dag\,L}\,
\a_{p_3}^{\dag\,M}\,\a_{p_4}^{\dag\,N}\, | {\a}_1\ra\, |
{\a}_2\ra\la {\a}_2|\,\la {\a}_1|\,\a_{p_4}^N\,\a_{p_3}^M
\,\a_{p_2}^L\,\a_{p_1}^K
\end{equation}

\ni where the sum over mode numbers is restricted by the level
matching condition $\sum_i p_i = 0$. Although there are many possible
contractions of this state with the oscillators in $|H_3\ra $, we will
only be concerned with those which lead to log divergent sums. These
are the ones where both oscillators in the prefactor of $|H_3\ra$
contract with the oscillators in ${\bf 1}_B$. We find this
contribution to $\delta E^{(2)}$ to be

\bea &&\delta E^{\mbox{div}}_{H_3}=\int_0^1
\frac{dr}{4!\,r\,(1-r)} \left(g_2 \frac{r(1-r)}{4} \right)^2 \,
\sum_{p_2\,p_3\,p_4} \frac{-\alpha_3\,r}{2\,\omega_n\,r -
\sum_{i=1}^4\omega_{p_i}} \times \cr &&\left(2\, \frac{\omega_{p_1} +
\omega_{p_2}}{-r\,\alpha_3} \, {\widetilde N}^{1\,1}_{-p_1\, p_2}
\right)^2 8\cdot12\,\left\{\left({\widetilde N}^{3\,1}_{n\, p_3}
{\widetilde N}^{3\,1}_{-n\, p_4} \right)^2 + {\widetilde
N}^{3\,1}_{n\, p_3} {\widetilde N}^{3\,1}_{-n\, p_3} {\widetilde
N}^{3\,1}_{n\, p_4} {\widetilde N}^{3\,1}_{-n\, p_4}\right\}
\label{divH34} 
\eea 

\ni where $p_1 = -(p_2+p_3+p_4)$. The factor of $12$ is combinatoric
and counts the number of ways equivalent contractions can be made. The
factor of $8$ comes from a sum over the spacetime indices of ${\bf
1}_B$. It is easy to see that in the above, the sum over $p_2$ is log
divergent. In fact, it is the very same form as appears in
(\ref{divH3}). In order to evaluate the leading $\m$ dependence of the
expression (\ref{divH34}), one must consider the forms of the Neumann
matrices given in appendix \ref{app:relations}. The matrices which have
one leg in the external string, i.e. $\wt N^{3\,r}_{n\,p}$ and $\wh
Q^{3\,r}_{n\,p}$, contain poles at $p = \b_r n$. The sums over mode
numbers involved with these Neumann matrices are then dominated by the
residues given in (\ref{howtodosums}). Specifically, they are ${\cal
O}(\m^0)$. This allows us to dispense with the sums over $p_3$ and
$p_4$ in (\ref{divH34}), as far as $\m$ power counting is
concerned. The remaining sum over $p_2$ is executed via replacing
$p_2$ with $z=\m|\a_3|p'$ and integrating over $p'$. One then finds
that the $\m$ dependence drops out completely from the squared term
involving $\wt N^{1\,1}_{-p_1\,p_2}$, while the measure of the
integration over $p'$ cancels the $\m^{-1}$ stemming from the energy
denominator. One then has that $\delta E^{\mbox{div}}_{H_3} \sim
\mbox{constant}$, and therefore $\delta E^{(2)}/\mu \sim
\sqrt{\lambda'}$. There are also contributions from intermediate
states which contain two bosonic and two fermionic impurities, however
these produce convergent sums and ${\cal O}(\lambda')$ contributions
to $\delta E^{(2)}/\mu$. The four-fermion channel is forbidden because
it produces a delta function (i.e. a trace) on the external state's
spacetime indices.

We now show that the contact term contribution stemming from the following
intermediate state,

\be
{\bf 1}_F= \int_0^1\frac{dr}{3!\,r(1-r)} \sum_{p_1\,p_2\,p_3\,p_4}
\beta_{p_1}^{\dag\,\S_1 \S_2}\,\a_{p_2}^{\dag\,L}\,
\a_{p_3}^{\dag\,M}\,\a_{p_4}^{\dag\,N}\, | {\a}_1\ra\, | {\a}_2\ra\la
{\a}_2|\,\la {\a}_1|\,\a_{p_4}^N\,\a_{p_3}^M
\,\a_{p_2}^L\,\beta_{p_1}^{\S_1 \S_2}
\ee

\ni cancels the divergent piece coming from the $H_3$ contribution,
leaving an ${\cal O}(\lambda')$ contribution to $\delta
E^{(2)}/\mu$. The log divergent piece comes from contractions where
the $\a^\dag$ in the prefactor of $|Q_3\ra$ is joined with one of the
bosonic oscillators in ${\bf 1}_F$. One finds,

\bsp \label{divCON}
\delta E^{\mbox{div}}_{H_4}=\int_0^1 \frac{dr}{3!\,r\,(1-r)}
\left( g_2 \frac{1}{4} \sqrt{\frac{r\,(1-r)\,
\alpha'}{-2\,\alpha_3^3}} \right)^2 \sum_{p_2\,p_3\,p_4} \left(2\,
G_{p_1} K_{-p_2} \right)^2 \\ \times 8\cdot6\,\left\{
\left({\widetilde N}^{3\,1}_{n\, p_3} {\widetilde
N}^{3\,1}_{-n\, p_4} \right)^2 +  {\widetilde N}^{3\,1}_{n\, p_3}
{\widetilde N}^{3\,1}_{-n\, p_3} {\widetilde N}^{3\,1}_{n\, p_4}
{\widetilde N}^{3\,1}_{-n\, p_4} \right\} 
\end{split}
\ee 

\ni In the above one sees the very same pattern as was seen in section
\ref{sec:trace} for the trace state. The sum over $p_2$ is divergent
on the positive side, and cancels the divergence in
(\ref{divH34}). 

The remaining (convergent) expression gives an ${\cal O}(\lambda')$
contribution to $\delta E^{(2)}/\mu$. To see this we note that the
poles in $p_3$ and $p_4$ set these variables to an ${\cal O}(\m^0)$
quantity. Then $p1 = \e - p_2$ by level matching, where $\e = 2\,r\,n$
represents this ${\cal O}(\m^0)$ quantity. Ignoring the common factor
involving the $\wt N^{3\,1}_{n\,p_i}$'s, the remaining expressions may
be expressed as (suppressing the integration over $r$)

\be
\delta E^{\mbox{div}}_{H_3} + \delta E^{\mbox{div}}_{H_4}
= g_2^2 \frac{(1-r)}{(4\pi)^2|\a_3|} \left[
\frac{2(\o_2 + p_2)}{\o_1 \o_2} + \frac{2(\o_1 \o_2 + r^2 \m^2 \a_3^2
  - p_1 p_2)}{(2 \,r\, \o_n - \o_1-\o_2-\o_3-\o_4)\,\o_1\o_2} \right].
\ee

\ni The next step is to notice that $2\,r\,\o_n - \o_3 -\o_4 = 0$ in
the large-$\m$ limit due to $p_3$ and $p_4$ being set to
$r\,n$. Combining the terms, the leading large-$\m$ behaviour is given
by

\be
\delta E^{\mbox{div}}_{H_3} + \delta E^{\mbox{div}}_{H_4}
= g_2^2 \frac{(1-r)}{(4\pi)^2|\a_3|} 2 \sum_{p_2=-\infty}^\infty
\frac{p_2}{\sqrt{p_2^2 + (r\m\a_3)^2}\sqrt{(p_2-\e)^2 + (r\m\a_3)^2}}
\sim \m^{-1}
\ee

\ni which results in the advertised large-$\m$ behaviour.  Again,
there is a non-divergent contribution from the intermediate state with
three fermionic and one bosonic impurity which we will ignore. There
are two other choices for distributing the intermediate-state
oscillators amongst the two strings. We may express them in pairs

\bea &&{\bf 1}_B= \int_0^1\frac{dr}{3!\,r(1-r)} \sum_{p_1\,p_2\,p_3}
\a_{p_1}^{\dag\,K}\,\a_{p_2}^{\dag\,L}\, \a_{p_3}^{\dag\,M}\, |
{\a}_1\ra\, \a_{0}^{\dag\,N}\,| {\a}_2\ra\la {\a}_2|\,\a_{0}^N\,\la
{\a}_1| \,\a_{p_3}^M \,\a_{p_2}^L\,\a_{p_1}^K \cr &&{\bf 1}_F=
\int_0^1\frac{dr}{2!\,r(1-r)} \sum_{p_1\,p_2\,p_3}
\beta_{p_1}^{\dag\,a}\,\a_{p_2}^{\dag\,L}\, \a_{p_3}^{\dag\,M}\,|
{\a}_1\ra\, \a_{0}^{\dag\,N}\,| {\a}_2\ra\la {\a}_2|\,\a_{0}^N\,\la
{\a}_1| \,\a_{p_3}^M \,\a_{p_2}^L\,\beta_{p_1}^a \eea

\ni where $\sum_{i=1}^3 p_i = 0$ and,

\bea &&{\bf 1}_B= \int_0^1\frac{dr}{2\cdot (2!)^2 \,r(1-r)}
\sum_{p_1\,p_2} \a_{p_1}^{\dag\,K}\,\a_{-p_1}^{\dag\,L}\, |
{\a}_1\ra\, \a_{p_2}^{\dag\,M}\,\a_{-p_2}^{\dag\,N}\, | {\a}_2\ra \la
{\a}_2|\,\a_{-p_2}^N\,\a_{p_2}^M \,\la
{\a}_1|\,\a_{-p_1}^L\,\a_{p_1}^K\cr &&{\bf 1}_F= \int_0^1\frac{dr}{2!
\,r(1-r)} \sum_{p_1\,p_2} \a_{p_1}^{\dag\,K}\,\a_{-p_1}^{\dag\,L}\, |
{\a}_1\ra\, \a_{p_2}^{\dag\,M}\,\beta_{-p_2}^{\dag\,a}\, | {\a}_2\ra
\la {\a}_2|\,\beta_{-p_2}^a\,\a_{p_2}^M \,\la
{\a}_1|\,\a_{-p_1}^L\,\a_{p_1}^K. \eea

\ni One may show the very same cancellation mechanism applies between
each pair here; essentially the difference is that the $\wt
N^{3\,r}_{n\,p}$ factors in (\ref{divH34}) and (\ref{divCON}) have
opposite string label $r$ to the remaining factors in those
expressions. Finally one may show that the intermediate state
with a $\b^\dag_0$ alone on string 2 does not lead to a divergent
contribution.

We therefore find that the entire contribution to $\delta E^{(2)}/\mu$
from the four impurity channel is convergent / leads as $\lambda'$. It
is not hard to generalize the above argument to ${\bf 1}_B$'s
containing an arbitrary number of bosonic impurities and no fermionic
impurities. The divergent expressions cancel against contact
interactions with ${\bf 1}_F$'s containing one fermionic and the same
number (less-one) of bosonic oscillators as ${\bf 1}_B$. Adding
fermionic impurities is far less trivial because of the complicated
nature of the prefactors of $|H_3\ra$ and $|Q_3\ra$. However, in the
next section a more elegant argument is presented which claims the
absence of log divergences for arbitrary impurity intermediate states.

It is important to note that the $|[{\bf 9}, {\bf 1}]\ra^{(ij)}$ state
receives no contributions to its energy shift from the zero impurity
channel and so we don't need to worry about $\sqrt{\l'}$ behaviour
hiding there.
% ========================================================================== %
\subsection{Generalizing to arbitrary impurities}
\label{sec:arbimp} 

It is possible to formally manipulate the contact term in such a way
that the $H_3$ portion of the energy shift is cancelled entirely,
leaving a convergent expression, which appears devoid of any
$\sqrt{\lambda'}$ contributions to $\delta E^{(2)}/\mu$. The
manipulation proceeds through the supersymmetry algebra; for
completeness we include both the ``dot-undot'' and ``undot-dot''
representations of the supercharges, see appendix
\ref{app:fermrep}. At order $g_2$ we have

\bea
&&\left\{Q_{2\alpha_1\dot\alpha_2},Q_{3\beta_1\dot\beta_2}\right\}+
\left\{Q_{3\alpha_1\dot\alpha_2},Q_{2\beta_1\dot\beta_2}\right\}
=-2\epsilon_{\alpha_1\beta_1}\epsilon_{\dot\alpha_2\dot\beta_2}
H_3~,\cr
&&\left\{Q_{2\dot\alpha_1\alpha_2},Q_{3\dot\beta_1\beta_2}\right\}+
\left\{Q_{3\dot\alpha_1\alpha_2},Q_{2\dot\beta_1\beta_2}\right\}
=-2\epsilon_{\dot\alpha_1\dot\beta_1}\epsilon_{\alpha_2\beta_2}
H_3 \label{Q2Q3} 
\eea 

\ni analogously to order $g^2_2$ one has 

\bsp
\left\{Q_{3\alpha_1\dot\alpha_2},Q_{3\beta_1\dot\beta_2}\right\}
+\left\{
Q_{2\alpha_1\dot\alpha_2},Q_{4\beta_1\dot\beta_2}\right\}+\left\{
Q_{4\alpha_1\dot\alpha_2},Q_{2\beta_1\dot\beta_2}\right\}=
-2\epsilon_{\alpha_1\beta_1}\epsilon_{\dot\alpha_2\dot\beta_2}
H_4, \\
\left\{Q_{3\dot\alpha_1\alpha_2},Q_{3\dot\beta_1\beta_2}\right\}+\left\{
Q_{2\dot\alpha_1\alpha_2},Q_{4\dot\beta_1\beta_2}\right\}+\left\{
Q_{4\dot\alpha_1\alpha_2},Q_{2\dot\beta_1\beta_2}\right\}
=-2\epsilon_{\dot\alpha_1\dot\beta_1}\epsilon_{\alpha_2\beta_2}
H_4. \label{Q3Q3} 
\end{split}
\ee

\ni In order to dispense with the $\e_{\a\,\b}$'s, we employ the
Hermitian conjugates of $Q_3$, see again appendix \ref{app:fermrep}

\be
\left\{Q_{2\beta_1\dot\beta_2},Q_3^{\beta_1\dot\beta_2}\right\}=+4
H_3,~~~
\left\{Q_{2\dot\beta_1\beta_2},Q_3^{\dot\beta_1\beta_2}\right\}=+4 H_3
\label{h3} 
\ee 

\ni and 

\bea H_4={1\over 8}
Q_{3\beta_1\dot\beta_2}Q_3^{\beta_1\dot\beta_2}+{1\over 8}
Q_{3\dot\beta_1\beta_2} Q_3^{\dot\beta_1\beta_2}+{1\over 8}
Q_{4\beta_1\dot\beta_2}Q_2^{\beta_1\dot\beta_2}+{1\over 8}
Q_{4\dot\beta_1\beta_2} Q_2^{\dot\beta_1\beta_2}\nonumber
\\+{1\over 8}
Q_{2\beta_1\dot\beta_2}Q_4^{\beta_1\dot\beta_2}+{1\over 8}
Q_{2\dot\beta_1\beta_2} Q_4^{\dot\beta_1\beta_2} \label{H4}. 
\eea

\ni Using these formula, the contribution of $H_4$ to $\delta
E^{(2)}$ can be rewritten as a sum of a term which cancels the $H_3$
contribution plus other pieces which all contain $Q_2$ acting on one
of the external states. Taking the expectation value of part of
(\ref{H4}), and introducing $P$ as a representation of unity, we have

\bea\label{okok} {1\over 8} \left< Q_{3\beta_1\dot\beta_2}
Q_3^{\beta_1\dot\beta_2} + Q_{3\dot\beta_1\beta_2}
Q_3^{\dot\beta_1\beta_2} \right>={1\over 8} \left<
Q_{3\beta_1\dot\beta_2}P
\frac{E_0-H_2}{E_0-H_2}Q_3^{\beta_1\dot\beta_2}\right> \\ +{1\over 8}
\left<
Q_{3\dot\beta_1\beta_2}P\frac{E_0-H_2}{E_0-H_2}Q_3^{\dot\beta_1\beta_2}\right>.
\eea

\ni It could be that the energy denominator which we have introduced
here will have a zero.  In that case, the projector $P$ is a reminder
to define the singularity using a principle value
prescription\footnote{ There is one additional subtlety, the
intermediate states must each obey the level-matching condition.  This
condition can be enforced by inserting a projection operator.  For
example, for two-string intermediate states, we can combine such a
projector with the energy denominator as

\begin{equation}
\frac{P}{E_0-H_2} = \int_0^\infty d\tau ~e^{E_0\tau}\int_{-\pi}^\pi
\frac{ d\theta_1}{2\pi} \int_{-\pi}^\pi \frac{d\theta_2}{2\pi}
~e^{-H_2^{(1)}\tau+i \theta_1 N^{(1)}}~e^{-H_2^{(2)}\tau+i\theta_2
N^{(2)}}
\end{equation}

\ni where 

\begin{equation} N^{(r)}=\sum_n n\left(
a^{I(r)\dagger}_n a^{I(r)}_n+b^{(r)\dagger}_{an}b^{(r)}_{an}\right)
\end{equation}

\ni with $r=1,2$ are the level number operators for the two
intermediate strings.  The net effect of the operators in the above
equation is to make the replacement
$\left(a_n^{(r)\dagger},b_n^{(r)\dagger}\right)\to\left(
e^{-\omega_n\tau+in\theta_{(r)}} a_n^{(r)\dagger},
e^{-\omega_n\tau+in\theta_{(r)}}b_n^{(r)\dagger}\right)$ for all
creation operators which lie to the right of the projector. Then,
after the matrix element is computed, we multiply it by $e^{E_0\tau}$
and integrate over $\tau$ and $\theta_r$.  Any potential divergences
come from the region near $\tau=0$. }. Equation (\ref{okok}) can be
written as 

\be =-{1\over 8} \left<
Q_{3\beta_1\dot\beta_2}\frac{P}{E_0-H_2}\left[H_2,Q_3^{\beta_1\dot\beta_2}
\right]\right> -{1\over 8} \left<
Q_{3\dot\beta_1\beta_2}\frac{P}{E_0-H_2}\left[H_2,Q_3^{\dot\beta_1\beta_2}
\right]\right> \label{com} 
\ee 

\ni where we remind the reader of (\ref{subtelty}), which ensures that
$H_2$ acting on the external state gives a positive $E_0$. Up to order
$g_2$ the following equation holds

\be
\left[H_2,Q_3^{\beta_1\dot\beta_2}\right]=\left[Q_2^{\beta_1\dot\beta_2},H_3\right]
\ee 

\ni so that (\ref{com}) becomes 

\be ={1\over 8} \left<
Q_{3\beta_1\dot\beta_2}\frac{P}{E_0-H_2}\left[H_3,Q_2^{\beta_1\dot\beta_2}\right]
\right> +{1\over 8} \left<
Q_{3\dot\beta_1\beta_2}\frac{P}{E_0-H_2}\left[H_3,Q_2^{\dot\beta_1\beta_2}\right]
\right>. 
\ee 

\ni Since $Q_2$ commutes with $H_2$ one has 

\bea =&+&{1\over 8} \left<
Q_{2\beta_1\dot\beta_2}Q_3^{\beta_1\dot\beta_2}\frac{P}{E_0-H_2}
H_3\right> +{1\over 8} \left<
Q_{2\dot\beta_1\beta_2}Q_3^{\dot\beta_1\beta_2}\frac{P}{E_0-H_2}
H_3\right> \cr &+&{1\over 8} \left<
Q_{3\beta_1\dot\beta_2}\frac{P}{E_0-H_2} H_3
Q_2^{\beta_1\dot\beta_2}\right> +{1\over 8} \left<
Q_{3\dot\beta_1\beta_2}\frac{P}{E_0-H_2} H_3
Q_2^{\dot\beta_1\beta_2}\right> \cr &-&\left< H_3 \frac{P}{E_0-H_2}
H_3\right> 
\eea 

\ni and the last term cancels the $H_3$ contribution to
the energy shift. The final expression for the energy shift is

\begin{eqnarray}\label{newnice}
\delta E^{(2)} =&+&{1\over 8} \left<
Q_{2\beta_1\dot\beta_2}Q_3^{\beta_1\dot\beta_2}\frac{P}{E_0-H_2}
H_3\right> +{1\over 8} \left<
Q_{2\dot\beta_1\beta_2}Q_3^{\dot\beta_1\beta_2}\frac{P}{E_0-H_2}
H_3\right> \cr &+&{1\over 8} \left<
Q_{3\beta_1\dot\beta_2}\frac{P}{E_0-H_2} H_3
Q_2^{\beta_1\dot\beta_2}\right> +{1\over 8} \left<
Q_{3\dot\beta_1\beta_2}\frac{P}{E_0-H_2} H_3
Q_2^{\dot\beta_1\beta_2}\right> \cr &+&{1\over 4} \left<
Q_{2\beta_1\dot\beta_2}Q_4^{\beta_1\dot\beta_2} \right> +{1\over
4} \left< Q_{2\dot\beta_1\beta_2}Q_4^{\dot\beta_1\beta_2} \right>
\cr &+&{1\over 4} \left< Q_{4\beta_1\dot\beta_2}
Q_2^{\beta_1\dot\beta_2}\right> +{1\over 4} \left<
Q_{4\dot\beta_1\beta_2} Q_2^{\dot\beta_1\beta_2}\right>.
\end{eqnarray}

\ni It is amusing to note that the vanishing energy correction for a
supersymmetric external state is manifest in (\ref{newnice}), since if
$Q_2$ annihilates the external state, all of the terms are identically
zero. As was mentioned in section \ref{sec:contact}, $Q_4$ is unknown
and it is consistent with the closure of the super-algebra to set it
to zero here. Further, for the calculations at hand here, the
``dot-undot'' terms are identical to the ``undot-dot'' terms and so we
continue to simply use double the latter.

Using the $\left|[{\bf 9}, {\bf 1}]\ra^{(ij)}\right.$ external state,
we can check that what is left is manifestly convergent for the four
impurity channel, and then show that the addition of impurities will
not disturb this, leaving $ {\cal O}(\lambda')$ contributions at every
order in impurities.  We have two sorts of terms in (\ref{newnice}),
which we can represent schematically as follows

\begin{equation}
\label{newstuf}
\delta E_1 = \sum_I \frac{ \Big( \la \Phi | \la I | Q_3 \ra \Big)
\Big( \la \Psi | \la I | H_3 \ra \Big)^* } {E_{\Phi} - E_{I}} \qquad
\delta E_2 = \sum_I\frac{ \Big( \la \Phi | \la I | H_3 \ra \Big) \Big(
\la \Psi | \la I | Q_3 \ra \Big)^* } {E_{\Phi} - E_{I}}
\end{equation}

\ni where $|\Phi\ra$ is the $\left|[{\bf 9}, {\bf
1}]\ra^{(ij)}\right.$ external state, $|\Psi\ra = Q_2 |\Phi\ra $, and
$|I\ra$ is a level-matched, two-string intermediate state. In order
to evaluate the convergence and large $\mu$ behaviour of these terms,
we can be entirely schematic. We take (see (\ref{Q2BMN}) for the
expression of $Q_2$ in the BMN basis)

\begin{equation}
| \Psi \ra \sim \sqrt{-\mu\alpha_3} \,\b^\dag_n \, \a^\dag_{-n} | \a_3 \ra
\qquad | \Phi \ra \sim \a^\dag_n \, \a^\dag_{-n} | \a_3 \ra
\end{equation}

\ni while for the purpose of evaluating convergence we can take

\begin{equation}\label{conv}
G^{(1)}_p \sim \frac{1}{\sqrt{p}} \qquad K^{(1)}_{-p} \sim \mbox{constant} \qquad
{\widetilde N}^{3\,r}_{n\, p} \sim \frac{1}{p} \qquad
{\widetilde N}^{r\,s}_{q\, p} \sim \frac{1}{p + q}
\end{equation}

\ni where we take all integers to be positive. Let us begin with $\d
E_1$ in (\ref{newstuf}), we have two choices for four impurity
intermediate states

\bea
\label{inter}
&&|I\ra \sim \a^\dag_{p_1}\b^\dag_{p_2}\a^\dag_{p_3}\a^\dag_{p_4} |
\a_1 \ra | \a_2 \ra \cr &&|I\ra \sim
\a^\dag_{p_1}\b^\dag_{p_2}\b^\dag_{p_3}\b^\dag_{p_4} | \a_1 \ra |
\a_2 \ra .
\eea

\ni We can proceed with the first one, which will give

\bsp
\delta E_1 \sim
\sqrt{x}
\sum_{p_1\,p_2\,p_3\,p_4}\frac{1}
{2\,r\,\omega_n - \sum_{i=1}^4\,\omega_{p_i}}\,
\la \a_3 | \a_n \, \a_{-n} \la \a_2 | \la \a_1
| \a_{p_1}\, \b_{p_2} \,\a_{p_3}\, \a_{p_4}\, |Q_3\ra\\
\times \Big( \la \a_3 | \b_n \, \a_{-n} \la \a_2 | \la \a_1
| \a_{p_1}\, \b_{p_2} \,\a_{p_3}\, \a_{p_4}\, |H_3\ra \Big)^*~~~~~~
\end{split}
\ee

\ni where $x = -\mu\alpha_3$ and $\sum_i p_i=0$.  There are two
general ways in which we can contract the $\b^{(r)}$'s. They can
connect to factors of $\sum_m G_m \b^\dag_m$ in the prefactors of
$|H_3\ra$ and $|Q_3\ra$, or they can pair-up to bring down a factor of
${\wh Q}^{r\,s}_{m\, p}$ from the exponential. As far as convergence
and large $x$ power-counting is concerned however, $G^{(r)}_m
G^{(s)}_p$ is equivalent to ${\wh Q}^{r\,s}_{m\, p}$, and so we will
simply use the former. When contracting $\b^{(3)}$'s there is a
fundamental difference between $G^{(3)}_n G^{(r)}_p$ and ${\wh
Q}^{3\,r}_{n\, p}$, as far as large $x$ behaviour is concerned, because
of the pole in the latter. In fact ${\wh Q}^{3\,r}_{n\, p}$ is
essentially equivalent to ${\widetilde N}^{3\,r}_{n\, p}$ and
therefore the two can be interchanged in this analysis.

Because $K_{-p}$ goes as a constant for large $p$, the worst
convergence will always be realized by contracting the
intermediate bosonic impurities with the prefactors of $|H_3\ra$ and
$|Q_3\ra$. These contractions will yield\footnote{ Note that any
contraction which would yield a delta function on the external
state's spacetime indices is naturally zero here because we have
chosen to analyze the traceless symmetric $|[{\bf 9}, {\bf
1}]\ra^{(ij)}$ state. It is a simple matter to analyze the trace
state of section \ref{sec:trace} here, and one finds convergence as
well, however the number of (inverse) powers of summation
variables will be 4 in the worst case, and thus the convergence is
marginal. In no case does $\sqrt{\lambda'}$ behaviour occur here.}

\bsp
\label{bb}
\delta E_1 \sim
\sqrt{x}\sum_{p_1\,p_2\,p_3\,p_4 }
\frac{
G^{(1)}_{p_2} {\wt N}^{3\,1}_{-n\, p_1} K^{(1)}_{-p_3} {\wt N}^{3\,1}_{n\, p_4}
\times  K^{(1)}_{-p_3} K^{(1)}_{p_4} {\wt N}^{3\,1}_{-n\, p_1}
\begin{cases}
{\widetilde Q}^{3\,1}_{n\, p_2} - {\widetilde Q}^{1\,3}_{p_2\, n} \\
G^{(3)}_n G^{(1)}_{p_2}
\end{cases}}
{2\,r\,\omega_n - \sum_{i=1}^4\,\omega_{p_i}}
\end{split}
\ee

\ni Taking $p_4 = -(p_1+p_2+p_3)$, and using (\ref{conv}) we see that

\bea
&&\delta E_1 \sim \sum_{p_1\,p_2\,p_3} \frac{1}{(p_1+p_2+p_3)^2} \frac{1}{p_1^2}
\begin{cases}
\frac{1}{p_2^{3/2}}\cr
\frac{1}{p_2}\end{cases}
\eea

\ni where all $p_i$ are considered absolute valued, or equivalently
the sum considered over positive integers. This is manifestly
convergent. Continuing on to evaluate the leading $x$ dependence, for
the top choice in (\ref{bb}) we have poles for all three summation
variables, while in the large $x$ limit the $K$'s go as constants,
$G\sim 1/\sqrt{x}$ and the energy denominator is linear in $x$, thus
giving $\delta E_1 \sim 1/x$. For the bottom choice in (\ref{bb}),
$p_1$ and $p_3$ have poles, while the sum over $p_2$ must be executed
using (\ref{nopoles}).  The scaling turns out identical however. Thus
$\delta E_1/\mu$ is convergent and ${\cal O}(\lambda')$. One can
repeat this argumentation for the second intermediate state in
(\ref{inter}) and find the same behaviour. Also the entire exercise
may be repeated for $\delta E_2$ in (\ref{newstuf}) using the
following intermediate states

\bea\label{meme} &&|I\ra \sim
\a^\dag_{p_1}\a^\dag_{p_2}\a^\dag_{p_3}\a^\dag_{p_4} | \a_1 \ra |
\a_2 \ra \cr &&|I\ra \sim
\a^\dag_{p_1}\a^\dag_{p_2}\b^\dag_{p_3}\b^\dag_{p_4} | \a_1 \ra |
\a_2 \ra \eea

\ni and one discovers the same behaviour. The essential point is that
we will always have at least 5 (inverse) powers of the summation
variables, while the number of summation variables is 3. Alternate
positionings of the oscillators in the intermediate states such as $
|I\ra \sim \a^\dag_{p_1}\a^\dag_{p_2} | \a_1 \ra \,
\a^\dag_{p_3}\a^\dag_{p_4} | \a_2 \ra $ only improves the
convergence, since level matching removes one more summation variable
in these cases.

We can now consider adding additional pairs of fermionic and bosonic
impurities to the intermediate state $|I\ra$. This will add two
factors of ${\wt N}^{1\,1}_{p_i\,p_j}$ or two factors of
$G^{(1)}_{p_i} G^{(1)}_{p_j}$ (or equivalently two factors of $\wh
Q^{1\,1}_{p_i p_j}$). Either way the number of powers of summation
variables increases in concert with the number of summation variables,
preserving the convergence. Similarly the leading behaviour in
$\lambda'$ is unaffected. So it would seem that there are ${\cal
O}(\lambda')$ contributions to $\delta E^{(2)}/\mu$ at every order in
impurities, however any non-perturbative $\sqrt{\lambda'}$ behaviour is
absent.

% ========================================================================== %
\subsection{Summary and conclusions}

We have presented an important set of results regarding the impurity
non-conserving channel in light-cone string field theory on the
plane-wave. The original expectation \cite{Roiban:2002xr} that the
four-impurity channel would lead as $\sqrt{\l'}$ has been
contradicted; we find that this behaviour (for any number of
intermediate impurities) is a manifestation of log-divergent mode
number sums present in equal and opposite amounts in the standard
$H_3$ and contact terms of the string field theory. This result is
pleasing for two reasons: 1) because string amplitudes must be finite
and 2) because $\sqrt{\l'}$ behaviour would present a serious
challenge for reproduction in the gauge theory. A further result of
our analysis is that, generically, all intermediate states contribute
to the leading $\l'$ term in the mass shift. This result is disturbing
because the prospects of calculating the full shift for all channels
is at least daunting, if not impossible. On the other hand it may
explain the discrepancy between the impurity conserving result and
that from the gauge theory. Physically, it seems non-sensical that an
intermediate string with an arbitrarily high energy is equally as
important as one whose energy is commensurate with the external state
whose mass is receiving the correction. Indeed, at some point one
would have to concern themselves with backreaction on the
geometry. The analysis in section (\ref{sec:arbimp}) is very generic; a
cancellation mechanism could be hiding in the vertices which kill off
powers of $\l'$ as the number of intermediate-state impurities is
raised. This is a very interesting direction to explore. 

The reader may be concerned about details having been swept under the
rug. We mentioned various intermediate states which we claimed,
without demonstration, were non-divergent. One might also hope to find
the coefficients of the leading $\l'$ term to be numerically
suppressed as the impurity non-conservation is increased. In section
\ref{sec:dlcq}, an honest four-impurity calculation is presented where
all contributions to the leading $\l'$ result have been properly
rendered. The result is non-divergent, non-zero, and not obviously
suppressed numerically.

Finally, we also find that the reflection symmetry factor of
$\frac{1}{2}$ dressing the $H_3$ term in \cite{Roiban:2002xr}, and
argued incorrect in \cite{Gutjahr:2004dv}, must indeed be incorrect:
dressing this term ruins the cancellation of log-divergences and
renders the theory in-finite.
 
% ************************************************************************** %
\section{Calculation of the mass-shift via alternate vertices}
\label{sec:altvert}

This section is a presentation of the author's original work published
in \textsf{arXiv:hep-th/0605080} \cite{Grignani:2006en}. \\

The construction of the light-cone string field theory given in
section \ref{sec:detintvert}, i.e. (\ref{H3andQ3full}), is not
unique. The construction was guided by ensuring conservation of
momentum, or that the string worldsheets touch at $\t=0$, which gave
the exponential factors $|V\ra$, and by the requirement that the
supersymmetry algebra was obeyed, which determined the prefactors. The
exponential factors are the unique method of ensuring (super)-locality
while the prefactors in (\ref{H3andQ3full}) are but one possible
solution. The literature contains two others, one due to Di Vecchia,
Petersen, Petrini, Russo, and Tanzini or DVPPRT
\cite{DiVecchia:2003yp} and another due to Dobashi and Yoneya or DY
\cite{Dobashi:2004nm}. The vertex we have used thus far was developed
by Spradlin, Volovich \cite{Spradlin:2002ar, Spradlin:2002rv}, and by
Stefanski and Pankiewicz \cite{Pankiewicz:2002tg, Pankiewicz:2003kj}
and so we refer to it as SVPS. In \cite{Grignani:2006en}, the author
of this thesis calculated the impurity-conserving channel contribution
to the mass shift stemming from these three choices of vertex. It was
found that all vertices respected the same divergence cancellation
mechanism and are therefore finite and lead as $\l'$. The DY vertex
produced the best agreement with gauge theory, correctly reproducing
the leading ${\cal O}(g_2^2\l')$ term.

% ========================================================================== %
\subsection{The DVPPRT vertex}

% -------------------------------------------------------------------------- %
\subsubsection{Introduction}

In the construction of the SVPS vertex (\ref{H3andQ3full}), an
important point was glossed over. The symmetry group of the plane-wave
background (\ref{ppmet}) is broken from $SO(8)$ to $SO(4)\times
SO(4)\times \bZ_2$ by the presence of the Ramond-Ramond field. The
presence of such a field is well-known for complicating the
quantization of worldsheet fermions; here it creates an ambiguity in
the $\bZ_2$-parity of the fermionic ground state. The trouble is found
in the strange re-organizing of the fermionic modes given in
(\ref{fnm}), (\ref{fnmm}). In fact, the original treatment
\cite{Metsaev:2002re} followed a more usual procedure, defining
creation and annihilation operators $\{\theta^a_n, \theta^{b\,\dag}_m\}
= \d^{ab}\d_{mn}$

\be
\vartheta^a_n = \sqrt{\frac{\a'|n|}{2\,\o_n |\a|}} \left(\theta^a_n + 
\theta^{a\,\dag}_{-n} \right), \qquad
\l_n^a =  \sqrt{\frac{\o_n |\a|}{2\,\a'|n|}} \left(\theta^a_{-n} + 
\theta^{a\,\dag}_{n} \right)
\ee

\ni and a ``vacuum'' state

\be
\theta_n^a |\wt 0\ra = 0, \qquad a_n^I |\wt 0\ra = 0
\ee

\ni which is precisely what one would do in flat space. In the
plane-wave background, this ``vacuum'' is not a zero-energy state,
indeed

\be
H_2 \,|\wt 0\ra = 4\m \,|\wt 0\ra.
\ee

\ni The vacuum (\ref{PANKvacc}) is related to this vacuum via

\be
|0;\a\ra = \theta_0^5 \,\theta_0^6 \,\theta_0^7 \,\theta_0^8 |\wt 0\ra
\ee

\ni i.e. the difference lies in the fermion zero modes. In
\cite{Chu:2002eu}, it was noted that the two vacuua have opposite
$\bZ_2$ parity for this reason. In attempting to preserve a smooth
limit to flat space, the SVPS construction chooses

\be
\bZ_2 \, |\wt 0\ra =  |\wt 0\ra, \qquad
\bZ_2 \, |0;\a\ra = - |0;\a\ra
\ee

\ni and so must use $\bZ_2$-odd prefactors in the interaction vertices
(\ref{H3andQ3full}). The DVPPRT vertex chooses the opposite, forsaking
the smooth continuation to flat space as $\m \rightarrow 0$, and
requiring $\bZ_2$-even prefactors. These they construct in the
simplest possible way, by directly employing the quadratic Hamiltonian
and supercharges

\be
\begin{split}
| H_3^{\text{DVPPRT}} \ra = \theta \, H_2 |V\ra = f\,\frac{g_2\, \a'}{16
  \,\a_3^3} \left( K^I K^I + \wt K^I \wt K^I + \text{fermions} \right) |V\ra\\ |
  Q_{3\,\b_1 \db_2}^{\text{DVPPRT}} \ra = \theta \, Q_{2\,\b_1\db_2}
  |V\ra = f\,\frac{g_2\,\eta}{4\,\a_3^3} \sqrt{-\frac{\a' \k}{2}} \left(
  K^{\dg_1}_{\b_1} \, Z_{\dg_1\,\db_2} - i K^{\g_2}_{\db_2} \,
  Y_{\b_1\,\g_2} \right)|V\ra
\end{split}
\ee

\ni where $\theta = -g_2 \,f\, r(1-r)/4$, and we have not explicitly
calculated the fermionic portion of the $H_3$ prefactor as it will not
concern us in the following calculations. It is obvious that these
vertices obey the superalgebra; they have inherited that property from
the free generators $H_2$ and $Q_2$.

% -------------------------------------------------------------------------- %
\subsubsection{Divergence cancellation}

\ni We would like to verify that the divergence cancellation mechanism
found in section \ref{sec:trace} for the SVPS vertex is also at play
here. Unlike the SVPS case, the $H_3$ divergence does not stem from
the two-bosonic-impurity intermediate state. There is, however,
another divergence that was not present in the SVPS case. It is due to
the contribution coming from matrix elements with two fermionic
impurities in the intermediate state. In particular, the relevant
matrix elements are given by

\bsp\label{ferm}
\langle \alpha_3 | \alpha^i_{n} \alpha^i_{-n} \,
\langle &\alpha_2 | \langle \alpha_1 | \beta_{p(1)}^{\a_1\a_2}
\beta_{-p(1)\b_1\b_2} |H^{\text{DVPPRT}}_3 \rangle =\\  
&4\,g_2 r\,(1-r)\left(
\frac{\omega_{n}^{(3)}}{\alpha_3} +
\frac{\omega_{p}^{(1)}}{\alpha_1} \right){\widehat
Q}_{-p\,p}^{1\,1}{\widetilde
N}_{-n\,n}^{3\,3}\delta_{\b_1}^{\a_1} \delta_{\b_2}^{\a_2} 
\end{split}
\ee

\noindent and similarly for the intermediate state with dotted
indices. The divergent contribution to the energy shift coming from
these matrix elements is found (by taking the large $p$ limits of the
summands) to be

\begin{equation}\label{diveccH3div}
\delta E^{\text{div}}_{H_3} \sim -\frac{1}{2}\int_0^1 dr\,
\,\frac{g_2^2\,r(1-r)}{r\,|\alpha_3|\,\pi^2} \left( {\widetilde
N}^{3\,3}_{n\,-n} \right)^2 \, \sum_p \frac{1}{|p|}.
\end{equation}

\ni The contribution from the contact term stems from the following
matrix element

\bea \label{contactmed} &&\left( g_2 \frac{\eta}{4}
\sqrt{\frac{r\,(1-r)\, \alpha'}{-2\,\alpha_3^3}} \right)^{-1}
\langle \alpha_3 | \alpha^i_{n} \alpha^i_{-n} \, \langle \alpha_2 |
\langle \alpha_1 | \alpha^{K\,(1)}_{p} \beta^{(1)\,
\Sigma_1\,\Sigma_2}_{-p} |Q^\text{DVPPRT}_{3\, \beta_1 {\dot \beta}_2} \rangle
=\cr && 2\biggl( G_{|p|}^{(1)} \, K_{-n}^{(3)} {\widetilde
N}^{3\,1}_{n\,p} + G_{|p|}^{(1)} \, K_{n}^{(3)} {\widetilde
N}^{3\,1}_{-n\,p}
 \biggr)( \sigma^k)^{\dot{\sigma_1}}_{ \beta_1 } \delta^{{\dot \sigma_2}}_{{\dot \beta_2}}
+8\, G_{|p|}^{(1)} \, K_{p}^{(1)} {\widetilde N}^{3\,3}_{n\,-n} (
\sigma^K)^{\Sigma}_{ \beta } \delta^{\Sigma}_{\beta}. \eea

\noindent The divergent contribution to the energy shift is found to
be

\begin{equation}
\delta E^{\text{div}}_{H_4} \sim +\int_0^1 dr\,
\,\frac{g_2^2\,r(1-r)}{r\,|\alpha_3|\,\pi^2} \left( {\widetilde
N}^{3\,3}_{n\,-n} \right)^2 \, \sum_{p>0} \frac{1}{p}.
\end{equation}

\noindent Noting that in the $H_3$ contribution the divergence is
found for both positive and negative $p$, while in the $H_4$
contribution the divergence occurs only for negative $p$, and hence
a relative factor of 2 is induced in the $H_3$ term, one sees that
the logarithmically divergent sums cancel identically between the
$H_3$ and contact terms, leaving a convergent sum. This result can
be generalized to arbitrary impurity channels, as was done for the
SVPS case in section \ref{sec:arbimp}.

% -------------------------------------------------------------------------- %
\subsubsection{Impurity-conserving mass-shift}

We now present the calculation of the impurity-conserving channel
contribution to the mass shift of the $|[{\bf 9},{\bf 1}]\ra$ state
(\ref{91}). The general method is outlined in detail in appendix
  \ref{app:example}, section \ref{appsec:example}. Beginning with the
  $H_3$ term of the mass-shift, we find\footnote{There is an implicit
    division of the energy-shift $\d E$ by the parameter $\m$ in the
    remainder of the text. We have also dropped the integration
    $\int_0^1 dr$, which is implied in all subsequent amplitudes.}

\be\label{h3diveccIN}
\begin{split}
\d E_{H_3}^{\text{DVPPRT}}=
\frac{2}{r(1-r)}\frac{g_2^2\,\a'^2}{64\,\a_3^6}
\sum_{r_1\,r_2}\sum_{q_1\,q_2} &\Biggl[ \left( \wt
L_{n\,q_1}^{3\,r_1} \right)^2 \left( \wt N_{-n\,q_2}^{3\,r_2}
\right)^2 + \wt L_{n\,q_1}^{3\,r_1} \wt L_{n\,q_2}^{3\,r_2} \wt
N_{-n\,q_2}^{3\,r_2} \wt N_{-n\,q_1}^{3\,r_1} \\ &+ \wt
L_{-n\,q_1}^{3\,r_1} \wt L_{n\,q_1}^{3\,r_1} \wt
N_{n\,q_2}^{3\,r_2} \wt N_{-n\,q_2}^{3\,r_2} + \wt
L_{-n\,q_1}^{3\,r_1} \wt L_{n\,q_2}^{3\,r_2} \wt
N_{n\,q_2}^{3\,r_2} \wt N_{-n\,q_1}^{3\,r_1} \Biggr] \\ &\times
\frac{-\a_3\left (\d^{r_1\,r_2}\d_{q_1+q_2}
+ (1 - \d^{r_1\,r_2})\d_{q_1}\d_{q_2}\right)}{2\,\o_{n}  - \beta_{r_1}^{-1} \o_{q_1} -
\beta_{r_2}^{-1} \o_{q_2}} + (n \lr -n)
\end{split}
\ee

\ni where the $L_{n\,q}^{3\,r}$'s are defined in (\ref{defLs}). The
sums are evaluated using (\ref{howtodosums}); the result is\footnote{A
systematic code was developed to take input of the form
(\ref{h3diveccIN}) and to produce output of the form
(\ref{h3diveccOUT}). The code was used for all calculations in this
section. It correctly reproduces by-hand calculations and so we are
confident it is accurate.}

\be\label{h3diveccOUT}
\begin{split}
\d E_{H_3}^{\text{DVPPRT}}=
\frac{g_2^2}{32\pi^2}&\Biggl[ -\left(\frac{2}{3}+\frac{5}{4\pi^2 n^2}\right)\l' 
+3\left(\frac{1}{\pi^2}+\frac{1}{2\pi}\right)\l'^{3/2}
+n^2\left(1-\frac{9}{8\pi^2 n^2}\right)\l'^2 \\
&-5 n^2\left(\frac{2}{\pi^2}+\frac{3}{4\pi}\right)\l'^{5/2}
-5n^4\left(\frac{1}{4}-\frac{21}{32\pi^2 n^2}\right)\l'^3\\
&+n^4\left(\frac{105}{16\pi}+\frac{94\,}{5\pi^2}\right)\l'^{7/2}
+{\cal O}(\l'^4)
\Biggr].
\end{split}
\ee

\ni Continuing with the contact term, we find

\be\label{q3diveccIN}
\begin{split}
\d E_{H_4}^{\text{DVPPRT}}=
-\frac{g_2^2\,\a'}{16\,\a_3^3} \sum_{r_1\,r_2}\sum_{q_1\,q_2}
&\Biggl[
 \left( K_{n} \right)^2  \left( G_{q_1} \right)^2  \left( \wt N_{-n\,q_2}^{3\,r_2}
\right)^2
+ K_{n} K_{-n}  \left( G_{q_1} \right)^2 \wt N_{n\,q_2}^{3\,r_2} 
\wt N_{-n\,q_2}^{3\,r_2} \Biggr] \\
&\times \left (\d^{r_1\,r_2}\d_{q_1+q_2}
+ (1 - \d^{r_1\,r_2})\d_{q_1}\d_{q_2}\right)
+ (n \lr -n)
\end{split}
\ee

\ni with result

\be\label{q3diveccOUT}
\begin{split}
\d E_{H_4}^{\text{DVPPRT}}=
\frac{g_2^2}{32\pi^2}&\Biggl[ \left( \frac{1}{3}+\frac{5}{8\pi^2
    n^2}\right) \l' 
-\frac{3}{2}\left(\frac{1}{\pi^2}+\frac{1}{2\pi}\right)\l'^{3/2}
-n^2\left(\frac{1}{6}-\frac{19}{16\pi^2 n^2}\right)\l'^2 \\
&+n^2\left(\frac{11}{4\pi^2}+\frac{9}{8\pi}\right)\l'^{5/2}
+\frac{n^4}{8}\left(1 - \frac{105}{8\pi^2 n^2}\right)\l'^3\\
&-n^4\left(\frac{45}{32\pi}+\frac{73\,}{20\pi^2}\right)\l'^{7/2}
+{\cal O}(\l'^4)
\Biggr].
\end{split}
\ee

\ni Adding the contributions from the $H_3$ and contact terms, we find
the complete shift to be \cite{Grignani:2006en}

\be\boxed{
\begin{split}
\d E^{\text{DVPPRT}} &=  \frac{g_2^2}{4\pi^2}\Biggl[
-\left(\frac{1}{24} + \frac{5}{64\pi^2 n^2}\right)\l'
+\frac{3}{16}\left(\frac{1}{\pi^2}+\frac{1}{2\pi}\right)\l'^{3/2}\\
&\qquad+ n^2\left(\frac{5}{48}+\frac{1}{128\pi^2 n^2}\right)\l'^2
-n^2\left(\frac{29}{32\pi^2}+\frac{21}{64\pi}\right)\l'^{5/2}\\
 & \qquad+ n^4\left(-\frac{9}{64}+\frac{105}{512\pi^2
n^2}\right)\l'^{3} + n^4 \left( \frac{303}{160\pi^2} +
\frac{165}{256\pi}\right)\l'^{7/2} +{\cal O}(\l'^4) \Biggr]. 
\end{split}}
\ee

\ni This result does not fare very well in agreeing with the gauge
theory result (\ref{gaugeresult}). It would seem that the DVPPRT
vertex is either not correct or not complete based on this test of
it. In the next section we will repeat the calculation using still
another vertex. There we will find the best agreement with gauge
theory yet found.

% ========================================================================== %
\subsection{The ``holographic'' DY vertex}

% -------------------------------------------------------------------------- %
\subsubsection{Introduction}

The basis of the AdS/CFT correspondence is the GKP-W relation
(\ref{GKPW}) discussed in section \ref{sec:GKPW}. In the plane-wave
background, a set of supergravity states is singled-out from the full
set of $AdS_5 \times S^5$ states. These plane-wave states have large
angular momentum $J$ on the five-sphere. From the perspective of the
$AdS_5$ space, these are very massive Kaluza-Klein states, with $m^2 =
J(J-4)$. The picture developed in section \ref{sec:GKPW} was that the
two point function of the dual CFT operators should be envisaged as a
process whereby the insertion of the first operator on the boundary of
$AdS_5$ causes the propagation of a supergravity mode into the bulk,
which then turns back again and joins the second operator. For a
sufficiently heavy supergravity mode, this propagation does not stray
far from the classical geodesic joining the insertion points of the
operators on the boundary. Technically, this means that the
semi-classical action is dominated by a saddle which is the geodesic
trajectory. Since in the plane-wave limit we are taking the mass of
the supergravity modes to be very large, such a semi-classical
treatment should be valid.

When we introduced the GKP-W relation in section \ref{sec:GKPW}, we
chose to analyze the equations in Euclidean signature. In fact, we
will see below that the picture of propagation from boundary to
boundary makes little sense without requiring this signature. Recall
that the metric of $AdS_5$ in global coordinates may be expressed as

\be\label{globads2}
ds^2 = \frac{1}{\cos^2\theta} \left(-dt^2 + d\theta^2 +
\sin^2\theta \,d \O_{3}^2\right) .
\ee

\ni To solve the massive geodesics (for pure radial motion, i.e. $d
\O_{3}^2=0$) is not terribly difficult. First we impose the constraint
that the five-velocity squares to -1, while we also note that
independence of $t$ gives $\dot t = \xi \cos^2 \theta$, then

\be\label{massgeo}
-\cos^2\theta = -\xi^2 \cos^4\theta + \dot\theta^2 \qquad \rightarrow
\qquad
\frac{\dot \theta}{\dot t} =\frac{d\theta}{d t} = \frac{\sqrt{\cos^2
    \theta - \xi^{-2}}}{\cos\theta}.
\ee

\ni Integrating we find 

\be\label{lorentz}
\sin \theta = \sqrt{1-\xi^{-2}} \, \sin t, \qquad \xi > 1.
\ee

\ni The boundary of $AdS_5$ sits at $\theta = \pi/2$, or $\sin \theta
= 1$. Thus a massive geodesic never reaches the boundary; it turns
back into the bulk at some $\theta < \pi/2$. Further, if we allowed
for some angular motion, we would find that at its closest approach,
the turning point, the particle's motion is {\it parallel} to the
boundary. This seems to contradict the picture of particles
originating from the boundary and propagating into the bulk.

Now consider the massive geodesic in the Euclidean picture; the
negative signs in the equation on the left-hand side of
(\ref{massgeo}) will be flipped to positive. The right-hand side
becomes

\be
\frac{\dot \theta}{\dot t} =\frac{d\theta}{d t} = \frac{\sqrt{-\cos^2
    \theta + \xi^{-2}}}{\cos\theta}, \qquad \xi < 1
\ee

\ni which has as solution

\be
\sin \theta = \sqrt{\xi^{-2} - 1} \cosh\left(t-t_0\right), \qquad t_0
= \ln \sqrt{\xi^{-2} - 1}.
\ee

\begin{figure}[ht]
\begin{center}
\includegraphics*[bb=0 110 590 685, height=2.0in,width=2.75in]{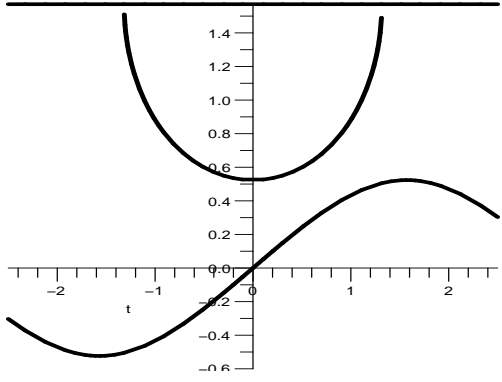}
\end{center}
\caption{Massive geodesics in $AdS$ are shown for Lorentzian
  (sinusoidal) and Euclidean (catenary) signatures. The axes are
  $\theta(t)$ vs. $t$, see (\ref{globads2}). The boundary of $AdS$ is
  the horizontal line at $\theta=\pi/2$. Only the Euclidean trajectory
  is consistent with the GKP-W picture.}
\label{fig:adsmassgeo}
\end{figure}

\ni This geodesic reaches to the boundary and terminates normal to it;
this is consistent with the GKP-W picture, see figure \ref{fig:adsmassgeo}.
\ni For this reason, the AdS/CFT correspondence is usually stated as a
relation between Euclidean CFT correlators and processes occurring in
Euclidean $AdS$. 

In \cite{Dobashi:2002ar}, these Euclidean geodesics were interpreted
as quantum mechanical tunnelling trajectories. Consider the
Poincar\'{e} patch of Lorentzian $AdS$

\be\label{ponyon}
ds^2 = R^2 \frac{dz^2}{z^2} + \frac{1}{R^2z^2} \left( d\vec x^2 - dt^2\right).
\ee

\ni The field equation for a scalar $\Phi(z,\vec x=0,t) = e^{i\o
t}\phi(z)$ of mass $m^2=J(J-4)$ is

\be\label{sceom}
\biggl( z^2 \p_z^2 - 3\,z\, \p_z + R^4 z^2 \o^3 - J(J-4) \biggr) \phi(z) = 0.
\ee

\ni Consider now applying a WKB approximation
$\phi(z)=G(z)\exp(iS(z))$; the leading function $G(z)$ is of order 1,
while the phase $S(z)$, being proportional to the potential is of
order $J$. Note also that in the plane-wave limit $J\sim R^2$. Keeping
the leading terms only, the result of plugging the WKB form into
(\ref{sceom}) is

\be
z^2\left(\frac{dS}{dz}\right)^2 - R^4 z^2 \o^2 + J^2 = 0,\qquad
\rightarrow \qquad \frac{dS}{dz} = \sqrt{R^4\o^2 - \frac{J^2}{z^2}}
\ee

\ni and therefore $S(z)$ is real only if $z^2 \geq J^2/(\o^2 R^4)$;
the boundary at $z=0$ is obviously excluded. However, the field is
free to tunnel to the boundary, i.e. we may let $S(z)$ become
imaginary. The tunnelling trajectory can then be found by noting

\be
\phi^*(z) P_z \phi(z) = \phi^*(z) (-i \p_z) \phi(z) 
= |\phi|^2 \frac{dS}{dz} 
\ee

\ni and therefore associating the momentum along $z$ to this via

\be
\frac{dS}{dz} = \frac{R^2}{z^2} \frac{dz}{d\t} 
\simeq \frac{J}{z^2} \frac{dz}{d\t} 
\ee

\ni where $\t$ is the affine parameter along the trajectory. This
gives (for the tunnelling solution)

\be
\frac{dz}{d\t} = \pm z \sqrt{1 - \frac{z^2\o^2R^4}{J^2}}
\ee

\ni which integrates to

\be\label{funy}
z = \frac{J}{R^2\o \cosh\t}
\ee

\ni reproducing the catenary shown in figure
\ref{fig:adsmassgeo}. Replacing $\t \rightarrow i\t$ in (\ref{funy})
reproduces the Lorentzian geodesic (\ref{lorentz}) or equivalently
corresponds to the propagating solution for $\phi(z)$. However, note
that since $J \sim d\psi/d\t$, where $\psi$ is an angle in $S^5$, and
$\o \sim d t/d \t$, where $t$ is the time coordinate of the boundary
CFT, such a rotation would need to be accompanied by a double Wick
rotation of the $AdS_5\times S^5$ metric in which both $t$ and $\psi$
become imaginary. This means that the tunnelling picture may be derived
from the standard Lorentzian $AdS_5\times S^5$ through double Wick
rotation of $\psi$ and $t$. 

In the work \cite{Dobashi:2004nm}, Dobashi and Yoneya continued this
picture of holography to a construction of the light-cone string field
theory vertices for the plane-wave background. They calculated the
effective action for a massive scalar field along the aforementioned
tunnelling trajectory. In their analysis, the first $SO(4)$ excitations
come directly from harmonic oscillator ground states where the
frequency of the harmonic oscillator is given by the mass of the
supergravity state (i.e. $m^2 = \D(\D-4)$). The other $SO(4)$,
associated with $D_i Z$ insertions in the BMN operators, stem from
excited states of these harmonic oscillators. The result is that the
cubic coupling of the excited states is dictated completely by the
cubic coupling of the ground states. This makes a definite prediction
for the zero-mode sector of the string field theory; the cubic
Hamiltonian $H_3$ ought to only count excitations of the first
$SO(4)$. Of course this explicitly breaks the $\bZ_2$ symmetry of the
plane-wave background. The perspective is that this symmetry is
``accidental'' from the point of view of holography, and indications
that it is not manifest at the level of CFT three-point functions were
discovered already in \cite{Chu:2003ji}. We will endeavour to give a
concise summary of the construction of the DY vertex. The effective
action for computing the three-point functions of SUGRA scalars was
worked-out in \cite{Lee:1998bx}. It is given by

\be\label{effsc}
S = \frac{4N^2}{(2\pi)^5} \int d^5 x \sqrt{-g} \left[ \frac{1}{2}\left(\nabla \phi^i
  \right)^2 + \frac{1}{2} m_i^2 \left(\phi^i\right)^2 - \frac{1}{3}
  G_{ijk} \phi^i \phi^j \phi^k \right]
\ee

\ni and leads to agreement (via the GKP-W relation) with three-point
functions of the dual CFT operators ${\cal O}_{\D_i}(x)$, where $m_i^2 =
\D_i(\D_i-4)$,

\be
\la {\cal O}_{\D_1}(x_1) {\cal O}_{\D_2}(x_2)  {\cal O}_{\D_3}(x_3)
\ra = \frac{C_{123}}{|x_1-x_2|^{2\a_3} |x_2-x_3|^{2\a_1}
  |x_3-x_1|^{2\a_2} }
\ee

\ni where $\a_1 = (\D_2 + \D_3 - \D_1)/2$, and similarly for $\a_2$
and $\a_3$. The coefficient $C_{123}$ is given by a
$\D_i$-dependent constant multiplied by the cubic coupling
$G_{123}$

\be
C_{123} = {\cal N}(\D_1,\D_2,\D_3) G_{123}.
\ee

\ni The strategy of Dobashi and Yoneya is to expand (\ref{effsc})
about the tunnelling trajectory, quantize the free part of the action
using creation/annihilation operators, and then to calculate the
matrix elements of the cubic Hamiltonian stemming from
(\ref{effsc}). Let $\t$ be the affine parameter along the tunnelling
trajectory, while $\vec y = (\vec x, z)$ are the fluctuations in the
given coordinates (see (\ref{ponyon})). The effective metric is then
\cite{Dobashi:2004nm}

\be
ds^2 = (1+ \vec y^2)d\tilde\t^2 + d\vec y ^2, \qquad \t = \tilde \t +
\frac{\vec y^2 }{2} \tanh\tilde \t 
\ee

\ni while the free part of the effective action becomes

\be
\frac{4 N^2}{(2\pi)^5} \int d\tilde \t d^4 y \left[ \left( 1 -
  \frac{1}{2} y^2 \right)\p_{\tilde\t} \bar
  \Phi_i \p_{\tilde\t}\Phi_i + \p_y \bar \Phi_i \p_y \Phi_i +  \left( 1 +
  \frac{1}{2} y^2 \right)\D_i(\D_i-4) \bar \Phi_i \Phi_i \right]
\ee

\ni where $\D = J + k_i$ is the dimension of the BMN operator with $k$
insertions of the ${\cal N}=4$ SYM scalar $\Phi^i_{{\cal
    N}=4}$. Rewriting the fields as

\be
\Phi_i = e^{-J\tilde\t} \phi_0^{(J)}(\vec y) \psi(\t), \qquad
\bar \Phi_i = e^{J\tilde\t} \phi_0^{(J)}(\vec y) \bar \psi(\t)
\ee

\ni where $\phi_0^{(J)}(\vec y)$ is the ground state wave function of
the operator $-\p_y^2 + J^2 \vec y^2$

\be
\phi_0^{(J)}(\vec y) = \left(\frac{J}{\pi}\right)^2
\exp\left(-\frac{1}{2}J\vec y^2\right)
\ee

\ni allows the $y$-directions to be integrated-out, leaving the
following free action for the $\psi(\t)$ fields

\be
\int d\t \sum_i \left[ \bar \psi_i \p_{\t}
  \psi_i - \p_{\t} \bar \psi_i \psi_i + k_i \bar \psi_i \psi_i \right]
\ee

\ni and the following form for the interaction

\be
\frac{1}{2} \int d\t \sum_{i_1,i_2,i_3} \l_{i_1,i_2,i_3} \left( \bar
\psi_{i_1} \psi_{i_2} \psi_{i_3} + \text{h.c.}\right)
\ee

\ni where $ \l_{i_1,i_2,i_3} = {\cal M}(\D_i) G_{i_1 i_2 i_3}$, where
$ {\cal M}(\D_i)$ is a constant dependent on the $\D_i$, and we take
$J_1+J_2 = J_3$ to conserve angular momentum.  By comparing $\l_{123}$
with $C_{123}$, a direct map may be made between the $AdS_5\times S^5$
couplings $C_{123}$ and the (what ought to be) plane-wave cubic
Hamiltonian coefficient $\l_{123}$. The result is that this coupling
is proportional to $k_2 + k_3 - k_1$, i.e. the quadratic Hamiltonian
counting the excitation energies of the BMN states

\be
\l_{123} \propto k_2 +k_3 -k_1.
\ee

\ni The other $SO(4)$'s worth of excitations, corresponding to
insertions of $D_iZ$ in the BMN operators, are conjectured to
correspond here to excited states of $\phi^{(J)}$

\be
\phi_n^{(J)} (\vec y) = \prod_{i=1}^4 \left( \frac{J}{\pi}
\right)^{1/4} \frac{2^{-n_i/2}}{\sqrt{n_i!}} H_{n_i}(\sqrt{J} \vec y)
  e^{-J y^2/2},
\ee

\ni where the excitation number $n_i$ corresponds to the insertion of
$n_i$ $D_iZ$'s in the BMN operator. The crucial element here is that
the couplings for the excited states are related directly to those
of the ground states via

\be
\l_{123}^{n_1 n_2 n_3} = \l_{123}^{000} \frac{\pi J_1}{J_2 J_3} \int d^4 y
\phi_{n_1}^{(J_1)}(\vec y) \phi_{n_2}^{(J_2)}(\vec y) \phi_{n_3}^{(J_3)}(\vec y).
\ee

\ni The result is that the cubic Hamiltonian is still proportional
only to the energies of the first $SO(4)$ excitations $k_2 + k_3 -
k_1$. 

At the end of the day, the vertex proposed by Dobashi and Yoneya must,
at the level of supergravity states (i.e. string zero modes), count
only the energies of the first $SO(4)$. This is accomplished by taking
an average of the $\bZ_2$-even prefactor of DVPPRT and the $\bZ_2$-odd
prefactor of SVPS. In this way the second $SO(4)$ zero modes
cancel-out. The proposal is then

\bsp
|H_3^{\text{DY}}\ra = \frac{1}{2} \left( |H_3^{\text{DVPPRT}}\ra +
|H_3^{\text{SVPS}}\ra \right)\\
|Q_3^{\text{DY}}\ra = \frac{1}{2} \left( |Q_3^{\text{DVPPRT}}\ra +
|Q_3^{\text{SVPS}}\ra \right)
\end{split}
\ee

% -------------------------------------------------------------------------- %
\subsubsection{Divergence cancellation}

The cancellation of divergences demonstrated in \cite{Grignani:2005yv}
for the SVPS vertex, was shown to extend to the DY vertex in
\cite{Grignani:2006en}. In fact, we now show that an arbitrary linear
combination of the SVPS and DVPPRT vertices,

\bea H^N_3 = \alpha \,H^{\text{SVPS}}_3 + \beta\, H^{\text{DVPPRT}}_3 \\
Q^N_3 = \alpha\, Q^{\text{SVPS}}_3 + \beta\, Q^{\text{DVPPRT}}_3 \eea

\noindent similarly yields a finite energy shift. We calculate the
mass shift of the trace state as in section \ref{sec:trace}. The
divergence stemming from the $H_3$ term is simply $\a^2$ times the
SVPS $H_3$ divergence (\ref{divH3}) plus $\b^2$ times the DVPPRT
divergence (\ref{diveccH3div}). The reason is simple - the SVPS
divergence stems from an entirely bosonic intermediate state, while
(\ref{diveccH3div}) results from an entirely fermionic one. This
precludes any divergences arising from cross-terms. We note that the
SVPS divergence (\ref{svpsdivH3}) is exactly equal to
(\ref{diveccH3div}), therefore we have

\be 
\delta E^{\mbox{div}}_{H^N_3} \sim -(\a^2+\b^2)
\frac{1}{2}\int_0^1 dr\,
\,\frac{g_2^2\,r(1-r)}{r\,|\alpha_3|\,\pi^2} \left( {\widetilde
N}^{3\,3}_{n\,-n} \right)^2 \, \sum_p \frac{1}{|p|}. 
\ee

\noindent The pieces of the SVPS $Q_3$ relevant to a two-impurity
channel calculation are exactly $Q_3^{\text{DVPPRT}}$ with $K \lr \wt
K$, therefore, from (\ref{divH4})

\bea \label{contactmey} &&\left( g_2 \frac{\eta}{4}
\sqrt{\frac{r\,(1-r)\, \alpha'}{-2\,\alpha_3^3}} \right)^{-1}
\langle \alpha_3 | \alpha^i_{n} \alpha^i_{-n} \, \langle \alpha_2 |
\langle \alpha_1 | \alpha^{K\,(1)}_{p} \beta^{(1)\,
\Sigma_1\,\Sigma_2}_{-p} |Q^{\text{DVPPRT}}_{3\, \beta_1 {\dot \beta}_2} \rangle
=\cr &&~~~ 2 G_{|p|}^{(1)} \,\biggl( \biggl[ \a
\,(K_{-n}^{(3)}{\widetilde N}^{3\,1}_{-n\,p}+K_{n}^{(3)}{\widetilde
N}^{3\,1}_{n\,p})+\b\,( K_{-n}^{(3)}{\widetilde N}^{3\,1}_{n\,p} +
K_{n}^{(3)}{\widetilde N}^{3\,1}_{-n\,p}) \biggr]
 ( \sigma^k)^{\dot{\sigma_1}}_{ \beta_1 } \delta^{{\dot \sigma_2}}_{{\dot
 \beta_2}}\cr
&&~~~~~+4\, (\b\,K_{p}^{(1)}+\a\,K_{-p}^{(1)}) {\widetilde
N}^{3\,3}_{n\,-n} ( \sigma^K)^{\Sigma}_{ \beta }
\delta^{\Sigma}_{\beta}\biggr) \eea

\noindent The last term in (\ref{contactmey}) gives rise to a
log-divergent sum, the large-$p$ behaviour of which is

\begin{equation}
\delta E^{\mbox{div}}_{H^N_4} \sim +(\a^2+\b^2)\int_0^1 dr\,
\,\frac{g_2^2\,r(1-r)}{r\,|\alpha_3|\,\pi^2} \left( {\widetilde
N}^{3\,3}_{n\,-n} \right)^2 \, \sum_{p>0} \frac{1}{p}.
\end{equation}

\noindent Thus, by the same arguments as section \ref{sec:trace}, the
energy shift is finite for arbitrary $\a$ and $\b$. The DY vertex uses
$\a=\b=1/2$, and this combination exclusively gives rise to the
agreement with gauge theory which will be presented in the next
sub-section. Again, as for the DVPPRT vertex, the generalization of
these arguments to the impurity non-conserving channels is a
straightforward application of the treatment given in section
\ref{sec:arbimp}.

% -------------------------------------------------------------------------- %
\subsubsection{Impurity-conserving mass-shift}
\label{sec:DYimpcons}

In order to verify the validity of our results, we use two different
methods for calculating the mass-shift. The first is straight-forward

\be\label{standard}
\d E =
\frac{\la H_3^{\text{DY}} | e \ra \la e | H_3^{\text{DY}} \ra}{E_0 - H_2^{\text{int}}} 
+ \frac{1}{4} \la Q_3^{\text{DY}} | e \ra \la e | Q_3^{\text{DY}}\ra 
\ee

\noindent where $|e\ra$ is the $|[{\bf 9, 1}]\ra$ external state
(\ref{91}), and where the superscript ``int'' refers to internal
states (i.e. strings number 1 and 2). For the second method, we recall
that

\be
|H_3^{\text{DY}}\ra = \frac{1}{2} \left(\theta H_2 \,|V\ra + |H_3\ra \right) \qquad
|Q_3^{\text{DY}}\ra = \frac{1}{2} \left(\theta Q_2 \,|V\ra + |Q_3\ra \right)
\ee

\noindent where $\theta = -g_2 r(1-r)/4$, and $|H_3\ra$ and $|Q_3\ra$
are the SVPS vertices (\ref{H3andQ3full}). Because of the simple form
of the DVPPRT vertices, a perhaps simpler form for the DY energy shift
can be derived. We begin by considering some matrix elements

\be
\la e | H_3^{\text{DY}} \ra = -\frac{\theta}{2}\,\Delta E \la e| V\ra 
+ \frac{1}{2} \la e | H_3\ra 
\ee

\be
\la e | Q_3^{\text{DY}} \ra = \frac{\theta}{2}\,\la e| Q_2^{(3)} |V\ra 
+\frac{\theta}{2}\,Q_2^{\text{int}} \la e |V\ra 
+ \frac{1}{2} \la e | Q_3\ra 
\ee

\noindent where $\Delta E = E_0 -
H_2^{\text{int}}$, where $E_0$ is the energy of the external
state. Plugging these into (\ref{standard}) we have, beginning with the
$H_3$ term

\be\label{H3yon}
\d E^{\text{DY}}_{H_3} =
\frac{\theta^2}{4} \, \Delta E \, \la V | e\ra\la e| V\ra
+\frac{1}{4}\,\d E^{\text{SVPS}}_{H_3} 
- \frac{\theta}{4} \Bigr( \la V | e \ra \la e | H_3 \ra + \text{h.c.} \Bigl)
\ee

\noindent and now for the contact term

\be\label{Q3yon}
\begin{split}
\d E^{\text{DY}}_{H_4} 
&= \frac{\theta^2}{16} \left| \la Q_2\, e| V \ra \right|^2
+ \frac{\theta^2}{16} \left| Q_2^{\text{int}} \la e| V \ra \right|^2
+\frac{1}{4}\,\d E^{\text{SVPS}}_{H_4}\\
&+ \frac{\theta^2}{16} \Bigr( \la V| Q_2 \,e \ra Q_2^{\text{int}} \la e | V \ra +
\text{h.c.}\Bigl)
+ \frac{\theta}{16} \Bigr( \la V | Q_2 \,e \ra \la e | Q_3\ra +
\text{h.c.}\Bigl)\\
&+ \frac{\theta}{16} \Bigr( \la Q_3 | e \ra Q_2^{\text{int}} \la e | V \ra +
\text{h.c.}\Bigl)
\end{split}
\ee

\noindent where $|Q_2\,e\ra = Q_2^{(3)} |e\ra$. The second term of
(\ref{Q3yon}) can be combined with the
first term of (\ref{H3yon}) by noting that 

\be
\frac{1}{4}Q_2^{\dag\,\text{int}} Q_2^{\text{int}} = H_2^{\text{int}} 
- \sum_{r=1}^2 \frac{1}{\a_r} \sum_q q\,N^{(r)}_q
\ee

\noindent where $N^{(r)}_q$ is the number operator for string $r$ and
mode $q$. However, the level matching is true independently on each
string, and so the extra term is zero. The result of adding
the second term of (\ref{Q3yon}) to the first term of (\ref{H3yon}) is
thus:

\be
\frac{\theta^2}{4} E_0 \la V | e\ra\la e| V\ra
\ee

\noindent The last terms of (\ref{Q3yon}) and (\ref{H3yon}) may also
be combined. We note that,

\be
\la Q_3 | e \ra Q_2^{\text{int}} \la e | V \ra = 
4 \la H_3 | e \ra \la e | V \ra 
- \la Q_3 |  Q_2\,e \ra \la e | V \ra
\ee

\noindent The first term on the RHS will cancel the last term of
(\ref{H3yon}). At the end of the day, the following expression for $\d
E^{\text{DY}}$ may be used:

\be\label{new}
\begin{split}
\d E^{\text{DY}} &= \frac{\theta^2}{4} E_0  \la V | e\ra\la
e| V\ra+ \frac{\theta^2}{16} \left| \la Q_2\, e| V \ra \right|^2
+ \frac{1}{4} \d E^{\text{SVPS}}\\
&+ \frac{\theta^2}{16} \Bigr( \la V| Q_2 \,e \ra Q_2^{\text{int}} \la e | V \ra +
\text{h.c.}\Bigl)+ \frac{\theta}{16} \Bigr( \la V | Q_2 \,e \ra \la e | Q_3\ra +
\text{h.c.}\Bigl)\\
&-\frac{\theta}{16} \Bigr( \la V | e \ra \la Q_2\,e | Q_3\ra +
\text{h.c.}\Bigl)
\end{split}
\ee

\noindent which is the second method we have used to do the
calculations. Both methods employ the general methodology of appendix
\ref{app:example}, section \ref{appsec:example}.\\

\ni \underline{First method}\\

The $H_3$ contributions are of three varieties, 

\be
\d E_{H_3} = \frac{1}{4} \d E^{\text{SVPS}}_{H_3} + \frac{1}{4} \d
E^{\text{DVPPRT}}_{H_3}  
+ \frac{1}{2} \frac{\la H_3^{\text{DVPPRT}} | e \ra \la e | H_3 \ra}{\Delta E}.
\ee

\ni We find

\be
\begin{split}\d E^{\text{SVPS}}_{H_3}=
\frac{2}{r(1-r)}\frac{g_2^2\,\a'^2}{64\,\a_3^6}
\sum_{r_1\,r_2}\sum_{q_1\,q_2} &\Biggl[ \left( L_{n\,q_1}^{3\,r_1}
\right)^2 \left( \wt N_{-n\,q_2}^{3\,r_2} \right)^2 +
L_{n\,q_1}^{3\,r_1} L_{n\,q_2}^{3\,r_2} \wt N_{-n\,q_2}^{3\,r_2}
\wt N_{-n\,q_1}^{3\,r_1} \\ &+ L_{-n\,q_1}^{3\,r_1}
L_{n\,q_1}^{3\,r_1} \wt N_{n\,q_2}^{3\,r_2} \wt
N_{-n\,q_2}^{3\,r_2} + L_{-n\,q_1}^{3\,r_1} L_{n\,q_2}^{3\,r_2}
\wt N_{n\,q_2}^{3\,r_2} \wt N_{-n\,q_1}^{3\,r_1} \Biggr] \\ &\times
\frac{-\a_3 \left(\d^{r_1\,r_2} \d_{q_1+q_2} + (1 -
\d^{r_1\,r_2})\d_{q_1}\d_{q_2}\right)}{2\,\o_{n}  - \beta_{r_1}^{-1} \o_{q_1} -
\beta_{r_2}^{-1} \o_{q_2}} + (n \lr -n)
\end{split}
\ee

\ni which evaluates to

\be
\begin{split}
\d E^{\text{SVPS}}_{H_3} = 
\frac{g_2^2}{32\pi^2}&\Biggl[ \frac{15}{2\pi^2 n^2}\l' 
+3\left(\frac{1}{\pi^2}+\frac{1}{2\pi}\right)\l'^{3/2}
-\frac{27}{4\pi^2}\l'^2 
-n^2\left(\frac{5}{\pi^2}+\frac{9}{4\pi}\right)\l'^{5/2}\\
&+\frac{111 n^2}{16\pi^2}\l'^3
+n^4\left(\frac{45}{16\pi}+\frac{33\,}{5\pi^2}\right)\l'^{7/2}
+{\cal O}(\l'^4)
\Biggr]
\end{split}
\ee

\ni while for the DVPPRT vertex, we use the result (\ref{h3diveccIN}),
(\ref{h3diveccOUT}). Next we have the cross-term, the expression is

\be
\begin{split}
\d E^{\text{S-DV}}_{H_3} = 2\frac{\la H_3^{\text{DVPPRT}} | e \ra \la
e | H_3 \ra}{\Delta E}&=\\
\frac{2}{r(1-r)}\frac{g_2^2\,\a'^2}{32\,\a_3^6}
\sum_{r_1\,r_2}\sum_{q_1\,q_2} &\Biggl[ \wt L_{n\,q_1}^{3\,r_1}
L_{n\,q_1}^{3\,r_1} \left( \wt N_{-n\,q_2}^{3\,r_2} \right)^2 + \wt
L_{n\,q_1}^{3\,r_1} L_{n\,q_2}^{3\,r_2} \wt N_{-n\,q_2}^{3\,r_2}
\wt N_{-n\,q_1}^{3\,r_1} \\ &+ \wt L_{-n\,q_1}^{3\,r_1}
L_{n\,q_1}^{3\,r_1} \wt N_{n\,q_2}^{3\,r_2} \wt
N_{-n\,q_2}^{3\,r_2} + \wt L_{-n\,q_1}^{3\,r_1}
L_{n\,q_2}^{3\,r_2} \wt N_{n\,q_2}^{3\,r_2} \wt
N_{-n\,q_1}^{3\,r_1} \Biggr] \\ &\times
\frac{-\a_3\left(\d^{r_1\,r_2} \d_{q_1+q_2} + (1 -
\d^{r_1\,r_2})\d_{q_1}\d_{q_2}\right)}{2\,\o_{n}  -
\beta_{r_1}^{-1} \o_{q_1} - \beta_{r_2}^{-1} \o_{q_2}} + (n \lr -n).
\end{split}
\ee

\noindent The result is

\be
\begin{split}
\d E^{\text{S-DV}}_{H_3} = 
\frac{g_2^2}{32\pi^2}&\Biggl[ \left(\frac{8}{3}+\frac{20}{\pi^2 n^2}\right)\l' 
-6\left(\frac{1}{\pi^2}+\frac{1}{2\pi}\right)\l'^{3/2}
-n^2\left(\frac{8}{3}+\frac{14}{\pi^2 n^2}\right)\l'^2 \\
&+ n^2\left(\frac{15}{\pi^2}+\frac{6}{\pi}\right)\l'^{5/2}
+n^4\left(\frac{8}{3}+\frac{41}{4\pi^2 n^2}\right)\l'^3\\
&-n^4\left(\frac{9}{\pi}+\frac{97\,}{4\pi^2}\right)\l'^{7/2}
+{\cal O}(\l'^4)
\Biggr].
\end{split}
\ee

\ni Adding the contributions together we have

\be
\d E^{\text{DY}}_{H_3} = \frac{1}{4} \left(
\d E^{\text{SVPS}}_{H_3} + \d E^{\text{DVPPRT}}_{H_3}
+\d E^{\text{S-DV}}_{H_3} \right)
\ee

\noindent and so the $H_3$ portion of the DY energy shift is given
by

\be
\begin{split}
\d E^{\text{DY}}_{H_3}= 
\frac{g_2^2}{4\pi^2}&\Biggl[ \frac{3}{4}\left(\frac{1}{12}+\frac{35}{32\pi^2 n^2}\right)\l' 
-5n^2\left(\frac{1}{96}+\frac{35}{256\pi^2 n^2}\right)\l'^2 \\
&+n^4\left(\frac{17}{384}+\frac{655}{1024\pi^2 n^2}\right)\l'^3
+n^4\left(\frac{3}{256\pi}+\frac{23\,}{640\pi^2}\right)\l'^{7/2}
+{\cal O}(\l'^4)
\Biggr].
\end{split}
\ee

The contact term contributions are similarly of three varieties,

\be
\d E_{H_4} = \frac{1}{4} \d E^{\text{SVPS}}_{H_4} + \frac{1}{4} \d
E^{\text{DVPPRT}}_{Q_3}  
+ \frac{1}{8} \la Q_3^{\text{DVPPRT}} | e \ra \la e | Q_3 \ra.
\ee 

\ni We find that

\be
\begin{split}\d E^{\text{SVPS}}_{H_4}=
-\frac{g_2^2\,\a'}{16\,\a_3^3} \sum_{r_1\,r_2}\sum_{q_1\,q_2}
&\Biggl[
 \left( K_{-n} \right)^2  \left( G_{q_1} \right)^2  \left( \wt N_{-n\,q_2}^{3\,r_2}
\right)^2
+ K_{-n} K_{n}  \left( G_{q_1} \right)^2 \wt N_{n\,q_2}^{3\,r_2} 
\wt N_{-n\,q_2}^{3\,r_2} \Biggr]\\
&\times \left(\d^{r_1\,r_2} \d_{q_1+q_2} + (1 -
\d^{r_1\,r_2})\d_{q_1}\d_{q_2}\right) + (n \lr -n)
\end{split}
\ee

\noindent with result

\be
\begin{split}
\d E^{\text{SVPS}}_{H_4}= 
\frac{g_2^2}{32\pi^2}&\Biggl[ \left( \frac{1}{3}+\frac{5}{8\pi^2
    n^2}\right) \l' 
-\frac{3}{2}\left(\frac{1}{\pi^2}+\frac{1}{2\pi}\right)\l'^{3/2}
-n^2\left(\frac{1}{6}-\frac{19}{16\pi^2 n^2}\right)\l'^2 \\
&+n^2\left(\frac{11}{4\pi^2}+\frac{9}{8\pi}\right)\l'^{5/2}
+\frac{n^4}{8}\left(1 - \frac{105}{8\pi^2 n^2}\right)\l'^3\\
&-n^4\left(\frac{45}{32\pi}+\frac{73\,}{20\pi^2}\right)\l'^{7/2}
+{\cal O}(\l'^4)
\Biggr].
\end{split}
\ee

\ni Again, for the DVPPRT contributions we refer to (\ref{q3diveccIN})
and (\ref{q3diveccOUT}). The expression for the cross-term is

\be
\begin{split}
\d &E^{\text{S-DV}}_{H_4}=\frac{1}{2} \la Q_3^D | e \ra \la e | Q_3 \ra=\\
&-\frac{g_2^2\,\a'}{8\,\a_3^3} \sum_{r_1\,r_2}\sum_{q_1\,q_2}
\Biggl[
 K_{n} K_{-n}  \left( G_{q_1} \right)^2  \left( \wt N_{-n\,q_2}^{3\,r_2}
\right)^2
+ K_{n} K_{n}  \left( G_{q_1} \right)^2 \wt N_{n\,q_2}^{3\,r_2} 
\wt N_{-n\,q_2}^{3\,r_2} \Biggr]\\
&\qquad\qquad\times \left(\d^{r_1\,r_2} \d_{q_1+q_2} + (1 -
\d^{r_1\,r_2})\d_{q_1}\d_{q_2}\right) + (n \lr -n)
\end{split}
\ee

\noindent with result

\be
\begin{split}
\d E^{\text{S-DV}}_{H_4} = 
\frac{g_2^2}{32\pi^2}&\Biggl[ -2\left( \frac{1}{3}+\frac{5}{8\pi^2
    n^2}\right) \l' 
+3\left(\frac{1}{\pi^2}+\frac{1}{2\pi}\right)\l'^{3/2}
+4n^2\left(\frac{1}{6}+\frac{1}{2\pi^2 n^2}\right)\l'^2 \\
&-2n^2\left(\frac{11}{4\pi^2}+\frac{9}{8\pi}\right)\l'^{5/2}
-4 n^4\left(\frac{1}{6} + \frac{5}{16\pi^2 n^2}\right)\l'^3\\
&+n^4\left(\frac{3}{\pi}+\frac{317\,}{40\pi^2}\right)\l'^{7/2}
+{\cal O}(\l'^4)
\Biggr].
\end{split}
\ee

\ni Adding the contributions together we have

\be
\d E^{\text{DY}}_{H_4} = \frac{1}{4} \left(
\d E^{\text{SVPS}}_{H_4} + \d E^{\text{DVPPRT}}_{Q_3}
+\d E^{\text{S-DV}}_{H_4} \right)
\ee

\noindent and so the $H_4$ portion of the DY energy shift is given
by

\be
\begin{split}
\d E^{\text{DY}}_{H_4} = 
\frac{g_2^2}{4\pi^2}&\Biggl[ n^2\left(\frac{1}{96}+\frac{35}{256\pi^2 n^2}\right)\l'^2 
-\frac{5 n^4}{128} \left(\frac{1}{3}+\frac{29}{8\pi^2 n^2}\right)\l'^3 \\
&+\frac{n^4}{256}\left(\frac{3}{2\pi}+\frac{5}{\pi^2}\right)\l'^{7/2}
+{\cal O}(\l'^4)
\Biggr].
\end{split}
\ee

\ni Assembling the $H_3$ and contact term results, we arrive at the final
expression for the Yoneya energy shift

\be\label{N4final}
\boxed{
\begin{split}
\d E^{\text{DY}} = 
\frac{g_2^2}{4\pi^2} \Biggl[ 
&\left(\frac{1}{12}+\frac{35}{32\pi^2 n^2}\right)
\left(\frac{3}{4}\l'-\frac{n^2}{2}\l'^2\right)
+\frac{n^4}{32}\left(1+\frac{255}{16\pi^2 n^2}\right) \l'^3\\
&+\frac{n^4}{512}\left(\frac{9}{\pi}+\frac{142}{5\pi^2}\right)\l'^{7/2}
+{\cal O}(\l'^4) \Biggl]
\end{split}}
\ee

\ni\underline{Second method}\\

Referring to (\ref{new}) we have five contributions to consider,
beyond the SVPS result. In this section we enumerate these results
and show that the final answers are in agreement with the first
method calculations. The terms of (\ref{new}) which are independent of
the SVPS vertices lead individually like a constant, however
together they lead as $\l'$. We therefore present the results for the
sum of these terms. The remaining terms are individually of ${\cal
O}(\l')$ and are presented individually. The terms independent of the
SVPS vertices are

\be
\begin{split}
\d E_1 = \frac{\theta^2}{16} E_0  \la V | e\ra\la
e| V\ra+ \frac{\theta^2}{16} \left| \la Q_2\, e| V \ra \right|^2
+ \frac{\theta^2}{16} \Bigr( \la V| Q_2 \,e \ra Q_2^{\text{int}} \la e
| V \ra + \text{h.c.} \Bigl)
\end{split}
\ee

\noindent where some of the relevant matrix elements can be found in
appendix \ref{app:example}. The resulting expressions are (for clarity
we suppress the level-matching factor of $\left(\d^{r_1\,r_2} \d_{q_1+q_2} + (1 -
\d^{r_1\,r_2})\d_{q_1}\d_{q_2}\right)$ in the remainder of this section)

\be
\begin{split}
\frac{g_2^2\,r(1-r)}{32\,(-\a_3)} \left(\o_{n}-n\right)\sum_{r_1\,r_2}\sum_{q_1\,q_2}
\Biggl[
\left(\wt N_{n\,q_1}^{3\,r_1} \right)^2 &\left( \wt N_{-n\,q_2}^{3\,r_2}
\right)^2
+ \wt N_{n\,q_1}^{3\,r_1} \wt N_{n\,q_2}^{3\,r_2} \wt N_{-n\,q_2}^{3\,r_2}
\wt N_{-n\,q_1}^{3\,r_1} \Biggr] + (n \lr -n)\\
\end{split}
\ee

\be
\begin{split}
\frac{g_2^2\,r(1-r)}{64\,(-\a_3)} \sum_{r_1\,r_2}\sum_{q_1\,q_2}
\Biggl[
\left|\O_{n}\,\wh Q_{n\,q_1}^{3\,r_1}\right|^2 &\left( \wt N_{-n\,q_2}^{3\,r_2}
\right)^2
+ \O_{n} \O_{-n} \wh Q_{n\,q_1}^{3\,r_1\,*} \wh
Q_{-n\,q_1}^{3\,r_1} \wt N_{n\,q_2}^{3\,r_2} \wt N_{-n\,q_2}^{3\,r_2}
\Biggr] + (n \lr -n)\\
\end{split}
\ee

\be
\begin{split}
-i\frac{g_2^2\,r(1-r)}{32\,(-\a_3)} \sum_{r_1\,r_2}\sum_{q_1\,q_2}
\Biggl[
\O_{n}\,\wh Q_{n\,q_1}^{3\,r_1\,*}
\frac{\O_{q_1}}{\sqrt{\b_{r_1}}} &\wt N_{n\,q_1}^{3\,r_1} \left( \wt N_{-n\,q_2}^{3\,r_2}
\right)^2\\
&+ \O_{n}\,\wh Q_{n\,q_1}^{3\,r_1\,*}
\frac{\O_{q_1}}{\sqrt{\b_{r_1}}} \wt N_{n\,q_1}^{3\,r_1} \wt
N_{-n\,q_2}^{3\,r_2} \wt N_{n\,q_2}^{3\,r_2}
\Biggr] + (n \lr -n)\\
\end{split}
\ee

\noindent respectively. The result is

\be
\begin{split}
\d E_1 = 
\frac{g_2^2}{32\pi^2}&\Biggl[ -\left( \frac{1}{12}+\frac{5}{32\pi^2
    n^2}\right) \l' 
+\frac{3}{8}\left(\frac{1}{\pi^2}+\frac{1}{2\pi}\right)\l'^{3/2}
+n^2\left(\frac{5}{24}+\frac{1}{64\pi^2 n^2}\right)\l'^2 \\
&-\frac{n^2}{16}\left(\frac{29}{\pi^2}+\frac{21}{2\pi}\right)\l'^{5/2}
+\frac{ n^4}{32}\left(-9 + \frac{105}{8\pi^2 n^2}\right)\l'^3\\
&+n^4\left(\frac{165}{128\pi}+\frac{303\,}{80\pi^2}\right)\l'^{7/2}
+{\cal O}(\l'^4)
\Biggr].
\end{split}
\ee

\ni The first remaining term is given by

\be
\d E_2 = \frac{\theta}{16} \Bigr( \la V | Q_2 \,e \ra \la e | Q_3\ra +
\text{h.c.}\Bigl)
\ee

\noindent which gives the following expression

\be
i \frac{g_2^2}{16 \a_3^3} \sqrt{\frac{\a' \k}{4 \a_3}} \Biggl[
\O_{n} K_{-n} G_{q_1} \wh Q^{3\,r_1\,*}_{n\,q_1} \left( \wt
N^{3\,r_2}_{-n\,q_2} \right)^2
+\O_{n} K_{n} G_{q_1} \wh Q^{3\,r_1\,*}_{n\,q_1} \wt
N^{3\,r_2}_{-n\,q_2}  \wt N^{3\,r_2}_{n\,q_2} \Biggr] + (n \lr -n)
\ee

\noindent yielding the result

\be
\begin{split}
\d E_2 = 
\frac{g_2^2}{32\pi^2}&\Biggl[ \left( \frac{1}{6}+\frac{35}{16\pi^2
    n^2}\right) \l' 
-n^2\left(\frac{1}{6}+\frac{5}{4\pi^2 n^2}\right)\l'^2 \\
&+\frac{n^2}{2}\left(\frac{1}{\pi^2}+\frac{3}{8\pi}\right)\l'^{5/2}
+n^4\left(\frac{1}{6} + \frac{31}{32\pi^2 n^2}\right)\l'^3\\
&-n^4\left(\frac{3}{8\pi}+\frac{21\,}{20\pi^2}\right)\l'^{7/2}
+{\cal O}(\l'^4)
\Biggr].
\end{split}
\ee

\ni The next term is

\be
\d E_3 = -\frac{\theta}{16} \Bigr( \la V | e \ra \la Q_2\,e | Q_3\ra +
\text{h.c.}\Bigl)
\ee

\noindent which gives the expression

\be
\frac{g_2^2\,\a'}{32 \a_3^3} \Biggl[
K_{n} K_{-q_1} \wt N^{3\,r_1}_{n\,q_1} \left( \wt
N^{3\,r_2}_{-n\,q_2} \right)^2
+ K_{n} K_{-q_2} \wt N^{3\,r_2}_{-n\,q_2}  \wt
N^{3\,r_1}_{n\,q_1} \wt N^{3\,r_1}_{-n\,q_1} \Biggr] + (n \lr
-n)
\ee

\noindent and results in 

\be
\begin{split}
\d E_3 = 
\frac{g_2^2}{32\pi^2}&\Biggl[ \left( \frac{1}{3}+\frac{5}{2\pi^2
    n^2}\right) \l' 
-\frac{3}{4}\left(\frac{1}{\pi^2}+\frac{1}{2\pi}\right)\l'^{3/2}
-n^2\left(\frac{1}{3}+\frac{7}{4\pi^2 n^2}\right)\l'^2 \\
&+\frac{n^2}{4}\left(\frac{15}{2\pi^2}+\frac{3}{\pi}\right)\l'^{5/2}
+n^4\left(\frac{1}{3} + \frac{41}{32\pi^2 n^2}\right)\l'^3\\
&-n^4\left(\frac{9}{8\pi}+\frac{97\,}{32\pi^2}\right)\l'^{7/2}
+{\cal O}(\l'^4)
\Biggr].
\end{split}
\ee

\ni Adding the contributions,

\be
\d E^{\text{DY}} = \frac{1}{4}\d E^{\text{SVPS}} + \d E_1 +\d E_2
+ \d E_3
\ee

\noindent we find the identical result (\ref{N4final}).

% ========================================================================== %
\subsection{Discussion}

The expression for the DY vertex calculation of the
impurity-conserving mass-shift (\ref{N4final}), represents the best
matching with the gauge theory result (\ref{gaugeresult}) yet
achieved. There is a leading factor of $3/4$ in the $\l'$ term which
we can scale away by employing the undetermined function $f$ which
appeared in the vertices. If we scale $f$ by $\sqrt{4/3}$ we will
achieve agreement of the leading $\l'$ term with gauge
theory. Although the $\l'^2$ term is of the correct form, the
coefficient is not in agreement. We also note the absence of
half-integer powers of $\l'$ up to (but not including) the $7/2$'s
power. This fairs much better than the SVPS, and one half-power better
than the DVPPRT results. 

We showed in section \ref{sec:divcan} that any ``reflection symmetry
factor'' which would effectively multiply the contact term by 2
relative to the $H_3$ term is incommensurate with finiteness of the
mass-shift. Mysteriously, however, if the contact terms are blindly
scaled by a factor of 2, the agreement with gauge theory is enhanced
for both the SVPS (\ref{pankN4}) and DY results,

\bsp
\d E^{\text{SVPS}}_{2H_4} = \frac{g_2^2}{4\pi^2} \left[
\left(\frac{1}{12}+\frac{35}{32\pi^2 n^2}\right)
\left(\l'-\frac{n^2}{2}\l'^2\right) +
\frac{n^2}{16\pi^2}\l'^{5/2}\right.\cr +
\left.n^4\left(\frac{1}{32}+\frac{117}{256\pi^2 n^2}\right) \l'^3
-\frac{7 n^4}{80\pi^2}\l'^{7/2} +{\cal O}(\l'^4) \right] 
\end{split}
\ee

\bsp
\d E^{\text{DY}}_{2H_4} = \frac{g_2^2}{4\pi^2}
\frac{3}{4}\left[ \left(\frac{1}{12}+\frac{35}{32\pi^2 n^2}\right)
\left(\l'-\frac{n^2}{2}\l'^2\right)
+n^4\left(\frac{7}{288}+\frac{365}{768\pi^2 n^2}\right)
\l'^3\right.\cr + \left.
n^4\left(\frac{1}{10\pi^2}+\frac{1}{32\pi}\right)\l'^{7/2}+{\cal
O}(\l'^4) \right] 
\end{split}
\ee

\noindent however, the DY result is still superior in that the
$\l'^{5/2}$ power is absent. The meaning (if any) of this coincidence
is not clear to us at this stage.

The DY vertex has thus produced the best match to gauge theory so far.
It matches the $\l'$ term, exhibits the correct form of the $\l'^2$
term, and displays the absence of half-integer powers of $\l'$ to a
rather high order. It is possible that higher-orders in intermediate
state impurities would correct the result to a complete match with
gauge theory; indeed a scheme whereby higher orders in impurities
somehow contribute only higher orders in $\l'$ would be very physical
and pleasing. Whether or not this is the case remains to be seen and
requires an honest calculation from these channels; as we have shown
in section \ref{sec:arbimp}, generically this is not the case.

% ************************************************************************** %
\section{Wrapping $x^-$: discrete light-cone quantization}
\label{sec:dlcq}

In an important paper by Mukhi, Rangamani, and Verlinde
\cite{Mukhi:2002ck}, a version of the plane-wave / BMN operator
correspondence was derived whereby the light-cone direction $x^-$ is
compactified leading to a discrete light-cone momentum $p^+$. The dual
gauge theory is no longer ${\cal N}=4$ supersymmetric Yang-Mills, but
an ${\cal N}=2$ quiver gauge theory corresponding to a stack of $N_1$
D-branes at a $\bC^3/\bZ_{N_2}$ orbifold point. The plane-wave is
obtained via a Penrose limit on $AdS_5\times S_5/\bZ_{N_2}$, where the
five-sphere is orbifolded into $N_2$ domains. 

The authors of \cite{DeRisi:2004bc} computed the non-planar
corrections to the anomalous dimensions of the gauge theory operators
corresponding to strings on the discrete light-cone plane-wave. It is
therefore interesting to consider light-cone string field theory in
this discrete light-cone quantization. This section presents original,
unpublished work of the author and collaborators of
\cite{Grignani:2005yv} concerning this DLCQ light-cone string field
theory.

% ========================================================================== %
\subsection{Introduction}

The space $AdS_5\times S_5/\bZ_{N_2}$ may be expressed as 

\bsp
ds^2 = R^2\Bigr[&-\cosh^2\r \, dt^2 + d\r^2 + \sinh^2\r \, d\O_3^2\\ 
& + d\a^2 + \sin^2 \a \, d\theta^2 + \cos^2 \a \bigl( d\g^2 + \cos^2 \g
  \, d\chi^2 + \sin^2\g \, d\phi^2 \bigr) \Bigl]
\end{split}
\ee

\ni where $\theta$, $\chi$, and $\phi$ are azimuthal angles
$\in[0,2\pi]$. The orbifold is realized by imposing the following
identifications

\be
\chi \sim \chi + \frac{2\pi}{N_2}, \qquad
\phi \sim \phi - \frac{2\pi}{N_2}.
\ee

\ni The Penrose limit is realized via the re-scalings $r = \r R$, $w =
\a R$, $y = \g R$, and the introduction of light-cone coordinates

\be
x^+ = \frac{1}{2} ( t + \chi), \qquad x^- = \frac{R^2}{2} ( t - \chi).
\ee

\ni Taking the limit $R\rightarrow \infty$, the plane-wave metric
(\ref{ppmet}) is obtained, albeit with compactifications

\be
x^+ \sim x^+ + \frac{\pi}{N_2}, \qquad x^- \sim x^- + \frac{\pi
  R^2}{N_2}.
\ee

\ni If we then take $N_2 \rightarrow \infty$, the periodicity in $x^+$ and
$\phi$ is removed, while scaling $N_2 \sim R^2$ causes a finite
compactification of $x^-$. The implication for string theory is very
simple; it is unchanged up to two important features

\begin{enumerate}
\item The light-cone momentum is quantized in units of inverse
  compactification radius $R_- = R^2/(2N_2)$

\be\label{dlcqp+}
2 p^+ = \frac{k}{R_-},\qquad k \in \bZ, ~ k > 0
\ee 

\item The string is free to wrap $x^-$, leading to a modified level
  matching

\be\label{dlcqm}
\prod_i a^{I\,\dag}_{n^b_i} \prod_j b^{a\,\dag}_{n^f_j}\,|0;k,m\ra
\rightarrow  \sum_i n^b_i + \sum_j n^f_j  = k m, \qquad m \in \bZ 
\ee

\ni where $m$ is the wrapping number.

\end{enumerate}

The dual gauge theory is constructed by considering $N_1$ coincident
D3-branes sitting at a $\bC^3/\bZ_{N_2}$ orbifold point. There are
thus $N_2$ copies of the $N_1$ branes. The gauge group of the
un-orbifolded theory is then broken as follows

\be SU(N_1N_2) \rightarrow SU(N_1)_1 \times SU(N_1)_2 \times \ldots
\times SU(N_1)_{N_2} \ee

\ni so that there are now $N_2$ separate $SU(N_1)$ gauge groups. The
action of the orbifold group generator $\G$ on the six scalars of the
parent ${\cal N}=4$ SYM is as follows

\bsp
\G : &\left( \frac{1}{\sqrt{2}} ( \Phi_1 + i \Phi_2 ), 
\frac{1}{\sqrt{2}} ( \Phi_3 + i \Phi_4 ), 
\frac{1}{\sqrt{2}} ( \Phi_5 + i \Phi_6 ) \right)\\ &\qquad = 
\left( \frac{\o}{\sqrt{2}} ( \Phi_1 + i \Phi_2 ), 
\frac{\o^{-1}}{\sqrt{2}} ( \Phi_3 + i \Phi_4 ), 
\frac{1}{\sqrt{2}} ( \Phi_5 + i \Phi_6 ) \right)
\end{split}
\ee

\ni where $\o = \exp(2\pi i /N_2)$. This leads to new bi-fundamental
fields $A_I$, $B_I$ which each have one leg each in $SU(N_1)_I$ and
$SU(N_1)_{I+1}$, corresponding to the first and second combinations of
the parent scalars, and complex scalars $\Phi_I$ in the adjoint
representation of $SU(N_1)_I$, corresponding to the remaining parent
scalar combination $\frac{1}{\sqrt{2}} ( \Phi_5 + i \Phi_6 )$. The
resulting gauge theory is known as a {\it quiver} theory (see
\cite{Douglas:1996sw}) and in this case carries half the supersymmetry
of the parent theory. The relation (\ref{radius}) is not modified,
i.e. we simply replace $N \rightarrow N_1 N_2$ so that

\be
R^2 = \sqrt{4 \pi g_s \,\a'^2 N_1 N_2}, \qquad g^2_{YM} = 4 \pi g_s.
\ee 

\ni Given the scaling $N_2 \sim R^2$, we are instructed to take $N_1
\sim N_2$ so that $g_s$ remains fixed. Each gauge group has a
coupling constant given by $(g^I_{YM})^2 = 4\pi g_s N_2$, so that the
relevant 't Hooft coupling is $\l = (g^I_{YM})^2 N_1 = 4\pi g_s N_1
N_2$. 

Following the treatment of BMN given in section \ref{sec:BMN},
we would like to identify the appropriate ``large-J'' limit of the
orbifolded theory in order to identify the operators dual to DLCQ
plane-wave strings. Two angular momenta are identified
\cite{Mukhi:2002ck}

\be
J = -\frac{i}{2 N_2} \left(\p_\chi - \p_\phi\right), \qquad 
J' = -\frac{i}{2} \left( \p_\chi + \p_\phi \right)
\ee

\ni so that the light-cone momenta are expressed as 

\bsp
2 p^- &= i (\p_t + \p_\chi ) = \D - N_2 J - J'\\
2 p^+ &= i\frac{(\p_t - \p_\chi)}{R^2} = \frac{\D+ N_2 J + J'}{R^2}.
\end{split}
\ee

\ni In analogy with the BMN case, we would like to take $\D$ and
$N_2J + J'$ to infinity as $R^2$, while keeping their difference
finite. The charges of the $A_I$, $B_I$, and $\Phi_I$ fields are as
follows \cite{Mukhi:2002ck}

\begin{center}
\begin{tabular}[t]{c|c|c|c|}
 & $\D$ & $N_2 J$ & $J'$ \\ \hline
$A_I$ & 1 & $1/ 2$ & $1/2$ \\ \hline
$B_I$ & 1 & $-1/2$ & $1/2$ \\ \hline
$\Phi_I$ & 1 & 0 & 0 \\ \hline
\end{tabular}
\end{center}

\ni which indicates that the desired operators are long chains of
$A_I$'s (which have $\D = N_2J + J'$), with insertions of $\Phi_I,\bar
\Phi_I, B_I, \bar B_I$ as the fundamental impurities which have $\D =
1$ while having $N_2 J+J' =0$. The other $SO(4)$ impurities are
constructed via insertions of derivatives of the $A_I$. In order that
the operator be gauge invariant, the product must be over all $N_2$
copies of $SU(N_1)$. The simplest state is the dual of the DLCQ string
vacuum. It is therefore not surprising to find

\be\label{k1vac}
|k=1,m=0 \ra \lr \frac{1}{\sqrt{N_1^{N_2}}} \Tr \left( A_1 A_2 \ldots
A_{N_2} \right)
\ee

\ni where we note that the string vacuum must have $m=0$ (a string
must exist in order to wrap a direction). For general $k$, the
operator is 

\be
|k,m=0 \ra \lr \frac{1}{\sqrt{N_1^{kN_2}}} \Tr \biggl( \left( A_1 A_2 \ldots
A_{N_2} \right)^k \biggr)
\ee

\ni so that $k$ copies of the string $A_1 \ldots A_{N_2}$ are traced
over. Adding impurities we see a novel feature as compared to the
standard BMN picture. Consider the addition of a single impurity to
the operator (\ref{k1vac})

\be
\left( a_n^{5\,\dag} + i a_n^{6\,\dag} \right) |k=1,m\ra \lr
\sum_{I=1}^{N_2} e^{2\pi i n I/N_2} \Tr \left( A_1 \ldots
A_{I-1} \Phi_I A_I \ldots A_{N_2}  \right).
\ee
%\left(A_1\ldots A_{N_2} \right)^{k-1}

\ni This would have been zero by cyclicity of the trace, but here each
insertion position is inequivalent to the next, so that this state is
non-zero. This is a wrapping state with $m = n$. For general $k$, we
have

\be
\left( a_n^{5\,\dag} + i a_n^{6\,\dag} \right) |k,m\ra \lr
\sum_{I=1}^{kN_2} e^{2\pi i n I/(kN_2)} \Tr \left( A_1 \ldots
A_{I-1} \Phi_I A_I \ldots A_{N_2}  \left(A_1\ldots A_{N_2} \right)^{k-1}\right)
\ee

\ni where, since cyclicity gives the same trace under $I \rightarrow
I+N_2$, $n$ must be $k$ times an integer. This is just the level
matching condition $n=km$. The construction of higher impurity states
is straightforward \cite{Mukhi:2002ck}. 

In \cite{DeRisi:2004bc}, the DLCQ analogue of (\ref{gaugeresult}) was
computed for one and two-impurity operators built upon $k=1,2$, and
$3$ vacuua. The couplings $\l'$ and $g_2$ may be expressed in terms of
$N_1$, $N_2$, and $k$ using (\ref{dlcqp+})

\bsp
\a' p^+ &= \frac{\a' k}{2 R_-} = \frac{\a' k N_2}{R^2} =
\frac{k}{g_{YM}} \sqrt{\frac{N_2}{N_1}} ~~\rightarrow~~
\l' = \frac{1}{(\a' p^+)^2} = \frac{g^2_{YM} N_1}{k^2 N_2}\\
g_2 &= g^2_{YM} (\a' p^+)^2 = \frac{k^2 N_2}{N_1}
\end{split}
\ee

\ni where we note that $\mu$ has been scaled out of the metric
here. The results of \cite{DeRisi:2004bc} may be summarized as
follows.

\begin{enumerate}

\item Single impurity operators receive only planar corrections to
  their anomalous dimension; this implies the absence of string-loop
  corrections to the masses of the dual string states. The planar loop
  corrections reproduce the expansion of the free string energy.

\item Two-impurity operators with $k=1$ similarly receive no
  non-planar corrections. The planar loop corrections also reproduce
  the expansion of the free string energy.

\item Two-impurity operators with $k=2$ receive the correct
  free-string planar corrections, but also receive a leading
  non-planar correction to the anomalous dimension

\be\label{dlcqk2}
\D - N_2 J - J' = \left( 2 + \frac{1}{2} (n_1^2 + n_2^2) \l' + \ldots
\right) + \begin{cases}
g_2^2 \left(\frac{1}{16\pi^2}\l' + \ldots \right)~ n_1,n_2 ~\text{odd}\\
0\qquad n_1,n_2 ~\text{even}
\end{cases} 
\ee

\ni which truncates at ${\cal O}(g_2^2)$. Note that $n_1$ and $n_2$
are the mode numbers of the dual string oscillators obeying the level
matching condition $n_1 + n_2 = 2 m$, where $m \in \bZ$.

\item Two-impurity operators with $k=3$ receive the correct
  free-string planar corrections, and also receive non-planar
  corrections to arbitrary order in $g_2^2$. The leading
  result is given by

\bsp\label{dlcqk3}
&\D - N_2 J - J' = \left( 2 + \frac{1}{2} (n_1^2 + n_2^2) \l' +
\ldots \right)\\ &+ \frac{g_2^2 \l'}{16\pi^2} \left[ 1 + \frac{6}{\pi
(n_1-n_2)} \left( \cos \left(\pi n_1\over 3 \right) \sin \left(\pi
n_1\over 3\right) - \cos \left(\pi n_2\over 3 \right) \sin \left(\pi
n_2\over 3 \right) \right)\right]\\
&+ \ldots
\end{split}
\ee

\ni where {\bf all} non-planar corrections (not just the leading term shown)
vanish for $n_1, n_2$ multiples of three. 

\end{enumerate}

\ni We also take $n_1 \neq n_2$ in all results shown here, i.e. $m\neq
0$. It is an interesting pursuit to attempt to calculate these
non-planar corrections using string loops as has been attempted for
the standard BMN operators in the previous sections. In the next
section we will endeavour to reproduce (\ref{dlcqk2}), (\ref{dlcqk3})
and the results discussed under items 1) and 2) above using DLCQ
light-cone string field theory on the plane-wave background.

% ========================================================================== %
\subsection{Results}

As we asserted in (\ref{dlcqp+}) and (\ref{dlcqm}), the light-cone
string field theory is unchanged in the DLCQ case, with the exception
of a modified level-matching condition and a discretized
$p^+$. Because $p^+$ is conserved and non-zero, the string with $k=1$
cannot split, as there is no lower $p^+$ strings to split into; this
is the dual-reflection of item 2) from the previous subsection. For
the same reason, we see that once the $k=2$ string is split into two
$k=1$ strings, the only choice is to re-join to a $k=2$
state. Therefore, the mass-shift of a $k=2$ string may not be of
higher than $g^2_2$ order; as was summarized in item 3). The string
theory manifestation of item 1) (that single-impurity states receive
no loop corrections) is in fact also responsible for the lack of $k=2$
corrections when both external mode numbers are even, or in the case
of $k=3$, when both are multiples of 3. The source is the factor of
$\sin n\pi r$ which occurs in each Neumann matrix and associated
quantity which has a leg in the external string (string \#3), see
appendix \ref{app:relations}. Recall that $r = \a_1/|\a_3|=k_1/k$, so that
if $n_i k_1/k$ is an integer for some external string excitation
$\a^\dag_{n_i}$, the entire amplitude will vanish. This factor comes
about from the decomposition of the modes of the external string into
those of the two internal strings at $\t=0$, see figure
\ref{fig:strcoords}. If the undulations of string \#3 at $\t=0$ are
orthogonal to those of strings \#1 or \#2, then obviously the string
worldsheets cannot be in contact, and therefore cannot interact. This
situation is realized if one of the $n_i$ is a multiple of the
external light-cone momentum $k$, i.e. for $k=2$ when at least one
$n_i$ is even, for $k=3$ when at least one $n_i$ is a multiple of
three, or, when only a single external impurity is present, always
since $n=km$ by level-matching, where $m$ is the external wrapping
number.

% -------------------------------------------------------------------------- %
\subsubsection{$k=2$ Impurity-conserving mass-shift}

The calculation of the specific one-loop mass-shift for $k=2$ proceeds
along the same lines as was performed in section
\ref{sec:altvert}. The difference is that the mode numbers of the
external $|[{\bf 9},{\bf 1}]\ra$ state have distinct, odd values
$n_1$ and $n_2$ satisfying

\be
n_1 + n_2 = 2 m
\ee

\ni where $m$ is the external wrapping number. For the
impurity-conserving channel, we may either place the two
intermediate-state impurities on the same string (say string \#1), or
one on each string. In the former case string \#2 is in its vacuum
state and necessarily has wrapping number $m_2 = 0$. The level-matching
condition for the excited string gives $q_1+q_2 = m_1$, where the
$q_i$ are the internal mode numbers; conservation of wrapping number
then gives $m_1 = m$. In the latter case we have $q_1 = m_1$, and $q_2
= m_2$ while $m_1+m_2=m$. Thus the two choices for distribution of
intermediate state impurities are indistinguishable, both leading to
the same condition which is introduced into the amplitudes via the
factor $\d_{q_1,m-q_2}$ where $m = (n_1+n_2)/2 \in \bZ$. We begin with
the SVPS result for the $H_3$ term

\be
\begin{split}
\d E^{\text{SVPS}}_{H_3}=
\frac{2}{r(1-r)}\frac{g_2^2\,\a'^2}{64\,\a_3^6}
\sum_{r_1\,r_2}\sum_{q_1\,q_2} &\Biggl[ \left( L_{n_1\,q_1}^{3\,r_1}
\right)^2 \left( \wt N_{n_2\,q_2}^{3\,r_2} \right)^2 +
L_{n_1\,q_1}^{3\,r_1} L_{n_1\,q_2}^{3\,r_2} \wt N_{n_2\,q_2}^{3\,r_2}
\wt N_{n_2\,q_1}^{3\,r_1} \\ &+ L_{n_2\,q_1}^{3\,r_1}
L_{n_1\,q_1}^{3\,r_1} \wt N_{n_1\,q_2}^{3\,r_2} \wt
N_{n_2\,q_2}^{3\,r_2} + L_{n_2\,q_1}^{3\,r_1} L_{n_1\,q_2}^{3\,r_2}
\wt N_{n_1\,q_2}^{3\,r_2} \wt N_{n_2\,q_1}^{3\,r_1} \Biggr] \\ &\times
\frac{-\a_3\,\d_{q_1,m-q_2}}{\o_{n_1} + \o_{n_2} - \beta_{r_1}^{-1}
\o_{q_1} - \beta_{r_2}^{-1} \o_{q_2}} + (n_1 \lr n_2)
\end{split}
\ee

\ni where now instead of an integration over a continuous $r\in[0,1]$,
$r$ is fixed at $1/2$. The result is  

\be
\begin{split}
\left(\d E^{\text{SVPS}}_{H_3}\right)_{k=2} = 
\frac{g_2^2}{16\pi^2}&\Biggl[ 
8 \left(\frac{1}{\pi^2}+\frac{1}{2\pi}\right)\l'^{3/2}
-\frac{1}{\pi}\left(\frac{13}{4}(n_1^2+n_2^2) 
+ \frac{1}{2} n_1 n_2 \right)\l'^{5/2}\\
&-\frac{1}{\pi^2}\left(\frac{22}{3}(n_1^2+n_2^2) 
+ \frac{4}{3} n_1 n_2 \right)\l'^{5/2}
+ \ldots \Biggr].
\end{split}
\ee

\ni The contact term contribution is as follows

\be
\begin{split}\d E^{\text{SVPS}}_{H_4}=
-\frac{g_2^2\,\a'}{16\,\a_3^3} \sum_{r_1\,r_2}\sum_{q_1\,q_2}
&\Biggl[
 \left( K_{-n_1} \right)^2  \left( G_{q_1} \right)^2  \left( \wt N_{n_2\,q_2}^{3\,r_2}
\right)^2
+ K_{-n_1} K_{-n_2}  \left( G_{q_1} \right)^2 \wt N_{n_1\,q_2}^{3\,r_2} 
\wt N_{n_2\,q_2}^{3\,r_2} \Biggr] \d_{q_1,m-q_2}\\ + (n_1 \lr n_2)
\end{split}
\ee

\ni giving

\be
\begin{split}
\left(\d E^{\text{SVPS}}_{H_4}\right)_{k=2} = 
\frac{g_2^2}{16\pi^2}&\Biggl[ 
\l' + \left( \frac{n_1+n_2}{2} 
- 4\left(\frac{1}{\pi^2}+\frac{1}{2\pi}\right)\right)\l'^{3/2}\\
&-\left(\frac{n_1^2+n_2^2}{4} +
2(n_1+n_2)\left(\frac{1}{\pi^2}+\frac{1}{2\pi}\right)\right)\l'^2\\
-\biggl( \frac{n_1^3+n_2^3}{2}
-\frac{1}{\pi}&\left(\frac{13}{8}(n_1^2+n_2^2)+\frac{1}{4} n_2 n_1
\right)
-\frac{1}{\pi^2}\left(\frac{23}{6} (n_1^2+n_2^2) + \frac{1}{3} n_1 n_2
\right) \biggr)\l'^{5/2}
+ \ldots \Biggr].
\end{split}
\ee

\ni Combining the results we find

\be\label{SVPSk2}\boxed{
\begin{split}
\left(\d E^{\text{SVPS}}\right)_{k=2} = 
\frac{g_2^2}{16\pi^2}&\Biggl[ \l' +  \left( \frac{n_1+n_2}{2} 
+ 4\left(\frac{1}{\pi^2}+\frac{1}{2\pi}\right)\right)\l'^{3/2}\\
&-\left(\frac{n_1^2+n_2^2}{4} +
2(n_1+n_2)\left(\frac{1}{\pi^2}+\frac{1}{2\pi}\right)\right)\l'^2\\
-\biggl( \frac{n_1^3+n_2^3}{2}
+\frac{1}{\pi}&\left(\frac{13}{8}(n_1^2+n_2^2)+\frac{1}{4} n_2 n_1
\right)
+\frac{1}{\pi^2}\left(\frac{7}{2} (n_1^2+n_2^2) +  n_1 n_2
\right) \biggr)\l'^{5/2}
+ \ldots \Biggr].
\end{split}}
\ee

\ni This result does display a leading agreement with the gauge theory
result (\ref{dlcqk2}). However, it also suffers maximally from
half-integer powers of $\l'$. We will see that the DY vertex will do
better, in analogy with the standard case. First, we present the
results for the DVPPRT vertex. The expression for the $H_3$ term is

\be
\begin{split}\d E^{\text{DVPPRT}}_{H_3}=
\frac{2}{r(1-r)}\frac{g_2^2\,\a'^2}{64\,\a_3^6}
\sum_{r_1\,r_2}\sum_{q_1\,q_2} &\Biggl[ \left( \wt
L_{n_1\,q_1}^{3\,r_1} \right)^2 \left( \wt N_{n_2\,q_2}^{3\,r_2}
\right)^2 + \wt L_{n_1\,q_1}^{3\,r_1} \wt L_{n_1\,q_2}^{3\,r_2} \wt
N_{n_2\,q_2}^{3\,r_2} \wt N_{n_2\,q_1}^{3\,r_1} \\ &+ \wt
L_{n_2\,q_1}^{3\,r_1} \wt L_{n_1\,q_1}^{3\,r_1} \wt
N_{n_1\,q_2}^{3\,r_2} \wt N_{n_2\,q_2}^{3\,r_2} + \wt
L_{n_2\,q_1}^{3\,r_1} \wt L_{n_1\,q_2}^{3\,r_2} \wt
N_{n_1\,q_2}^{3\,r_2} \wt N_{n_2\,q_1}^{3\,r_1} \Biggr] \\ &\times
\frac{-\a_3\,\d_{q_1,m-q_2}}{\o_{n_1} + \o_{n_2} - \beta_{r_1}^{-1}
\o_{q_1} - \beta_{r_2}^{-1} \o_{q_2}} + (n_1 \lr n_2)
\end{split}
\ee

\ni with result

\be
\begin{split}
\left(\d E^{\text{DVPPRT}}_{H_3}\right)_{k=2} = 
\frac{g_2^2}{16\pi^2}&\Biggl[ 
-2 \l' + 8 \left(\frac{1}{\pi^2}+\frac{1}{2\pi}\right)\l'^{3/2}
+\frac{3}{2}(n_1^2+n_2^2)\l'^2\\
-\frac{1}{\pi}&\left(\frac{21}{4}(n_1^2+n_2^2) 
+ \frac{1}{2} n_1 n_2 \right)\l'^{5/2}
-\frac{1}{\pi^2}\left(\frac{38}{3}(n_1^2+n_2^2) 
- \frac{4}{3} n_1 n_2 \right)\l'^{5/2}
+ \ldots \Biggr]
\end{split}
\ee

\ni while the contact term gives 

\be
\begin{split}\d E^{\text{DVPPRT}}_{H_4}=
-\frac{g_2^2\,\a'}{16\,\a_3^3} \sum_{r_1\,r_2}\sum_{q_1\,q_2}
&\Biggl[
 \left( K_{n_1} \right)^2  \left( G_{q_1} \right)^2  \left( \wt N_{n_2\,q_2}^{3\,r_2}
\right)^2
+ K_{n_1} K_{n_2}  \left( G_{q_1} \right)^2 \wt N_{n_1\,q_2}^{3\,r_2} 
\wt N_{n_2\,q_2}^{3\,r_2} \Biggr]\d_{q_1,m-q_2}\\ + (n_1 \lr n_2)
\end{split}
\ee

\ni with result

\be
\begin{split}
\left(\d E^{\text{DVPPRT}}_{H_4}\right)_{k=2} = 
\frac{g_2^2}{16\pi^2}&\Biggl[ 
\l' - \left( \frac{n_1+n_2}{2} 
+ 4\left(\frac{1}{\pi^2}+\frac{1}{2\pi}\right)\right)\l'^{3/2}\\
&-\left(\frac{n_1^2+n_2^2}{4} -
2(n_1+n_2)\left(\frac{1}{\pi^2}+\frac{1}{2\pi}\right)\right)\l'^2\\
+\biggl( \frac{n_1^3+n_2^3}{2}
+\frac{1}{\pi}&\left(\frac{13}{8}(n_1^2+n_2^2)+\frac{1}{4} n_2 n_1
\right)
+\frac{1}{\pi^2}\left(\frac{23}{6} (n_1^2+n_2^2) + \frac{1}{3} n_1 n_2
\right) \biggr)\l'^{5/2}
+ \ldots \Biggr].
\end{split}
\ee

\ni Combining these results we obtain the mass-shift for the DVPPRT
vertex

\be\label{DVPPRTk2}\boxed{
\begin{split}
\left(\d E^{\text{DVPPRT}}\right)_{k=2} = 
\frac{g_2^2}{16\pi^2}&\Biggl[ 
-\l' - \left( \frac{n_1+n_2}{2} 
- 4\left(\frac{1}{\pi^2}+\frac{1}{2\pi}\right)\right)\l'^{3/2}\\
&+\left(\frac{5}{4}(n_1^2+n_2^2) +
2(n_1+n_2)\left(\frac{1}{\pi^2}+\frac{1}{2\pi}\right)\right)\l'^2\\
+\biggl( \frac{n_1^3+n_2^3}{2}
-\frac{1}{\pi}&\left(\frac{29}{8}(n_1^2+n_2^2)+\frac{1}{4} n_2 n_1
\right)
-\frac{1}{\pi^2}\left(\frac{53}{6} (n_1^2+n_2^2) - \frac{5}{3} n_1 n_2
\right) \biggr)\l'^{5/2}
+ \ldots \Biggr]
\end{split}}
\ee

\ni which fails to agree with the gauge theory result even at the
leading order, as the sign is incorrect. Finally, we compute the extra
cross-terms required to assemble the DY result. The $H_3$ cross-term
is given by

\be
\begin{split}
\d E^{\text{S-DV}}_{H_3} = 2\frac{\la H_3^{\text{DVPPRT}} | e \ra \la
e | H_3^{\text{SVPS}} \ra}{\Delta E}&=\\
\frac{2}{r(1-r)}\frac{g_2^2\,\a'^2}{32\,\a_3^6}
\sum_{r_1\,r_2}\sum_{q_1\,q_2} &\Biggl[ \wt L_{n_1\,q_1}^{3\,r_1}
L_{n_1\,q_1}^{3\,r_1} \left( \wt N_{n_2\,q_2}^{3\,r_2} \right)^2 + \wt
L_{n_1\,q_1}^{3\,r_1} L_{n_1\,q_2}^{3\,r_2} \wt N_{n_2\,q_2}^{3\,r_2}
\wt N_{n_2\,q_1}^{3\,r_1} \\ &+ \wt L_{n_2\,q_1}^{3\,r_1}
L_{n_1\,q_1}^{3\,r_1} \wt N_{n_1\,q_2}^{3\,r_2} \wt
N_{n_2\,q_2}^{3\,r_2} + \wt L_{n_2\,q_1}^{3\,r_1}
L_{n_1\,q_2}^{3\,r_2} \wt N_{n_1\,q_2}^{3\,r_2} \wt
N_{n_2\,q_1}^{3\,r_1} \Biggr] \\ &\times \frac{-\a_3\,\d_{q_1,m-q_2}}{\o_{n_1} +
\o_{n_2} - \beta_{r_1}^{-1} \o_{q_1} - \beta_{r_2}^{-1} \o_{q_2}} +
(n_1 \lr n_2)
\end{split}
\ee

\ni resulting in

\be
\begin{split}
\left(\d E^{\text{S-DV}}_{H_3}\right)_{k=2} = 
\frac{g_2^2}{16\pi^2}&\Biggl[ 
8 \l' - 16 \left(\frac{1}{\pi^2}+\frac{1}{2\pi}\right)\l'^{3/2}
-4(n_1^2+n_2^2)\l'^2
+\frac{1}{\pi}\left(\frac{17}{2}(n_1^2+n_2^2) 
+ n_1 n_2 \right)\l'^{5/2}\\
&+\frac{20}{\pi^2} (n_1^2+n_2^2) \l'^{5/2}
+ \ldots \Biggr].
\end{split}
\ee

\ni The contact cross-term is given by

\be
\begin{split}
\d &E^{\text{S-DV}}_{H_4}=\frac{1}{2} \la Q_3^{\text{DVPPRT}} | e \ra
\la e | Q_3^{\text{SVPS}} \ra=\\ &-\frac{g_2^2\,\a'}{8\,\a_3^3}
\sum_{r_1\,r_2}\sum_{q_1\,q_2} \Biggl[ K_{n_1} K_{-n_1} \left( G_{q_1}
\right)^2 \left( \wt N_{n_2\,q_2}^{3\,r_2} \right)^2 + K_{n_1}
K_{-n_2} \left( G_{q_1} \right)^2 \wt N_{n_1\,q_2}^{3\,r_2} \wt
N_{n_2\,q_2}^{3\,r_2} \Biggr]\d_{q_1,m-q_2} + (n_1 \lr n_2)
\end{split}
\ee

\ni with result

\be
\begin{split}
\left(\d E^{\text{S-DV}}_{H_4}\right)_{k=2} = 
\frac{g_2^2}{16\pi^2}&\Biggl[ 
-2\l' + 8\left(\frac{1}{\pi^2}+\frac{1}{2\pi}\right)\l'^{3/2}\\
&+\left(n_1^2+n_2^2\right)\l'^2\\
-\biggl(\frac{1}{\pi}&\left(\frac{15}{4}(n_1^2+n_2^2)+\frac{3}{2} n_2 n_1
\right)
+\frac{1}{\pi^2}\left(\frac{26}{3} (n_1^2+n_2^2) + \frac{8}{3} n_1 n_2
\right)\biggr)\l'^{5/2}
+ \ldots \Biggr].
\end{split}
\ee

\ni Assembling the final result $\d E^{\text{DY}} = (\d E^{\text{DVPPRT}} + \d
E^{\text{S-DV}} + \d E^{\text{SVPS}})/4$, we find

\be\label{DYk2}
\boxed{
\begin{split}
\left(\d E^{\text{DY}}\right)_{k=2} = 
\frac{g_2^2}{16\pi^2} \frac{3}{2}\Biggl[
\l' -\frac{n_1^2+n_2^2}{3}\l'^2
-\frac{(n_1+n_2)^2}{6}\left(\frac{1}{\pi^2}+\frac{1}{2\pi}\right)\l'^{5/2}+\ldots
\Biggr]
\end{split}}
\ee

\ni which we have verified using the so-called ``second method''
outlined in section \ref{sec:DYimpcons}. This result matches the
leading order gauge theory result (\ref{dlcqk2}) if we re-scale the
undetermined function $f$ (appearing in front of the vertices) by
$\sqrt{2/3}$. The result is superior to the SVPS result (\ref{SVPSk2})
as it does not contain the $3/2$'s power of $\l'$. It would be
interesting to know whether the $\l'^2$ term also agrees with gauge
theory, however the gauge theory computation of this term has yet to
be done.

% -------------------------------------------------------------------------- %
\subsubsection{$k=3$ Impurity-conserving mass-shift}

For the $k=3$ string, the splitting and level-matching are more
involved. There are two distinct cases, the first is when string \#1
has $k_1=1$. We can then distribute the two intermediate state
impurities both on string \#1, both on string \#2, or one impurity
per string (of which there are two equivalent configurations). The
next case is when the assignments of light-cone momenta are reversed,
so that string \#1 has $k_1 = 2$ (and so string \#2 has $k_2 =
1$). This just counts the $k_1 = 1$ case again, leading to a factor of
two. The level-matching is therefore achieved via the insertion of the
following operator    

\be\label{k3levelm}
r = \frac{1}{3},~~~~ 2\biggl( \d^{r_1,1}\d^{r_2,1} \d_{q_1,m-q_2}
+ \d^{r_1,2} \d^{r_2,2} \d_{q_1,2m-q_2} + 2\d^{r_1,1} \d^{r_2,2}
\d_{q_1,2(m-q_1)} \biggr)
\ee

\ni where the intermediate-state impurities have mode-number/string
label configurations $(q_1,r_1)$ and $(q_2,r_2)$, and $m=(n_1+n_2)/3
\in \bZ$ is the external state winding number while $n_1$ and $n_2$ are
integers and not multiples of three. 

The expressions given for the $k=2$ case in the previous subsection
are equally valid here, however with the replacement of the $k=2$
delta function with (\ref{k3levelm}). The results are difficult to
obtain for high order in $\l'$, and so we present leading order
results only. Since the calculations are straightforward, we will be
brief and simply state the results

\be\label{SVPSk3}\boxed{
\left(\d E^{\text{SVPS}}\right)_{k=3} = 
\frac{g_2^2\l'}{16\pi^2}\Biggl[ 1 +  \frac{9}{2} \frac{ \left[
 \cos \left(\pi n_1\over 3 \right) \sin \left(\pi
n_1\over 3\right) - \cos \left(\pi n_2\over 3 \right) \sin \left(\pi
n_2\over 3 \right) \right]}{\pi(n_1-n_2)} \Biggr] + \ldots}
\ee 

\be\label{DVPPRTk3}\boxed{
\left(\d E^{\text{DVPPRT}}\right)_{k=3} = 
\frac{g_2^2\l'}{16\pi^2}\Biggl[ -1 +  \frac{3}{2} \frac{ \left[
 \cos \left(\pi n_1\over 3 \right) \sin \left(\pi
n_1\over 3\right) - \cos \left(\pi n_2\over 3 \right) \sin \left(\pi
n_2\over 3 \right) \right]}{\pi(n_1-n_2)} \Biggr] + \ldots}
\ee 

\be\label{DYk3}\boxed{
\left(\d E^{\text{DY}}\right)_{k=3} = 
\frac{g_2^2\l'}{16\pi^2}\frac{3}{2}\Biggl[ 1 +  \frac{3}{2} \frac{ \left[
 \cos \left(\pi n_1\over 3 \right) \sin \left(\pi
n_1\over 3\right) - \cos \left(\pi n_2\over 3 \right) \sin \left(\pi
n_2\over 3 \right) \right]}{\pi(n_1-n_2)} \Biggr] + \ldots}
\ee 

\ni Comparing with the gauge theory result (\ref{dlcqk3}), we see that
although the dependence on the external mode numbers is of the correct
form, the coefficient of the second term is not matched by any of the
vertices. Further, the first term of the DVPPRT does not match on
account of the sign.  

% -------------------------------------------------------------------------- %
\subsubsection{$k=2$ Four impurity channel mass-shift}

We have had success in matching the leading $k=2$ mass-shift to gauge
theory using both the SVPS and DY vertices and the impurity conserving
channel. It is therefore interesting to see whether or not a
miraculous cancellation appears at the four-impurity channel, such
that it leads as $\l'^2$ or higher. The $k=2$ setting makes the
calculation simpler than it would be for the standard, continuous $p^+$
case. The reasons for this are as follows.

\begin{enumerate}

\item The intermediate strings have only one possible distribution of
  $p^+$: each string must have $k =1$. This gives the same
  level-matching condition regardless of the distribution of the four
  impurities amongst the two strings; thus one may be chosen and the
  result multiplied by 16.

\item The leading $\l'$ term comes only from those expressions
  containing a double-pole in one of the intermediate mode number
  sums. This allows us to discard many complicated terms from the
  calculation.

\item The result will be independent of $n_1$ and $n_2$, and therefore
  just a number. The expressions are therefore simple and easy
  to manipulate. 

\end{enumerate}

The calculation was performed by the author using two methods ``in
parallel'' as a check on the results. The methods used are the
standard $H_3$ and contact term we have been using all along, and the
manifestly convergent method (\ref{newstuf}) developed in section
\ref{sec:arbimp}. We can relate the quantities appearing in these two
methods by exploiting the superalgebra. We have

\be
[H_2,Q_3] = [Q_2,H_3].
\ee

\noindent Let $|\phi\ra = |[{\bf 9,1}]\ra$, $|I\ra=$ be a generic two
string intermediate state, $\la \psi | = \la \phi |\,Q_2$, and
$|\Lambda\ra = Q_2 \,|I\ra$. Then

\bsp
&\la \phi | \,H_2\,Q_3 - Q_3\,H_2 \,|I\ra = \la \phi| \,Q_2\,H_3 - H_3\,Q_2 \,|I\ra\\
&\left( \frac{\o_n^{(3)}}{-\a_3} -\sum_{s=1}^2 \, \frac{\o_{p_s}^{(s)}}{-\b_s\a_3} \right)
\, \la \phi | Q_3 \,|I\ra =
\la \psi|\,H_3 |I\ra - \la \phi|\,H_3 \,|\Lambda\ra\\
\end{split}
\ee

\noindent therefore

\be\label{pardE1}
\delta E_1 = 
\frac{\la \phi |\la I|\,Q_3\ra ~\la H_3 | I\ra|\psi\ra}{4\,\Delta E} = 
\frac{1}{4}\left| \la \phi|\la I|Q_3\ra \right|^2 
+\frac{ \la \phi |\la I|\,Q_3\ra ~\la \phi|\la \Lambda| H_3\ra^* }{4\,\Delta E}.
\ee

\ni Further, we have

\be
\{ Q_2, Q_3 \} = 4 H_3.
\ee

\ni Taking the expectation value in the same way, we find

\bsp
\la \phi | \,Q_2\,Q_3 + Q_3\,Q_2 \,|I\ra &= 4\,\la \phi| \,H_3\,|I\ra\\
\, \la \psi | Q_3 \,|I\ra +\la \phi|\,Q_3 \,|\Lambda\ra &= 
4\, \la \phi|\,H_3 |I\ra 
\end{split}
\ee

\noindent and therefore

\be\label{pardE2}
\delta E_2 = 
\frac{\la \psi |\la I|\,Q_3\ra ~\la H_3 | I\ra|\phi\ra}{4\,\Delta E} = 
\frac{\left| \la \phi|\la I|H_3\ra \right|^2 }{\Delta E}
-\frac{ \la \phi |\la I|\,H_3\ra ~\la \phi|\la \Lambda| Q_3\ra^* }{4\,\Delta E}.
\ee

\ni We have calculated all three terms in (\ref{pardE1}) and in
(\ref{pardE2}), for the four impurity channel, at leading order in
$\l'$, checking that the two methods give the same result. Many of the
matrix elements are to be found in appendix \ref{app:example}. The
various intermediate states may be classified by the number of
$\a_q^i$, $\a_q^{i'}$, $\b^{\a_1\a_2}_q$, and $\b^{\da_1\da_2}_q$
impurities. As an example we will show the $\d E_1$ calculation of the
$\a'\,\a\, \a\, \b$ channel - i.e. one undotted fermion, two bosons
from the first $SO(4)$ and one from the second. We begin by finding
the $\a'\,\a\, \a\, \b$ contributions from $\la H_3 | I\ra|\psi\ra$,
found in (\ref{Q2onH3}). The only source for a pole in an intermediate
state mode number is the Neumann matrix $\wt N^{3\,r}_{n\,q}$,
therefore we ignore any contribution which does not contain this
matrix. Further, as per usual, we are only interested in those
contributions which do not result in a delta function on the external
state's spacetime indices. We find the contribution to be

\bsp 
\la Q_2 : \a_{n_1}^k \a_{n_2}^l | \la I | \, &H_3 \ra = \\
\frac{g_2 \a'}{8 \a_3^3}&\frac{\bar \eta}{\sqrt{2|\a_3|}}
\s^k_{\b_1\dg_1} \O_{n_1} G_{n_1} \wt N^{3\,s_2}_{n_2\,p_2}
\a^{\dag\,(s_2)\,l}_{p_2} \left[ K^{\dr_1 \r_1} \wt K^{\dr_2 \r_2} +
\wt K^{\dr_1 \r_1} K^{\dr_2 \r_2} \right] Y_{\r_1 \r_2}
\d^{\dg_1}_{\dr_1} \e_{\db_2 \dr_2}\\ 
&\qquad\qquad\qquad\qquad\qquad + (n_1 \lr n_2)\\ 
= \frac{g_2 \a'}{8 \a_3^3}\frac{\bar \eta}{\sqrt{2|\a_3|}}
\s^k_{\b_1\dg_1}& \s^{i_1\,\dg_1\r_1}  \s^{i'\,\r_2}_{~~\db_2}   
\O_{n_1} G_{n_1} \wt N^{3\,s_2}_{n_2\,p_2} G_{p_1} \left[ K_{p_3}
  K_{-p_4} + K_{-p_3} K_{p_4} \right] 
\a_{p_2}^{\dag\,l} \a_{p_3}^{\dag\,i_1} \a_{p_4}^{\dag\,i'}
\b_{p_1\,\r_1\r_2}^{\dag\,l} \\
&\qquad\qquad\qquad\qquad\qquad+ (n_1 \lr n_2).
\end{split}
\ee

\ni The contribution to the $\a'\,\a\, \a\, \b$ channel from $\la
\phi |\la I|\,Q_3\ra$ is read-off from (\ref{NQ3elem}), it is

\be \left( \la \a_{n_1}^i \a_{n_2}^j |\la I|\,Q_3\ra \right)^\dag =
\frac{g_2 \bar \eta}{4 \a_3^3} \sqrt{\frac{\a' \k}{2}} (-i) \,
\s^{j'}_{\l_2 \dl_2} G_{q_4} \e^{\dl_2 \db_2} K_{-q_3} \wt
N^{3\,r_1}_{n_1\,q_1} \wt N^{3\,r_2}_{n_2\,q_2} \a^{j'}_{q_3}
\a_{q_1}^{i} \a_{q_2}^{j} \b_{q_4}^{\b_1 \l_2}.  \ee

\ni The next step is to calculate

\bsp\label{aaab}
&\left( \la \a_{n_1}^i \a_{n_2}^j |\la I|\,Q_3\ra \right)^\dag
\la Q_2 : \a_{n_1}^k \a_{n_2}^l | \la I | \, H_3 \ra =
-\frac{g_2^2 \a'^{3/2}}{128 \a_3^6} (-\a_3) \left(2 \d^{ik} \d^{jl}\right)
\left( 2 \d^{i'j'} \d^{i'j'} \right) \O_{n_1} G_{n_1}\\
 &\qquad\qquad\qquad \times K_{-q_3}
\left(G_{q_4}\right)^2 \left( \wt N^{3\, r_2}_{n_2\,q_2} \right)^2 \wt
N^{3\, r_1}_{n_1\,q_1} \left[ K_{q_1} K_{-q_3} + K_{-q_1} K_{q_3}
  \right] + (n_1 \lr n_2).
\end{split}
\ee

\ni Finally, we must level-match and sum. This is accomplished via

\be
\sum_{r_1,r_2,r_3,r_4=1}^2 \,\sum_{\substack{q_1, q_2, q_3, q_4\\\sum
    q_i = m}} = 16 \,\sum_{\substack{q_1, q_2, q_3, q_4\\\sum
    q_i = m}} 
\ee

\ni reflecting the fact that all distributions of intermediate-state
impurities over the internal strings are equivalent. The factor of
$\left( \wt N^{3\, r_2}_{n_2\,q_2} \right)^2$ in (\ref{aaab}) plays a
very important r\^{o}le. It provides a double pole in the sum over $q_2$,
fixing it to $n_2/2$ and causing $\left( \wt N^{3\, r_2}_{n_2\,q_2}
\right)^2$ to evaluate to $\frac{1}{2}$ in the large-$\m$ limit. Since
the remaining mode numbers will effectively be order-$\m$, one can
simply set $n_1$ and $n_2$ to zero, leaving a sum over two
mode-numbers, $q_1$ and $q_3$ say, while $q_4 = -(q_1+q_3)$. Taking
the large-$\m$ limit of the remaining expressions, we find

\be\label{intform}
\d E_1 = \frac{g_2^2 \l'}{16 \pi^4} \int_{-\infty}^\infty d q_1 
 \int_{-\infty}^\infty d q_3 
\frac{\left( \L^+_3 + \L^-_3
  \right) \L_1^+ \left[ \L_1^+ \L_3^+ - \L^-_1 \L^-_3 \right] }{q_1
  \o_1 \o_3 \o_4 \left( 1 - \o_1 - \o_3 - \o_4 \right) }
\ee

\ni where we have scaled $\m\a_3$ out of all quantities so that

\be
\o_i = \sqrt{q_i^2 + 1}, \qquad \L^+_i = \sqrt{\o_i + 1},\qquad \L^-_i
=e(q_i) \sqrt{\o_i - 1}.
\ee

\ni Unfortunately (\ref{intform}) is as far as we can go, integrals of
this form do not have closed analytical solutions. However, we can
still calculate the total four-impurity channel shift and express it
in terms of integrals of this form. We have done this, and verified
our results as indicated previously, by ensuring that (\ref{pardE1})
and (\ref{pardE2}) are satisfied for each channel (i.e. combinations
of impurities), and finally that both the standard method and $(\d E_1
+ \d E_2)/4$ give the same result for the complete four-impurity
mass-shift. The cancellation of divergences discussed in section
\ref{sec:divcan} are found explicitly; this is a confirmation of the
work in that section. 

A complete presentation of the calculation would fill many pages and
we will opt not to do this. The result is however, that neither the
SVPS, DVPPRT, nor the DY vertices give a zero result. These vertices
contribute to the leading $\l'$ order for the four-impurity
channel. The results may be expressed by approximate numerical results
for the integrals

\be\label{k24imp}\boxed{
\d E_{k=2}^{\text{4-imp.}} = \frac{g_2^2 \l'}{\pi^4} \times
\begin{cases}
-0.68 \qquad \text{SVPS}\\
0.29 ~~(0.22 ~~L.R.) \qquad \text{DY}\\
-0.42 \qquad \text{DVPPRT}
\end{cases}}
\ee  

\ni where we have indicated in brackets the result of a suggested
extension of the DY vertices proposed by Lee and Russo
\cite{Lee:2004cq}, affecting only the impurity non-conserving
channels.

% -------------------------------------------------------------------------- %
\subsection{Discussion}

The DLCQ light-cone string field theory has been investigated at the
impurity conserving level for the $k=2$ and $k=3$ two impurity
external state. Further the four-impurity channel has also been
investigated for the $k=2$ state. The results for $k=2$ are
inconclusive. We have available only an external mode number
independent prediction from gauge theory (\ref{dlcqk2}) to compare
to. The impurity conserving channel gives such a number for any of the
three vertices considered (\ref{SVPSk2}), (\ref{DVPPRTk2}), (\ref{DYk2}). 
The four-impurity channel also contributes at this leading order
(\ref{k24imp}). The total shift from the impurity-conserving and
four-impurity channel is negative for SVPS and DVPPRT, which is a
mismatch with gauge theory, but since we have no evidence of a
truncation of $\l'$ terms above four impurities, higher channels may
correct this. It is reassuring that the DY result is free of
half-powers of $\l'$ at the impurity-conserving level; however the
higher orders of the four-impurity channel could easily contain
half-integer powers; our analysis was only able to capture the leading
term.

The results for $k=3$ (\ref{SVPSk3} - \ref{DYk3}) fail to reproduce
the gauge theory result (\ref{dlcqk3}). The dependence on the external
mode numbers is correct, it is the coefficients which are
mismatched. What effect higher impurity channels may have on these
results remains a mystery. 

The broad outlines of the gauge theory results are captured here -
truncation of the $k=2$ spectrum, protection of the $k=1$ spectrum,
and the absence of corrections when the external mode numbers are
multiples of the external light-cone momenta. The general form of
corrections also seems correct, however the precise details continue
to be lacking. The main issue is the effect of higher impurity
channels. Until these can be brought under control, the validity of
the vertices cannot truly be known. 

% ************************************************************************** %
\section{Conclusions}

In the plane-wave limit, the AdS/CFT correspondence stands the best
chance of being systematically tested beyond the classical
(i.e. planar) level. Light-cone string field theory on the plane-wave
background is the tool for carrying out such tests. As it stands, the
correct form of the string interaction vertices is ambiguous. Various
proposals are put forward, but at the base of all of them is the
fundamental construction on the foundation of (super)-locality: the
strings must touch (in superspace) where they interact. Unfortunately
symmetry alone is not enough to completely fix the interactions. There
are two main issues as regards the light-cone string field theory on
the plane-wave 1) The lack of a construction for the quartic
supercharge $Q_4$, and 2) The lack of tools to analyze the higher
impurity channels. Without these ingredients, the validity of the
proposed vertices will likely remain unknown. 

It is troublesome that a
correspondence conjectured to be valid on the basis of symmetries
fails to be tested due to a lack of (string theory) information beyond
those symmetries. Indeed, we do not have a first principles approach
to constructing {\it the} light-cone string field theory; and so
attempting to match gauge theory results takes on an air of
predetermined conclusions. On the other hand many features of the
gauge theory treatment are manifested in the light-cone string field
theory and  the question of agreement essentially comes down to one of
coefficients. 

The structure of this string field theory deserves to be explored
further. One would not be too surprised to find a cancellation
mechanism limiting the order of the results in $\l'$ as the number of
intermediate state impurities is increased. Further, a cancellation
mechanism for the half-powers of $\l'$, as was shown in this chapter
for $\sqrt{\l'}$, seems possible and worth looking for. An explicit
construction involving a non-zero $Q_4$ would also go a long way in
elaborating the theory. Testing the AdS/CFT correspondence at the
``quantum'' level, that is, the non-planar/string-loop level, remains
one of the most important pursuits in fleshing-out and comprehending
the duality.

% ************************************************************************** %
% ************************************************************************** %
% ************************************************************************** %

\chapter{Free energy and phase transition of the matrix model on a
  plane-wave}
\label{sec:matrixmodel}

{\small
\begin{quote}
Double, double toil and trouble;\\ Fire burn, and caldron bubble.\\
\rightline{--- Shakespeare's {\it Macbeth}, Act IV, Scene 1
$\;\;\;\;\;\;\;\;\;\;\;\;\;\;\;\;\;\;\;\;\;\;\;\;\;\;\;\;\;\;\;\;\;\;\;\;\;\;\;\;\;\;\;
\;\;\;\;\;\;\;\;\;\;\;\;\;\;\;\;\;$}
\end{quote}}

In section \ref{sec:clos} we mentioned that 11-dimensional
supergravity (\ref{S11}) plays a privileged r\^{o}le. Eleven is the
maximum spacetime dimension before massless particles of spin greater
than 2 are introduced by supergravity. The lower dimensional
supergravities are (essentially) derivable via dimensional reduction
from this master theory. Superstring theory is a quantization of
10-dimensional gravity; finding a supersymmetric quantization of
11-dimensional gravity might then produce a master theory from which
all string theories are derivable. It was in this effort that
``M-theory'' or ``Matrix'' theory was developed. The path was to
attempt the quantization of a membrane (a 2-spatial dimensional
object) in an 11-dimensional target space. Working in the light-cone
gauge and promoting spatial worldvolume coordinates to matrices, a
regularization or discretization was achieved, resulting in a theory of
$N\times N$ matrices which depend on a single time-like
parameter. This matrix quantum mechanics was shown by Banks, Fischler,
Shenker, and Susskind (BFSS) \cite{Banks:1996vh} to also describe a
collection of $N$ D0-branes in type-IIA superstring theory. It was
then found that various classical and quantum mechanical processes in
11-d supergravity were captured by the matrix model, leading to the
conjecture that the full second-quantized theory containing 11-d SUGRA
as its low-energy limit was encoded by the matrix model.
\begin{figure}[ht]
\begin{center}
\includegraphics*[bb=0 0 280 220, height=2.5in]{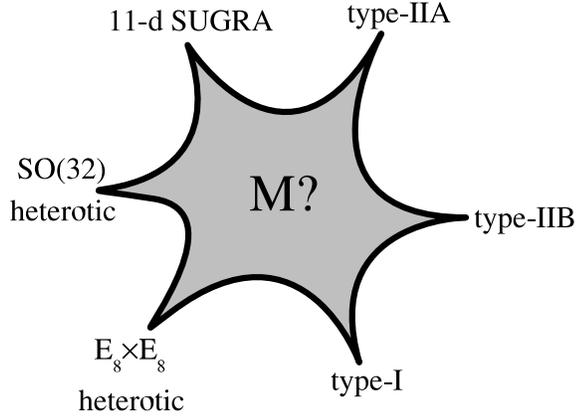}
\end{center}
\caption{The five types of string theories: type-IIA, type-IIB,
  type-I, and the two heterotic string theories are related via
  various dualities. The low-energy limit of the BFSS matrix model,
  which may be understood as a collection of D0-branes in type-IIA
  string theory, gives 11-dimensional supergravity. Could this matrix
  model also represent a master theory from which all string theories
  arise as limits?}
\label{fig:starfish}
\end{figure}
A more ambitious proposal is that the BFSS matrix model describes a
{\it master} theory containing within it, as limits, all known string
theories as well as 11-d SUGRA, see figure \ref{fig:starfish}. 

The BFSS matrix model suffers from the drawback that it does not
contain a perturbative coupling constant. The plane-wave background
(\ref{ppmet}) introduced by Berenstein, Maldacena, and Nastase
\cite{Berenstein:2002jq} also has a cousin in 11-dimensional SUGRA,
and the membrane can be quantized in the presence of this background.
Alternatively, one may consider the collection of $N$ D0-branes on the
10-dimensional type-IIA plane-wave. These approaches both lead to the
plane-wave matrix model which does have a perturbative coupling
\cite{Dasgupta:2002hx}. Essentially, this matrix model (and the BFSS
model) is a 1-dimensional gauge theory whose large-$N$ limit
corresponds to the low-energy 11-dimensional SUGRA limit. In this
sense, it is a manifestation of a gauge/gravity duality like the
AdS/CFT correspondence. In the standard AdS/CFT duality, considering
the CFT at finite temperature is dual to a gas of gravitons in the AdS
space, characterized by the same temperature. As the temperature is
raised, the AdS space undergoes a phase transition leading to the
production of a large black-hole, an object which is thermally
stable. This is known as the Hawking-Page phase transition
\cite{Hawking:1982dh}. The transition on the gauge theory side is
conjectured to be the analogue of the deconfinement transition in QCD
\cite{Witten:1998zw}. At low energies, the degrees of freedom are
singlets of the gauge group $SU(N)$, and so the free energy is of
order one. As the temperature is raised, charged states are liberated,
so that at high enough temperature every possible state is excited. In
this phase the free energy scales as $N^2$, the total number of
fundamental excitations of the theory. As we will see, the plane-wave
matrix model also shares a deconfinement transition. The dual gravity
interpretation, however, is less clear. Determining the order of this
transition is therefore an interesting endeavour, as it should shed
some light on the dual process. Notwithstanding that, it is of general
interest to understand deconfinement transitions wherever they arise,
as this information should help us to eventually understand the QCD
deconfinement transition, a subject of paramount importance in physics
and cosmology.

% ************************************************************************** %
\section{M-theory and the BFSS matrix model}

In section \ref{sec:clos}, we showed how type-IIA supergravity could
be derived from 11-dimensional supergravity via dimensional
reduction. When a theory is dimensionally reduced, the extra
dimensions are taken to be compact with radius $R$. This gives the
familiar Kaluza-Klein mechanism, where momentum in the compact
direction becomes mass in the dimensionally reduced theory. The mass
comes in units of $R^{-1}$, and so as $R\rightarrow 0$, the
zero-momentum modes become decoupled from the infinitely more massive
Kaluza-Klein states, which can be ignored. Beginning with type-IIA
string theory, we can actually follow this process in reverse, and
watch while this theory {\it grows} an extra dimension, becoming a
theory whose low-energy limit is 11-d SUGRA, i.e. M-theory. The trick
is to consider the D0-branes of type-IIA superstring theory. These
objects have a mass given by

\be
\t_0 = \frac{1}{g_s \sqrt{\a'}}
\ee 

\ni where $g_s$ is the string coupling. These objects are charged
under the Ramond-Ramond vector potential $A_1$, see (\ref{IIA}). We
discussed in section \ref{sec:sugrapbranes} that parallel D-branes do
not interact as a consequence of their gravitational attraction
balancing their form-field repulsion exactly. A stable ground state of
a system of parallel Dp-branes was then to have them coincident, as
excitations consisting of open strings stretched between separated
pairs would tend to pull them together. A collection of point-like
objects are always parallel, and so a coincident arrangement of $n$
D0-branes counts simply $n$ times the D0-brane mass

\be\label{Mspec}
\t_n = n\t_0 = \frac{n}{g_s \sqrt{\a'}}.
\ee 

\ni This is immediately reminiscent of a tower of Kaluza-Klein states
on a compact direction of radius $R = g_s \sqrt{\a'}$. Indeed, as $g_s
\rightarrow \infty$, the spectrum (\ref{Mspec}) becomes continuous, $R
\rightarrow \infty$, and type-IIA string theory grows a new
decompactified direction out of its non-perturbative, point-like
D0-branes. This strong-coupling limit of type-IIA superstring theory,
whatever its true description may be, is given the name
``M-theory''. At the supergravity level, M-theory is just
11-dimensional supergravity, whose action is (\ref{S11}).

The action of $N$ D0-branes may be derived as a dimensional reduction
of 10-dimensional supersymmetric Yang-Mills theory with gauge group
$SU(N)$ to 0+1-dimensions. Nine of the ten components of the gauge
field $A_\m$ become the scalar fields $X^I$, while the remaining gauge
field is zero-dimensional and is called $A_0$

\be\label{BFSS}
S = \frac{1}{2R} \Tr \int dt \left[ D_0 X^I D_0 X^I +i\,\theta^T
  D_0 \,\theta + \frac{1}{2} \left[ X^I, X^J \right]^2 -\theta^T
  \g_I \left[ \theta, X^I \right] \right]
\ee

\ni where $D_0 = \p_t - i [A_0,\ldots]$, the fermionic superpartners
$\theta$ have been included, and all fields are $N\times N$
matrices. The scalars $X^I$ have a very pretty interpretation. We know
from section \ref{sec:dbranes} that the VEV's of these fields describe
the transverse shape of a general D-brane. In fact, here, for
D0-branes, all spatial directions are transverse. The ``position'' of
the D0-branes may become non-commutative or ``fuzzy''. The lowest
energy configuration is to take the $\la X^I \ra$ constant, and have
the commutator vanish, thus allowing them to be simultaneously
diagonalized. The eigenvalues are precisely the positions of the $N$
D0-branes. Turning on off-diagonal elements of $\la X^I \ra$ gives a
non-commutative geometry, where the ``positions'' of the branes are
matrix valued. In two important works, \cite{Banks:1996vh} and
\cite{Susskind:1997cw}, convincing arguments were given that
(\ref{BFSS}) indeed describes the discrete light-cone quantization or
DLCQ of M-theory. The infinite $N$ limit should then correspond to
decompactified M-theory viewed in the infinite momentum frame
\cite{Weinberg:1966jm}. Perhaps the most convincing evidence is that
11-dimensional supergravity scattering amplitudes are readily
retrieved using (\ref{BFSS}), as are the extended objects of 11-d
SUGRA (see \cite{Taylor:2001vb} for a review). The significance of the
BFSS model is that (\ref{BFSS}) was obtained previous to those
authors' work, in a very different context. In the 1980's there was a
campaign to attempt the quantization of 11-dimensional gravity via a
two-dimensional membrane, in much the same way that a one-dimensional
string led to the quantization of 10-dimensional gravity
\cite{deWit:1988ig} (see \cite{Taylor:2001vb} for a modern review and
more references). This work led to the promotion of the membrane
embedding functions to matrices, in order to provide a regularization
to the theory. The action was then found to be precisely (\ref{BFSS}).

The membrane/BFSS theory has flat-directions (those in which the
commutator vanishes), leading to a continuous spectrum. Prior to BFSS
this was interpreted as an instability in the dynamics of the
membrane. Adding a long spike, of vanishing area, to a membrane
incurs a vanishing energy cost. A ``sea-urchin'' picture of the
membrane then emerges, with large and wild fluctuations in membrane
shape which cost nearly no energy, leading to a continuous spectrum.
This instability was a stumbling block for membrane research, and
stymied its progress. BFSS provided a natural interpretation for this
continuous spectrum. The theory ought to be considered {\it second}
quantized, as it is capable of describing multi-particle states
(i.e. multiple D0-branes); ergo a continuous spectrum. Indeed, as we
have mentioned above, BFSS showed that (\ref{BFSS}) was capable of
describing the scattering of multi-particle states in 11-d SUGRA.

The continuous spectrum of the BFSS model, though turned from a
liability to an asset, still makes calculations challenging compared
to a model with a discrete spectrum. The other drawback of
(\ref{BFSS}) is that it has no tunable coupling constant. The coupling
$R$ is essentially the 11-dimensional Newton's constant, leading to a
rather peculiar quantum-classical correspondence. The non-linear terms
of 11-dimensional Einstein gravity are reproduced through quantum loop
corrections stemming from (\ref{BFSS}). The full classical 11-d
gravity therefore requires the all-loop results of the matrix quantum
mechanics. For a general process, these loop corrections are not
perturbative; indeed there is no sense in which $R$ is small.

% ************************************************************************** %
\section{The plane-wave matrix model}

In the seminal work by Berenstein, Maldacena, and Nastase
\cite{Berenstein:2002jq}, a deformation of the BFSS model was given
which may be understood as the action of N D0-branes on the type-IIA
plane-wave background, or equivalently as the quantization of the
supermembrane in an 11-d SUGRA plane-wave \cite{Dasgupta:2002hx}. The
trough of the plane-wave background (see figure \ref{fig:trough})
causes the previously flat directions to become massive, leading to a
discrete spectrum, while the parameter $\m$ leads to a tunable coupling
constant. The plane-wave matrix model thus cures the two drawbacks of
the BFSS model, and presents itself as an instance of M-theory which
readily lends itself to exploration. The 11-dimensional plane-wave can
be obtained via a Penrose limit either of $AdS_4 \times S^7$ or $AdS_7
\times S^4$, both maximally symmetric solutions of 11-dimensional
supergravity. The result is a plane-wave with different masses for
three of the transverse directions as compared to the remaining six

\bsp\label{11ppmet} 
ds^2 = -2 dx^+ dx^- + dx^I dx^I - \left[ \left(
\frac{\m}{3} \right)^2 x^{\bar a} x^{\bar a}
+\left(\frac{\m}{6}\right)^2 x^i x^i \right] dx^+ dx^ +\\ F_{123+} =
\m
\end{split}
\ee

\ni where $I=1,\ldots,9$, $\bar a=1,\ldots 3$, and $i = 4,\ldots,9$. The
action of the plane-wave matrix model is then given by
\cite{Berenstein:2002jq}, \cite{Dasgupta:2002hx}

\bsp\label{PWMM} 
S &=\frac{1}{2R} \int d\tau \Tr \Biggl( D X^{\bar a}
D X^{\bar a} +D {X}^i D {X}^i + i \psi^{\dagger I \alpha} D \psi_{I
\alpha}\\ &\qquad-\left(\frac{\mu}{3}\right)^2 (X^{\bar a})^2 -
\left(\frac{\mu}{6} \right)^2 (X^i)^2 - \frac{\mu}{4} \psi^{\dagger I
\alpha} \psi_{I \alpha}\\ &\qquad+ \frac{R^2}{2} [X^{\bar a}, X^{\bar
b}]^2 + R^2 [ X^{\bar a}, X^i]^2 +\frac{R^2}{2} [X^i , X^j]^2 -
i\mu\frac{2R}{3 } \epsilon_{\bar a \bar b \bar c} X^{\bar a} X^{\bar
b} X^{\bar c}\\ &-R\, \psi^{\dagger I \alpha} \sigma^{\bar a}_\alpha
{}^\beta \left[X^{\bar a}, \psi_{I \beta}\right] + \frac{R}{2}
\epsilon_{\alpha \beta} \psi^{\dagger \alpha I} {\sf g}^i_{IJ}
\left[X^i, \psi^{\dagger \beta J}\right] - \frac{R}{2}
\epsilon^{\alpha \beta} \psi_{\alpha I} ({\sf g}^{i \dagger})^{IJ}
\left[X^i, \psi_{\b J}\right] \Biggr)
\end{split}
\ee

\ni where all variables transform in the adjoint representation of the
gauge group $X^i\to UX^iU^{\dagger}$, etc. The time derivatives are
covariant, $D = \del_\tau - i\left[A,...\right]$ with an $N\times N$
Hermitian gauge field $A$. The fermions have $8$ complex components
with $I,J=1,\ldots,4$ and $\alpha,\beta=1,2$. The spin matrix has the
property $ {\sf g}^i ({\sf g}^j)^\dagger + {\sf g}^j ({\sf
g}^i)^\dagger = 2 \delta^{ij} {\bf 1}_{4\times 4} $.
$\epsilon_{\alpha\beta}$ and $\epsilon_{\bar a \bar b \bar c}$ are
antisymmetric tensors.

The classical supersymmetric vacuua of (\ref{PWMM}), and the
perturbation theory about those vacuua, were discovered by Dasgupta,
Sheikh-Jabbari, and Van Raamsdonk \cite{Dasgupta:2002hx}. They noted
that the bosonic potential is given by

\be
V = \frac{R}{2} \Tr \left[ \left( \frac{\m}{3R} X^{\bar a} + i
  \e^{\bar a \bar b \bar c} X^{\bar b} X^{\bar c} \right)^2 -
  \frac{1}{2}\left[X^i,X^j\right]^2 - \left[X^i, X^{\bar a} \right]^2
  + \left(\frac{\m}{6R}\right)^2 X^i X^i \right]
\ee

\ni where, for supersymmetric solutions, each term must vanish
independently. The solutions are simple and beautiful

\be
X^{\bar a} = \frac{\m}{3R} J^{\bar a},\qquad X^i = 0
\ee

\ni where $J^{\bar a}$ are an $N$-dimensional representation of
$SU(2)$

\be
[J^{\bar a}, J^{\bar b}] = i \e^{\bar a \bar b \bar c}  J^{\bar c}.
\ee 

\begin{figure}[ht]
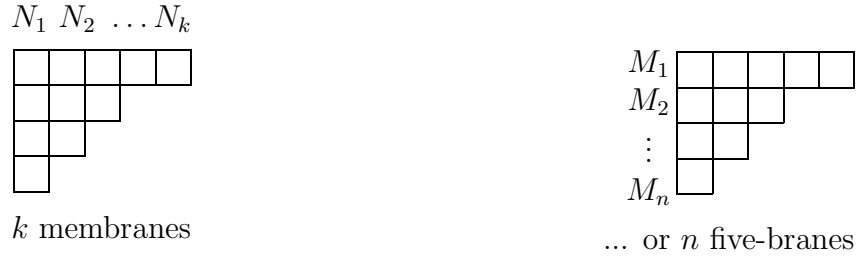

\begin{minipage}{0.4\textwidth}
\bsp\nonumber
N_1~N_2~\ldots N_k\\
\yng(5,3,2,1)\\
\text{$k$ membranes}
\end{split}
\ee
\end{minipage}
\begin{minipage}{0.6\textwidth}
\vspace{0.75cm}
\bsp\nonumber
\begin{matrix}
M_1\cr M_2 \cr \vdots \cr M_n 
\end{matrix}
\begin{aligned}
\yng(5,3,2,1)
\end{aligned}\\
\text{... or $n$ five-branes} 
\end{split}
\ee
\end{minipage}
\caption{The vacuua of the plane-wave matrix model (\ref{PWMM}) are
  given by $N$-dimensional representations of $SU(2)$. These may
  in turn be pictured as Young tableaux. A given representation may be
  interpreted as a collection of membranes (shown on the left) where
  each column corresponds to a single membrane whose radius is
  proportional to its size $N_i$; or as a collection of five-branes,
  where the r\^{o}les of column and row are reversed (shown on the
  right).}
\label{fig:youngtab}
\end{figure}

\ni The M-theory interpretation of these vacuua was given in
\cite{Dasgupta:2002hx}, and in a subsequent paper
\cite{Maldacena:2002rb}. The extended objects of 11-dimensional
supergravity are of two varieties. The action (\ref{S11}) contains a
three-form potential indicating that objects with 3-dimensional or
6-dimensional worldvolumes can couple to it electrically or
magnetically, respectively. These are the membranes or ``M2-branes''
and fivebranes or ``M5-branes'' of the theory. A general
$N$-dimensional representation of $SU(2)$ has a block-diagonal
structure where the size of the blocks is given by a partition
$\{N_1, \ldots, N_k\}$ of $N$, i.e. $\sum_{i=1}^k N_i = N$. Let $N_1
\geq N_2 \geq \ldots \geq N_k$, then the partition may be represented
by a Young tableau, see figure \ref{fig:youngtab}, with $k$ columns
whose depths are given by $\{N_1, \ldots, N_k\}$. 
These are naturally
interpreted as a collection non-commutative ``fuzzy-spheres'', which
approach, in the large-$N$ limit, a collection of spherical M2-branes
with radii

\be r_i = \sqrt{\frac{1}{N_i} \Tr X_i^2 } = \frac{\m N_i}{6 R} =
\frac{\m \,p^+_i}{6} \ee

\ni where we have indicated the blocks of $X^{\bar a}$ via the index
  $i$, and have used the fact that $J^{\bar a} J^{\bar a} = {\bf
  1}_{N\times N} (N^2-1)/4$. We have also noted that $N_i/R$ is to be
  interpreted as the amount of light-cone momentum $p^+$ in the DLCQ
  of M-theory. However, if every vacuum is interpreted as M2-branes,
  this leaves the question of how the M5-brane vacuua are encoded in
  the theory. The work of Maldacena, Van Raamsdonk, and Sheikh-Jabbari
  \cite{Maldacena:2002rb} cleared-up this riddle. For finite $N$, a
  given vacuum is ambiguous. In addition to the M2-brane
  interpretation, it may also be viewed as a collection of $n$
  five-branes, where $n$ is the size of the largest irreducible
  representation (depth of the deepest column in the Young
  tableau). The number of units of $p^+$ (and therefore, in the
  appropriate limit, the radius) of the five-branes are then given by
  lengths $\{M_1,\ldots,M_n\}$ of the rows in the Young tableau, see
  figure \ref{fig:youngtab}. In other words, the number of units of
  $p^+$ of the $i$-th five-brane is given by $M_i$, the number of
  irreducible representations of size greater than or equal to
  $i$. These interpretations are disambiguated via different large-$N$
  limits. To obtain a classical configuration of M2-branes, one takes
  all $N_i\rightarrow \infty$, while keeping $k$ fixed. The classical
  M5-branes are obtained by taking all $M_i\rightarrow \infty$, while
  keeping $n$ fixed. This corresponds to having an infinite number of
  repetitions of each of the $k$ irreducible representations, while
  keeping the sizes of those representations fixed. The M2-brane limit
  is just the opposite: infinite-sized representations with fixed
  repetitions. The trivial vacuum, $X^{\bar a} = 0$, being $N$ copies
  of the trivial one-dimensional representation of $SU(2)$,
  corresponds to a single five-brane whose (one-loop corrected) radius
  is \cite{Maldacena:2002rb}

\be
r_5 =  \sqrt{\frac{1}{N} \Tr X^2 } =\sqrt{\frac{18 N^2}{\m p^+}}.
\ee

The coupling constant arising from perturbation theory about the
M2-brane and M5-brane vacuua (i.e. at large-$N$) are different. The
representations of $SU(2)$ break the $U(N)$ gauge symmetry of
(\ref{PWMM}) down to a residual $U(n_k)$ symmetry, where $n_k$ is the
number of repetitions of the representation with size $k$. The
M2-brane limit has finite $n_k$ and the coupling constant is 

\be
g_{\text{eff}} = \left(\frac{3 R}{\m} \right)^{\frac{3}{2}}.
\ee

\ni The M5-brane limit, due to its enhanced gauge symmetry, picks-up
factors of $n=\sum n_k$ in index loops, leading to a 't Hooft coupling 

\be
\l_{\text{eff}} =g_{\text{eff}}^2 n 
= \left(\frac{3 R}{\m} \right)^3 n.
\ee

\ni In the next section, we will present work of the author of this
thesis concerning the thermodynamics of the plane-wave matrix model
about the single five-brane vacuum. 

% ************************************************************************** %
\section{Free energy and phase transition in the single five-brane vacuum}
\label{sec:FEPTPW}

This section is a presentation of the author's original work published
in \textsf{arXiv:hep-th/0409318} \cite{Hadizadeh:2004bf}.\\

The thermodynamics of the plane-wave matrix model was investigated by
Furuchi, Schre-\\iber, and Semenoff in \cite{Furuuchi:2003sy}. They
discovered that the theory expanded about the five-brane vacuua
demonstrates a first-order phase transition (\`{a} la Gross-Witten
\cite{Gross:1980he}) corresponding to the deconfinement of the
plane-wave matrix model. This transition was found to be unique to the
five-brane vacuua; the membrane vacuua do not demonstrate a phase
transition. They showed that matrix models generally posses Hagedorn
transitions \cite{Hagedorn:1965st}, and associated this first-order
transition with the Hagedorn temperature of M-theory in the five-brane
background. That analysis was based an a one-loop calculation of the
effective action expanded about the single five-brane vacuum. It was
therefore important to understand whether or not the first-order
nature of the transition remained at higher loop-order. As we will
describe, the two-loop effective action is not sufficient to answer
this question, the three-loop effective action (or at least portions
thereof) must be obtained. This calculation was undertaken by the
author of this thesis and his collaborators in
\cite{Hadizadeh:2004bf}, where it was found that the transition
remains first order. In the following subsections the details of this
work will be presented.

% ========================================================================== %
\subsection{Introduction}

The thermodynamics of a quantum field theory (see \cite{Gross:1980br}
for a discussion) are investigated by considering a Euclideanized
path-integral in which the time is compactified on a circle of
circumference $\b \sim T^{-1}$, where $T$ is the temperature of the
resulting ensemble. As an example, consider a quantum field theory of
a single scalar field $\phi(\vec x, t)$. The transition amplitude
between two states $|\phi_0\ra$ (at time $t=0$) and $|\phi_1\ra$ (at
time $t=t'$) defined by

\bsp
\phi(\vec x,0) | \phi_0 \ra = \phi_0(\vec x)\,
|\phi_0\ra\\
\phi(\vec x,0) | \phi_1 \ra = \phi_1(\vec x)\,
|\phi_1\ra
\end{split}
\ee

\ni is given in the path-integral formalism by (see
\cite{Peskin:1995ev}, pg. 282)

\be
\la \phi_1 | e^{-iH t'} |\phi_0\ra = \int_{\substack{\phi(\vec x, 0) =
  \phi_0(\vec x) \\
\phi(\vec x, t') = \phi_1(\vec x)}}
 \left[
  d\phi \right] \exp \left( i \int_0^{t'} dt\, {\cal L}[\phi(\vec x, t)] \right)
\ee

\ni where ${\cal L}$ is the Langrangian of the system. The partition
function for an ensemble defined by the (inverse) temperature $\b$ is
given by

\be
Z = \Tr e^{-\b H} = \sum_{\phi} \la \phi | e^{-\b H}| \phi \ra
\ee

\ni where $\{|\phi\ra\}$ is a complete set of states spanning the
configuration space of the theory. Thus by taking $it = \t$,
and by setting $it'=\b$, we find that

\be
Z = \sum_{\phi} \la \phi | e^{-\b H } |\phi\ra = \int_{\phi(\vec x, \t) =
  \phi(\vec x,\t+\b)} 
 \left[
  d\phi \right] \exp \left( \int_0^{\b} d\t\, {\cal L}[\phi(\vec x,\t)] \right)
\ee

\ni where now the functional integration proceeds over the space of
periodic fields $\phi(\vec x,\t) = \phi(\vec x, \t + \b)$. For
fermions, a similar treatment shows that anti-periodic boundary
conditions must be imposed, i.e. $\psi(\t) = - \psi(\t+\b)$. The
reason may be traced back to the grassman nature of the fermionic
fields. The Euclidean action is then defined as $S_E \equiv -
\int_{0}^{\b} d\t\, {\cal L}(\t)$, so that, for example, the partition
function for the plane-wave matrix model is, schematically

\be
Z = \int \left[d \psi\right] \left[ d X \right] \frac{\left[ d A
    \right]}{\text{gauge orbits}} e^{-S_E}.
\ee

\ni The free energy is then given simply by $F=-S_E$. 

As mentioned at the beginning of this chapter, we will find a
``deconfinement'' transition in this theory. The concept of
confinement is usually associated with spatial separation of quarks in
QCD, however the plane-wave matrix model has no spatial dimensions,
and so the concept of confinement in this context must be
clarified. In a confined phase, all states are singlets of the gauge
group. The number of such states is order 1 as compared to the rank
$N$ of the gauge group. In this phase we therefore expect that the
free energy would not scale with $N$. In the deconfined phase, the
singlet states decompose into liberated, charged states, of which
there are as many as the number of elements in the group, i.e. $\sim
N^2$. We therefore expect to find

\bsp
\lim_{N\rightarrow \infty} \frac{F}{N^2} &= 0 \qquad \text{confined}\\
\lim_{N\rightarrow \infty} \frac{F}{N^2} &\neq 0 \qquad
\text{deconfined}.
\end{split}
\ee

\ni It requires an infinite amount of energy to insert a charged,
fundamental particle, i.e. a quark, into the confined phase. In the
deconfined phase, this chemical potential is finite, a reflection of
the fact that the quarks are liberated. The difference in the free
energy when a quark is added is given by \cite{Polyakov:1978vu,Susskind:1979up}

\be\label{ploop}
F_q [T] - F_0[T] = -T \ln \la P \ra,\qquad P = \frac{1}{N} \Tr \left(
e^{i\oint d\t\,A}\right)
\ee

\ni where $P$, the Wilson loop about the Euclidean time circle, is
known as the Polyakov loop. One then has that

\bsp
\la P \ra &= 0 \qquad \text{confined}\\
\la P \ra &\neq 0  \qquad \text{deconfined}.
\end{split}
\ee

\ni The Polyakov loop is therefore an order parameter for the
deconfinement transition. We will proceed by calculating the effective
action for the Polyakov loop, in order to determine the critical
temperature, and the order of the phase transition in the plane-wave
matrix model.

% ========================================================================== %
\subsection{Gauge fixing and 1-loop results}

The gauge fixing and 1-loop effective action was worked out in
\cite{Furuuchi:2003sy}. We provide a summary of these results, taken
directly from \cite{Hadizadeh:2004bf}.
 
The partition function is given by the functional integral
\begin{equation} Z=\int [dA][dX^i][d\psi]e^{-\int_0^\beta d\tau
L[A,X^i,\psi]}\end{equation} where $L$ is the Euclidean time
Lagrangian
\begin{equation}
\begin{split}
L &=\frac{1}{2R} \Tr \left( D {X}^i D {X}^i +
 D X^{\bar a} D X^{\bar a} -
\psi^{\dagger I \alpha} D \psi_{I \alpha}
\right)\\
 & \!   + \frac{1}{ 2R}  \Tr \left(
 \left(\frac{\mu}{3}\right)^2 (X^{\bar a})^2
+ \left(\frac{\mu}{6} \right)^2 (X^i)^2  + \frac{\mu}{4}
\psi^{\dagger I \alpha} \psi_{I \alpha} + i\mu\frac{2R}{3}
\epsilon_{\bar a \bar b \bar c} X^{\bar a} X^{\bar b} X^{\bar c}
\right. \\
& \qquad + R\,\psi^{\dagger I \alpha} \sigma^{\bar a}_\alpha
{}^\beta [X^{\bar a}, \psi_{I \beta}] - \frac{R}{2}
\epsilon_{\alpha \beta} \psi^{\dagger \alpha I} {\sf g}^i_{IJ}
[X^i, \psi^{\dagger \beta J}] + \frac{R}{2} \epsilon^{\alpha
\beta}
\psi_{\alpha I} ({\sf g}^{i \dagger})^{IJ} [X^i, \psi_{\alpha J}] \\
& \left. \qquad -  \frac{R^2}{2} [X^i , X^j]^2 - \frac{R^2}{2}
[X^{\bar a}, X^{\bar b}]^2 - R^2 [ X^{\bar a}, X^i]^2 \right)
\end{split}
\end{equation}
  The bosonic and fermionic variables have periodic
and antiperiodic boundary conditions, respectively
$$A(\tau+\beta)=A(\tau) ~~,~~X^i(\tau+\beta)=X^i(\tau)
~~,~~\psi(\tau+\beta)=-\psi(\tau). $$ Since the boundary conditions
for fermions and bosons are different, supersymmetry is broken
explicitly. Of course this is expected at finite temperature where
bosons and fermions have different thermal distributions.
Supersymmetry is restored in the zero temperature limit.  We will
see the results of this explicitly in the following. 

To begin, we must fix the gauge.  It is most convenient to use the
gauge freedom to make the variable $A$ static and diagonal,
$$
\frac{d}{d\tau}A_{ab}=0 ~~,~~ A_{ab}=A_a\delta_{ab}
$$
Once this is done, the remaining degrees of freedom of $A$ are the
time-independent diagonal components, $A_a$. We shall see that
they eventually appear in the form $\exp\left( i\beta A_a\right)$.

The Faddeev-Popov determinant for the first of these gauge fixings
is\footnote{Using zeta-function regularization,
$${\det}'\left(-\frac{d}{d\tau}\right)=\beta$$.}
\begin{equation}\label{fp1}
{\det}'\left(-
\frac{d}{d\tau}\left(-\frac{d}{d\tau}+i(A_a-A_b)\right)\right) =
{\det}'\left(-
\frac{d}{d\tau}\right){\det}'\left(-\frac{d}{d\tau}+i(A_a-A_b)\right)
\end{equation}
where the boundary conditions are periodic with period $\beta$.
The prime means that the zero mode of time derivative operating on
periodic functions is omitted from the determinant. Once the gauge
field is time-independent, we do the further gauge fixing which
makes it diagonal. The Faddeev-Popov determinant for diagonalizing
it is the familiar Vandermonde determinant,
$$
\prod_{a\neq b}|A_a-A_b|.
$$
This is also just the factor that the time independent zero mode
would contribute to the second of the determinants in (\ref{fp1}).
Including it gives the determinant
\begin{equation}\label{fp2}
\prod_{a\neq b} {\det}'\left(-
\frac{d}{d\tau}\right)\det\left(-\frac{d}{d\tau}+i(A_a-A_b)\right)
\end{equation}
where there is now no prime on the second factor.  These
determinants can be found explicitly. We will do this shortly.

If we expand about the classical vacuum  $X_{\rm cl}^a=0=X_{\rm
cl}^i$, we find the partition function in the 1-loop approximation
is
\begin{equation}
Z= \int dA_a\prod_{a\neq b} \frac{ {\det}'\left(
-d/d\tau\right){\det}\left(-D_{ab}\right)
 \det^8\left(
-D_{ab}+\frac{\mu}{4}\right) }{ \det^{3/2}\left(
-D_{ab}^2+\frac{\mu^2}{9}\right) \det^{3} \left(
-D_{ab}^2+\frac{\mu^2}{36}\right) }
\end{equation}
where $D_{ab}=\frac{d}{d\tau}-i(A_a-A_b)$. The first two terms in
the numerator are the Faddeev-Popov determinant.  The third term
comes from fermions whereas the denominator is from bosons.
Using the formula
$$
\det\left(
-\frac{d}{d\tau}+\omega\right)=2\sinh\frac{\beta\omega}{2}
$$
with periodic boundary conditions and
$$
\det \left(
-\frac{d}{d\tau}+\omega\right)=2\cosh\frac{\beta\omega}{2}
$$
with antiperiodic boundary conditions, we can
write\footnote{Because the matrix model action
(\ref{PWMM}) is invariant under replacing $A$ by $A$
plus a constant times the unit matrix, we see that the integrand
in (\ref{almostpartf}) is indeed invariant under translating
all values of $A_a$ by the same constant.}
\begin{equation}\label{almostpartf}
Z= \int_{-\pi}^{\pi} \prod_{a=1}^N \frac{d\left( \beta
A_a\right)}{2\pi} \prod_{a\neq b} \frac{
[1-e^{i\beta(A_a-A_b)}][1+e^{-\beta\mu/4+i\beta(A_a-A_b)}]^8
 }{ [1-e^{-\beta\mu/3+i\beta(A_a-A_b)}
 ]^3[1-e^{-\beta\mu/6+i\beta(A_a-A_b)}]^6  }
\end{equation}

 Note that, because of supersymmetry, the zero temperature
 ($\beta\to\infty$) limit of the partition function is one.
 It also has a symmetry under replacing $e^{-\beta\mu}$ by
 $1/e^{-\beta\mu}$.

We must now do the remaining integral when $N\to\infty$. There are
$N$ integration variables $A_a$ and the action, which is the
logarithm of the integrand is generically of order $N^2$ which is
large in the large $N$ limit. For this reason, the integral can be
done by saddle point integration. This amounts to finding the
configuration of the variables $A_a$ which minimize the effective
action:
\begin{eqnarray}
S_{\rm eff}=\sum_{a\neq b} \left( -\ln
[1-e^{i\beta(A_a-A_b)}]-8\ln[1+e^{-\beta\mu/4+i\beta(A_a-A_b)}]
 +\right.\nonumber\\ \left.~~~~+3\ln[1-e^{-\beta\mu/3+i\beta(A_a-A_b)}
 ]+6\ln[1-e^{-\beta\mu/6+i\beta(A_a-A_b)}]\right)
 \label{effectiveaction}
\end{eqnarray}
To study the minima, it is illuminating to Taylor expand the
logarithms in the phases (this requires some assumptions of
convergence for the first log)
\begin{eqnarray}
S_{\rm eff}= \sum_{n=1}^\infty \frac{
1-8(-)^{n+1}r^{3n}-3r^{4n}-6r^{2n}}{n}~\phi_{-n}\phi_n \label{eff}
\end{eqnarray}
Here, $$ r=\exp\left( -\beta\mu/12\right) $$ and
\begin{equation}\label{momenta}
\phi_n~=~\frac{1}{N}\sum_{a=1}^N e^{in\beta A_a} \end{equation}
Recalling (\ref{ploop}), we note that $\phi_n$ are multiply wound
Polyakov loop operators evaluated in the static, diagonal gauge.
The zeroth moment is normalized
\begin{equation}\label{constraints1}
\phi_0=1
\end{equation}
The other elements are constrained by sum rules.  The density
defined by
\begin{eqnarray} \rho(\chi)&=&\frac{1}{N}\sum_{a=1}^\infty
\delta(\chi-\beta A_a)~\geq 0 \nonumber \\&=&\sum_n e^{-2\pi i
n\chi}\phi_n \label{constraints2}\end{eqnarray} is a non-negative
function. For example, if only $\phi_0$ and $\phi_{\pm 1}$ are
nonzero, (\ref{constraints2}) implies that $|\phi_1|\leq 1/2$.

In this one-loop approximation, the action is quadratic in the
Polyakov loops. When all coefficients of the quadratic terms are
positive, the action is minimized by $\phi_n=0$ for $n\neq 0$.
This is the confining phase.  When a coefficient becomes negative,
the effective action is minimized with one of the loops nonzero.
The result is a condensation of the loops.

As we raise the temperature from zero (and lower $\beta$ from
infinity), the first mode to condense is $n=1$. This occurs when
$$r_c=1/3~~\to ~~ T_c=\frac{\mu}{12\ln3}\approx .0758533 \mu$$
and $\phi_1\neq 0$ when $T>T_c$. 

 Note that this condensation breaks a U(1) symmetry. This is associated with
the center of the gauge group $U(1)\in U(N)$.  It arises from the
fact that all variables are in the adjoint representation. In the
Euclidean path integral,   gauge transformations $X(\tau)\to
U(\tau)X(\tau)U^{\dagger}(\tau)$ must preserve the periodicity of
the dynamical variables.  They therefore must be periodic up to an
element of the center, $U(\beta)=e^{i\theta}U(0)$. The Polyakov
loop, on the other hand, being the holonomy on the time circle,
does transform as $P\to e^{i\theta}P$.

Even once the static, diagonal gauge is fixed, there is a vestige
of this symmetry where $\beta A_a\to \beta A_a+\theta$ or
$\phi_n\to e^{in\theta}\phi_n$. This symmetry restricts the form
of the effective action for Polyakov loops, so that the term with
$\phi_{k_1}\ldots\phi_{k_n}$ must have $\sum{k_i}=0$. It is a good
symmetry of the confined phase and it is spontaneously broken in
the deconfined phase.  The Polyakov loop operator is an order
parameter for this symmetry breaking.

% ========================================================================== %
\subsection{Three-loop effective action}
\label{sec:threeeff}

We now present the original work of the author of this thesis which
was reported in \cite{Hadizadeh:2004bf}. The effective action is
calculated up to three-loop order. The two and three-loop pieces of
the effective action are given by the sum of connected vacuum diagrams
of that loop-order. In order to calculate these, the relevant
propagators must be determined.

% -------------------------------------------------------------------------- %
\subsubsection{Propagators}

Our strategy here will be to construct a Euclidean Green function
which obeys the equation

\begin{equation}
\left(-\left(
\frac{d}{d\tau}-iA_{ab}\right)^2+\omega^2\right)G_{ab}(\tau)=\delta(\tau)
\end{equation}
which has the periodicity
\begin{equation}
G(\tau+\beta)=G(\tau)
\end{equation}

\ni We will begin by constructing $G(\tau)$ in the domain
$-\beta\leq\tau\leq\beta$ and then continuing it periodically
outside of this domain.  For this we use the Heaviside function

\begin{eqnarray}
\theta(\tau)=\left\{ \begin{matrix} 1 & 0<\tau<\beta \cr 0 &
-\beta < \tau < 0 \cr \end{matrix} \right.
\\
\theta(-\tau)=\left\{ \begin{matrix} 0 & 0<\tau<\beta \cr 1 &
-\beta < \tau < 0 \cr \end{matrix} \right.
\end{eqnarray}

\ni Then our ans\"{a}tz for the Green function is

\begin{equation}
G(\tau)= \eta^{\tau/\beta}\left( g_+(\tau)
\theta(\tau)+g_-(\tau)\theta(-\tau)\right)
\end{equation}

\ni where

\begin{equation}
\eta=e^{-i \beta A_{ab}}=e^{i\beta A_b}/e^{i\beta A_a}\equiv
z_b/z_a
\end{equation}

\ni The Green function equation is obeyed if

\begin{equation}
\left( -\frac{d^2}{d\tau^2}+\omega^2\right)g_{\pm}(\tau)=0
~\longrightarrow~
g_{\pm}(\tau)=a_{\pm}e^{\omega\tau}+b_{\pm}e^{-\omega\tau}
\end{equation}

\ni and

\begin{equation}
g_+(\tau)=g_-(\tau) ~~,~~
\frac{d}{d\tau}g_+(\tau)-\frac{d}{d\tau}g_-(\tau)=-1
\end{equation}

\ni And the green function is periodic within the domain
$-\beta<\tau<\beta$ if

\begin{equation}
\tau<0~~~\eta g_-(\tau)= g_+(\tau+\beta)
\end{equation}

\ni The unique solution of these equations is

\begin{equation}
G(\tau)= \frac{ \eta^{-\frac{\tau}{\beta}+1} }{2\omega} \left(
\frac{  e^{\omega(\tau-\beta)} }{1-\eta e^{-\omega\beta}} + \frac{
e^{-\omega\tau} }{ \eta - e^{-\omega\beta}}\right)\theta(\tau) +
\frac{ \eta^{-\frac{\tau}{\beta}} }{2\omega} \left( \frac{
e^{\omega\tau} }{1-\eta e^{-\omega\beta}} + \frac{
e^{-\omega(\tau+\beta)} }{ \eta -
e^{-\omega\beta}}\right)\theta(-\tau)
\end{equation}

\ni If needed, this Green function should be extended periodically to
all values of $\tau$.

We note that this Green function is a sum of two green functions
for linear differential operators,

\begin{equation}
(\tau| \frac{1}{-D^2+\omega^2} |\tau')=\frac{1}{2\omega} (\tau|
\frac{1}{D+\omega} + \frac{1}{-D+\omega} |\tau')
\end{equation}

\ni which implies

\begin{equation}
G(\tau)=\frac{1}{2\omega} \left( g_1(\tau)+g_2(\tau) \right)
\end{equation}

\ni where

\begin{equation}
\left( D+\omega\right)g_1(\tau)=\delta(\tau) ~~~,~~~\left(
-D+\omega\right)g_2(\tau)=\delta(\tau)
\end{equation}

\ni with the same periodic boundary condition that is satisfied by
$G(\tau)$. We then have that

\begin{eqnarray}
g_1(\tau)=   \eta^{-\frac{\tau}{\beta}+1}   \left( \frac{
e^{\omega(\tau-\beta)} }{1-\eta e^{-\omega\beta}}
\right)\theta(\tau) +   \eta^{-\frac{\tau}{\beta}} \left( \frac{
e^{\omega\tau} }{1-\eta e^{-\omega\beta}}  \right)\theta(-\tau)
\\
g_2(\tau)=   \eta^{-\frac{\tau}{\beta}+1}   \left(   \frac{
e^{-\omega\tau} }{ \eta - e^{-\omega\beta}}\right)\theta(\tau) +
 \eta^{-\frac{\tau}{\beta}}   \left(   \frac{
e^{-\omega(\tau+\beta)} }{ \eta -
e^{-\omega\beta}}\right)\theta(-\tau).
\end{eqnarray}

\ni Note that $g_1^*(-\t) = g_2(\t)$, so that from now on we will use
$g(\t) \equiv g_1(\t)$ only. Similarly, a fermionic propagator obeys

\begin{equation}
 \left(
-D+\omega\right)g_f(\tau)=\delta(\tau)
\end{equation}
with the anti-periodic boundary condition
\begin{equation}
g_f(\tau+\beta)=-g_f(\tau)
\end{equation}

\ni We can similarly construct it in the interval $-\beta<\tau<\beta$
and continue it anti-periodically to the real line. The fermionic
propagator is 

\begin{equation} g_f(\tau)=-
\eta^{-\frac{\tau}{\beta}+1} \left( \frac{ e^{-\omega\tau} }{ \eta
+ e^{-\omega\beta}}\right)\theta(\tau) +
 \eta^{-\frac{\tau}{\beta}}   \left(   \frac{
e^{-\omega(\tau+\beta)} }{ \eta +
e^{-\omega\beta}}\right)\theta(-\tau)
\end{equation}

\ni The full propagators are then given by

\begin{equation}
\left< X^i_{ab}(\tau)\, X^j_{cd}(\tau') \right>
= \frac{R}{2\omega} \,\delta^{ij}\delta_{ad}\delta_{bc}  \left[ g(\tau'-\tau) +
g^*(\tau-\tau') \right]_{ab}
\label{scalar_prop}
\end{equation}

\noindent where, for this expression only, we can take $i,j$ to be either flavour of
scalar. For the fermions, we have

\begin{equation}
\begin{split}
 \left<(\psi_{I\alpha})_{ab}(\tau)\, (\psi^{\dagger J\beta})_{cd}(\tau')\right> =
2R \,\delta_{ad}\delta_{bc}\, \delta_{\alpha}^{\beta}\, \delta_I^J\, g_f{}_{ab}(\tau'-\tau)\\
\left<(\psi^{\dagger I\alpha})_{ab}(\tau)\, (\psi_{J\beta})_{cd}(\tau')\right> =
-2R\,\delta_{ad}\delta_{bc}\, \delta^I_J \,\delta^{\alpha}_{\beta} \,g^*_f{}_{ab}(\tau-\tau')
\end{split}
\label{ferm_prop}
\end{equation}

\noindent where $\langle\psi\psi\rangle=\langle\psi^
\dagger\psi^\dagger\rangle=0$.

% -------------------------------------------------------------------------- %
\subsubsection{2-loop diagrams}

\begin{figure}[ht]
\begin{center}
\includegraphics*[bb=170 645 425 725, height=1.25in]{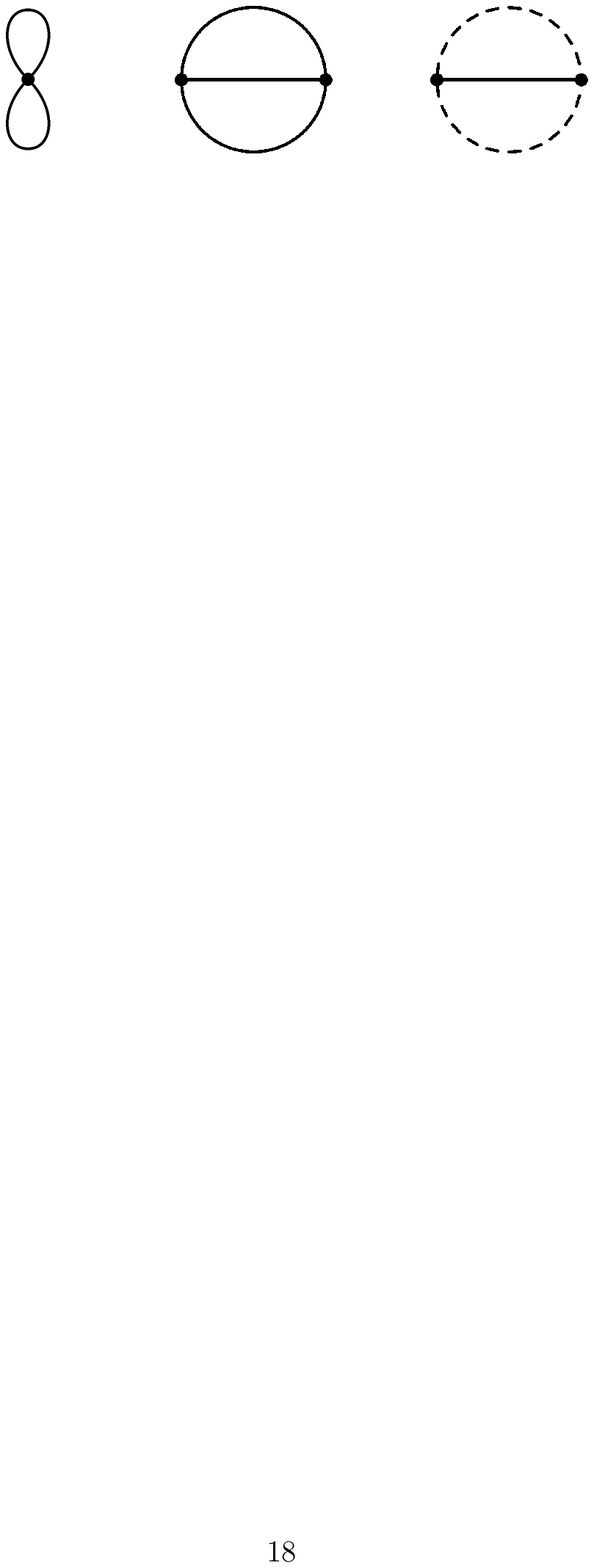}
\end{center}
\caption{The two-loop connected vacuum diagrams for the plane-wave
  matrix model at finite temperature about the single five-brane
  vacuum. Dashed lines refer to fermions, solid lines to scalars.}
\label{fig:pwmm2loop}
\end{figure}

The connected vacuum diagrams at two loops may be divided into three
forms, as shown in figure \ref{fig:pwmm2loop}, where we use a dashed
line to represent a fermion, while a solid line is used to indicate a
scalar propagator.
As an example of how these calculations are performed, we will
explicitly present the calculation of the third diagram in figure
\ref{fig:pwmm2loop}.\\

\ni\underline{Last diagram of figure \ref{fig:pwmm2loop}}\\

The relevant term in the action is

\begin{equation}
\frac{1}{2}\Tr\left( +\psi^{\dagger I\alpha} {\sigma^{\bar a}}_{\alpha}^{\beta}
\left[ X^{\bar a}, \psi_{J\beta} \right]
- \frac{1}{2} \epsilon_{\alpha\beta}\,\psi^{\dagger I\alpha}
{\sf g}^{i}_{IJ} [ X^{i}, \psi^{\dagger J\beta} ]
+ \frac{1}{2} \epsilon^{\alpha\beta}\,\psi_{I\alpha}
({\sf g}^{i \dagger})^{IJ} [ X^{i}, \psi_{J\beta} ] \right).
\end{equation}

\noindent The diagram comes from expanding $\exp(-S)$ to second order
in the path integral. Since $\langle\psi\psi\rangle=\langle\psi^
\dagger\psi^\dagger\rangle=0$,
the surviving terms are

\begin{equation}
\begin{split}
\Biggl< \frac{1}{2}\frac{1}{4} \int d\tau \, d\tau' \, \biggl\{ &\Tr
\left( \psi^{\dagger}{\sigma^{\bar a}}[ X^{\bar a}, \psi
]\right)(\tau)\, \Tr \left( \psi^{\dagger}{\sigma^{\bar b}}[ X^{\bar
b}, \psi ]\right)(\tau')\\ + &\Tr \left( -\frac{1}{2} \,\epsilon\,
\psi^{\dagger}\,{\sf g}^{i}[ X^{i}, \psi^\dagger ]\right)(\tau)\, \Tr
\left( \frac{1}{2} \,\epsilon\,\psi\,{\sf g}^{j}[ X^{j}, \psi
]\right)(\tau')\\ + &\Tr \left( \frac{1}{2} \,\epsilon\,\psi\,{\sf
g}^{i}[ X^{i}, \psi ]\right)(\tau)\, \Tr \left( -\frac{1}{2}
\,\epsilon\,\psi^{\dagger}\,{\sf g}^{j}[ X^{j}, \psi^\dagger
]\right)(\tau') \biggr\} \Biggr>.
\end{split}
\label{terms}
\end{equation}

\noindent Considering the first term first, and writing it in terms of
matrix indices we have

\begin{equation}\label{thisting}
\frac{1}{8 } \int d\tau \, d\tau' \, \left< \psi^\dagger_{ab}
\sigma^{\bar a} \left( X^{\bar a}_{bc} \psi_{ca} - \psi_{bc} X^{\bar
a}_{ca} \right)(\tau)\, \psi^\dagger_{de} \sigma^{\bar b} \left(
X^{\bar b}_{ef} \psi_{fd} - \psi_{ef} X^{\bar b}_{fd} \right)(\tau')
\right>.
\end{equation}

\noindent Keeping only the planar contributions, and noting again that
$\langle\psi\psi\rangle=\langle\psi^\dagger\psi^\dagger\rangle=0$ this
becomes

\begin{equation}
\frac{1}{8} \int d\tau \, d\tau' \, \sigma^{\bar a}\sigma^{\bar b}
\left\{ \left< X^{\bar a}_{ca} X^{\bar b}_{fd} \right> \left<\psi_{bc}
\psi^\dagger_{de}\right> \left< \psi^\dagger_{ab} \psi_{ef} \right> +
\left< X^{\bar a}_{bc} X^{\bar b}_{ef} \right> \left< \psi_{ca}
\psi^\dagger_{de}\right> \left< \psi^\dagger_{ab} \psi_{fd} \right>
\right\}
\end{equation}

\noindent where the first field in each expectation value is evaluated at
$\tau$ and the second at $\tau'$. Recalling the form of the
propagators (\ref{scalar_prop}) and (\ref{ferm_prop}) we have:

\begin{equation}
-8\frac{(2R)^2}{8 } \sum_{abc} \int d\tau \, d\tau' \,
 \left[ \frac{3R}{2\omega_1}(g+g^*_{-})_{ca}^{\omega_1}
 \right]\left[
(g_f)_{bc} (g_{f-}^*)_{ab} + (g_f)_{ab} (g_{f-}^*)_{bc} \right]
\label{ferm_thingy}
\end{equation}

\noindent where the subscript ``$-$'' indicates time reversal.  The
factor of 8 comes from $\delta^I_I \Tr \sigma \sigma$. The factor of 3
from the fact that there are three scalars of the first flavour.  Now,
noting that $\omega_1 = \mu/3$ we have

\begin{equation}
-\frac{2 R^3}{\mu} \sum_{abc} \int d\tau \, d\tau' \,
\left[ 9\,(g+g^*_{-})_{ca}^{\omega_1}
\right]\left[
(g_f)_{bc} (g_{f-}^*)_{ab} + (g_f)_{ab} (g_{f-}^*)_{bc} \right].
\label{ferm_gs}
\end{equation}

\noindent Now we attack the fermion propagator terms

\begin{equation}
(g_f)_{bc} (g_{f-}^*)_{ab} = \phi_{bc}^{(\tau'-\tau)}
\left[ \frac{-\phi_{bc}^{-\beta}}{1+\phi_{bc}^{-\beta}} \theta
+ \frac{1}{1+\phi_{bc}^{-\beta}} \bar\theta \right] \,
\phi_{ab}^{*(\tau-\tau')}
\left[ \frac{-\phi_{ab}^{*-\beta}}{1+\phi_{ab}^{*-\beta}} \bar\theta
+ \frac{1}{1+\phi_{ab}^{*-\beta}} \theta \right]
\end{equation}

\noindent where $\phi_{ab} = e^{iA_{ab}+\omega}$, $\theta =
\theta(\tau'-\tau)$, and $\bar \theta \equiv \theta(\tau-\tau')
$. Using the fact that $\theta^2=\theta$ and that $\theta \bar\theta =
0$, and that $A_{ab} \equiv A_a-A_b$, we have:

\begin{equation}
(g_f)_{bc} (g_{f-}^*)_{ab} = -e^{iA_{ac}(\tau'-\tau)}
\frac{\phi_{bc}^{-\beta}\theta
+\phi_{ab}^{*-\beta}\bar\theta}{(1+\phi_{bc}^{-\beta})(1+\phi_{ab}^{*-\beta})}.
\end{equation}

\noindent Therefore

\begin{equation}
\begin{split}
&(g_f)_{bc} (g_{f-}^*)_{ab} + (g_f)_{ab} (g_{f-}^*)_{bc}=\\
&-e^{iA_{ac}(\tau'-\tau)}  \left\{\left[ \frac{\phi_{bc}^{-\beta}}
{(1+\phi_{bc}^{-\beta})(1+\phi_{ab}^{*-\beta})} +
\frac{\phi_{ab}^{-\beta}}
{(1+\phi_{ab}^{-\beta})(1+\phi_{bc}^{*-\beta})} \right] \theta
+ \Biggl[ \text{c.c.} \Biggr] \bar \theta \right\}
\end{split}
\end{equation}

\noindent while for the scalar propagators we have

\begin{equation}
\left[g+g^*_{-}\right]_{ca}^{\omega} =
\frac{1}{|1-\phi_{ca}^{-\beta}|^2} \left\{
\left[ {\cal A}_{ca} \phi_{ca}^{(\tau'-\tau)} +
{\cal B}_{ca} \phi_{ca}^{*(\tau-\tau')} \right] \theta
+\left[ {\cal A}^*_{ca} \phi_{ca}^{*(\tau-\tau')} +
{\cal B}^*_{ca} \phi_{ca}^{(\tau'-\tau)} \right]\bar \theta
\right\}
\end{equation}

\noindent where,

\begin{equation}
{\cal A}_{ca} = (1-\phi_{ca}^{*-\beta}) \phi_{ca}^{-\beta}
\qquad\qquad
{\cal B}_{ca} = (1-\phi_{ca}^{-\beta}).
\label{ABC}
\end{equation}

\noindent Now it can be seen that,

\begin{equation}
\left[(g_f)_{bc} (g_{f-}^*)_{ab} + (g_f)_{ab} (g_{f-}^*)_{bc}\right]
\left[g+g^*_{-}\right]_{ca}^{\omega} = G \theta + H\bar\theta
\label{pre_ferm_Re}
\end{equation}

\noindent and by changing variables in the second term such that $\tau$ and $\tau'$
are interchanged, one notes that $H \rightarrow G^*$, so that

\begin{equation}
\left[(g_f)_{bc} (g_{f-}^*)_{ab} + (g_f)_{ab} (g_{f-}^*)_{bc}\right]
\left[g+g^*_{-}\right]_{ca}^{\omega} = (G + G^*)\,\theta = 2\text{Re}(G)\, \theta.
\label{ferm_Re}
\end{equation}

\noindent Now notice that the $e^{iA_{ac}(\tau'-\tau)}$ term from the
fermion propagators kills the gauge field dependence of the scalar propagators

\begin{equation}
e^{iA_{ac}(\tau'-\tau)} \phi_{ca}^{(\tau'-\tau)} = e^{\omega(\tau'-\tau)}
\qquad\qquad
e^{iA_{ac}(\tau'-\tau)} \phi_{ca}^{*(\tau-\tau')} = e^{-\omega(\tau'-\tau)}.
\end{equation}

\noindent Thus yielding the following form for $G$

\begin{equation}
G = -\frac{
\left( {\cal A}_{ca} e^{\omega(\tau'-\tau)}
+ {\cal B}_{ca} e^{-\omega(\tau'-\tau)} \right)
\left(\bar{\cal A}_{bc}\bar{\cal B}_{ab} + \bar{\cal A}_{ab}\bar{\cal B}_{bc} \right)}
{C_{ca}\bar C_{ab} \bar C_{bc}}
\label{ferm_G}
\end{equation}

\noindent where

\begin{equation}
\begin{split}
&C_{ab} = | 1- \phi_{ab}^{-\beta}(\omega_i) |^2 \qquad
\bar C_{ab} = | 1 + \phi_{ab}^{-\beta}(\omega_f) |^2\\
&{\cal A}_{ab} = \left[1-\phi_{ab}^{*-\beta}(\omega_i)\right] \phi_{ab}^{-\beta}(\omega_i)
\qquad \bar{\cal A}_{ab} = \left[1+\phi_{ab}^{*-\beta}(\omega_f)\right]
\phi_{ab}^{-\beta}(\omega_f)\\
&{\cal B}_{ab} = 1-\phi_{ab}^{-\beta}(\omega_i)
\qquad \bar{\cal B}_{ab} = 1+\phi_{ab}^{-\beta}(\omega_f).
\end{split}
\end{equation}

\noindent The integrations over $\tau$ and $\tau'$ are performed using

\begin{equation}
\int_{-\beta/2}^{\beta/2} d\tau \int_{-\beta/2}^{\beta/2} d\tau'\,
\theta(\tau'-\tau)\, e^{\omega(\tau'-\tau)} =
\frac{e^{\omega\beta}-1-\beta\omega}{\omega^2}.
\end{equation}

\noindent The contribution (\ref{thisting}) may then be reduced to the
following form

\begin{equation}
\begin{split}
\frac{-2}{C^{\omega_1}_{ca}\bar C_{ab} \bar C_{bc}} \frac{3\beta}{\mu}
&\Biggl\{ \pm\left[ \cos \beta A_{ab} + \cos \beta A_{bc} \right]\\
&\times \left[e^{-\beta\mu/4} + e^{-3\beta\mu/4} + e^{-11\beta\mu/12}
+ e^{-17\beta\mu/12} - 2 e^{-7\beta\mu/12} - 2 e^{-13\beta\mu/12} \right]\\
&+\left[2+2\cos\beta A_{ca}\right]
\left[e^{-\beta\mu/2}+e^{-7\beta\mu/6}-2e^{-5\beta\mu/6} \right]
\Biggr\}\left(-9\,\frac{2 R^3}{\mu}\right)
\end{split}
\end{equation}

\noindent where the negative sign comes from (\ref{ferm_G}), the factor of
$2$ from (\ref{ferm_Re}), the factor of $3\beta/\mu$ from the final
reduction, and the final factor of $-2\cdot 9R^3/\mu$ from (\ref{ferm_gs}).

\noindent The form of the $C$'s is

\begin{equation}
C_{ab} = 1 - 2e^{-\beta\omega} \cos\beta A_{ab} + e^{-2\beta\omega}
\qquad
\bar C_{ab} = 1 + 2e^{-\beta\omega} \cos\beta A_{ab} + e^{-2\beta\omega}.
\end{equation}

\noindent Now for the second and third terms in (\ref{terms}),
starting with the second term

\begin{equation}
-\frac{1}{32} \int d\tau \, d\tau' \, \left<
\,\epsilon\,\psi^\dagger_{ab}\, {\sf g}^{i} \left( X^{i}_{bc}
\psi^\dagger_{ca} - \psi^\dagger_{bc} X^{i}_{ca} \right)(\tau)\,
\,\epsilon\,\psi_{de} \,{\sf g}^{\dagger j} \left( X^{j}_{ef} \psi_{fd} - \psi_{ef}
X^{j}_{fd} \right)(\tau')  \right>.
\end{equation}

\noindent There are more planar contributions here than for the first
term of (\ref{terms}), we have

\begin{equation}
\begin{split}
-\frac{1}{32} \int d\tau \, d\tau' \, \,\epsilon\,\,\epsilon\,
{\sf g}^{i}{\sf g}^{\dagger j} &\left\{
\left< X^{i}_{ca} X^{j}_{fd} \right> \left<\psi^\dagger_{bc} \psi_{de}\right>
\left< \psi^\dagger_{ab}  \psi_{ef} \right> +
\left< X^{i}_{bc} X^{j}_{ef} \right> \left< \psi^\dagger_{ca} \psi_{de}\right>
\left< \psi^\dagger_{ab} \psi_{fd} \right> \right.\\
&+\left.\left< X^{i}_{ca} X^{j}_{ef} \right> \left<\psi^\dagger_{bc} \psi_{fd}\right>
\left< \psi^\dagger_{ab}  \psi_{de} \right> +
\left< X^{i}_{bc} X^{j}_{fd} \right> \left< \psi^\dagger_{ca} \psi_{ef}\right>
\left< \psi^\dagger_{ab} \psi_{de} \right>.
\right\}
\end{split}
\end{equation}

\noindent Not surprisingly each term contributes the same quantity,
and a factor of four is gained, the result is

\begin{equation}
+\frac{8\cdot 4}{32} (2R)^2 \cdot 6  \sum_{abc} \int d\tau \, d\tau' \,
 \left[ \frac{R}{2\omega_2}(g+g^*_{-})_{ca}^{\omega_2}
 \right]\left[
 (g_{f-}^*)_{ab}  (g_{f-}^*)_{bc}\right].
\end{equation}

\noindent Here we encounter the structure
$\Tr \epsilon^2 \Tr {\sf g}^i {\sf g}^\dagger{}^i = -8\cdot6$, where the sign comes
from the fact that $\epsilon = i \sigma^2$. The factor of $6$ counts the six
scalars of the second flavour.
The third term in (\ref{terms}) is identical except that
$\langle \psi^\dagger \psi \rangle \rightarrow \langle \psi \psi^\dagger \rangle$,
and thus the full expression is

\begin{equation}
\frac{8\cdot 4}{32} (2R)^2 \cdot 6  \sum_{abc} \int d\tau \, d\tau' \,
 \left[ \frac{R}{2\omega_2}(g+g^*_{-})_{ca}^{\omega_2}
 \right]\left[
 (g_{f-}^*)_{ab}  (g_{f-}^*)_{bc} + (g_f)_{ab} (g_f)_{bc} \right].
\end{equation}

\noindent or more concisely,

\begin{equation}
\frac{2 R^3}{\mu} \sum_{abc} \int d\tau \, d\tau' \,
\left[ 36\,(g+g^*_{-})_{ca}^{\omega_2}
\right]\left[(g_{f-}^*)_{ab} (g_{f-}^*)_{bc} + (g_f)_{ab} (g_f)_{bc}\right].
\label{ferm_gs2}
\end{equation}

\noindent We continue as before by evaluating the fermion propagators

\begin{equation}
\begin{split}
(g_f)_{ab} (g_f)_{bc} &= e^{(iA_{ac}+2\omega_f)(\tau'-\tau)}
\frac{\phi_{ab}^{-\beta}\phi_{bc}^{-\beta}\theta
+\bar\theta}{(1+\phi_{bc}^{-\beta})(1+\phi_{ab}^{-\beta})} \\
(g_{f-}^*)_{ab}(g_{f-}^*)_{bc} &= e^{(iA_{ac}-2\omega_f)(\tau'-\tau)}
\frac{\phi_{ab}^{*-\beta}\phi_{bc}^{*-\beta}\bar\theta
+\theta}{(1+\phi_{bc}^{*-\beta})(1+\phi_{ab}^{*-\beta})}.
\end{split}
\end{equation}

\noindent Now the same integration variable switch employed in
(\ref{pre_ferm_Re}) and (\ref{ferm_Re}) can be used here, yielding

\begin{equation}
G = \frac{ \left( {\cal A}_{ca} e^{\omega_2(\tau'-\tau)} + {\cal
B}_{ca} e^{-\omega_2(\tau'-\tau)} \right) \left(\bar{\cal
A}_{ab}\bar{\cal A}_{bc} e^{2\omega_f(\tau'-\tau)} + \bar{\cal
B}_{ab}\bar{\cal B}_{bc} e^{-2\omega_f(\tau'-\tau)}\right)}
{C_{ca}\bar C_{ab} \bar C_{bc}}.
\end{equation}

\noindent One may then reduce to obtain the final result

\begin{equation}
\begin{split}
\frac{2}{C^{\omega_2}_{ca}\bar C_{ab} \bar C_{bc}} \frac{3\beta}{2\mu}
&\Biggl\{ \pm\left[ \cos \beta A_{ab} + \cos \beta A_{bc} \right]\\
&\times \left[e^{-\beta\mu/4} + e^{-5\beta\mu/12} + e^{-11\beta\mu/12}
+ e^{-13\beta\mu/12} -2 e^{-3\beta\mu/4} - 2 e^{-7\beta\mu/12}
\right]\\ &+\cos\beta A_{ca} \left[e^{-7\beta\mu/6}
+e^{-\beta\mu/6}-e^{-\beta\mu/2}-e^{-5\beta\mu/6}\right]\\ &+ 1
+e^{-4\beta\mu/3}+ 2e^{-2\beta\mu/3}- 2e^{-\beta\mu/3}- 2e^{-\beta\mu}
\Biggr\}\left(36\,\frac{2 R^3}{\mu}\right)
\end{split}
\end{equation}

\noindent where the factor of $2$ comes from (\ref{ferm_Re}), the
factor of $3\beta/2\mu$ from the final reduction, and the final factor
of $2\cdot 36R^3/\mu$ from (\ref{ferm_gs2}).\\

\ni\underline{Final 2-loop effective action result}\\

The other diagrams in figure \ref{fig:pwmm2loop} are similarly
calculated. The details are presented in appendix
\ref{app:pwmm2loop}. The final result may be stated in a compact form
using the variable $r \equiv \exp(-\beta\mu/12)$

\begin{equation}\label{seff2loop}
\begin{split}
&S_\text{eff}^\text{2-loops} = \frac{27\beta R^{3}}{4\mu^{2}} \Biggl(
-\frac{(1-r^8)}{C_{ab}^{\omega_{1}}C_{ca}^{\omega_{1}}}
-20\frac{(1-r^4)}{C_{ab}^{\omega_{2}}C_{ca}^{\omega_{2}}}
-12\frac{(1-r^8)(1-r^4)}{C_{ab}^{\omega_{1}}C_{ca}^{\omega_{2}}}\\
+&\frac{(r^8+4r^4+1)(r^4-1)^4 +
\left[\cos\beta A_{a b}+\cos\beta  A_{b c}+\cos\beta A_{ca}\right]
2r^4(r^4-1)^4}{C_{ab}^{\omega_{1}}C_{bc}^{\omega_{1}}C_{ca}^{\omega_{1}}}\\
+&16 \frac{r^3(r^4-r^2+1)(r^4-1)^2(r^2+1)
\left[\cos\beta A_{a b}+\cos\beta  A_{b c}\right]
+ r^6(r^4-1)^2\left[2+2 \cos\beta A_{ca}\right] }
{\bar C_{ab}\bar C_{bc}C_{ca}^{\omega_{1}}}\\
+&32\frac{r^3(r^4-1)^2(r^2+1)\left[\cos\beta A_{a b}+\cos\beta  A_{b c}\right]
+r^2(r^8-1)(r^4-1)\cos\beta A_{ca} + (r^4-1)^2(r^8+1) }
{\bar C_{ab}\bar C_{bc}C_{ca}^{\omega_{2}}} \Biggr)
\end{split}
\end{equation}

\ni where we have

\begin{equation}
\begin{split}
C_{ab}^{\omega_1} &= 1 - 2\,r^4\,\cos{\beta A_{ab}} + r^8\\
C_{ab}^{\omega_2} &= 1 - 2\,r^2\,\cos{\beta A_{ab}} + r^4\\
{\bar C}_{ab} &= 1 + 2\,r^3\,\cos{\beta A_{ab}} + r^6.
\end{split}
\end{equation}

\noindent In the zero temperature limit, $r\rightarrow 0$ and $C,\bar C \rightarrow 1$.
We then see that the free energy is $1+20+12-1-32=0$, and so there is a SUSY
cancellation at zero temperature. In order restate this result in
terms of the multiply-wound Polyakov loops, we use the following
identities

\bsp
\left( C_{ab}^{\omega_1} \right)^{-1} &= \sum_n \frac{r^{4|n|}}{1-r^8}
  \eta_a^{-n} \eta^n_b\\
\left( C_{ab}^{\omega_2}\right)^{-1} &= \sum_n \frac{r^{2|n|}}{1-r^4}
  \eta_a^{-n} \eta^n_b\\
\left( {\bar C}_{ab}\right)^{-1} &= \sum_n \frac{r^{3|n|}\,(-1)^n}{1-r^6}
  \eta_a^{-n} \eta^n_b
\end{split}
\ee

\ni where $\eta_a = \exp(i\b A_a)$. So that, for example, we may
express quantities such as the following

\bsp
\left( {\bar C}_{ab} {\bar C}_{ab} C_{ab}^{\omega_2} \right)^{-1} = 
\sum_{\substack{nmp\\abc}} \frac{(-1)^n r^{3|n|}}{1-r^6}  \frac{(-1)^m
  r^{3|m|}}{1-r^6}  \frac{r^{2|p|}}{1-r^4} \eta^{-n}_a \eta^n_b
\eta^{-m}_b \eta^m_c \eta^{-p}_c \eta^p_a \\
= \frac{1}{(1-r^6)^2}\frac{1}{1-r^4} \sum_{pn}\sum_{abc} \eta_a^p
\eta_b^n \eta^{-p-n}_c (-1)^n \sum_m r^{\left(3|m| + 3 |m+n| +
  2|m+n+p|\right)}\\
= \frac{N^3}{(1-r^6)^2}\frac{1}{1-r^4} \sum_{pn} \phi_p
\phi_n \phi_{-p-n} (-1)^n \sum_m r^{\left(3|m| + 3 |m+n| +
  2|m+n+p|\right)}
\end{split}
\ee 

\ni where the sum over $m$ is a straightforward, if tedious,
application of geometric series, and can be performed
analytically. We may use this method to re-express (\ref{seff2loop})
as

\begin{equation}\label{seffphiform}
\begin{split}
\frac{1}{N^2}\,S_{\text{eff}}^{\text{2-loops}} 
&= 3\lambda \ln(r) \, \sum_{mn} \phi_n \, \phi_m \, \phi_{-m-n} \Biggl[
-4 \frac{ r^8-r^4+1 }{ (r^4+1)^2 } r^{4|n|+4|m|}\\
&+16 \frac{ (r^4-r^2+1)(r^2+1)^2 }{ (r^4+r^2+1)(r^4+1) } \,(-1)^n \, r^{3|n|+4|m|} 
-16 \frac{ (r^2+1)^2 }{ r^4+r^2+1 } \, (-1)^{n+m}\, r^{3|n|+3|m|} \\
&+32 \frac{ (r^2+1)^2 }{ r^4+r^2+1 } \, (-1)^n \,r^{3|n|+2|m|}
-12 \, r^{4|n|+2|m|} -20\, r^{2|n|+2|m|}\\
&-4\frac{ (r^4-1)(r^8+r^4+1) }{ (r^4+1)^3 } \, F_{mn}(4,4)
+16 \frac{ (r^2+1)(r^{10}-1) }{ (r^4+1)(r^4+r^2+1)^2 } \, (-1)^m\,F_{mn}(3,4)\\
&-16 \frac{(r^2+1)(r^6-r^4+r^2-1) }{ (r^4+r^2+1)^2 } \, (-1)^m\,F_{mn}(3,2)
\Biggr]
\end{split}
\end{equation}

\noindent where we have used the coupling $\l = (3R/\m)^3N$, and
define the function $F_{mn}(a,b)$ in the following manner

\begin{equation}
F_{mn}(a,b) = \left\{ \begin{aligned}
F^1_{mn}(a,b) &\qquad m,n \geq 0 \quad \text{or} \quad m,n < 0\\
F^2_{mn}(a,b) &\qquad n < 0, m \geq -n\quad \text{or} \quad n \geq 0, m < -n\\
F^3_{mn}(a,b) &\qquad m < 0, m \geq -n \quad \text{or} \quad m \geq 0, m < -n
\end{aligned}
\right.
\end{equation}

\noindent where

\begin{equation}
\begin{split}
F^1_{mn}(a,b) = r^{a(2+n+m) + b} &\Biggl[
\frac{ r^{b(n+m)-an} }{ 1-r^{2a+b} } + \frac{ r^{-b-2a+an} }{ 1-r^{2a+b} }
+\frac{ r^{-2a-an+b(n+m)} }{ r^b-1 } \\
&- \frac{ r^{-2a-an+bn} }{ r^b-1 } 
- \frac{ r^{-2a-an+bn} }{ -r^b+r^{2a} }  + \frac{ r^{-2a+an} }{ -r^b+r^{2a} }  
\Biggr]
\end{split}
\end{equation}

\begin{equation}
\begin{split}
F^2_{mn}(a,b) = r^{a(n+m) + b} &\Biggl[
\frac{ r^{2a - an +b(n+m)} }{ 1-r^{2a+b} } + \frac{ r^{-b-an-bn} }{ 1-r^{2a+b} }\\
&+\frac{ r^{-an+b(n+m)} +r^{-b-an}-r^{-an}}{ r^b-1 }
 - \frac{ r^{-bn-b-an} }{ 1 - r^b }
\Biggr]
\end{split}
\end{equation}

\begin{equation}
\begin{split}
F^3_{mn}(a,b) = r^{a(n-m) + b} &\Biggl[
\frac{ r^{-an+2a+bn} }{ 1-r^{2a+b} } + \frac{ r^{-b+an+2am} }{ 1-r^{2a+b} }
+\frac{ r^{-an+bn} }{ r^b-1 }\\
& - \frac{ r^{-an+b(n+m)} }{ r^b-1 } 
- \frac{ r^{-an+b(n+m)} }{ -r^b+r^{2a} }  + \frac{ r^{an+2am} }{ -r^b+r^{2a} }  
\Biggr].
\end{split}
\end{equation}
% -------------------------------------------------------------------------- %
\subsubsection{3-loop diagrams}

As one can imagine, the complexity at the three-loop level is far
greater than at 2-loops. An extensive C++ code was written by the
author to produce all three loop diagrams, and their associated
combinatoric prefactors. The output of this code may be summarized as
follows. We introduce some new notation to simplify the presentation

\begin{equation}
\begin{split}
P_{ab}(t_{21}) &= \frac{R}{2\omega_1}[g(t_2-t_1) + g^*(t_1-t_2)]_{ab}^{\omega_1}\\
Q_{ab}(t_{21}) &= \frac{R}{2\omega_2}[g(t_2-t_1) + g^*(t_1-t_2)]_{ab}^{\omega_2}\\
F_{ab}(t_{21}) &= 2R\,{g_f}_{ab}(t_2-t_1)\\
G_{ab}(t_{21}) &= -2R\,{g^*_f}_{ab}(t_1-t_2).
\end{split}
\end{equation} 

\ni\underline{Cat's eye diagram}\\

Results should be multiplied by $R^2$.

\begin{center}
\[
\parbox{20mm}{ 
\includegraphics*[bb= 215 483 385 627,width=1in, height=0.847in]{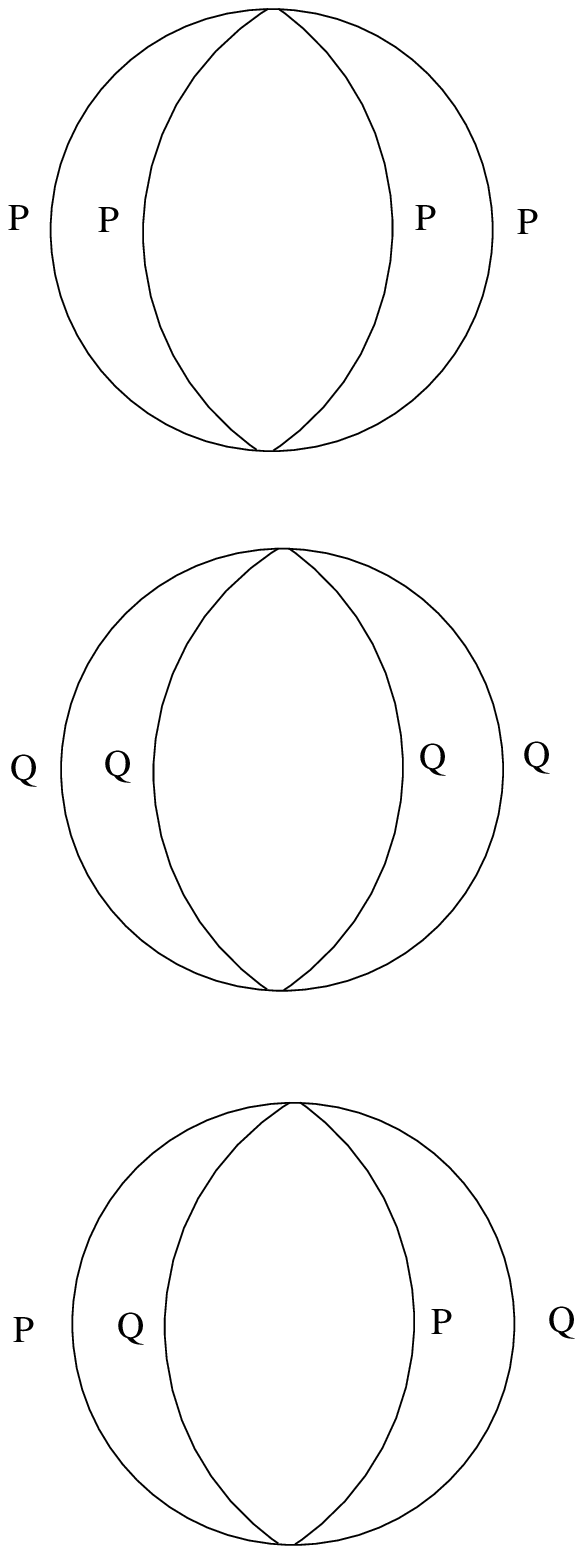}}
\qquad\qquad 
3\,\frac{3}{2} P_{ab}(t_{10}) P_{bc}(t_{10}) P_{cd}(t_{10})
P_{da}(t_{10})\]
\[  
\parbox{20mm}{ 
\includegraphics*[bb= 215 327 385 470,width=1in,height=0.847in]{catsi.ps}}
\qquad\qquad 
15\,\frac{3}{2} Q_{ab}(t_{10}) Q_{bc}(t_{10})
Q_{cd}(t_{10}) Q_{da}(t_{10})\]
\[
\parbox{20mm}{ 
\includegraphics*[bb= 215 168 395 310,width=1.074in,height=0.847in]{catsi.ps}}
\qquad\qquad
9 \,P_{ab}(t_{10}) P_{bc}(t_{10}) Q_{cd}(t_{10}) Q_{da}(t_{10})
+ 18\, P_{ab}(t_{10}) Q_{bc}(t_{10}) P_{cd}(t_{10}) Q_{da}(t_{10})
\]
\end{center}

\ni\underline{Triple bubble diagram}\\

Results should be multiplied by $R^2$.

\begin{center}
\[
\parbox{20mm}{
\includegraphics*[bb= 180 530 418 616,width=1.31in, height=0.473in]{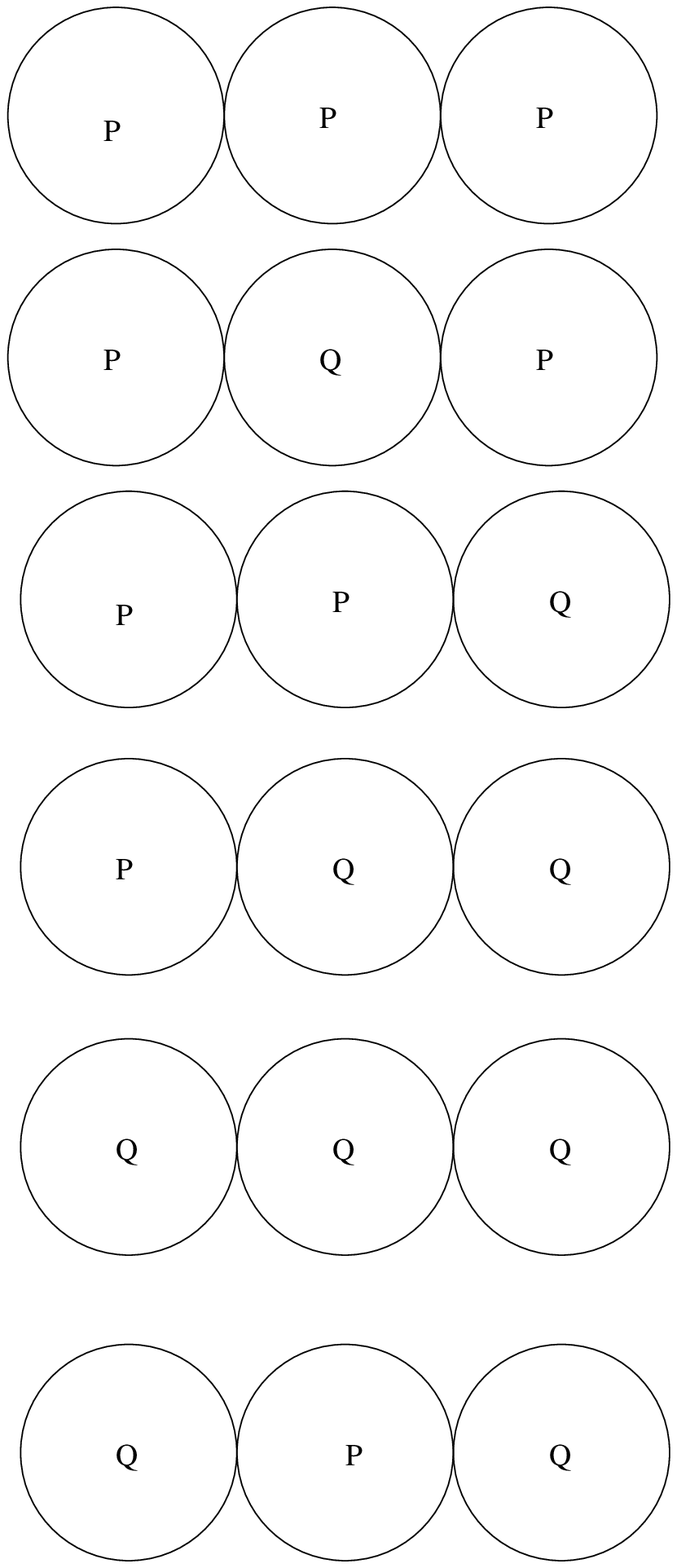}}
\qquad\qquad
27 \biggl[ P_{ab}(0) P_{cd}(0) + P_{ab}(0) P_{ad}(0)\biggr] Q_{ac}(t_{10}) Q_{ca}(t_{10})\]
\[
\parbox{20mm}{
\includegraphics*[bb= 180 143 418 229,width=1.31in, height=0.473in]{trip_bubble.ps}}
\qquad\qquad
54 \biggl[ Q_{ab}(0) Q_{cd}(0) + Q_{ab}(0) Q_{ad}(0)\biggr] P_{ac}(t_{10}) P_{ca}(t_{10})\]
\[
\parbox{20mm}{
\includegraphics*[bb= 180 445 418 531,width=1.31in, height=0.473in]{trip_bubble.ps}}
\qquad\qquad
36 \biggl[ P_{ab}(0) Q_{cd}(0) + P_{ab}(0) Q_{ad}(0)\biggr] P_{ac}(t_{10}) P_{ca}(t_{10})\]
\[
\parbox{20mm}{
\includegraphics*[bb= 180 350 418 436,width=1.31in, height=0.473in]{trip_bubble.ps}}
\qquad\qquad
90 \biggl[ P_{ab}(0) Q_{cd}(0) + P_{ab}(0) Q_{ad}(0)\biggr] Q_{ac}(t_{10}) Q_{ca}(t_{10})\]
\[
\parbox{20mm}{
\includegraphics*[bb= 180 615 418 701,width=1.31in, height=0.473in]{trip_bubble.ps}}
\qquad\qquad
12 \,P_{ab}(0) P_{cd}(0) P_{ac}(t_{10}) P_{ca}(t_{10})\]
\[
\parbox{20mm}{
\includegraphics*[bb= 180 250 418 336,width=1.31in, height=0.473in]{trip_bubble.ps}}
\qquad\qquad
150 \,Q_{ab}(0) Q_{cd}(0) Q_{ac}(t_{10}) Q_{ca}(t_{10})\]
\end{center}

\ni\underline{Theta-bubble diagram}\\

Results should be multiplied by $R$.

\begin{center}
\[
\parbox{20mm}{
\includegraphics*[bb= 310 385 460 490,width=0.726in, height=0.508in]{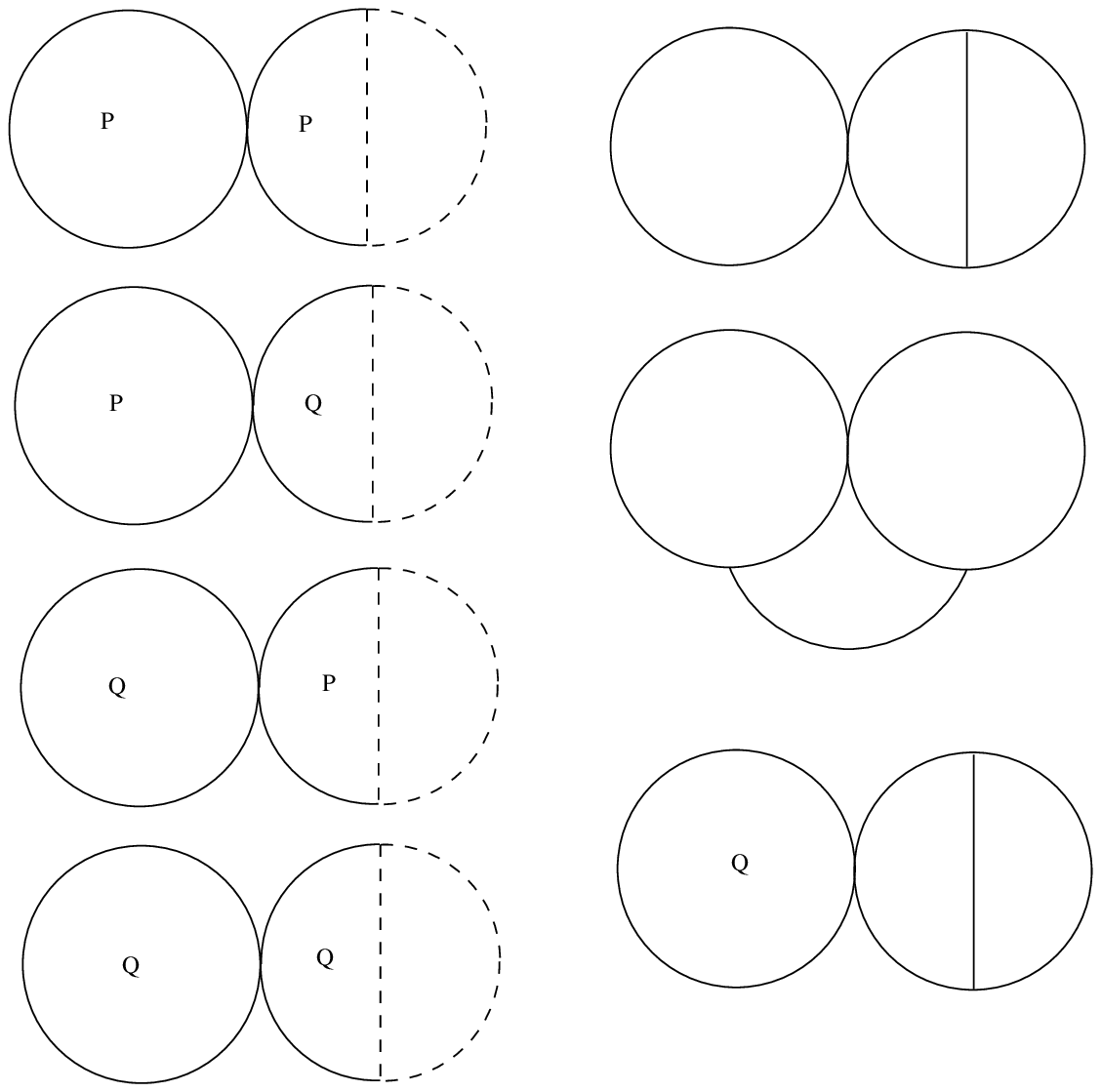}}
\qquad
-9\, P_{ab}(t_{10}) P_{bc}(t_{10}) P_{ac}(t_{21}) P_{cd}(t_{20}) P_{da}(t_{20})\]
\[
\parbox{20mm}{
\includegraphics*[bb= 310 500 460 578,width=0.907in, height=0.473in]{th_bubble.ps}}
\qquad
-12\, P_{ab}(0) P_{ca}(t_{10}) P_{ac}(t_{20}) P_{cd}(t_{21}) P_{da}(t_{21})\]
\[
\parbox{20mm}{
\includegraphics*[bb= 310 287 460 365,width=0.907in, height=0.473in]{th_bubble.ps}}
\qquad
-36\, Q_{ab}(0) P_{ca}(t_{10}) P_{ac}(t_{20}) P_{cd}(t_{21}) P_{da}(t_{21})\]
\[
\parbox{20mm}{
\includegraphics*[bb= 132 505 282 583,width=0.907in, height=0.473in]{th_bubble.ps}}
\qquad
-12\, P_{ab}(0) P_{ca}(t_{10}) P_{ac}(t_{20}) 
\biggl[ F_{cd}(t_{21}) G_{da}(t_{21}) + F \leftrightarrow G\biggr] \]
\[
\parbox{20mm}{
\includegraphics*[bb= 133 424 283 502,width=0.907in, height=0.473in]{th_bubble.ps}}
\qquad
-36\, P_{ab}(0) Q_{ca}(t_{10}) Q_{ac}(t_{20}) 
\biggl[ F_{cd}(t_{21}) F_{da}(t_{21}) + F \leftrightarrow G\biggr] \]
\[
\parbox{20mm}{
\includegraphics*[bb= 136 341 286 419,width=0.907in, height=0.473in]{th_bubble.ps}}
\qquad
-36\, Q_{ab}(0) P_{ca}(t_{10}) P_{ac}(t_{20}) 
\biggl[ F_{cd}(t_{21}) G_{da}(t_{21}) + F \leftrightarrow G\biggr] \]
\[
\parbox{20mm}{
\includegraphics*[bb= 136 260 286 338,width=0.907in, height=0.473in]{th_bubble.ps}}
\qquad
-60\, Q_{ab}(0) Q_{ca}(t_{10}) Q_{ac}(t_{20}) 
\biggl[ G_{cd}(t_{21}) G_{da}(t_{21}) + F \leftrightarrow G\biggr] \]
\end{center}

\ni\underline{Circle-T Diagram}

\begin{center}
\[
\parbox{20mm}{
\includegraphics*[bb= 193 260 296 360,width=0.7in, height=0.688in]{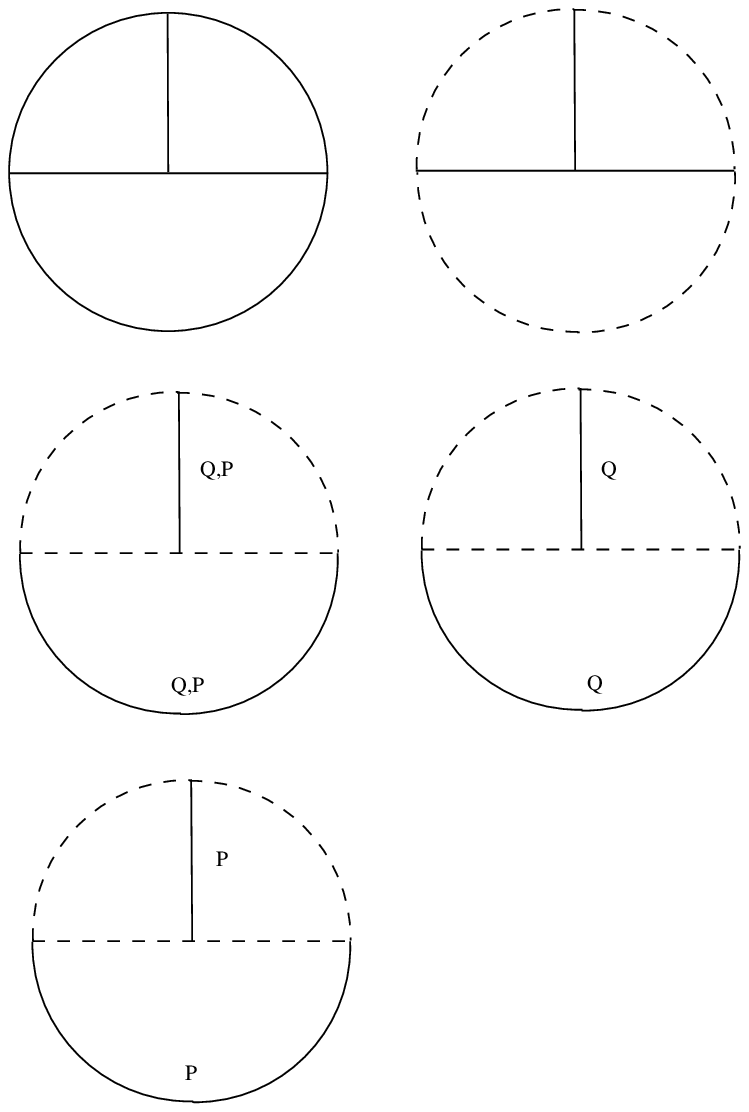}}
\qquad
-\frac{3}{8}\, P_{ab}(t_{20}) P_{cd}(t_{31}) 
\biggl[ F_{bd}(t_{10}) F_{bc}(t_{21}) F_{ac}(t_{32}) G_{da}(t_{30})
+F \leftrightarrow G \biggr] \]
\[
\parbox{20mm}{
\includegraphics*[bb= 190 372 293 472,width=0.7in, height=0.688in]{T.ps}}
\qquad
-\frac{9}{2}\, P_{ab}(t_{20}) Q_{cd}(t_{31}) 
\biggl[ F_{bd}(t_{10}) G_{bc}(t_{21}) G_{ac}(t_{32}) G_{da}(t_{30})
+F \leftrightarrow G\biggr] \]
\[
\parbox{20mm}{
\includegraphics*[bb= 304 372 407 472,width=0.7in, height=0.688in]{T.ps}}
\qquad
-3\, Q_{ab}(t_{20}) Q_{cd}(t_{31}) 
\biggl[ G_{bd}(t_{10}) F_{bc}(t_{21}) G_{ac}(t_{32}) G_{da}(t_{30})
+F \leftrightarrow G\biggr] \]
\[
\parbox{20mm}{
\includegraphics*[bb= 304 482 407 582,width=0.7in, height=0.688in]{T.ps}}
\qquad
-2\, P_{ab}(t_{30}) P_{ca}(t_{31}) P_{bc}(t_{32}) 
\biggl[ F_{bd}(t_{20}) G_{dc}(t_{21}) G_{da}(t_{10})
- F \leftrightarrow G\biggr] \]
\[
\parbox{20mm}{
\includegraphics*[bb= 185 482 288 582,width=0.7in, height=0.688in]{T.ps}}
\qquad
\frac{1}{2}\, P_{ab}(t_{10}) P_{bc}(t_{20}) P_{ca}(t_{30}) P_{db}(t_{21}) 
P_{ad}(t_{31}) P_{dc}(t_{32}) 
\]
\end{center}

\ni\underline{Two-rung ladder diagrams}\\

\begin{center}
\[
\parbox{20mm}{
\includegraphics*[bb= 164 300 244 420,width=0.6in, height=0.9in]{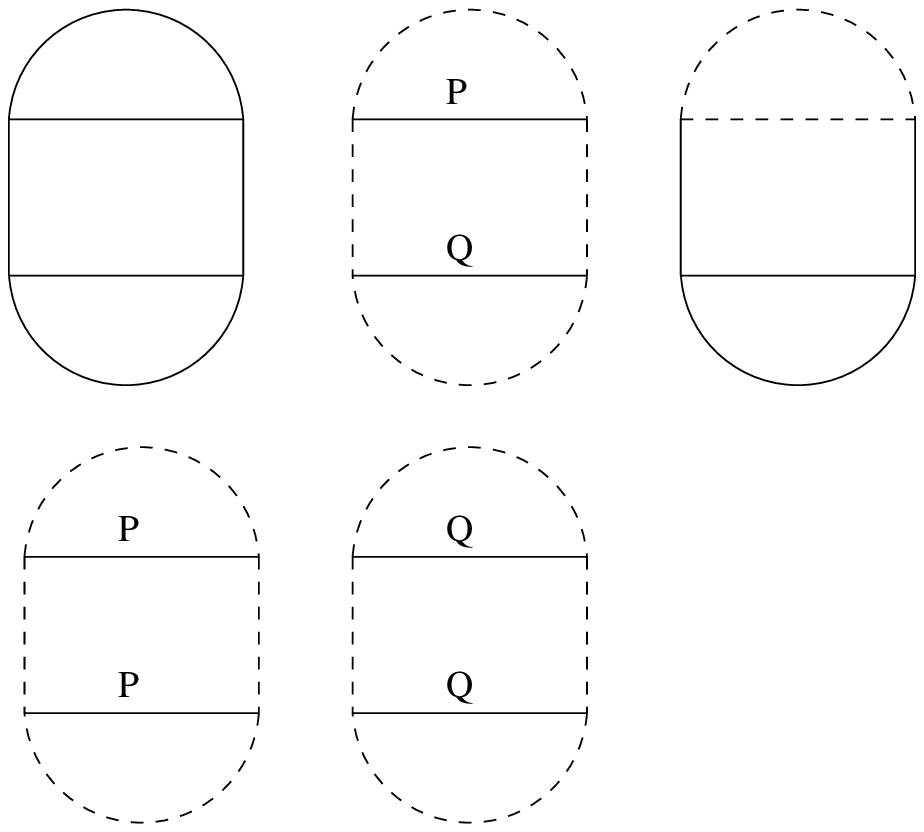}}
\frac{9}{4}\, P_{ab}(t_{10}) \biggl[ P_{cd}(t_{32}) G_{bc}(t_{32})
+ P_{bc}(t_{32}) G_{cd}(t_{32}) \biggr] 
G_{bd}(t_{20}) G_{bd}(t_{13}) F_{da}(t_{10}) + (F \leftrightarrow G) \]
\[
\parbox{20mm}{
\includegraphics*[bb= 256 300 337 420,width=0.6in, height=0.9in]{dblbar.ps}}
9\, Q_{ab}(t_{10}) \biggl[ Q_{cd}(t_{32}) G_{bc}(t_{32})
+ Q_{bc}(t_{32}) G_{cd}(t_{32}) \biggr] 
F_{bd}(t_{20}) F_{bd}(t_{13}) F_{da}(t_{10}) + (F \leftrightarrow G) \]
\[
\parbox{20mm}{
\includegraphics*[bb= 313 360 393 480,width=0.6in, height=0.9in]{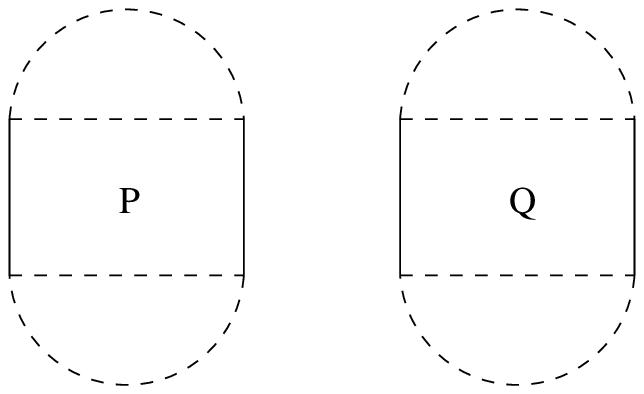}}
6 \, Q_{ab}(t_{10}) Q_{ba}(t_{32}) F_{bc}(t_{20}) F_{ca}(t_{20}) 
\biggl[ F_{ad}(t_{31}) F_{db}(t_{31}) + G_{db}(t_{31}) G_{ad}(t_{31})\biggr] 
+ (F \leftrightarrow G)\]
\[
\parbox{20mm}{
\includegraphics*[bb= 200 360 280 480,width=0.6in, height=0.9in]{dblbar2.ps}}
3 \, P_{ab}(t_{10}) P_{ba}(t_{32}) F_{bc}(t_{20}) G_{ca}(t_{20}) 
\biggl[ F_{ad}(t_{31}) G_{db}(t_{31}) + F_{db}(t_{31}) G_{ad}(t_{31})\biggr] 
+ (F \leftrightarrow G)\]
\[
\parbox{20mm}{
\includegraphics*[bb= 257 425 337 545,width=0.6in, height=0.9in]{dblbar.ps}}
\qquad\qquad
-9\, P_{ab}(t_{10}) Q_{cd}(t_{32})  
\biggl[F_{bd}(t_{20}) G_{da}(t_{10}) G_{db}(t_{31}) G_{bc}(t_{32})
+ (F \leftrightarrow G) \biggr]\]\[
\qquad\qquad\qquad\qquad\qquad
-9\, P_{ab}(t_{10}) Q_{ca}(t_{32})  
\biggl[F_{da}(t_{20}) G_{bd}(t_{10}) G_{ad}(t_{31}) G_{dc}(t_{32})
+ (F \leftrightarrow G) \biggr]
\]
\[
\parbox{20mm}{
\includegraphics*[bb= 355 425 435 545,width=0.6in, height=0.9in]{dblbar.ps}}
\qquad\qquad\qquad
6 \, P_{ab}(t_{20}) P_{ba}(t_{31}) P_{ac}(t_{32}) P_{cb}(t_{32}) 
\biggl[ F_{bd}(t_{10}) G_{da}(t_{10})+ (F \leftrightarrow G)\biggr] 
\]
\[
\parbox{20mm}{
\includegraphics*[bb= 160 425 240 545,width=0.6in, height=0.9in]{dblbar.ps}}
\qquad\qquad\qquad\qquad\qquad\qquad
3 \, P_{ab}(t_{10}) P_{ca}(t_{10}) P_{bc}(t_{20}) P_{cb}(t_{31}) 
P_{bd}(t_{32}) P_{dc}(t_{32})
\]
\end{center}

\ni\underline{Processing the 3-loop diagrams}\\

The diagrams must be processed and integrated as is done in appendix
\ref{app:pwmm2loop} for the 2-loop diagrams. Obviously, this process
is horribly complicated and was achieved via computer algebra
systems. The output of each diagram was obtained in a form analogous
to (\ref{seff2loop}). It would be far too cumbersome to display those
results here. The zero temperature limit was then taken in order to
verify the SUSY cancellation. The results of this zero temperature
limit are as follows.

\[
\parbox{20mm}{
\includegraphics*[bb= 215 483 385 627,width=1in, height=0.847in]{catsi.ps}}
\qquad
=\frac{2187}{64} \frac{\beta R^6}{\mu^5}\quad \parbox{20mm}{
\includegraphics*[bb= 215 327 385 470,width=1in, height=0.847in]{catsi.ps}}
\qquad
=\frac{10935}{2} \frac{\beta R^6}{\mu^5}  \]

\[\parbox{20mm}{
\includegraphics*[bb= 215 168 395 310,width=1.074in, height=0.847in]{catsi.ps}}
\qquad
=\frac{2187}{2} \frac{\beta R^6}{\mu^5}
\]

\begin{center}
\[
\parbox{20mm}{
\includegraphics*[bb= 180 530 418 616,width=1.31in, height=0.473in]{trip_bubble.ps}}
\qquad\qquad =6561\, \frac{\beta R^6}{\mu^5}\qquad \parbox{20mm}{
\includegraphics*[bb= 180 143 418 229,width=1.31in, height=0.473in]{trip_bubble.ps}}
\qquad\qquad =6561\, \frac{\beta R^6}{\mu^5}
\]
\[
\parbox{20mm}{
\includegraphics*[bb= 180 445 418 531,width=1.31in, height=0.473in]{trip_bubble.ps}}
\qquad\qquad =2187\, \frac{\beta R^6}{\mu^5}\qquad \parbox{20mm}{
\includegraphics*[bb= 180 350 418 436,width=1.31in, height=0.473in]{trip_bubble.ps}}
\qquad\qquad =43740\, \frac{\beta R^6}{\mu^5}
\]
\[
\parbox{20mm}{
\includegraphics*[bb= 180 615 418 701,width=1.31in, height=0.473in]{trip_bubble.ps}}
\qquad\qquad =\frac{729}{4} \frac{\beta R^6}{\mu^5}\qquad \parbox{20mm}{
\includegraphics*[bb= 180 250 418 336,width=1.31in, height=0.473in]{trip_bubble.ps}}
\qquad\qquad =72900\, \frac{\beta R^6}{\mu^5}\]
\end{center}

\begin{center}
\[
\parbox{20mm}{
\includegraphics*[bb= 310 385 460 490,width=0.726in, height=0.508in]{th_bubble.ps}}
\qquad =-\frac{10935}{32} \frac{\beta R^6}{\mu^5}  \qquad \qquad \parbox{20mm}{
\includegraphics*[bb= 310 500 460 578,width=0.907in, height=0.473in]{th_bubble.ps}}
\qquad =-729\, \frac{\beta R^6}{\mu^5}  \]

\[
\parbox{20mm}{
\includegraphics*[bb= 132 505 282 583,width=0.907in, height=0.473in]{th_bubble.ps}}
\qquad \qquad \qquad =0 \qquad \qquad\parbox{20mm}{
\includegraphics*[bb= 310 287 460 365,width=0.907in, height=0.473in]{th_bubble.ps}}
\qquad  =-4374\, \frac{\beta R^6}{\mu^5}\]

\[
\parbox{20mm}{
\includegraphics*[bb= 133 424 283 502,width=0.907in, height=0.473in]{th_bubble.ps}}
\quad   =-43740\, \frac{\beta R^6}{\mu^5} \quad 
\parbox{20mm}{
\includegraphics*[bb= 136 341 286 419,width=0.907in, height=0.473in]{th_bubble.ps}}
\quad =0 \quad \parbox{20mm}{
\includegraphics*[bb= 136 260 286 338,width=0.907in, height=0.473in]{th_bubble.ps}}
\quad  =-145800\, \frac{\beta R^6}{\mu^5}\]

\end{center}

\begin{center}
\[
\parbox{20mm}{
\includegraphics*[bb= 193 260 296 360,width=0.7in, height=0.688in]{T.ps}}
\qquad \qquad = 0 \qquad\qquad \parbox{20mm}{
\includegraphics*[bb= 190 372 293 472,width=0.7in, height=0.688in]{T.ps}}
\qquad =2916\, \frac{\beta R^6}{\mu^5} \]

\[
\parbox{20mm}{
\includegraphics*[bb= 304 372 407 472,width=0.7in, height=0.688in]{T.ps}}
\quad =7776\, \frac{\beta R^6}{\mu^5}\quad 
\parbox{20mm}{
\includegraphics*[bb= 304 482 407 582,width=0.7in, height=0.688in]{T.ps}}
\quad =0  \quad \parbox{20mm}{
\includegraphics*[bb= 185 482 288 582,width=0.7in, height=0.688in]{T.ps}}
\quad =\frac{6561}{64} \frac{\beta R^6}{\mu^5} \]

\end{center}

%\begin{center}
\[
\parbox{20mm}{
\includegraphics*[bb= 164 300 244 420,width=0.6in, height=0.9in]{dblbar.ps}}
\qquad \qquad = 0 \qquad\qquad \parbox{20mm}{
\includegraphics*[bb= 256 300 337 420,width=0.6in, height=0.9in]{dblbar.ps}}
\qquad  =-34992\, \frac{\beta R^6}{\mu^5}\]

\[
\parbox{20mm}{
\includegraphics*[bb= 313 360 393 480,width=0.6in, height=0.9in]{dblbar2.ps}}
\qquad \quad =73872\, \frac{\beta R^6}{\mu^5} \qquad \qquad\parbox{20mm}{
\includegraphics*[bb= 200 360 280 480,width=0.6in, height=0.9in]{dblbar2.ps}}
\qquad \qquad =0\]

\[
\parbox{20mm}{
\includegraphics*[bb= 257 425 337 545,width=0.6in, height=0.9in]{dblbar.ps}}
\quad =5832\, \frac{\beta R^6}{\mu^5}\quad
\parbox{20mm}{
\includegraphics*[bb= 355 425 435 545,width=0.6in, height=0.9in]{dblbar.ps}}
\quad =0 \quad \parbox{20mm}{
\includegraphics*[bb= 160 425 240 545,width=0.6in, height=0.9in]{dblbar.ps}}
\quad =\frac{24057}{32} \frac{\beta R^6}{\mu^5}
\]

\noindent One can check that the sum of the above factors is zero, as
guaranteed by SUSY. The next step is to extract the $|\phi_1|^4$
terms in the expression analogous to (\ref{seffphiform}). These ${\cal
O}(\l^2)$ terms are then combined with the one-loop (${\cal O}(\l^0)$) and
two-loop (${\cal O}(\l)$) results, in order to assemble

\begin{equation}
\frac{1}{N^2}S_{\text{eff}} =   \Delta_1 (r) |\phi_1|^2
+\Delta_2(r)|\phi_2|^2+ ~\lambda ~P_1(r) \, \biggl( \phi_1 \phi_1
\phi_{-2} + \text{c.c.} \biggl) + ~\lambda^2~ P_2(r) \, |\phi_1|^4
+ \ldots \label{approxaction}\end{equation}

\noindent The coefficients in (\ref{approxaction}) are found to be

\begin{eqnarray}
\Delta_1(r) =\left[1-8r^3-3r^4-6r^2\right] - 24\lambda~\left[
\ln(r) \, r^2 (r^2+1)
(r+1)^4 \right]- ~~~~~~~~~~~~~~\nonumber \\
-3\lambda^2r^2\,\left[ ~ \ln(r)^2~ \left( 68\, r^{10}+352\,
r^9+904\, r^8+1536\, r^7+2256\, r^6 +~~~~~\right. \right.
\nonumber
\\ \left. \left. +3104\, r^5+ 4120\, r^4+2304\, r^3+928\,
r^2+192\, r+16\right) -~~~\right. \nonumber
\\
-\left.   \ln(r)~\left( 27\, r^{10}+152\, r^9+390\, r^8+640\, r^7+
915\, r^6+1232\, r^5 +\right.\right. \nonumber \\ \left. \left.
+1748\, r^4+1184\, r^3+466\, r^2+440\, r+102 \right)~ \right]+...
\end{eqnarray}
\begin{equation}
\Delta_2(r) = \frac{1}{2} \left(1+  8 r^6 - 3 r^8 - 6
r^4\right)+\ldots~~~~~~~~~~~~~~~~~~~~~~~~~~~~~~~~~~~~~~~~~~~~~
\end{equation}

\begin{equation}
P_1(r) =-12 \ln(r) \, r^4 (2 r^4-4 r^3+3 r^2-4 r+5) (r+1)^4
\end{equation}

\begin{eqnarray}
 P_2(r)  =3 r^4 \left[~-\ln(r)^2~\left(~136\,
r^{12}+512\, r^{11}+704\, r^{10} -1308\, r^8-1376\, r^7+1560\, r^6
\right. \right. \nonumber \\ \left. \left. +6400 \, r^5+10896\,
r^4+8096\, r^3+2136\, r^2+1536\, r+240~\right) \right. + \nonumber
\\ \left.
+ \ln(r)~\left(~27\, r^{12} + 120\, r^{11} + 166\, r^{10} - 32\,
r^9 - 271\, r^8 - 16\, r^7 + 1044\, r^6 + \right. \right. \nonumber \\
\left. \left. + 2624\, r^5 + 4036\, r^4 + 3256\, r^3 + 774\, r^2 +
768\, r  + 944~\right)~ \right]+\ldots
\end{eqnarray}

\noindent Eliminating $\phi_2$ using its equation of motion, we
obtain the effective action for $\phi_1$, in the large $N$ limit,
and to order $\lambda^2$

\begin{equation}\label{final}\frac{1}{N^2}
S_{\text{eff}} =   \Delta_1 (r)|\phi_1|^2 + \lambda^2\left( P_2(r)
- \frac{ \left[ P_1(r)  \right]^2 } { \Delta_2(r) } \right) \,
|\phi_1|^4 + \ldots
\end{equation}

% ========================================================================== %
\subsection{Phase transition}

\begin{figure}[ht]
\begin{center}
\includegraphics*[bb=65 190 525 675, height=3.5in]{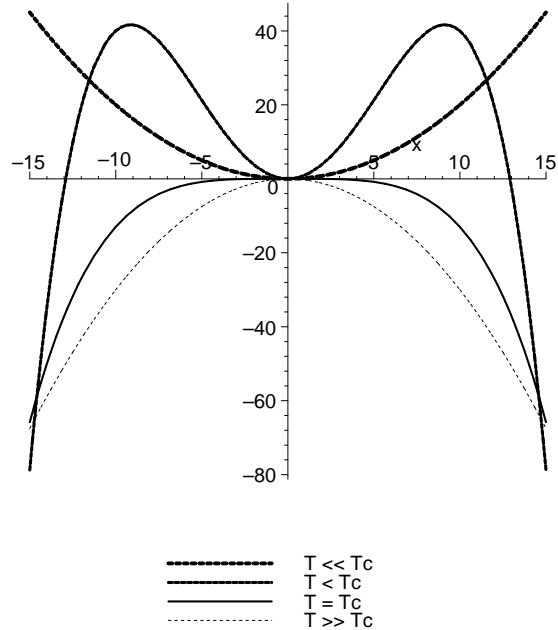}
\end{center}
\caption{The effective action (\ref{final}) for the first Polyakov
  loop $\phi_1$ vs. $\phi_1$. At low temperatures, $\phi_1=0$, but
  as the temperature is increased the negative quartic term grows,
  making this only a local minimum, and leading to bubble nucleation
  of the deconfined state. Above the critical temperature $T_c$ the
  quadratic term becomes negative and $\phi_1$ condenses to
  $\phi_1 \neq 0$ indicating total deconfinement.}
\label{fig:effact}
\end{figure}

Inspecting (\ref{approxaction}) and (\ref{final}) we see why it was
necessary to compute the effective action to three-loop order in order
to discover the first correction to the 1-loop transition temperature
$r_c=1/3$. It is because the next highest term in the action for the
lowest Polyakov loop $\phi_1$ is quartic, and its coefficient at
leading order receives contributions both from the 2-loop order
squared, i.e. $(\l P_1(r))^2$, and the 3-loop order, i.e. $\l^2
P_2(r)$. One can check that this quartic coefficient is negative over
the entire range of $r\in[0,1]$. We therefore have the following
picture of the effective potential for $\phi_1$. At low temperatures,
the action is quadratic and positive, as the temperature increases
(i.e. $r$ increases) the quartic term begins to become important and
we have a ``cowboy hat'' shape (see figure \ref{fig:effact}). At this
temperature, $\phi_1=0$ is no longer a global minimum and bubble
nucleation of the deconfined phase sets in. At a critical temperature
$T_c$, the quadratic term in the effective action vanishes, leaving an
inverted quartic, and the transition to the deconfined phase is
unimpeded. Finally, at temperatures above $T_c$, the quadratic term
reappears, but now with a negative sign.
The negative value of the quartic term indicates that the phase
transition is indeed first order. The critical temperature is found to
be

\begin{equation}\label{tc}
T_c = \frac{\mu}{12\ln(3)} \left[ 1 + \lambda\frac{2^6\cdot 5}{3^4}
-\lambda^2 \left( \frac{23\cdot 19927}{2^2\cdot 3^7}
+ \frac{1765769}{2^4 \cdot 3^8}\ln(3) \right) + \ldots \right].
\end{equation}

\ni The zeroth order term in the critical temperature is the
one found in \cite{Furuuchi:2003sy}. The term of first order in
$\lambda$ agrees with the result quoted in \cite{Spradlin:2004sx}.

 When $r$ is just less than the (1-loop) critical value $r_{c,0}=1/3$, the
 second zero of the effective action is found at
$$\vert\phi_1\vert^2=-\frac{1}{\lambda^2}\frac{
\Delta_1(r)}{P_2(r)-P_1^2(r)/\Delta_2(r)}.$$ Higher order terms in the
effective action are individually small at this value of $|\phi_1|^2$
when $-\frac{ \Delta_1(r)}{P_2(r)-P_1^2(r)/\Delta_2(r)}<<1$.  We are
further constrained by the fact that $\vert\phi_1\vert\leq 1$.  This
requires that $-\frac{
\Delta_1(r)}{P_2(r)-P_1^2(r)/\Delta_2(r)}<\lambda^2<<1$.  The number
$-\frac{ \Delta_1(r)}{P_2(r)-P_1^2(r)/\Delta_2(r)}$ is less than
$0.10$ in the range $0.2555<r\leq 1/3$ and is less than 0.001 in the
range $0.3174<r\leq 1/3$. If $r$ is sufficiently close to $r_{c,0}$, we
can reliably say that the absolute minimum of the potential is not at
$\phi_1=0$ but is elsewhere.  This sets an upper bound on the
transition temperature
$$T_{\rm trans.}  <T_{c,0}=\frac{\mu}{12\ln(3)}.$$ The tunnelling barrier
for bubble nucleation during the first order phase transition is of
order $1/\lambda^2$.

% ========================================================================== %
\subsection{Conclusions}

We have found that the phase transition in the weakly coupled plane
wave matrix model is indeed of first order.  As the temperature is
raised from zero, the curvature contained in the quadratic term in the
effective action still vanishes at some critical temperature.
However, before that point is reached, when there is still an energy
barrier between the two phases, the deconfined phase becomes the lower
energy state.  This is the generic behaviour at a first order phase
transition.  In fact, this behaviour is seen in other adjoint matrix
models \cite{Semenoff:1996xg}-\cite{Gattringer:1996fi}. It is also the
behaviour that is seen in the collapse of Anti de Sitter space to a
black hole, which is thought to be the analog of this phase transition
in supergravity of a similar deconfinement in ${\cal N}=4$
supersymmetric Yang-Mills theory \cite{Witten:1998zw}. It is difficult
to speculate what the dual 11-dimensional gravity process may be
here. We are working at weak coupling, so that the radius of the
five-brane is small compared to the string scale. In other words, the
limit of small $\l$ corresponds to a highly curved or ``stringy''
five-brane geometry \cite{Furuuchi:2003sy}. It also corresponds to
small $R$, meaning that we are far from the decompactified M-theory
limit. One may speculate that the first order transition persists at
strong coupling; if so one would not be surprised to find that it
corresponds to the nucleation of black-holes in a classical 11-d SUGRA
M5-brane background.

Our analysis does not allow us to compute the first order phase transition
temperature accurately, only to deduce that it is of first order.
It does, however, allow us to compute corrections to the Hagedorn
temperature. This is the temperature at which, if the confining
phase is superheated beyond where it is a global minimum of the
free energy, it eventually becomes perturbatively unstable.  It is
just the place where the corrected curvature of the effective
action vanishes, i.e. at $T_c$.

% ************************************************************************** %
% ************************************************************************** %
% ************************************************************************** %
\chapter{Exact 1/4 BPS Wilson loop: chiral primary correlator}
\label{sec:wilsonloop}

{\small
\begin{quote}
Show me slowly what I only\\ know the limits of\\
\rightline{--- Leonard Cohen {\it Dance me to the end of love}
$\;\;\;\;\;\;\;\;\;\;\;\;\;\;\;\;\;\;\;\;\;\;\;\;\;\;\;\;\;\;\;\;\;\;\;\;\;\;\;\;\;\;\;
\;\;\;\;\;\;\;\;\;\;\;\;\;\;\;\;\;$}
\end{quote}}

One of the most important class of operators in a gauge theory are the
Wilson loops. These are non-local, gauge invariant operators which may
be thought of as being responsible for the parallel transport of a
particle excitation of a field $\psi(x)$, in the fundamental
representation, about a closed path $C$

\be
\psi(x + C) = {\cal W} \,\psi(x).
\ee

\ni The Wilson loop $W$ is given by, for example in a standard
non-abelian gauge theory

\be
W = \Tr {\cal W} = \Tr P \exp \left[ -i \oint_C dx^\m A_\m \right]
\ee

\ni where the trace is over the fundamental representation of the
gauge group, and $P$ is the path-ordering symbol, which indicates that
in the Taylor expansion of the exponential, higher values of the
parameter along the path stand to the left. In section
\ref{sec:FEPTPW}, we saw that the Wilson loop about the compact
thermal circle served as an order parameter for deconfinement. Indeed,
the Wilson loop encodes information about the gauge theory
generally. If we take $C$ to be a rectangular path in the $(t,x)$
plane, of dimensions $T\times R$, so that the two sides of the
rectangle are separated in $x$ by $R$, then in the limit that $T \gg
R$, it is found that

\be
\la W \ra = A(R) \,e^{-T\, V(R)}
\ee

\ni where $A(R)$ is a numerical pre-factor, and $V(R)$ is the static
potential between a fundamental representation particle $\psi$ and its
anti-particle, i.e. in QCD, the quark-anti-quark potential. Indeed,
the temporal sides of the rectangle may be thought of as the
worldlines of the two particles. The fact that they remain straight,
despite the interaction between them, indicates that they are
effectively treated as infinitely massive.

The string-dual of the Wilson loop in ${\cal N}=4$ SYM was one of the
first correspondences worked out after the discovery of AdS/CFT
\cite{Maldacena:1998im}. Since then there has been much work on the
subject. One of the most significant developments was the work of
Erickson, Semenoff, and Zarembo \cite{Erickson:2000af}. They
discovered that for a circular Wilson loop, an infinite class of
diagrams contributing to $\la W \ra$ could be summed analytically. The
remaining diagrams show a strong promise of cancelling amongst
themselves, so that a function is obtained which interpolates smoothly
between weak and strong coupling. This is a powerful tool for probing
the AdS/CFT correspondence; typically (outside of the BMN
double-scaling limit) direct comparison of string results (strong
coupling) and CFT results (typically weak coupling) is
impossible. This chapter will introduce the Wilson loop in the AdS/CFT
correspondence and present original work of the author of this thesis
\cite{Semenoff:2006am} concerning the correlator of a certain $1/4$
BPS Wilson loop with chiral primary operators.

% ************************************************************************** %
\section{Introduction}

\begin{figure}[ht]
\begin{center}
\includegraphics*[bb=0 0 375 290, height=2.5in]{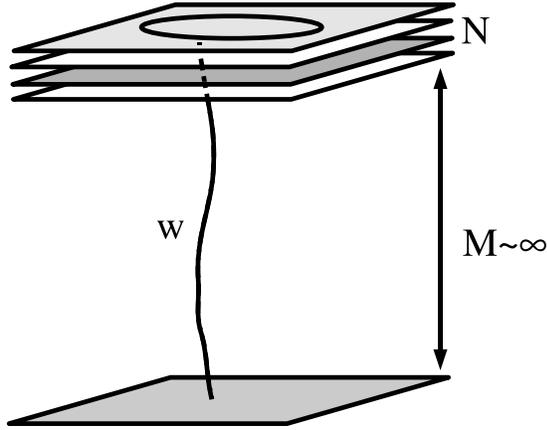}
\end{center}
\caption{Heavy, fundamental particles $w$ may be introduced into
  ${\cal N}=4$ SYM via a Higgsed $SU(N+1)$ theory built from a
  stack of $N+1$ coincident D3-branes. One brane is placed far from
  the remaining $N$, resulting in the introduction of stretched string
  modes, corresponding to the $w$ field, into the $SU(N)$ theory. The
  holonomy of such fields about a closed path gives the Wilson loop.}
\label{fig:wboson}
\end{figure}

In order to construct a Wilson loop in ${\cal N}=4$ SYM, we must find
a way to naturally introduce very massive, fundamental particles into
the theory. Recall from figure \ref{fig:stack} that ${\cal N}=4$ SYM
is the low-energy worldvolume theory of a stack of $N$ D3-branes. Now
consider a stack of $N+1$ D3-branes, where the extra brane is moved
far away from the remaining $N$, see figure \ref{fig:wboson}.
This of course corresponds to giving a VEV to the scalar which encodes
the transverse position of the separated brane. In fact
\cite{Drukker:1999zq}, the six scalars of the $SU(N+1)$ theory $\hat
\Phi^I$ may be expressed as the following $(N+1)\times(N+1)$ matrix

\be
\hat \Phi^I = \begin{pmatrix}  
\Phi^I & w^I\\  {w^I}^\dag & M\theta^I\\
\end{pmatrix}
\ee

\ni where $\theta$ is a constant normalized to $\theta^I \theta^I=1$,
$M$ represents the separation distance of the extra brane, while
$\Phi^I$ are the scalars of the $SU(N)$ theory. The length-$N$ column
field $w^I$ is evidently in the fundamental representation of
$SU(N)$. The VEV given to the lower right-hand corner element of $\hat
\Phi^I$ acts as a Higgs mechanism, imbuing $w^I$ with mass $M$. From
the geometric point of view afforded by the brane construction, the
$w$ field is simply those strings stretching from the stack of $N$
D3-branes to the separated brane; their tension is proportional to
$M$. Considering the propagation of a $w$ field about a closed loop
(from $x^\m$ to $y^\m$ and then back again) in the $SU(N)$ theory
gives the following path-integral representation
\cite{Drukker:1999zq}\footnote{As is typical in AdS/CFT, we work in
  Euclidean signature.}

\be\label{holonomy}
\int dy \la w(x) \, w^\dag(x) \, w(y) \, w^\dag(y) \ra
\sim \int {\cal D} x_\m \int {\cal D}A_\m {\cal D}\Phi^I
e^{-S_{SU(N)} - ML(x_\m)} W(x_\m) 
\ee

\ni where on the RHS, the $x_\m(\t)$ are closed paths of length
$L(x_\m)$, $S_{SU(N)}$ is the action of ${\cal N}=4$ SYM with gauge
group $SU(N)$, and the Wilson loop $W(x_\m)$ is given by

\be\label{wilsonloop}
W(x_\m) = \frac{1}{N} \Tr P \exp \left[ \oint_{C} d\t
\biggl(i \dot x_\m(\t) \, A_\m (x) + |\dot x(\t) | \theta^I \, \Phi^I(x) 
\biggr) \right].
\ee
 
\ni Thus, the Wilson loop measures the holonomy of the $w$ field about
a closed path. The scalars of the theory are coupled via $|\dot x(\t)|
\theta^I$, where, since $\theta^I\theta^I=1$, we may interpret
$\theta^I$ as a point on the five-sphere $S^5$. More generally,
researchers have considered paths on the five-sphere, given by
$\theta^I(\t)$.

\begin{figure}[ht]
\begin{center}
\includegraphics*[bb=0 85 525 330, height=1.75in]{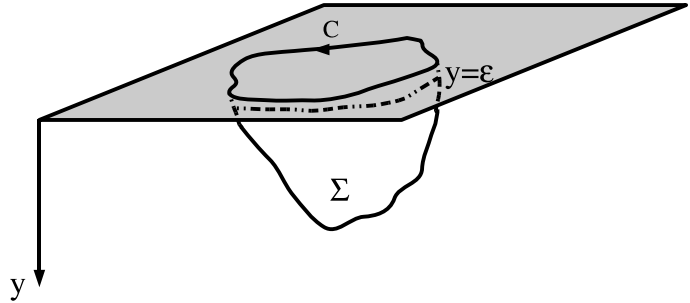}
\end{center}
\caption{The AdS/CFT dual of the Wilson loop (\ref{wilsonloop}) is
  given by a macroscopic string whose worldsheet is coincident with
  the Wilson loop at the boundary of $AdS_5$.}
\label{fig:wloop}
\end{figure}

The string dual of the ${\cal N}=4$ SYM Wilson loop (\ref{wilsonloop})
was suggested in an early paper by Maldacena
\cite{Maldacena:1998im}. The picture is very intuitive and is
summarized in figure \ref{fig:wloop}.
The Wilson loop is dual to the semi-classical partition function of
a macroscopic string in $AdS_5 \times S^5$, whose worldsheet falls
on the path of the Wilson loop at the boundary of $AdS_5$. More
precisely, at strong coupling $\l$, $\la W(x_\m) \ra$ is given by

\be
Z = \int {\cal D} X^\m \, {\cal D} Y^I\, {\cal D} h_{ab} 
{\cal D} \vartheta^\a \exp \biggl( -\frac{\sqrt{\l}}{2\pi} \int_{\S}
d^2\s \sqrt{h}\,h_{ab} \frac{\del_a X^\m \del_b X^\m + \del_a Y^I
  \del_b Y^I }{Y^2} + \text{fermions} \biggr) 
\ee

\ni where

\be 
X^\m |_{\del \S} = x_\m(\t), \qquad Y^I|_{\del \S} = \theta^I(\t)
Y|_{\del \S}, \qquad Y|_{\del\S} = 0. 
\ee

\ni The saddle-point is obtained when the string worldsheet $\S$ describes
a surface of minimal area ${\cal A}$

\be
{\cal A} =
\int_{\S}
d^2\s \frac{1}{Y^2} \sqrt{ \det \left( \del_a X^\m \del_b X^\m + \del_a Y^I
  \del_b Y^I \right) }
={\cal A}_{\text{reg.}} + \frac{L(C)}{\e}
\ee

\ni where we have noted that such an area is infinite, due to the
diverging nature of the area element in $AdS_5$ as the boundary is
approached. In fact, this area may be regulated by placing a cut-off
at $Y=\e$. The result is a finite regulated area ${\cal
  A}_{\text{reg.}}$, and an infinite piece proportional to the length
of the Wilson loop $L(C)$. This is perfectly consistent with the
gauge theory result (\ref{holonomy}), and allows us to associate

\be
M \lr \frac{1}{\e}.
\ee

\ni We then have that the expectation value of the Wilson loop is
given by the exponential of the regulated area of the minimal surface

\be
\la W \ra = e^{-\frac{\sqrt{\l}}{2\pi} {\cal A}_{\text{reg.}}}.
\ee

% ========================================================================== %
\subsection{Supersymmetric Wilson loops}

Consider as an example the straight line Wilson loop\footnote{The line
  is infinite and may be thought of as closing at infinity.} $x_\m(\t)
  = (\t,0,0,0)$, with $\theta^I$ constant. If we calculate the
  expectation value in perturbation theory, we find 

\begin{equation}
\begin{split}
\la W \ra &= 1 + \frac{1}{N} \Biggl< \Tr \int_0^\infty d\t_1 \int_0^{\t_1} d\t_2
\left( i A_0 + \theta \cdot \Phi \right)(\t_1) 
\left( i A_0 + \theta \cdot \Phi \right)(\t_2) + \ldots \Biggr> \\
&= 1 + \frac{-1 + \theta \cdot \theta}{4\pi^2 \left( x(\t_1) -
  x(\t_2) \right)^2} + \ldots = 1+ 0 + \ldots
\end{split}
\end{equation}

\noindent so the (combined scalar and gauge field) loop-to-loop
propagator vanishes for the straight line. In fact this is a special
case of a class of supersymmetric Wilson loops, due to Zarembo
\cite{Zarembo:2002an}, which have 

\be \theta^I(\t) =
\frac{\dot x_\m(\t)}{|\dot x(\t)|} \,M_\m^I, \qquad\text{where}\qquad
M_\m^I M_\n^I = \d_{\m\n} 
\ee

\noindent For these loops, the loop-to-loop propagator always
vanishes

\begin{equation}\nonumber
\begin{split}
\biggl< \left( i \dot x_\m A_\m + |\dot x| \theta \cdot \Phi \right)(\t_1)
       \left( i \dot x_\m A_\m + |\dot x| \theta \cdot \Phi
       \right)(\t_2) \biggr> &= \frac{-\dot x(\t_1) \cdot \dot x(\t_2) +
	 M_\m^I M_\n^I \dot x_\m(\t_1) \dot x_\n(\t_2)}{4\pi^2 \left( x(\t_1) -
  x(\t_2) \right)^2 } \\
&=0. 
\end{split}
\end{equation}

\ni As the name implies, this is a result of supersymmetry. As
$d=4,\,{\cal N}=4$ super-Yang-Mills theory is just a dimensional
reduction of $d=10,\,{\cal N}=1$ super-Yang-Mills theory, we may use
the ten dimensional supersymmetry transformations

\begin{equation}
\delta A^\mu = \frac{i}{2} {\bar \epsilon} \gamma^\mu \psi,
\quad \,
\delta \Phi^I = \frac{i}{2} {\bar \epsilon} \Gamma^I \psi,
\quad \,
\delta \psi = -\frac{1}{4} \Gamma^{MN} F_{MN} \epsilon,
\quad \,
\Gamma^{MN} = \frac{1}{2} [ \Gamma^M, \Gamma^N ]
\end{equation}

\noindent where $M = (\m,I)$ so that $I = 4,...,9$, and $\mu =
0,...,3$. The 10-d gamma matrices are $\Gamma^M = (\gamma^\mu,
\Gamma^I)$ and $\psi$ is a 10-d Majorana-Weyl fermion. The generalized
field strength $F^{MN}$ is understood as being built from the 10-d
gauge field $A^M = (A^\mu, \Phi^I)$. Further, the structure of $\e$, a
16-component spinor, is as follows

\be\label{fulleps}
\e = \e_0 + x_\m \g^\m \e_1
\ee

\ni where $\e_0$ corresponds to the Poincar\'{e}, and $\e_1$ to the
superconformal supersymmetries. Taking a supersymmetry variation of
the Wilson loop gives

\be
\d_{\epsilon} W = \frac{1}{N} \Tr P \int d\t \bar \psi
\left(i \dot x_\m \g^\m + |\dot x| \theta \cdot \G \right) \epsilon
\exp \biggl( \int d\t' \left( i \dot x_\m A_\m + |\dot x| \theta \cdot
\Phi \right) \biggr).
\ee

\ni Thus, if $(i \dot x_\m \g^\m + |\dot x| \theta \cdot \G ) \epsilon
= 0$ for some constant $\epsilon$, the Wilson loop will enjoy some
amount of supersymmetry. In fact this operator is nilpotent, potentially
indicating a halving of the supersymmetry

\be (i \dot x_\m
\g^\m + |\dot x| \theta \cdot \G )^2 = -\dot x^2 + \dot x^2 \,\theta
\cdot \theta =0 
\ee

\noindent but, in general solutions will require $\epsilon(\t)$ which
is {\it local} SUSY - not a symmetry of ${\cal N}=4$ SYM. In the case
of the supersymmetric loop, the path dependence factorizes 

\be \dot x_\m(\t)
\left(i\g^\m + M_\m^I \G^I \right)\bigl(\epsilon_0 + x_\n(\t) \g^\n
\epsilon_1\bigr) = 0 
\ee

\noindent which gives one halving for each non zero component of $\dot
x_\mu(\t)$, and which acts independently on the Poincar\'{e} and
superconformal supersymmetries. Thus we have that 

\be\label{zarembo}
 \text{ d-dimensional loop} \rightarrow
\left(\frac{1}{2}\right)^d ~~\text{ BPS}.
\ee

These supersymmetric loops, as suggested by the vanishing of
the loop-to-loop propagator, have a protected expectation value of
exactly unity, independent of the contour $x_\m(\t)$

\be
\bigl< W_{\text{SUSY}} \bigr> = 1 \qquad \text{(independent of contour)}.
\ee

\ni This has been proven using superspace techniques
\cite{Guralnik:2003di} up to $1/8$ BPS loops, while the remaining case of
$1/16$ BPS was proven in \cite{Kapustin:2006pk}. On the string side of
the AdS/CFT duality, this protection has also been proven
\cite{Dymarsky:2006ve} for the most general case. As a simple example,
we may consider the straight line. We expect ${\cal
A}_{\text{reg.}}=0$, since $\la W \ra =
\exp(-\sqrt{\l}{\cal A}_{\text{reg.}}/(2\pi))$. The string
worldsheet sits at a point on the $S^5$, so we just need the $AdS_5$
piece

\be
{\cal A} = \int d\t d\s \frac{1}{Y^2}
\sqrt{\left({X'}^2 + {Y'}^2 \right)  \left({\dot X}^2 + {\dot Y}^2
  \right) - \left( X'\cdot \dot X + Y'\dot Y \right) }.
\ee

\noindent If we set $X^{1,2,3} = 0$, we have two embedding functions
to worry about, these are $X^0(\s,\t)$ and $Y(\s,\t)$. But we also
have this much gauge invariance. Actually the choice 

\be X^0(\s,\t) =
\s \qquad Y(\s,\t) = \t 
\ee

\noindent solves the equations of motion trivially and obeys the
B.C.'s

\be
X^\m(\s,0) = x_\m = (\s,0,0,0) \qquad Y(\s,0) = 0
\ee

\noindent so then we have

\be
{\cal A} = \int d\s \int_\e^\infty \frac{1}{\t^2} = \frac{\int
  d\s}{\e} = \frac{L(C)}{\e} \rightarrow {\cal A}_{\text{reg.}} = 0.
\ee

% ========================================================================== %
\subsection{The 1/2 BPS circle: The straight line's conformal
  half-brother}

One of the most significant Wilson loops to have been considered is
the $1/2$ BPS circle. This is not a supersymmetric Wilson loop, indeed
it cannot be, since a circle has dimension 2, and would therefore be
$1/4$ BPS by (\ref{zarembo}). The $1/2$ BPS circle is given by

\be
x_\m(\t)=R\,(\cos \t, \sin
\t,0,0), \qquad \theta^I=\text{const.}.
\ee

\begin{figure}[ht]
\begin{center}
\includegraphics*[bb=0 0 510 250, height=2.0in]{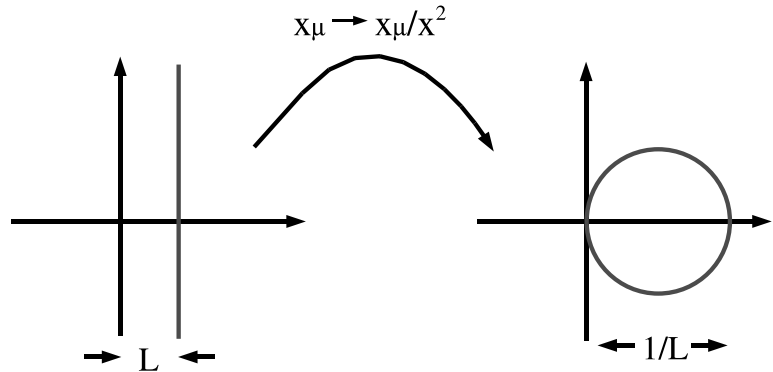}
\end{center}
\caption{The infinite straight line is related to the circle via a
  conformal inversion. This turns out to be a singular conformal
  transformation due to the infinite nature of the straight line.}
\label{fig:inversion}
\end{figure}

\ni This Wilson loop has an intimate connection with the
supersymmetric straight-line, considered in the last
subsection. Indeed the conformal inversion $x_\m \rightarrow x_\m/x^2$
maps the straight-line to the circle, as shown in figure
\ref{fig:inversion}.
We know that the gauge theory is a conformal field theory, and
therefore one might assume that a conformal transformation such as
$x_\m \rightarrow x_\m/x^2$ would not be detectable. However we will
see that $\la W \ra \neq 1$ for the $1/2$ BPS circle. This may be
traced to the supersymmetry, which has evidently changed compared to
the straight-line

\be 
\left(i \dot x_\m \g^\m + \theta \cdot \G \right) 
\left(\epsilon_0 + x_\m \g^\m \epsilon_1 \right)=0\qquad
\rightarrow\qquad i\g^1\g^0\epsilon_1 = -\theta\cdot\G \epsilon_0.
\ee

\ni We therefore see that the circle is indeed $1/2$ BPS, but as a result
of the superconformal and Poincar\'{e} supersymmetries being
related. In the case of the straight-line, we had two independent
conditions on each of $\e_0$ and $\e_1$. We will see below the
resolution of this apparent paradox. However, before we do this we
will introduce some very important work on the gauge theory
calculation of $\la W \ra$ for the $1/2$ BPS circle.

% -------------------------------------------------------------------------- %
\subsubsection{Erickson, Semenoff, and Zarembo}

In the seminal work \cite{Erickson:2000af}, Erickson, Semenoff, and
Zarembo succeeded in summing an infinite class of Feynman diagrams
contributing to $\la W \ra$, for the $1/2$ BPS circle. They noted that
the loop-to-loop propagator on the circle is a constant

\begin{equation}\label{halfbpsloop2loop}
\begin{split}
\biggl< \bigl(i\dot x_\m A_\m + \theta\cdot \Phi\bigr)%(\t_1)
 \bigl(i\dot x_\m A_\m + \theta\cdot \Phi\bigr) \biggr> =
\frac{1-\cos\t_1\cos\t_2 - \sin\t_1\sin\t_2}{4\pi^2\left[
(\cos\t_1-\cos\t_2)^2+(\sin\t_1-\sin\t_2)^2 \right]} = \frac{1}{8\pi^2}
\end{split}
\end{equation}

\begin{figure}[ht]
\begin{center}
\includegraphics[bb=0 0 530 130, width=5.75in]{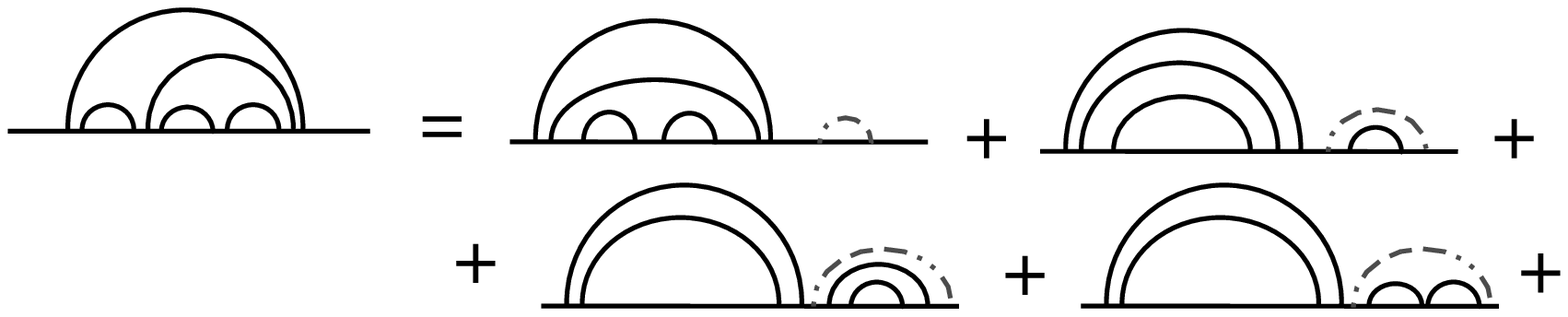}
\includegraphics[bb=0 0 350 75,width=3.25in]{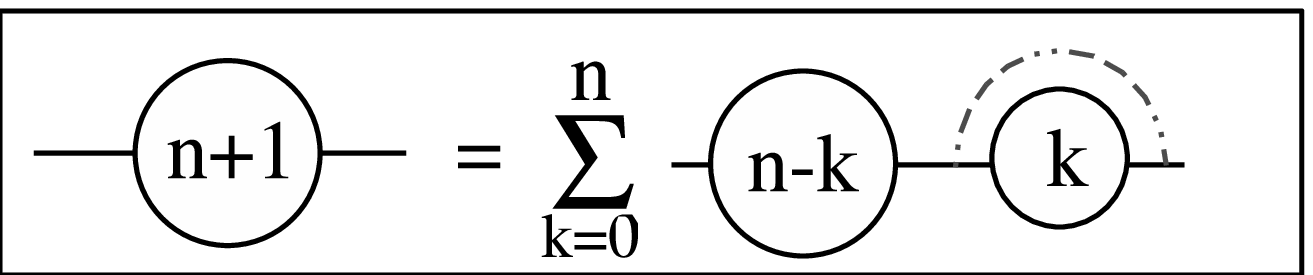}
\end{center}
\caption{Summing the planar ladder diagrams contributing to $\la W
  \ra$ for the $1/2$ BPS circle is reduced to a counting exercise
  owing to the constancy of the loop-to-loop propagator
  (\ref{halfbpsloop2loop}).}
\label{fig:planar}
\end{figure}

\noindent So that summing planar ladder diagrams becomes a counting
exercise. In figure \ref{fig:planar}, the circular Wilson loop is
opened to a horizontal line which is periodically identified.
The arches represent the loop-to-loop propagators. For example, as
shown, all five-propagator (planar, ladder) diagrams may be generated
by taking all four-propagator diagrams with a single separated arch
(dashed grey line), plus all three propagator diagrams in which all
one propagator diagrams are inserted under the separated arch, and so
on. This gives a recursion relation for the number of diagrams with
$n$ propagators which may then be solved

\be
N_{n+1} = \sum_{k=0}^n N_{n-k} N_k \qquad \rightarrow \qquad N_n =
\frac{(2n)!}{(n+1)!n!}
\ee

\noindent Taking care of factors from the path-ordered integration,
one finds

\be\label{bridge}
\la W \ra_{\text{ladders}} = \sum_{n=0}^\infty
\frac{(\l/4)^n}{(n+1)!n!} = \frac{2}{\sqrt{\l}} I_1(\sqrt{\l}).
\ee

The diagrams neglected in this treatment are those with internal
vertices. These were shown to cancel amongst themselves, up to two
loop order in \cite{Erickson:2000af} and \cite{Plefka:2001bu}, and to
three loop order in \cite{Arutyunov:2001hs}. The minimal area surface
of the string dual was found in \cite{Berenstein:1998ij}, with the
result

\be\label{halfbpsstring}
\la W \ra_{\text{string}} = e^{\sqrt{\l}}.
\ee

\ni Taking the large-$\l$ limit of (\ref{bridge}), one finds

\be\label{halfbpsgauge}
\la W \ra_{\text{gauge}}\simeq
\sqrt{\frac{2}{\pi}} \frac{e^{\sqrt{\l}}}{\l^{3/4}}
\ee

\ni and so the same exponential behaviour as the string result. In
fact, the presence of the prefactors may also be explained from the
string theory perspective. This leads one to suspect that
(\ref{bridge}) is in fact exact, and represents a continuous bridge
connecting weak and strong coupling. 

% -------------------------------------------------------------------------- %
\subsubsection{Drukker and Gross}

In their paper \cite{Erickson:2000af}, Erickson, Semenoff, and Zarembo
also noted that their results could be obtained from a Hermitian
matrix model 

\be
\la W_{\text{circle}} \ra = \frac{1}{Z} \int DM \frac{1}{N} \Tr \exp M
\exp\left(-\frac{2N}{\l} \Tr M^2 \right)
\ee

\ni In \cite{Drukker:2000rr}, Drukker and Gross went further with the
matrix model and solved also for arbitrary N

\be\label{laguerre}
\la W_{\text{circle}} \ra = \frac{1}{N} L^1_{N-1}
\left(-\sqrt{\l}/4N\right) e^{-\l/8N} = \frac{2}{\sqrt{\l}}
I_1(\sqrt{\l}) + \frac{\l}{48 N^2} I_2(\sqrt{\l}) + \ldots
\end{equation}

\begin{figure}[ht]
\begin{center}
\includegraphics[bb=0 0 310 230, height = 1.75in]{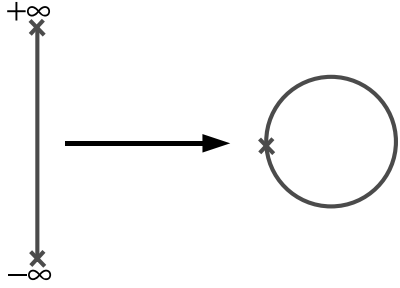}
\end{center}
\caption{Under the conformal inversion $x_\m \rightarrow x_\m/ x^2$,
  the ``end points'' of the straight line are mapped to a point on the
  circle. The dynamics of the straight line Wilson loop trivially
  vanish by supersymmetry, so that the discrepancy between the two
  Wilson loops is captured by a 0-dimensional theory living at a point
  on the circle.}
\label{fig:pointinf}
\end{figure}

\ni where $L_n^m$ is the Laguerre polynomial
$L_n^m(x)=1/n!\exp[x]x^{-m}(d/dx)^n (\exp[-x]x^{n+m})$. They also
understood that the inversion $x_\m \rightarrow x_\m/x^2$ is a
singular one, which gives a sort of {\it conformal anomaly}. The
dynamics are captured by a 0-dimensional theory at the point mapped
from infinity (see figure \ref{fig:pointinf}),
and this is why the matrix model works. In fact, the result is general

\be 
\la W_{\text{closed}} \ra = F(\l,N) \la W_{\text{open}} \ra
\ee

\ni for the relation of any ``open'' Wilson loop such as the
straight-line, to its conformally inverted, closed cousin. The
apparent breakdown of conformal invariance is then seen as a
consequence of a non-physical infinite Wilson loop, which does not
close explicitly.

The discrepancy between (\ref{halfbpsstring}) and (\ref{halfbpsgauge})
was also resolved in \cite{Drukker:2000rr}. They argued that a proper
treatment of the semi-classical string partition function should give
three powers of $\l^{-1/4}$, which should dress the main saddle-point
result (\ref{halfbpsstring}). These are associated with the
fluctuation determinants of three zero modes associated with the
relevant disk amplitude. Drukker and Gross also argued that the disk
can be decorated by degenerate handles, which gives an expansion in $1/N$

\be
\la W \ra_{\text{string}} = \sum_p \frac{C_p}{N^{2p}} \frac{\l^{(6p-3)/4}}{p!}
e^{\sqrt{\l}} \biggl( 1 + {\cal O}(1/\sqrt{\l}) \biggr)
\ee 

\noindent although the coefficients $C_p$ cannot be easily determined.
In fact, a large $\l$ expansion of their matrix model result gives
exactly this, with 

\be
C_p = \sqrt{\frac{2}{\pi}} \frac{1}{96^p}.
\ee

% ========================================================================== %
\subsection{Correlator with a chiral primary operator}

When viewed from a large distance, a compact Wilson loop (following
the closed path $C$) should look like an assembly of local operators
${\cal O}_{\Delta_i}(x)$ with conformal dimensions $\D_i$

\begin{equation}
W[C]=\la W[C] \ra \biggl(1+\sum_{\Delta_i > 0}{\cal O}_{\Delta_i}
(0)~L[C]^{\Delta_i} \xi_{\Delta_i}[C] + \ldots\biggr)
\end{equation}

\ni where $L[C]$ is the length of the Wilson loop, and $
\xi_{\Delta_i}[C]$ are some coefficients. The leading behaviour of the
correlator is given by the operators of smallest conformal dimension -
the chiral primaries (c.f. \cite{Minwalla:1997ka}), which we normalize as

\be
\la {\cal O}_\D(x) {\cal O}_{\D'}(0) \ra = \frac{\d_{\D\D'}}{(4\pi^2 x^2)^\D}.
\ee

\ni We then expect

\be\label{ldcorr}
\frac{ \la W[C]~
{\cal O}_{\Delta}(x) \ra }{\la W[C] \ra }
=\frac{L[C]^\Delta}{\left(4\pi^2|x|^2\right)^{\Delta}}
\xi_\Delta+\ldots
\ee

\noindent As an example, consider the 1/2 BPS circle with
$\theta^I=(1,0,...,0)$

\be
\begin{split}
W &= \la W \ra \biggl( \sum_k (2\pi R)^k \frac{1}{Nk!}\frac{1}{2^k}
\Tr \left( Z(0)+\bar Z(0) \right)^k + \ldots \biggr) \\
&= \la W \ra \biggl( 1 + \sum_{J\geq 2} {\cal O}_J(0) (2\pi R)^J \xi_J +
\ldots\biggr)
\end{split}
\end{equation}

\noindent where 
${\cal O}_J(x) = \frac{1}{\sqrt{\l^J J}} \Tr Z^J$, 
$Z=\Phi_1+i\Phi_2$. We then have, at leading order in $\l$ 

\be\label{xij}
\frac{\la {\cal O}_J(x) W(0) \ra}{\la W(0) \ra} = \left(\frac{2\pi R}{4\pi^2
  x^2}\right)^J \xi_J \qquad \xi_J = \frac{1}{N}\frac{1}{2^J J!}\sqrt{J\l^J}.
\ee

\noindent In fact, for the $1/2$ BPS circle, all planar (loop-to-loop)
ladders can also be summed for the calculation of $\xi_J$. This was
accomplished by Semenoff and Zarembo in \cite{Semenoff:2001xp}, where
leading non-ladder corrections were also found to vanish. The result
is

\be\label{bridge2}
\xi_J = \frac{1}{N}\frac{1}{2} \sqrt{\l J}
\frac{I_J(\sqrt{\l})}{I_1(\sqrt{\l})}
\ee

\ni representing another interpolating bridge between weak and strong
coupling, assuming non-ladder diagrams cancel at all orders in
perturbation theory.

\begin{figure}[ht]
\begin{center}
\includegraphics[bb=0 0 425 140, height = 1.25in]{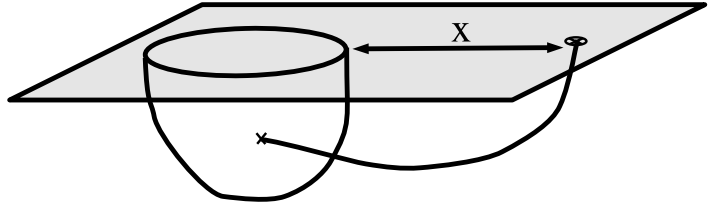}
\end{center}
\caption{The correlator of a Wilson loop with a chiral primary
  operator (\ref{ldcorr}) is dual to the exchange of a supergravity
  mode between the string worldsheet describing the Wilson loop and
  the boundary of $AdS_5$.}
\label{fig:exch}
\end{figure}

\begin{figure}[ht]
\begin{center}
\includegraphics[bb=0 0 555 160, height = 1.25in]{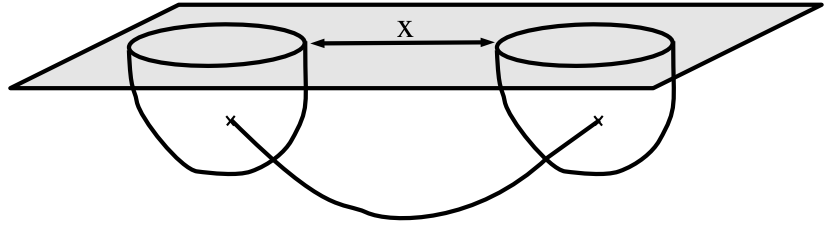}
\end{center}
\caption{A simpler method of calculating $\xi_J$ in (\ref{xij}) is to
  consider the exchange of the same supergravity mode pictured in
  figure \ref{fig:exch} between two widely separated Wilson loops.}
\label{fig:wloopexch}
\end{figure}

It is also possible to calculate $\xi_J$ at strong coupling using the
string side of the AdS/CFT duality. This was accomplished by
Berenstein, Corrado, Fischler, and Maldacena in
\cite{Berenstein:1998ij}. The result agrees precisely with the
large-$\l$ limit of (\ref{bridge2}). The chiral primaries are dual to
supergravitons propagating in $AdS_5\times S^5$. The large distance
correlator (\ref{ldcorr}) may be thought of as an exchange of such a
mode, between the loop's worldsheet and the boundary of $AdS_5$, see
figure (\ref{fig:exch}).
For the purpose of calculations, there is an easier method to obtain
$\xi_J$. Berenstein, Corrado, Fischler, and Maldacena pointed out that
the leading interaction between a pair of identical but widely
separated Wilson loops was mediated by the same supergravitons (see
figure \ref{fig:wloopexch}), leading to 

\be\label{loop2loopcorr}
\frac{\la W(x) W(0)\ra}{\la W(x) \ra\la W(0)\ra} = \sum_J \xi_J^2
\biggl(\frac{R}{x} \biggr)^{2J} + \ldots
\ee

\ni In practice, this is calculated by coupling the relevant
supergravitons to the string worldsheets and using the appropriate
bulk-to-bulk propagator. We will give the specific details below,
where we demonstrate this calculation for a special class of $1/4$ BPS
circular loops.

% ************************************************************************** %
\section{Exact 1/4 BPS loop: chiral primary correlator}
\label{sec:wilsonquartersection}

This section is a presentation of the author's original work published
in \textsf{arXiv:hep-th/0609158} \cite{Semenoff:2006am}.\\

In a recent paper \cite{Drukker:2006ga}, Drukker proposed and studied
the following circular Wilson loop

\be\label{quartcircle}
x_\m(\t) = R\,(\cos \t, \sin \t,0,0), \qquad 
\theta^I(\t)=(\sin\theta_0\cos
\t, \sin\theta_0\sin \t, \cos\theta_0,0,0,0).
\ee

\noindent When $\theta_0 = \pi/2$, we have the 1/4 BPS SUSY circle of
Zarembo, while when $\theta_0 = 0$, the 1/2 BPS circle is recovered.
For general $\theta_0$, there is one condition each on $\epsilon_{0}$
and $\epsilon_{1}$ (see (\ref{fulleps})), and one more condition
relating them

\be\label{Wsusy1}
\begin{split}
\sin\theta_0 (\g^1\G^2 +\g^2\G^1) \epsilon_0 = 0\\
\sin\theta_0 (\g^1\G^2 +\g^2\G^1) \epsilon_1 = 0\\
\cos\theta_0 \epsilon_0 = R(-i\g^1+\sin\theta_0\G^2) \G^3\g^2\epsilon_1
\end{split}
\ee

\begin{figure}[ht]
\begin{center}
\includegraphics[bb=0 0 190 190,height=1.75in]{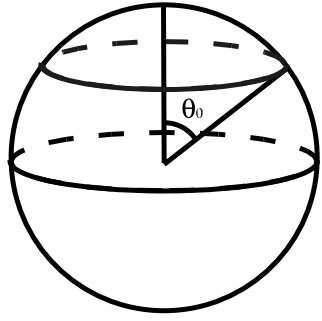}
\end{center}
\caption{The path $\theta^I(\t)$ (\ref{quartcircle}) describes a
  circle of latitude $\theta_0$ on an $S^2 \subset S^5$. }
\label{fig:quart}
\end{figure}

\ni and so the loop is generally $1/4$ BPS. The path $\theta^I(\t)$
describes a circle of latitude $\theta_0$ on an $S^2 \subset S^5$, see
figure \ref{fig:quart}.
Drukker discovered that, like for the case of the $1/2$ BPS circle,
the loop-to-loop propagator is a constant
$\cos^2\theta_0/8\pi^2$. This is just $\cos^2\theta_0$ times the $1/2$
BPS circle propagator. Therefore, the planar ladder diagrams can be
summed in exactly the same way they were for the $1/2$ BPS
circle. Further, leading internal vertex diagrams cancel in the
calculation of $\la W \ra$ by the same mechanism as for the $1/2$ BPS
circle. The only difference is that $\l \rightarrow \l' =
\cos^2\theta_0\l$. On the string side, the minimal surface for this
$1/4$ BPS circle was found by Drukker and it yields $\la W \ra =
\exp(\sqrt{\l'})$. It would thus seem that the results of the $1/2$
BPS circle are applicable here, albeit with the rescaled coupling
$\l'$. One therefore expects the matrix model result (\ref{laguerre}) 
to be applicable here, i.e.

\be
\la W_{1/4} \ra = \frac{1}{N} L^1_{N-1}
\left(-\sqrt{\l'}/4N\right) e^{-\l'/8N}.
\ee

In the work \cite{Semenoff:2006am}, the author of this thesis and
Semenoff expanded the $\l \rightarrow \l'$ correspondence to include
correlators with chiral primary operators. That work is described in
the balance of this chapter. Certain passages are taken from that
publication \cite{Semenoff:2006am}.

% ========================================================================== %
\subsection{Supersymmetry}

In the case of the $1/2$ BPS circle, the planar ladder diagrams
contributing to the correlator with a chiral primary operator are
summable and produce (\ref{bridge2}). The remaining diagrams appear to
cancel out. The reason for this cancellation is most likely the shared
supersymmetry between the chiral primary operator (CPO) and the Wilson
loop itself. It is therefore interesting to understand the degree of
shared SUSY between the $1/4$ BPS circle (\ref{quartcircle}) and a generic
CPO. We will consider a chiral operator which has an arbitrary $SO(6)$
orientation, beginning with

\be\label{chiralprim}
{\cal O}(0) = \frac{1}{\sqrt{J \l^J}}{\rm Tr}\left( u\cdot\Phi(0)\right)^J
\ee

\ni where $u$ is a complex 6-vector, satisfying the constraint that
$u^2=0$.   Being a scalar operator, conformal supersymmetries are
automatic.  This operator has some Poincar\'{e} supersymmetry if there
exist some non-zero constant spinors $\epsilon_0$ which solve the
equation
\begin{equation}\label{susy}
u\cdot\Gamma \epsilon_0=0
\end{equation}
There are solutions only when $\left(u\cdot\Gamma\right)^2=u^2=0$
which, as we have assumed, is the case. Then $u\cdot\Gamma$ is
half-rank and there are exactly eight independent non-zero solutions
of (\ref{susy}).

Now we can ask the question as to whether the eight independent
$\epsilon_0$ which solve (\ref{susy}) have anything in common with the
solutions of $\e_0$ arising from (\ref{Wsusy1}), i.e. are there
spinors which solve both of them? Before we answer this question, let
us backtrack to the case of the 1/2 BPS loop geometry.  There the top
two lines of (\ref{Wsusy1}) are absent and the spinors must solve the
last relation with $\theta_0=0$. This simply relates $\epsilon_1$ to
$\epsilon_0$, eliminating half of the possible spinors.  There are 16
independent solutions of this equation -- it is 1/2 BPS.  Now,
consider a chiral primary operator.  Without loss of generality, we
can consider the operator ${\rm Tr}\left(\Phi_1+i\Phi_2\right)^J$. It
is supersymmetric if $\epsilon_0$ satisfies the equation $$\left(
\Gamma^1+i\Gamma^2\right)\epsilon_0=0$$ The matrix
$\Gamma_1+i\Gamma_2$ has half-rank, so this requirement eliminates
half of the supersymmetries generated by $\epsilon_0$.  This leaves
eight supersymmetries which commute with both the 1/2 BPS Wilson loop
and the 1/2-BPS chiral primary operator. As we mentioned, this high
degree of residual joint supersymmetry is thought to be responsible
for the fact that, apparently, only ladder diagrams contribute to the
asymptotic limit of their correlator.

Returning to the 1/4 BPS loop and chiral primary with general
orientation, it is easy to see that there is a simultaneous solution
of (\ref{Wsusy1}) and (\ref{susy}) only when one of
the following holds:

\begin{itemize}

\item{}$u_1=u_2=0$.  We can always do an $SO(6)$ rotation
which commutes with the loop operator and sets
$(u_4,u_5,u_6)\to(u_4,0,0)$. Then,  there will be simultaneous
solutions of (\ref{Wsusy1}) and (\ref{susy}) only
when $u_3= iu_4$ or when $u_3=-iu_4$. In both of these cases, there
are four solutions, corresponding to 1/8 supersymmetry in common
between the chiral primary and the Wilson loop.  Up to a constant,
the chiral primary operator is ${\rm Tr}\left( \Phi_3 +
i\Phi_4\right)^J$ or the complex conjugate ${\rm Tr}\left( \Phi_3 -
i\Phi_4\right)^J$.

\item{}$u_3=u_4=0$. There is a solution when $u_1=\pm iu_2$ and there
is also 1/8 supersymmetry. The chiral primary is ${\rm Tr}\left(
\Phi_1+ i\Phi_2\right)^J$ or its complex conjugate. In this case, we
show in Appendix \ref{app3} that the coefficient $\xi_J$ which is
extracted from the long range part of the correlator of this
operator and the loop vanishes due to R-symmetry. Thus, for all
$J> 0$, the coefficients of ${\rm Tr}\left( \Phi_1+ i\Phi_2\right)^J$
or ${\rm Tr}\left( \Phi_1- i\Phi_2\right)^J$ in the operator
expansion of the 1/4 BPS loop are zero.

\item{}$u_1=\pm iu_2$. There are two non-zero solutions when $u_3=
iu_4$ or when $u_3=-iu_4$.  This corresponds to 1/16 supersymmetry.
There are essentially four operators,
$$ {\rm Tr}\left( \chi\left(\Phi_1+i\Phi_2\right)+
\left(\Phi_3+i\Phi_4\right)\right)^J $$ plus others with
substitutions of $\Phi_1-i\Phi_2$ or $\Phi_3-i\Phi_4$. In this case
too, because of R-symmetry the contribution with any non-zero power
of $\left( \Phi_1\pm i\Phi_2\right)$ will be zero.  The coefficient
$\xi_J[C_{\tiny 1/4}]$ for these operators is therefore the same as
those for the operator ${\rm Tr}\left(\Phi_3\pm i\Phi_4\right)^J $.
\end{itemize}

\ni Thus we see that the interesting quantity where there is some degree
of supersymmetry common to both the loop operator and the primary is

\begin{equation}\label{quartbps}
\xi_J[C_{\tiny 1/4}]=\lim_{|x|\to\infty} \left(
\frac{4\pi^2|x|^2}{2\pi R}\right)^J
\frac{1}{\sqrt{J\lambda^J}}\frac{\la 0|~ W[C_{\tiny 1/4}] ~{\rm
Tr}\left( 
\Phi_3(x)+i\Phi_4(x)\right)^J~|0\ra }{\la 0|~ W[C_{\tiny
1/4}]~|0\ra }
\end{equation}

\ni It is these partially supersymmetric configurations which we
expect to have some level of protection from quantum
corrections. Indeed, we shall find evidence for this.  All other
possibilities either vanish, are equivalent to (\ref{quartbps}) or
have no supersymmetry at all. The cases with no supersymmetry at all
are apparently not protected.

% ========================================================================== %
\subsection{Gauge theory calculation}

We will present arguments that the sum of planar ladder diagrams
contributing to the correlation function in (\ref{quartbps}) gives a
contribution which differs from the one for the 1/2 BPS loop quoted
in (\ref{bridge2}) by the simple replacement
$\lambda\to\lambda\cos^2\theta_0$, so that the total result is

\begin{equation}\label{quarterbps}
\xi_J[C_{\tiny 1/4}]=\frac{1}{N}\frac{1}{2}
~\sqrt{\lambda\cos^2\theta_0 J}~\frac{
I_J(\sqrt{\lambda\cos^2\theta_0})}{I_1(\sqrt{\lambda\cos^2\theta_0})}.
\end{equation}

\begin{figure}[ht]
\begin{center}
   \includegraphics[bb=419 279 190 508,height=2in]{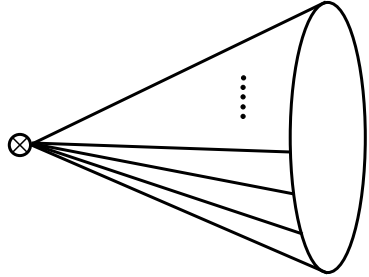}
\end{center}
 \caption{The leading planar contribution to $\la
   W[C_{\tiny 1/4}]~{\rm Tr}(\Phi_3+i\Phi_4)^J \ra$.
   There are $J$ lines connecting the chiral
   primary on the left with the circular Wilson loop on the right. }
\label{fig:leading}
\end{figure}

\ni To find this result using Feynman diagrams, we begin with the
lowest order diagrams, depicted in figure \ref{fig:leading}.
There, each occurrence of the scalar $\Phi_3$ in the composite
operator contracts with a scalar $\Phi_3$ in the Wilson loop. We
consider only the planar diagrams.  Each scalar $\Phi_3$ from the
Wilson loop carries a factor of $\cos\theta_0$, leading to an overall
factor of $(\cos\theta_0)^J$.  We are taking the convention for
Feynman rules where each line in the Feynman diagram results in a
factor of $\lambda$, totalling $\lambda^J$ for the diagram in figure
\ref{fig:leading}. With this convention, the chiral primary operator
has normalization $\lambda^{-J/2}$, as in (\ref{chiralprim}). The net
result is a factor of $\lambda^{J/2}$ which combines with the
$(\cos\theta_0)^J$ to give a coupling constant dependence in the form
$(\lambda\cos^2\theta_0)^{J/2}$.  This is identical to what one would
have obtained by taking the same diagram for the 1/2 BPS loop and
simply replacing $\lambda$ by $\lambda\cos^2\theta_0$.

\begin{figure}[ht]
\begin{center}
\includegraphics[height=1.2in,angle=180]{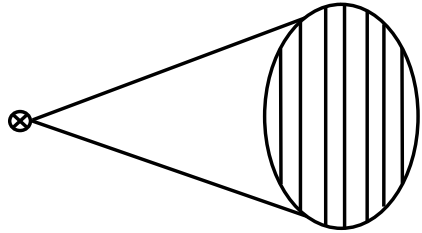}
\end{center}
\caption{A ladder diagram of $\la W[C_{\tiny 1/4}]~{\rm
Tr}(\Phi_3+i\Phi_4)^J \ra$. The ``rungs'' represent the combined gauge
field and scalar propagator. For clarity, $J$ has been set to 2.}
\label{fig:ladder2}
\end{figure}

To compute the next orders, we must decorate the diagram in figure
\ref{fig:leading} with propagators. The simplest are ladder diagrams,
see figure \ref{fig:ladder2}, which go between two points on the
periphery of the loop. 
They are described by summing the contribution of the vector and the
scalar field. Recall that the sum of scalar and vector propagators
connecting two points on arcs of the same circle is the constant
$\cos^2 \theta_0 /(8\pi^2)$. This is what makes ladder diagrams easy
to sum. We note that this propagator is accompanied by a factor of
$\lambda$, so the total $\lambda$ and $\theta_0$-dependence again
comes in the combination $\lambda\cos^2\theta_0$. Further, the only
difference from the analogous quantity for the 1/2 BPS loop is the
factor $\cos^2\theta_0$. Thus we see that the sum of ladders for this
1/4 BPS loop will be identical to that for the 1/2 BPS loop with the
replacement $\lambda\to \lambda \cos^2\theta_0$.  
\begin{figure}[ht]
\begin{center}
\includegraphics[height=3in]{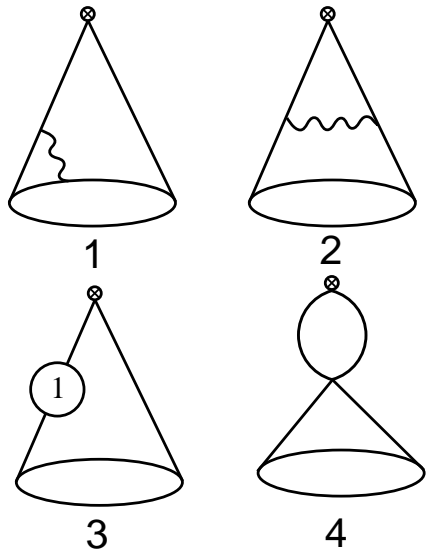}
\end{center}
\caption{The one-loop radiative corrections to $\la W[C_{\tiny
1/4}]~{\rm Tr}(\Phi_3+i\Phi_4)^J \ra$. Only an adjacent pair of the
$J$ scalar lines is shown.}
\label{fig:next}
\end{figure}

Finally, there are the diagrams that have not yet been included so
far. The conjecture is that they vanish.  The leading order are
depicted in figure \ref{fig:next}. By a simple generalization of the
argument obtained in \cite{Semenoff:2001xp} and explained in more
detail in \cite{Pestun:2002mr}, they can be shown to cancel
identically.  Assuming that this cancellation occurs to higher orders
as well, the result for the summation of all planar Feynman diagrams
is summarized in the formula (\ref{quarterbps}).

% ========================================================================== %
\subsection{String theory calculation}

The connected loop-loop correlator (\ref{loop2loopcorr}) has an
extremal surface whose boundary is the two loops.  When the loops have
large separation, this surface degenerates to two disc geometry
worldsheets whose boundaries are each loop with an infinitesimal tube
connecting them, see figure \ref{fig:wloopexch}. In the limit of large separation,
this tube is described by the propagator of the lightest gravity
modes, which at large $\lambda$ are 1/2 BPS supergravitons, the string
theory duals of the chiral primary operators. The connection between
the graviton propagator and the worldsheet is through a vertex
operator which must be identified and the connection point with the
vertex operator must be integrated over the worldsheet. The resulting
amplitude is proportional to the square of the desired operator
expansion coefficient, see (\ref{loop2loopcorr}).

To begin, the first step is to identify the minimal surface in
$AdS_5\times S^5$ whose boundary is the 1/4 BPS circle $C_{\tiny
1/4}$. This was done in \cite{Drukker:2006ga}. We will summarize it
here in more convenient coordinates.  We take the metric of $AdS_5
\times S^5$

\bea\label{metric} ds^2 &=& \sqrt{\lambda}\left( \frac{ dy^2 + d
r_1^2 + r_1^2 d\phi_1^2 + d r_2^2 + r_2^2 d \phi_2^2  }{y^2} \right.
\nonumber \\  &+& \left.
  d\theta^2 + \sin^2 \theta d\phi^2 + \cos^2
\theta \left( d\r^2 + \sin^2\r\, d\hat\phi^2 + \cos^2 \r\, d \tilde
\phi^2 \right)  \right). \eea  The string worldsheet is then
embedded as follows,
\begin{eqnarray}y = R \tanh \s \qquad r_1 =
\frac{R}{\cosh \s} \qquad \phi_1 = \t \qquad r_2=0 \qquad \phi_2={\rm const.}
\nonumber \\
 \sin \theta = \frac{1}{\cosh(\s_0 \pm \s)} \qquad \phi = \tau
 \qquad \rho=\frac{\pi}{2} \qquad \hat\phi=0 \qquad \tilde\phi={\rm const}.
\label{embed} 
\end{eqnarray}  

\ni where $\s \in [0,\infty]$ and $\t\in [0, 2\pi]$ are the
worldsheet coordinates. The contour $C_{\tiny 1/4}$ is the boundary
of the worldsheet at $\s=0$, which in turn sits at $y=0$, the boundary
of $AdS_5\times S^5$. The parameter
$\cos\theta_0=\frac{1}{\cosh\sigma_0}$.  The choice of $\pm$ sign in
the embedding of $\theta$ arises because there are two saddle points
in the classical action corresponding to wrapping the north or south
pole of the $S^5$. Of course the sign should be chosen to minimize the
classical action, which corresponds to choosing +. The other saddle
point is unstable, and the string worldsheet will slip-off the
unstable pole.

The supergravity modes that we are interested in are fluctuations of
the RR 5-form as well as the spacetime metric. They are by now very
well known, and details can be found in \cite{Kim:1985ez},
\cite{Berenstein:1998ij}, \cite{Lee:1998bx}, \cite{Semenoff:2004qr},
and \cite{Giombi:2006de}. The fluctuations are

\bea\label{fluct} \d
g_{\alpha \beta} &=& \left[-\frac{6\,J}{5}\,g_{\alpha \beta} +
\frac{4}{J+1} \, D_{(\alpha} D_{\beta)} \right] \,s^J(X)\,Y_J(\O),\cr
\d g_{IK } &=& 2\,k\,g_{IK } \,s^J(X)\,Y_J(\O)
\eea 

\ni where $\alpha,\beta$ are $AdS_5$ and $I,K$ are $S^5$ indices. The
symbol $X$ indicates coordinates on $AdS^5$ and $\O$ coordinates on
the $S^5$. The $D_{(\alpha} D_{\beta)}$ represents the traceless
symmetric double covariant derivative. The $Y_J(\O)$ are the spherical
harmonics on the five-sphere, while $s^J(X)$ have arbitrary profile
and represent a scalar field propagating on $AdS_5$ space with mass
squared $=J(J-4)$, where $J$ labels the representation of $SO(6)$ and
must be an integer greater than or equal to 2. (This is the
representation of $SO(6)$ which contains the chiral primary operators
that we are interested in.)

The supergravity field dual to the operator
$\Tr\left(u\cdot\Phi\right)^J$ is
obtained by choosing the combination of spherical harmonics with the
same quantum numbers and evaluating them on the worldsheet using
(\ref{embed}) (see appendix \ref{app2}) so that

\be
Y_J(\theta,\phi)= {\cal N}_J(u) \biggl[ u_1\sin \theta \,\cos\phi + u_2 \sin
\theta \,\sin\phi + u_3 \cos \theta \biggr]^J  
\ee  

\ni The worldsheets will be connected by the propagator for the scalar
supergravity mode $s^J(X)$.  The asymptotic form of this propagator
for large separation $x$ is 

\be\label{prop} P(X,\bar X ) = \la
s^J(X)\, s^J(\bar X) \ra \simeq \L_J \, \left( \frac{1}{x}
\right)^{2J} \, y^J\, \bar y^J 
\ee 

\ni where $\L_J = 2^J (J+1)^2 /(16\, N^2 J)$.  The barred quantities
are coordinates on the second Wilson loop worldsheet. Then, in the
large $\lambda$ limit, the Wilson loop correlator is 

\be \frac{\la 0|~
W[C_{\tiny 1/4},x]~ W^*[C_{\tiny 1/4},0]~|0\ra }{ \left|\la 0|~
W[C_{\tiny 1/4}]~|0\ra \right|^2 } = \int_{\S} \int_{\bar \S} \del_a
X^M \del^a X^N \,\d g_{MN}\, P(X,\bar X) \, \d \bar g_{\bar M \bar N}
\,\del_{\bar a} X^{\bar M} \del^{\bar a} X^{\bar N}, 
\ee 

\ni where $M,N=1,...,10$ and the $\delta g_{MN}$ are given in
(\ref{fluct}), except now we have removed the fluctuating parts,
$s^J(X)$ and replaced them by the propagator $P$.  The pullback of the
fluctuations (\ref{fluct}) to the worldsheet are found in appendix
\ref{app1}. Using them we have, 

\bea\label{integr} \frac{\la 0|~
W[C_{\tiny 1/4},x]~ W^*[C_{\tiny 1/4},0]~|0\ra }{ \left|\la 0|~
W[C_{\tiny 1/4}]~|0\ra \right|^2 } =
\frac{\L_J}{x^{2J}}\frac{\lambda}{16 \pi^2 } \Biggl[ 2J \int d\s d\t
y'^2 y^{J-2} Y_J(\theta,\phi)- ~~~~~~~~~~\cr ~~~~~~~ - 2J \int d\s d\t
(r_1'^2+r_1^2) y^{J-2} Y_J(\theta,\phi) + 2J \int d\s d\t
(\theta'^2+\sin^2 \theta) y^J Y_J(\theta,\phi) \Biggr]^2 
\eea 

\ni Each of the terms inside the square on the right-hand-side of the
above expression has a common factor of

\bea\label{str} \int_0^{2\pi}d\tau \,Y_J(\theta,\phi)= 
{\cal N}_J(u) \int_0^{2\pi}
d\t  \Bigl[ u_1\sin\theta \cos\t +u_2\sin\theta \sin\t +
u_3\cos\theta \Bigr]^J
\eea 

\ni From this expression we see that, consistent with our
expectations using R-symmetry on the gauge theory side, for the at
least 1/16 supersymmetric combination of loop and primary when
$u_2=\pm iu_1$, the dependence on $u_1$ and $u_2$ integrates to
zero.  If these parameters are chosen more arbitrarily, so that
there is no supersymmetry at all, the loop depends on them. In that
case the contributions proportional to powers of $u_1$ and $u_2$ in
the final result for the operator expansion coefficients do not
follow the rule that they are related to the 1/2 BPS loop ones by
the replacement of $\lambda$ by $\lambda\cos^2\theta_0$.  We
attribute this to absence of supersymmetry.  From here, we will
proceed with the supersymmetric case only by putting $u_1=u_2=0$ and
$u_3=1$.

We will now compute the integrals in (\ref{integr}) with this
assumption. We note that the embedding (\ref{embed}) has some nice
properties. For instance $y'^2 + r_1'^2 = r_1^2 = y'$ and also
$\sin^2 \theta = \theta'^2$. Using these, we can express the
integrals in (\ref{integr}) as follows 

\bea \frac{2^{-J/2}}{R^J} \int d\s y'^2 y^{\small J-2} \cos^J \theta =
2^{-J/2}\int_0^\infty d\s \frac{(\tanh \s)^{J-2}}{\cosh^4 \s} \tanh^J
(\s_0 \pm \s) \cr = 2^{-J/2}\int_0^1 dz (1-z^2)z^{J-2}\left(\frac{\pm
z + \cos \theta_0}{1 \pm z \cos \theta_0}\right)^J \\
\frac{2^{-J/2}}{R^J} \int d\s (r_1'^2+r_1^2) y^{\tiny J-2} \cos^J
\theta =2^{-J/2} \int_0^1 dz (1+z^2)z^{J-2}\left(\frac{\pm z + \cos
\theta_0}{1 \pm z \cos \theta_0}\right)^J \\ \frac{2^{-J/2}}{R^J} \int
d\s (\theta'^2+\sin^2 \theta) y^J \cos^J \theta = -2^{1-J/2}\int_{\mp
\cos \theta_0}^{-1} dz \left(\frac{ \pm z + \cos \theta_0}{1 \pm z
\cos \theta_0}\right)^J z^J \eea
  
\ni Putting everything together,

\bea\label{Qfinal} &&  \frac{\la 0|~ W[C_{\tiny 1/4},x]~ W^*[C_{\tiny
1/4},0]~|0\ra  }{ \left|\la 0|~
  W[C_{\tiny 1/4}]~|0\ra \right|^2 }
 = \cr
&&~~=16\,J^2\,
 \frac{\L_J}{2^J}\,\left(\frac{R}{x}\right)^{2J}\,\frac{\lambda}{4
}\Biggl[ \left\{ \int_{-1}^{\mp \cos \theta_0} dz - \int_0^1 dz
\right\} \left( \frac{\pm z + \cos \theta_0}{1\pm z\cos \theta_0}
\right)^J \, z^J\Biggl]^2\cr &&~~= 16\,J^2\,
\frac{\L_J}{2^J}\,\left(\frac{R}{x}\right)^{2J}\,\frac{\lambda}{4 }\left[
\frac{-(\pm)^{J+1} \cos \theta_0}{J+1} \right]^2 =
\frac{1}{4N^2}\,J\,\lambda\cos^2
\theta_0\left(\frac{R}{x}\right)^{2J} 
\eea  

\ni which is just the result for the 1/2 BPS circle
\cite{Berenstein:1998ij} with $\l \rightarrow \l \cos^2
\theta_0$. Using the prescription (\ref{loop2loopcorr}) to obtain from
the loop-to-loop correlator the overlap with the chiral primary in
question, we find $\xi_J[C_{\tiny
1/4}]=\sqrt{J\lambda\cos^2\theta_0}/2N$.  This is identical to the
large $\lambda$ limit of (\ref{quarterbps}). We have thus confirmed
that the sum of planar ladder diagrams agrees with the prediction of
AdS/CFT in the strong coupling limit. The emergence of this structure
on the supergravity side of the duality is non-trivial.  The
integrations over the $AdS_5$ and $S^5$ portions of the string
worldsheet conspire in a complicated way in (\ref{Qfinal}) to give the
$\l \rightarrow \cos^2\theta_0\,\l$ result.

It is instructive to consider  this calculation where both saddle
points of the classical action are kept in the path integral, as is
discussed in \cite{Drukker:2006ga}. There it was noted that the
semi-classical result for the expectation value of the Wilson loop
is a sum of two terms; one proportional to $\exp(\sqrt{\l'})$ and
the other to $\exp(-\sqrt{\l'})$, where $\l'=\cos^2 \theta_0\, \l$.
This was mirrored in the asymptotic expansion \cite{GR} of the
modified Bessel function of (\ref{laguerre})

\be
\begin{split}
&I_1(\sqrt{\l'}) =\\ &\frac{e^{\sqrt{\l'}}}{\sqrt{2\pi\sqrt{\l'}}}\,
\sum_{k=0}^{\infty} \left( \frac{-1}{2 \sqrt{\l'}} \right)^k\,
\frac{ \G(3/2+k) }{ k!\,\G(3/2-k) } \pm
i\, \frac{e^{-\sqrt{\l'}}}{\sqrt{2\pi\sqrt{\l'}}}\,
\sum_{k=0}^{\infty} \left( \frac{1}{2 \sqrt{\l'}} \right)^k\,
\frac{ \G(3/2+k) }{ k!\,\G(3/2-k) }
\end{split}
\ee  

\ni where the sign of the $i$ is ambiguous due to the {\it Stokes'
Phenomenon} \cite{watson}. The factor of $i$ was associated with the
fluctuation determinant of the three tachyonic modes associated with
the worldsheet slipping off the unstable pole of the five-sphere.

Due to the sign structure found in (\ref{Qfinal}) before squaring, the
analogous structure for the connected correlator of the primary with
the loop is a sum of a term proportional to $\exp(\sqrt{\l'})$ and of
another proportional to $(-1)^{J+1}\,\exp(-\sqrt{\l'})$. The sum of
these two terms should then be normalized by the expectation value of
the Wilson loop. If we employ the asymptotic expansions of the
modified Bessel functions in (\ref{quarterbps}), we have

\be\begin{split}
&\frac{I_J(\sqrt{\l'})}{I_1(\sqrt{\l'})} =\\
&\frac{ e^{\sqrt{\l'}}\,
\sum_{k=0}^{\infty} \left( \frac{-1}{2 \sqrt{\l'}} \right)^k\,
\frac{ \G(J+k+1/2) }{ k!\,\G(J-k+1/2) } \mp
i\, (-1)^J\,e^{-\sqrt{\l'}}\,
\sum_{k=0}^{\infty} \left( \frac{1}{2 \sqrt{\l'}} \right)^k\,
\frac{ \G(J+k+1/2) }{ k!\,\G(J-k+1/2) } }
{ e^{\sqrt{\l'}}\,
\sum_{k=0}^{\infty} \left( \frac{-1}{2 \sqrt{\l'}} \right)^k\,
\frac{ \G(3/2+k) }{ k!\,\G(3/2-k) } \pm
i\, e^{-\sqrt{\l'}}\,
\sum_{k=0}^{\infty} \left( \frac{1}{2 \sqrt{\l'}} \right)^k\,
\frac{ \G(3/2+k) }{ k!\,\G(3/2-k) } }.
\end{split}
\ee 

\ni This clearly reflects the presence of two saddle points in the
functional integrals in both the numerator and denominator.

% ========================================================================== %
\subsection{Summary}

The $1/4$ BPS circle is quite attractive as it provides a continuous,
one parameter family of circular Wilson loops which interpolate
between the supersymmetric circle of Zarembo and the celebrated $1/2$
BPS circle. Surprisingly, at the level of the Wilson loop expectation
value, this entire family of loops seem to be described by the $1/2$
BPS circle matrix model, with a rescaled coupling $\l' = \cos^2
\theta_0 \l$, which vanishes for the SUSY circle.

We have presented equal arguments that this correspondence holds for
the correlator of the $1/4$ BPS circle with a chiral primary operator,
as long as that operator shares the minimal 1/16 supersymmetry with
the loop. We have found that on the gauge theory side, the planar
ladders sum as they do for the 1/2 BPS correlator with a chiral
primary. Further, we have shown that the remaining diagrams cancel at
leading order. The result is that the $\l \rightarrow \l'$
prescription remains valid. At strong coupling, using string theory,
we recover the large-$\l'$ limit of our gauge theory result, as long
as the chiral primary in question shares SUSY. We find that when it
does not, the $\l \rightarrow \l'$ prescription breaks down. We
interpret this as an indication that the correlator is not protected
in this case. We therefore expect that higher-order gauge theory
calculations will display this lack of protection. It would be very
interesting to verify this.

Finally, we note that the double saddle points in the semi-classical
action for the string worldsheet describing the 1/4 BPS circle are
reflected in our gauge theory results, as was noted in
\cite{Drukker:2006ga} for the expectation value of the loop. It would
seem that, as long as a minimum of supersymmetry is maintained, the
$\l \rightarrow \l'$ prescription may be extended to include two point
functions with chiral primary operators.   

% ************************************************************************** %
% ************************************************************************** %
% ************************************************************************** %

\appendix

\chapter{Fermion representations}
\label{app:fermrep}

The fermionic normal modes (\ref{fnm}, \ref{fnmm}) break the $SO(8)$
symmetry to $SO(4)\times SO(4)$.  To make this symmetry manifest it is
convenient to label representations of $SO(4)_1\times SO(4)_2$ through
$(SU(2)\times SU(2))_1\times (SU(2)\times SU(2))_2$ spinor indices.
With this decomposition of the R-charge index, the fermionic fields
$\vartheta^a$ and $\lambda^a$, are expressed in terms of creation
operators $b^{\dagger}_{\a_1\a_2}$ and
$b^{\dagger}_{\dot\a_1\dot\a_2}$ which transform in the
$(1/2,0,1/2,0)$ and $(0,1/2,0,1/2)$ representations of $(SU(2)\times
SU(2))_1\times (SU(2)\times SU(2))_2$, respectively;
$\alpha_k$,$\da_k$ being two-component Weyl indices of $SO(4)_k$.  The
$SO(8)$ vector index $I$ splits into two $SO(4)\times SO(4)$ vector
indices $(i,i')$ so that we use vector index $i=1,\dots,4$ and
bi-spinor indices $\a_1,\da_1=1,2$ for the first $SO(4)$ and
$(i',\a_2,\da_2)$ for the second $SO(4)$.  Vectors are constructed in
terms of bi-spinor indices as $(\a_n)_{\a_1\dot
\a_1}=\sigma^i_{\a_1\dot\a_1} \a^i_n/\sqrt{2}$, $(\a_n)_{\a_2\dot
\a_2}=\sigma_{\a_2\da_2}^{i'} \a^{i'}_n/\sqrt{2}$ and transform as
$(1/2,1/2,0,0)$ and $(0,0,1/2,1/2)$, respectively. Here the
$\s$-matrices consist of the usual Pauli-matrices together with the 2d
unit matrix 

\be \s^i_{\a\da}=\bigl(i\t^1,i\t^2,i\t^3,-1\bigr)_{\a\da}
\ee 

\ni and satisfy the reality properties
$\bigl[\s^i_{\a\da}\bigr]^{\dag} = {\s^i}^{\da\a}$\,,
$\bigl[{\s^i}_{\a}^{\da}\bigr]^{\dag} = -{\s^i}^{\a}_{\da}$. These
properties are also satisfied by the fermionic oscillators, so that
$\left( \b_{n\,\a_1\,\a_2}\right)^\dag = \b^{\dag\,\a_1\,\a_2}_n$ and
$\left( \b_{n\,\a_1}^{\a_2}\right)^\dag = -\b^{\dag\,\a_1}_{n\,\a_2}$;
the same relations are obeyed for the dotted-index fermions.

Spinor indices are raised and lowered with the two-dimensional Levi-Civita symbols,
$\e_{\a\b}=\e_{\da\db}\equiv\left(\begin{matrix} 0
& 1 \cr -1 & 0 \end{matrix}\right)$, $( \e^{\a\,\b} )^\dag =
\e_{\b\,\a}$, for example

\be
A^{\a} = A_{\b}\, \e^{\a\,\b} \qquad A_{\a} = A^{\b} \,\e_{\a\,\b}
\ee

\ni and

\begin{equation}
\s^i_{\a\da} = \e_{\a\b}\e_{\da\db}\,{\s^i}^{\db\b}
\equiv \e_{\a\b}\,{\s^i}^{\b}_{\da} \equiv \e_{\da\db}\,{\s^i}^{\db}_{\a}\,.
\end{equation}

\ni The $\s$-matrices satisfy the relations

\begin{equation}
\s^i_{\a\da}{\s^j}^{\da\b}+\s^j_{\a\da}{\s^i}^{\da\b}
=2\d^{ij}\d_{\a}^{\b}\,,\qquad
{\s^i}^{\da\a}\s^j_{\a\db}+{\s^j}^{\da\a}\s^i_{\a\db}=
2\d^{ij}\d^{\da}_{\db}\,.
\end{equation}

\ni Some other properties satisfied by these matrices are

\bea
\label{rel1}
&&\e_{\a\b}\e^{\g\d}  = \d_{\a}^{\d}\d_{\b}^{\g}-\d_{\a}^{\g}\d_{\b}^{\d}\,,\\
&&\s^i_{\a\db}{\s^j}^{\db}_{\b}  = -\d^{ij}\e_{\a\b}+\s^{ij}_{\a\b}\,,\qquad
(\s^{ij}_{\a\b}\equiv \s^{[i}_{\a\da}{\s^{j]}}^{\da}_{\b}=\s^{ij}_{\b\a})\\
&&\s^i_{\a\da}{\s^j}^{\a}_{\db} = -\d^{ij}\e_{\da\db}+\s^{ij}_{\da\db}\,,
\qquad (\s^{ij}_{\da\db}\equiv \s^{[i}_{\a\da}{\s^{j]}}^{\a}_{\db}=\s^{ij}_{\db\da})\\
&&\s^k_{\a\da}\s^{k}_{\b\db}  = 2\e_{\a\b}\e_{\da\db}\,,\\
&&\s^{kl}_{\a\b}\s^{kl}_{\g\d}  = 4(\e_{\a\g}\e_{\b\d}+\e_{\a\d}\e_{\b\g})\,,\\
&&\s^{kl}_{\a\b}\s^{kl}_{\dg\dd}  = 0\,,\\
\label{rel7}
&&2\s^i_{\a\da}\s^{j}_{\b\db} = \d^{ij}\e_{\a\b}\e_{\da\db}
+\s^{k(i}_{\a_1\b_1}\s^{j)k}_{\da_1\db_1}
-\e_{\a\b}\s^{ij}_{\da\db}-\s^{ij}_{\a\b}\e_{\da\db}\,.
\eea

\be
(\s^{ij}_{\da\db})^\dag = {\s^{ij}}^{\da\db}
\ee

\be
\s^{k\,\db}_\g \, \s^{i\,j}_{\da\,\db} = \d^{ik} \s^j_{\g\,\da} - \d^{jk} \s^i_{\g\,\da}
\ee

\ni In this basis the gamma matrices have the following representation

\bea 
&&\gamma^i_{a\dot{a}} = \left(\begin{matrix} 0 &
\s^i_{\a_1\db_1}\d_{\a_2}^{\b_2} \cr
{\s^i}^{\da_1\b_1}\d^{\da_2}_{\db_2} & 0\end{matrix} \right)\
,\qquad~~ \g^i_{\dot{a}a} =\left(
\begin{matrix}
0 & \s^i_{\a_1\db_1}\d^{\da_2}_{\db_2} \cr
{\s^i}^{\da_1\b_1}\d_{\a_2}^{\b_2} & 0\end{matrix} \right)\ ,\\
&&\g^{i'}_{a\dot{a}} =\left(
\begin{matrix}
-\d_{\a_1}^{\b_1}\s^{i'}_{\a_2\db_2} & 0 \cr0 &
\d^{\da_1}_{\db_1}{\s^{i'}}^{\da_2\b_2}\end{matrix} \right)\ ,\qquad~~
 \g^{i'}_{\dot{a}a} =\left( 
\begin{matrix}
-\d_{\a_1}^{\b_1}{\s^{i'}}^{\da_2\b_2} & 0 \cr 0 &
\d^{\da_1}_{\db_1}\s^{i'}_{\a_2\db_2}\end{matrix} \right)\ .  \eea 

\ni and the projector reads 

\begin{equation}
\Pi_{ab} =
\left(\begin{matrix}\bigl(\s^1\s^2\s^3\s^4\bigr)_{\a_1}^{\b_1}\d_{\a_2}^{\b_2}
& 0 \cr 0 &
\bigl(\s^1\s^2\s^3\s^4\bigr)^{\da_1}_{\db_1}\d^{\da_2}_{\db_2}\end{matrix}\right)
= \left(\begin{matrix} \d_{\a_1}^{\b_1}\d_{\a_2}^{\b_2} & 0 \cr 0 &
-\d^{\da_1}_{\db_1}\d^{\da_2}_{\db_2}\end{matrix}\right)\,,
\end{equation}

\ni so that $(1\pm\Pi)/2$ projects onto $(1/2,0,1/2,0)$ and
$(0,1/2,0,1/2)$, respectively.

The supercharge $Q^-_{\a_1\dot \b_2}$ is a $(1/2,0,0,1/2)$ and
$Q^-_{\dot\a_1\b_2}$ is a $(0,1/2,1/2,0)$ representation. In this
notation it is convenient to define the linear combinations of the
free supercharges 

\be \sqrt{2}\eta Q\equiv
Q^-+i\bar{Q}^-~~~,~~~\sqrt{2}\bar{\eta} \widetilde{Q} \equiv
Q^--i\bar{Q}^- \ee 

\ni where $\eta=e^{i\pi/4}$, and $\bar{Q}^{\pm} =
e(\a)(Q^{\pm})^\dag$.  On the space of physical states they satisfy the
dynamical constraints

\bea
&&\left\{Q_{\a_1\da_2},Q_{\b_1\db_2}\right\}=
\left\{\widetilde{Q}_{\a_1\da_2},\widetilde{Q}_{\b_1\db_2}\right\}=
-2\epsilon_{\a_1\b_1}\epsilon_{\da_2\db_2}H\cr
&&\left\{Q_{\a_1\da_2},\widetilde{Q}_{\b_1\db_2}\right\}= -\mu
\epsilon_{\da_2\db_2}\left(\sigma^{ij}\right)_{\a_1\b_1}J^{ij}+ \mu
\epsilon_{\a_1\b_1}\left(\sigma^{i'j'}\right)_{\da_2\db_2}J^{i'j'}
\label{dynamconst}
\eea

\ni and similarly for $Q_{\da_1\a_2}$ and
$\widetilde{Q}_{\db_1\b_2}$. The free supercharge with raised
indices is understood as

\be
Q_2^{\a_1 \da_2} \equiv e(\a) \left( Q_{2\,\a_1 \da_2} \right)^\dag, \qquad 
Q_2^{\da_1 \a_2} \equiv e(\a) \left( Q_{2\,\da_1 \a_2} \right)^\dag
\ee

\ni and this gives

\be
Q_2^{\a_1 \da_2} Q_{2\,\a_1 \da_2}  = + 4 H_2 = Q_{2\,\a_1 \da_2} Q_2^{\a_1 \da_2}
\ee

\ni for these operators in the single string Hilbert space ${\cal
  H}_1$. For states in the three-string Hilbert space ${\cal H}_3$,
  i.e. $|Q_3\ra$, the $e(\a)$ is already encoded into the construction
  so that it should be dropped in the adjoint

\be
Q_{2\,\a_1 \da_2}  | Q_3^{\a_1 \da_2} \ra =
Q_2^{\a_1 \da_2}  | Q_{3 \a_1 \da_2} \ra\equiv
 \left( Q_{2\,\a_1 \da_2} \right)^\dag | Q_{3 \a_1 \da_2} \ra = + 4 |H_3\ra
\ee 

\ni and similarly $Q_3^{\a_1 \da_2} \equiv (Q_{3\,\a_1
\da_2})^\dag$. In the BMN basis, the full expression for the quadratic
supercharge $Q_{2\,\a_1\,\da_2}$ is\footnote{Note that
  ${\a^\dag_k}_{\a_1}^{\db_1} = - {\s^i}^{\db_1}_{\a_1} \a^{i\,\dag}_k /\sqrt{2}$, and
  similarly for the other $SO(4)$ since $[{\s^i}^{\a_1}_{\db_1}]^\dag =
- {\s^i}^{\db_1}_{\a_1}$.}

\be\label{Q2BMN}
\begin{split}
Q_{2\,\a_1\,\da_2} =
\frac{\bar \eta}{\sqrt{|\a|}}\,\sum_{k \neq 0} \O_k \Bigl( 
  &{\a^{\dag}_k}^{\db_1}_{\a_1} \, {\b_k}_{\db_1\,\da_2} 
+ i \, e(\a) \, {\a_k}^{\db_1}_{\a_1} \, {\b^{\dag}_k}_{\db_1\,\da_2} \\
+i\, &{\a^{\dag}_k}^{\b_2}_{\da_2} \, {\b_k}_{\a_1\,\b_2} 
+e(\a) \, {\a_k}^{\b_2}_{\da_2} \, {\b^{\dag}_k}_{\a_1\,\b_2} \Bigr) \\
+{\bar \eta}^{\,e(\a)}\, \sqrt{2\,\mu} \, \Bigl( 
 &{\a^{\dag}_0}^{\db_1}_{\a_1} \, {\b_0}_{\db_1\,\da_2}
+ i\,{\a_0}^{\db_1}_{\a_1} \, {\b^{\dag}_0}_{\db_1\,\da_2} \\
 + i \, e(\a) \,&{\a^{\dag}_0}^{\b_2}_{\da_2} \, {\b_0}_{\a_1\,\b_2} 
+e(\a) \,{\a_0}^{\b_2}_{\da_2} \, {\b^{\dag}_0}_{\a_1\,\b_2} \Bigr)
\end{split}
\ee

\ni where $\O_k$ is defined in (\ref{bigomega}).

Among states that are created by two oscillators, the state with
quantum numbers $(1,1,0,0)$ and $(0,0,1,1)$ which are created by two
bosons have no analogues amongst the two oscillator states containing
either one or two fermions.  Thus, they are not mixed with other
members of the supermultiplet.  These states in the main text are
denoted $\left|[{\bf 9}, {\bf 1}]\ra^{(ij)}\right.$ and $|[{\bf 1},
{\bf 9}]\ra^{(i'j')}$ in SO(8) notation.

% ************************************************************************** %
% ************************************************************************** %
% ************************************************************************** %
\chapter{Neumann matrices and associated quantities}
\label{app:neumann}

In this section we present the explicit expressions for the quantities
appearing in the prefactors and exponential part of $|H_3\rangle$ and
$|Q_3\rangle$ (\ref{H3andQ3full}).  Following the notation of \cite{Gutjahr:2004dv}, the
Neumann matrices can be written as

\bea
&&\wt{N}^{st}_{mn}=
\begin{cases}
\frac{1}{2}\bar{N}^{st}_{|m||n|}\left(1+U_{m(s)}U_{n(t)}\right)~~~~,m,n\neq0 \cr
\frac{1}{\sqrt{2}}\bar{N}^{st}_{|m|0}~~~~ m\neq 0 \cr
\bar{N}^{st}_{00}\,\end{cases}
\eea
with\footnote{To have a manifest symmetry in $1\leftrightarrow 2$ we
additionally redefined the oscillators as $(-1)^{s(n+1)}\a_{n(s)} \to
\a_{n(s)}$ for $n\in{\bZ}$, $s=1,2,3$ and analogously for the
fermionic oscillators.}
\be
\label{mn}
\bar{N}^{st}_{mn} =-(1-4\m\k K)^{-1}\frac{\k}{\a_s\omega_{n(t)}+
\a_t\omega_{m(s)}}\left[CU_{(s)}^{-1}C_{(s)}^{1/2}\bar{N}^s\right]_m
\left[CU_{(t)}^{-1}C_{(t)}^{1/2}\bar{N}^t\right]_n\,
\ee
\be
\label{m0}
\bar{N}^{st}_{m0} =
\sqrt{-2\m\k(1-\b_t)}\sqrt{\omega_{m(s)}}\bar{N}^s_m\,,\qquad t\in\{1,2\}
\ee
\be
\label{00a}
\bar{N}^{st}_{00} =
(1-4\m\k K)\left(\d^{st}+\sqrt{\b_s\b_t}\right)\,,\qquad  s,t\in\{1,2\}\,
\ee
\be
\label{00b} \bar{N}^{s3}_{00} = -\sqrt{\b_s}\,,\qquad s\in\{1,2\}
\ee
while 
\be
\wt Q^{rs}_{mn} = \begin{cases}
\frac{i}{2} e(m) \bar Q^{rs}_{|m||n|}, ~~~m,n\neq0\cr
\frac{i}{\sqrt{2}} e(m) \bar Q^{rs}_{|m|0}, ~~~m\neq0\cr
\bar Q^{rs}_{00}
\end{cases}
\ee
where \cite{Pankiewicz:2003kj}
\bsp
\bar Q^{rs}_{mn} = e(\a_r) \sqrt{\left|\frac{\a_s}{\a_r}\right|}
\left[ U^{1/2}_{(r)} C^{1/2} \bar N^{rs} C^{-1/2} U^{1/2}_{(s)}
  \right]_{mn},\qquad m,n>0\\
\bar Q^{sr}_{m0} = -\a_3 (1-\b_r) \sqrt{\a_r} 
\frac{e(\a_s)}{\sqrt{|\a_s|}} \left[\left(U_{(s)} C_{(s)} C
  \right)^{1/2} \bar N^s \right]_m,\qquad m>0\\
\bar Q^{3r}_{00} = -\bar Q^{r3}_{00} = 
\frac{1}{2}\sqrt{-\frac{\a_r}{\a_3}},\qquad \bar
Q^{rs}_{00}=0,\qquad r,s=\{1,2\}
\end{split}
\ee
and we note that $\bar Q^{sr}_{0m} = 0$, while
\be C_n = n\,,\qquad C_{n(s)} =
\omega_{n(s)}\equiv\sqrt{n^2+\bigl(\m\a_s\bigr)^2}\,,\qquad \k\equiv
\a_1\a_2\a_3 \ee \be U_{n(s)}
=\frac{1}{n}(\omega_{n(s)}-\m\a_s)\,,\qquad
U^{-1}_{n(s)}=\frac{1}{n}(\omega_{n(s)}+\m\a_s) \ee
and~\cite{He:2002zu}
\be
\label{K}
1-4\m\k K \approx -\frac{1}{4\pi r(1-r)\m\a_3}\,
\ee
\be
\label{N3}
\a_3\bar{N}^3_n \approx -\frac{\sin(n\pi r)}{\pi r(1-r)}\frac{1}{\omega_{n(3)}
\sqrt{-2\m\a_3(\omega_{n(3)}+\m\a_3)}}\,
\ee
\be
\label{Nr}
\a_3\bar{N}^s_n \equiv \a_3\bar{N}_n(\b_s) \approx -\frac{\sqrt{\b_s}}{2\pi r(1-r)}\frac{1}
{\omega_{n(s)}\sqrt{-2\m\a_3(\omega_{n(s)}-\m\a_3\b_s)}}
\ee
up to exponential corrections $\sim{\mc O}(e^{-\m\a_3})$
\footnote{To compare with the definition used in~\cite{He:2002zu} note
that $\bar{N}^s_{n\,\mbox{here}}=(-1)^{s(n+1)}
U_{n(s)}C_{n(s)}^{-1/2}\bar{N}^s_{n\,\mbox{there}}$.}. For the
bosonic constituents of the prefactor one has
\be
K^I = \sum_{s=1}^3\sum_{n\in{\bZ}}K_{n(s)}\a_{n(s)}^{I\,\dagger}\,,~~~
\wt{K}^I= \sum_{s=1}^3\sum_{n\in{\bZ}}K_{n(s)}\a_{-n(s)}^{I\,\dagger}\,
\ee
where
\be K_{0(s)}= (1-4\m\k
K)^{1/2}\sqrt{-\frac{2\m\k}{\a'}\bigl(1-\b_s\bigr)}\,,\qquad K_{0(3)}
= 0\, \ee and \be K_{n(s)}= -\frac{\k}{\sqrt{2\a'}\a_s}(1-4\m\k
K)^{-1/2}(\omega_{n(s)}+\m\a_s)
\sqrt{\omega_{n(s)}}\bar{N}^s_{|n|}\bigl(1-U_{n(s)}\bigr)\, \ee
For the fermionic constituents of the prefactor one has
\be Y^{\a_1\a_2} =
\sum_{s=1}^3\sum_{n\in{\bZ}}G_{|n|(s)}\b^{\dag\,\a_1\a_2}_{n(s)}\,,\qquad
Z^{\da_1\da_2} =
\sum_{s=1}^3\sum_{n\in{\bZ}}G_{|n|(s)}\b^{\dag\,\da_1\da_2}_{n(s)}\,,
\ee
where
\be
G_{0(s)} = (1-4\m\k K)^{1/2}\sqrt{1-\b_s}\,,\qquad G_{0(3)} = 0\,
\ee
and
\be
G_{n(s)}= \frac{e(\a_s)}{\sqrt{2|\a_s|}}\frac{\sqrt{-\k}}{(1-4\m\k K)^{1/2}}
\sqrt{(\omega_{n(s)}+\m\a_s)\omega_{n(s)}}\bar{N}^s_{|n|}\,
\ee
where in the above expressions we have used $\b_1\equiv r$ and $\b_2 \equiv 1-r$
(with $\b_t\equiv-\a_t/\a_3$ and $\a_3 < 0$).

% ************************************************************************** %
% ************************************************************************** %
% ************************************************************************** %
\chapter{Simpler forms and relations}
\label{app:relations}

We find a simpler expression for the Neumann matrices and associated
quantities

\be
\wt N^{3 \,r}_{n\,q} = -\frac{\sin(n\pi r) \sqrt{\b_r} \left(\L^+_n
  \L^+_q+ \L^-_n \L^-_q \right) }{2\pi\sqrt{\o_n \o_q} \left( q-\b_r n \right)}
\qquad
\wt N^{r\,s}_{q\,p} = \frac{ \sqrt{\b_r \b_s} \left( \L^+_q \L^+_p + \L^-_q \L^-_p \right) }
{4\pi \sqrt{ \o_q \o_p } \left( \b_s \o_q + \b_r \o_p \right) }
\ee

\be
\wh Q^{3 \,r}_{n\,q} = \frac{i\sin(|n|\pi r) 
  \left(\o_q+\b_r \o_n\right) }{2\pi\sqrt{\o_n \o_q} \left( q-\b_r n \right)}
\qquad
\wh Q^{r\,s}_{q\,p} = \frac{i\left( \b_s q - \b_r p \right) }
{4\pi \sqrt{ \o_q \o_p } \left( \b_s \o_q + \b_r \o_p \right) }
\ee

\ni where $\wh Q = \wt Q - \wt Q^T$. We also find

\be
K_n = +\a_3 \sin(n\pi r)  \sqrt{\frac{r(1-r)}{\pi\a'}} 
\frac{\L^-_n - \L^+_n}{\sqrt{\o_n}}
\ee

\be
K_q = -\a_3  \sqrt{\frac{r(1-r)}{\pi\a'\b_r}} 
\frac{\L^+_q - \L^-_q}{2\sqrt{\o_q}}
\ee

\be
G_q = \frac{1}{\sqrt{4\pi\o_q}}
\qquad G_n = -\frac{\sin(|n|\pi r)}{\sqrt{\pi\o_n}}
\ee

\be\label{bigomega}
\O_q = \L^+_q - \L^-_q \qquad \O_n = e(n) (\L^-_n - \L^+_n)
\ee

\noindent where,

\be
\L^+_q = \sqrt{\o_q - \b_r \m \a_3} \qquad \L^-_q = e(q) \sqrt{\o_q + \b_r \m \a_3}
\ee

\be
\L^+_n = \sqrt{\o_n - \m \a_3} \qquad \L^-_n = e(n) \sqrt{\o_n + \m \a_3}
\ee

\noindent We will also find use for

\be\label{defLs}
L^{3\,r}_{n\,q} \equiv K_n K_{-q} + K_{-n} K_{q} \qquad 
\wt L^{3\,r}_{n\,q} \equiv K_n K_{q} + K_{-n} K_{-q} .
\ee

\ni The following relations  may also be proven

\be\label{L2N}
K_p^{(s)}\,K_q^{(r)} + K_{-p}^{(s)}\,K_{-q}^{(r)} =
\frac{2\,\a_3^2\,r\,(1-r)}{\a'} \, 
\left( \frac{\o_q^{(r)}}{\b_r} + \frac{\o_p^{(s)}}{\b_s} \right)\,
\wt N_{q\,p}^{r\,s}
\ee

\be\label{L2N3}
K_n^{(3)}\,K_q^{(r)} + K_{-n}^{(3)}\,K_{-q}^{(r)} =
\frac{2\,\a_3^2\,r\,(1-r)}{\a'} \, 
\left( \frac{\o_q^{(r)}}{\b_r} - \o_n^{(3)} \right)\,
\wt N_{n\,q}^{3\,r}
\ee

\be
\O_q^{(r)} \, G_q^{(r)} = \sqrt{ \frac{\b_r\,\a'}{r\,(1-r)}} \,
\frac{1}{-\a_3} \, K_q^{(r)}
\ee

\be
\O_n^{(3)} \, G_n^{(3)} = \sqrt{ \frac{\a'}{r\,(1-r)}} \,
\frac{1}{-\a_3} \, K_n^{(3)}
\ee

\be
i\,\frac{\O_q^{(r)}}{\sqrt{\b_r}} \, \wh Q^{r\,s}_{q\,p} + 
\frac{\O_p^{(s)} }{\sqrt{\b_s}} \, \wt N^{r\,s}_{q\,p} =
 \sqrt{ \frac{\a'}{r\,(1-r)}} \, \frac{1}{-\a_3} K_q^{(r)} G_p^{(s)}
\ee

\be
i\,\O_n^{(3)} \, \wh Q^{3\,r}_{n\,q} + 
\frac{\O_q^{(r)} }{\sqrt{\b_r}} \, \wt N^{3\,r}_{n\,q} =
 \sqrt{ \frac{\a'}{r\,(1-r)}} \, \frac{1}{-\a_3} K_n^{(3)} G_q^{(r)}
\ee

\be
-i\,\frac{\O_q^{(r)}}{\sqrt{\b_r}} \, \wh Q^{3\,r}_{n\,q} + 
\O_n^{(3)} \, \wt N^{3\,r}_{n\,q} =
 \sqrt{ \frac{\a'}{r\,(1-r)}} \, \frac{1}{-\a_3} K_q^{(r)} G_n^{(3)}
\ee

\be
\left( \wt N^{r\, s}_{q\,p} \right)^2 - \left( \wh Q^{r\,s}_{q\,p} \right)^2
= \left(G^{(r)}_{|q|} \, G^{(s)}_{|p|} \right)^2 
\ee

\be
\left( \wt N^{3\, r}_{n\,q} \right)^2 + \left( \wh Q^{3\,r}_{n\,q} \right)^2
= -\left(G^{(r)}_{|q|} \, G^{(3)}_{|n|} \right)^2 
\ee

% ************************************************************************** %
% ************************************************************************** %
% ************************************************************************** %
\chapter{Calculational method}
\label{app:example}

% ************************************************************************** %
\section{Vertices and definitions}

We remind the reader of the construction of $|H_3\ra$ and $|Q_3\ra$ in (\ref{H3andQ3full})

\begin{equation}
|V \rangle = |E_{\alpha}\rangle |E_{\beta}\rangle\delta(\sum_{r=1}^3 \alpha_r)
\ee

\ni where $|E_{\alpha}\rangle$ and $|E_{\beta}\rangle$ are
exponentials of bosonic and fermionic oscillators respectively

\be
|E_{\alpha}\rangle=
\exp \left( \frac{1}{2} \sum_{r,s=1}^{3} \sum_{m,n = -\infty}^{\infty}
\alpha^{\dagger K}_{m\,(s)} {\widetilde N}^{st}_{mn} \alpha^{\dagger K}_{n\,(t)} \right )
| \alpha \rangle _{123}
\end{equation}

\ni and

\begin{equation}
|E_{\beta}\rangle=\exp\left(
\sum_{r,s=1}^3\sum_{m,n= -\infty}^{\infty}
\bigl(\b^{\a_1\a_2\,\dag}_{m(r)}\b^{\dag}_{n(s)\,\a_1\a_2}-
\b^{\da_1\da_2\,\dag}_{m(r)}\b^{\dag}_{n(s)\,\da_1\da_2}\bigr)
\wt{Q}^{rs}_{mn}\right)| \alpha \rangle _{123}
\end{equation}

\ni where $| \alpha \rangle _{123}=|0;\alpha_1 \rangle \otimes|0;\alpha_2
\rangle\otimes|0;\alpha_3 \rangle$. We then have

\bsp
&|H_3\ra =
g_2\,f(\m\a_3\,,\,\frac{\a_1}{\a_3})\frac{\alpha'}{8\,\a_3^3}
\Bigl[\bigl(K_i\K_j-\frac{\m\k}{\a'}\d_{ij}\bigr)v^{ij}
-\bigl(K_{i'}\K_{j'}-\frac{\m\k}{\a'}\d_{i'j'}\bigr)v^{i'j'}\\
&\qquad-K^{\da_1\a_1}\K^{\da_2\a_2}s_{\a_1\a_2}(Y)s^*_{\da_1\da_2}(Z)
-\K^{\da_1\a_1}K^{\da_2\a_2}s^*_{\a_1\a_2}(Y)s_{\da_1\da_2}(Z)\Bigr]|V\ra\,,\\
&|Q_{3\,\b_1\db_2}\ra =
 g_2\,\eta\,f(\m\a_3\,,\,
\frac{\a_1}{\a_3})\frac{1}{4\, \a_3^3}\,\sqrt{-\frac{\a'\k}{2}}
\Bigl(s_{\dg_1\db_2}(Z)t_{\b_1\g_1}(Y)\K^{\dg_1\g_1}\\
&\qquad\qquad\qquad\qquad\qquad\qquad\qquad\qquad\qquad\qquad
+ is_{\b_1\g_2}(Y)t^*_{\db_2\dg_2}(Z)\K^{\dg_2\g_2}\Bigr)|V\ra\,,\\
&|Q_{3\,\db_1\b_2}\ra = g_2\,{\bar \eta}\,f(\m\a_3\,,\,
\frac{\a_1}{\a_3})\frac{1}{4\, \a_3^3}\,\sqrt{-\frac{\a'\k}{2}}
\Bigl(s^*_{\g_1\b_2}(Y)t^*_{\db_1\dg_1}(Z)\K^{\dg_1\g_1}\\
&\qquad\qquad\qquad\qquad\qquad\qquad\qquad\qquad\qquad\qquad
+is^*_{\db_1\dg_2}(Z)t_{\b_2\g_2}(Y)\K^{\dg_2\g_2}\Bigr)|V\ra\,.
\end{split}
\ee

\ni where 

\be
K^I = \sum_{s=1}^3\sum_{n\in{\bZ}}K_{n(s)}\a_{n(s)}^{I\,\dagger}\,,~~~ \wt{K}^I=
\sum_{s=1}^3\sum_{n\in{\bZ}}K_{n(s)}\a_{-n(s)}^{I\,\dagger}\,
\ee

\be Y^{\a_1\a_2} =
\sum_{s=1}^3\sum_{n\in{\bZ}}G_{|n|(s)}\b^{\dag\,\a_1\a_2}_{n(s)}\,,\qquad
Z^{\da_1\da_2} = \sum_{s=1}^3\sum_{n\in{\bZ
}}G_{|n|(s)}\b^{\dag\,\da_1\da_2}_{n(s)}\,, \ee

\ni and

\begin{equation}
\K^{\dg_1\g_1} \equiv \K^i{\s^i}^{\dg_1\g_1}\,,\qquad
\K^{\dg_2\g_2} \equiv \K^{i'}{\s^{i'}}^{\dg_2\g_2}\,,
\end{equation} 

\ni where the $\s$-matrices are defined in appendix
\ref{app:fermrep}. We also have

\bsp\label{thisisv}
&v^{ij}  =
\d^{ij}\Bigl[1+\frac{1}{12}\bigl(Y^4+Z^4\bigr)+\frac{1}{144}Y^4Z^4\Bigr]\nonumber\\
&\qquad-\frac{i}{2}\Bigl[{Y^2}^{ij}\bigl(1+\frac{1}{12}Z^4\bigr)
-{Z^2}^{ij}\bigl(1+\frac{1}{12}Y^4\bigr)\Bigr]
+\frac{1}{4}\bigl[Y^2Z^2\bigr]^{ij}\,,\\
&v^{i'j'}  =
\d^{i'j'}\Bigl[1-\frac{1}{12}\bigl(Y^4+Z^4\bigr)+\frac{1}{144}Y^4Z^4\Bigr]\nonumber\\
&\qquad-\frac{i}{2}\Bigl[{Y^2}^{i'j'}\bigl(1-\frac{1}{12}Z^4\bigr)
-{Z^2}^{i'j'}\bigl(1-\frac{1}{12}Y^4\bigr)\Bigr]
+\frac{1}{4}\bigl[Y^2Z^2\bigr]^{i'j'}\,.
\end{split}
\ee

\ni Here we defined

\begin{equation}
{Y^2}^{ij} \equiv \s^{ij}_{\a_1\b_1}{Y^2}^{\a_1\b_1}\,,\quad
{Z^2}^{ij} \equiv \s^{ij}_{\da_1\db_1}{Z^2}^{\da_1\db_1}\,,\quad
\bigl(Y^2Z^2\bigr)^{ij} \equiv {Y^2}^{k(i}{Z^2}^{j)k}
\end{equation}

\ni and analogously for the primed indices.  We have also introduced
the following quantities quadratic and cubic in $Y$ and symmetric in
spinor indices

\begin{equation}
Y^2_{\a_1\b_1} \equiv Y_{\a_1\a_2}Y^{\a_2}_{\b_1}\,,\qquad
Y^2_{\a_2\b_2} \equiv Y_{\a_1\a_2}Y^{\a_1}_{\b_2}\,,
\end{equation}

\begin{equation}
Y^3_{\a_1\b_2} \equiv Y^2_{\a_1\b_1}Y^{\b_1}_{\b_2}=-Y^2_{\b_2\a_2}Y^{\a_2}_{\a_1}\,,
\end{equation}

\ni and quartic in $Y$ and antisymmetric in spinor indices

\begin{equation}
Y^4_{\a_1\b_1} \equiv
Y^2_{\a_1\g_1}{Y^2}^{\g_1}_{\b_1}=-\frac{1}{2}\e_{\a_1\b_1}Y^4\,,\qquad
Y^4_{\a_2\b_2} \equiv
Y^2_{\a_2\g_2}{Y^2}^{\g_2}_{\b_2}=\frac{1}{2}\e_{\a_2\b_2}Y^4\,,
\end{equation}

\ni where

\begin{equation}
Y^4 \equiv Y^2_{\a_1\b_1}{Y^2}^{\a_1\b_1}=-Y^2_{\a_2\b_2}{Y^2}^{\a_2\b_2}\,.
\end{equation}

\ni The spinorial quantities $s$ and $t$ are defined as

\begin{equation}
s(Y) \equiv Y+\frac{i}{3}Y^3\,\ ,~~~~~t(Y) \equiv \e+iY^2-\frac{1}{6}Y^4\,.
\end{equation}

\ni Analogous definitions can be given for $Z$. The normalization of
the dynamical generators is not fixed by the superalgebra at order
${\cal O}(g_2)$ and can be an arbitrary (dimensionless) function
$f(\m\a_3\,,\,\frac{\a_1}{\a_3})$ of the light-cone momenta and $\m$
due to the fact that $P^+$ is a central element of the algebra.

% ************************************************************************** %
\section{Commutation relations}

Rules for (anti)commutation of $\beta$ annihilation operator with $Y$
and $Z$ elements in the pre-factor:

\begin{equation}
Y^{\alpha_{1}\alpha_{2}}=\sum_{r=1}^{3}\sum_{n}G_{|n|\,(r)}\beta^{\dagger
\alpha_{1} \alpha_{2}}_{n\,(r)}
\end{equation}

\begin{equation}
\{\beta_{\gamma_{1}\,\gamma_{2}\,m\,(s)},Y^{\alpha_{1}\,\alpha_{2}} \}=G_{|m|\,(s)}
\delta^{\alpha_{1}}_{\gamma_{1}}\delta^{\alpha_{2}}_{\gamma_{2}}
\end{equation}

\begin{equation}
\{\beta_{\gamma_{1}\,\gamma_{2}\,m\,(s)},Y_{\alpha_{1}\,\alpha_{2}} \}=G_{|m|\,(s)}
\epsilon_{\alpha_{1}\,\gamma_{1}}\epsilon_{\alpha_{2}\,\gamma_{2}}
\end{equation}

\begin{equation}
\{\beta_{\gamma_{1}\,\gamma_{2}\,m\,(s)},Y_{\alpha_{1}\,\beta_{1}}^{2}\}=G_{|m|\,(s)}
\left(\epsilon_{\gamma_{1}\,\alpha_{1}}Y_{\beta_{1}\,\gamma_{2}}
+\epsilon_{\gamma_{1}\,\beta_{1}}Y_{\alpha_{1}\,\gamma_{2}}\right)
\end{equation}

\begin{equation}
\{\beta_{\gamma_{1}\,\gamma_{2}\,m\,(s)},Y_{\alpha_{2}\,\beta_{2}}^{2}\}=G_{|m|\,(s)}
\left(\epsilon_{\gamma_{2}\,\alpha_{2}}Y_{\gamma_{1}\,\beta_{2}}
+\epsilon_{\gamma_{2}\,\beta_{2}}Y_{\gamma_{1}\,\alpha_{2}}\right)
\end{equation}

\begin{equation}
\{\beta_{\gamma_{1}\,\gamma_{2}\,m\,(s)},Y_{\alpha_{1}\,\beta_{2}}^{3}\}=G_{|m|\,(s)}
\left(\epsilon_{\gamma_{1}\,\alpha_{1}}Y_{\gamma_{2}\,\beta_{2}}^{2}
-\epsilon_{\gamma_{2}\,\beta_{2}}Y_{\alpha_{1}\,\gamma_{1}}^{2}
+Y_{\alpha_{1}\,\gamma_{2}} \,Y_{\gamma_{1}\,\beta_{2}}\right)
\end{equation}

\begin{equation}
\{\beta_{\gamma_{1}\,\gamma_{2}\,m\,(s)},Y^{4}\}=-4\,G_{|m|\,(s)}
Y_{\gamma_{1}\,\gamma_{2}}^{3}
\end{equation}

And exactly the same for $Z$ and the dotted indices

% ************************************************************************** %
\section{Matrix elements}

Some useful matrix elements are (where $\la 3 | \equiv \la \a_3 |$)

\be
\la 3 | \,\a_n^{(3)\,i} \, \wt K^{\dg_1 \g_1} |V\ra
= \left( K_{-n}^{(3)} \, {\s^i}^{\dg_1 \g_1} + \wt K^{\dg_1 \g_1} \, \wt N^{3\,s}_{n\,p} \,
{\a_p^\dag}^{(s)\,i}
\right) \la 3 | V\ra
\ee

\noindent Where $s$ and any other internal string index is restricted to run over $1,2$ only.
We also have,

\be
\la 3 | \,\a_n^{(3)\,i} \, K_k \wt K_l |V\ra = \left(
K_n^{(3)} \, \wt K_l\, \d^{ik} + K_{-n}^{(3)} \, K_k\, \d^{il} 
+ K_k \wt K_l \, \wt N_{n\,p}^{3\,s}\, {\a_p^\dag}^{(s)\,i} \right) \la 3 | V\ra
\ee

\be
\begin{split}
\la 3 | \,\a_{n_1}^{(3)\,i}\,\a_{n_2}^{(3)\,j} \, K_k \,\wt K_l |V\ra
&= \Biggl( K_{n_1}^{(3)}\,K_{-n_2}^{(3)}\,\d^{ik}\,\d^{jl} +
K_{-n_1}^{(3)}\,K_{n_2}^{(3)}\,\d^{il}\,\d^{jk}\\ &+ K_{n_1}^{(3)}
\,\wt K_l \,\wt N_{n_2\,p}^{3\,s}\, {\a_p^\dag}^{(s)\,j} \,\d^{ik}
+K_{n_2}^{(3)} \,\wt K_l \,\wt N_{n_1\,p}^{3\,s}\,
{\a_p^\dag}^{(s)\,i} \,\d^{jk}\\ &+ K_{-n_1}^{(3)} \,K_k \,\wt
N_{n_2\,p}^{3\,s}\, {\a_p^\dag}^{(s)\,j} \,\d^{il} +K_{-n_2}^{(3)}
\,K_k \,\wt N_{n_1\,p}^{3\,s}\, {\a_p^\dag}^{(s)\,i} \,\d^{jl}\\ &+
K_k \,\wt K_l \left( \wt N_{n_1\,n_2}^{3\,3} \d^{ij} + \wt
N_{n_1\,p}^{3\,s}\,\wt N_{n_2\,q}^{3\,r}\, {\a_p^\dag}^{(s)\,i} \,
{\a_q^\dag}^{(r)\,j} \right)\Biggl) \la 3 | V\ra
\end{split}
\ee

\be
\begin{split}
\la 3 | \, \b_{n\,\,\s_1\s_2}^{(3)} \, t_{\b_1 \g_1} (Y) \,|V\ra &=
\Biggl\{ \left( \wt Q_{n\,p}^{3\,s} - \wt Q_{p\,n}^{s\,3} \right) \,
\b_{p\,\,\s_1\s_2}^{\dag\,(s)} \left( \e_{\b_1 \g_1} + iY^2_{\b_1
\g_1} + \frac{1}{12}\,\e_{\b_1 \g_1} Y^4 \right)\\ & + i G_{|n|}^{(3)}
\left( \e_{\s_1 \b_1} \,Y_{\g_1 \s_2} + \e_{\s_1 \g_1} \, Y_{\b_1
\s_2} \right) - \frac{1}{3} G_{|n|}^{(3)} \, \e_{\b_1 \g_1} \,
Y^3_{\s_1 \s_2} \Biggr\} \la 3 | V\ra
\end{split}
\ee

\be
\begin{split}
\la 3 | \, \b_{n\,\,\s_1\s_2}^{(3)} \, t_{\b_2 \g_2} (Y) \,|V\ra &=
\Biggl\{ \left( \wt Q_{n\,p}^{3\,s} - \wt Q_{p\,n}^{s\,3} \right) \,
\b_{p\,\,\s_1\s_2}^{\dag\,(s)} \left( \e_{\b_2 \g_2} + iY^2_{\b_2
\g_2} - \frac{1}{12}\,\e_{\b_2 \g_2} Y^4 \right)\\ & + i G_{|n|}^{(3)}
\left( \e_{\s_2 \b_2} \,Y_{\s_1 \g_2} + \e_{\s_2 \g_2} \, Y_{\s_1
\b_2} \right) + \frac{1}{3} G_{|n|}^{(3)} \, \e_{\b_2 \g_2} \,
Y^3_{\s_1 \s_2} \Biggr\} \la 3 | V\ra
\end{split}
\ee

\be
\begin{split}
\la 3 | \, \b_{n\,\,\s_1\s_2}^{(3)} \, s_{\b_1 \g_2} (Y) \,|V\ra &=
\Biggl\{ \left( \wt Q_{n\,p}^{3\,s} - \wt Q_{p\,n}^{s\,3} \right) \,
\b_{p\,\,\s_1\s_2}^{\dag\,(s)} \left( Y_{\b_1 \g_2} + \frac{i}{3}
Y^3_{\b_1 \g_2} \right) + G_{|n|}^{(3)} \, \e_{\s_1 \b_1} \, \e_{\s_2
\g_2}\\ &+ \frac{i}{3} G_{|n|}^{(3)} \left( \e_{\s_1 \b_1}\, Y^2_{\s_2
\g_2} - \e_{\s_2 \b_2}\,Y^2_{\s_1 \b_1} + Y_{\b_1 \s_2} \, Y_{\s_1
\g_2} \right ) \Biggr\} \la 3 | V\ra
\end{split}
\ee

\be \la \a_3 | \,\a_{n_1}^{(3)\,i}\,\a_{n_2}^{(3)\,j} \,
Q_{2\,\,\a_1\,\da_2} = \frac{\bar \eta}{\sqrt{2\,|\a_3|}} \la \a_3
| \, \left( \,\O_{n_2} \, {\s^j}_{\a_1}^{\db_1} \, \a_{n_1}^i \,
\b_{n_2\,\,\db_1 \da_2} + ~~ (i \leftrightarrow j, \,n_1
\leftrightarrow n_2 ) \right) \ee

We are now prepared to construct the matrix elements we need 
certain calculations, for instance,

\be\label{23}
\begin{split}
\e^{\a_1 \b_1} \, &\e^{\da_2 \db_2} \, \la 3 |\,
\a_{n_1}^{(3)\,i}\,\a_{n_2}^{(3)\,j} \, Q_{2\,\,\a_1\,\da_2} \, |
Q_{3\,\,\b_1\,\db_2} \ra = \frac{g_2}{4\,\a_3^3} \,
\sqrt{\frac{-\a'\,\kappa}{-4\,\a_3}} \,\O_{n_2}\,{\s^j}^{\dg_1}_{\a_1}
\e^{\a_1 \b_1} \, \e^{\da_2 \db_2} \\ &\times \Biggl[ \left\{
K_{-n_1}^{(3)}\,{\s^i}^{\ds_1 \s_1} + \wt K^{\ds_1 \s_1} \, \wt
N_{n_1\,p}^{3\,s}\, {\a_p^\dag}^{(s)\,i} \right\} \, t_{\b_1 \s_1} (Y)
\, \left\{ \left( Z_{\ds_1 \db_2} + \frac{i}{3} Z^3_{\ds_1 \db_2}
\right)\, \wh Q_{n_2\,q}^{3\,r} \, \b_{q\,\,\dg_1 \da_2}^{\dag\,
(r)}\right.\\ &\qquad\qquad\left.+ \,G_{|n_2|}^{(3)} \, \e_{\dg_1
\ds_1}\,\e_{\da_2 \db_2} + \frac{i}{3} G_{|n_2|}^{(3)} \left(
\e_{\dg_1 \ds_1} Z^2_{\da_2 \db_2} - \e_{\da_2 \db_2} \, Z^2_{\dg_1
\ds_1} + Z_{\ds_1 \da_2} \, Z_{\dg_1 \db_2} \right) \right\}\\ &+i\,
\wt K^{\ds_2 \s_2} \wt N_{n_1\,p}^{3\,s}\, {\a_p^\dag}^{(s)\,i} \,
s_{\b_1 \s_2} (Y) \biggl\{ \wh Q_{n_2\,q}^{3\,r} \,\b_{q\,\,\dg_1
\da_2}^{\dag\, (r)} \left( \e_{\db_2 \ds_2} -iZ^2_{\db_2 \ds_2} -
\frac{1}{12} \,\e_{\db_2 \ds_2} \,Z^4 \right) - \frac{1}{3} \,
G_{|n_2|}^{(3)} \, \e_{\db_2 \ds_2}\,Z^3_{\dg_1 \da_2} \\
&\qquad\qquad + \, i\, G_{|n_2|}^{(3)} \, \left( \e_{\da_2 \db_2} \,
Z_{\dg_1 \ds_2} + \e_{\da_2 \ds_2} \, Z_{\dg_1 \db_2} \right) \biggr\}
\Biggr] \la 3 | V\ra \quad + \quad(i \leftrightarrow j, n_1
\leftrightarrow n_2)
\end{split}
\ee

\noindent where $\wh Q = \wt Q - \wt Q^T$.

\be\label{NH3elem}
\begin{split}
\la 3 |\, \a_{n_1}^{(3)\,i}\,&\a_{n_2}^{(3)\,j} \, | H_3 \ra =
\frac{g_2\,\a'}{8\,\a_3^3} \Biggl[ \left( \wt N_{n_1\,n_2}^{3\,3} \,
\d^{ij} + \wt N_{n_1\,p}^{3\,s}\,\wt N_{n_2\,q}^{3\,r}\,
{\a_p^\dag}^{(s)\,i} \, {\a_q^\dag}^{(r)\,j} \right)\\ &
\qquad\qquad\qquad\qquad\times \biggl( \left[ K_k \wt K_l -
\frac{\m\,\a}{\a'}\, \d_{kl} \right]\,v^{kl} - \left[ K_{k'} \wt
K_{l'} - \frac{\m\,\a}{\a'}\, \d_{k'l'} \right]\,v^{k'l'} \\
&\qquad\qquad\qquad- K^{\dr_1 \r_1}\, \wt K^{\dr_2 \r_2} \, s_{\r_1
\r_2}(Y) \,s^*_{\dr_1 \dr_2}(Z) - \wt K^{\dr_1 \r_1}\, K^{\dr_2 \r_2}
\, s^*_{\r_1 \r_2}(Y) \,s_{\dr_1 \dr_2}(Z) \biggr)\\ +\biggl(
&K_{n_1}^{(3)}\,K_{-n_2}^{(3)}\,\d^{ik}\,\d^{jl} +
K_{-n_1}^{(3)}\,K_{n_2}^{(3)}\,\d^{il}\,\d^{jk} + K_{n_1}^{(3)} \,\wt
K_l \,\wt N_{n_2\,p}^{3\,s}\, {\a_p^\dag}^{(s)\,j} \,\d^{ik}
+K_{n_2}^{(3)} \,\wt K_l \,\wt N_{n_1\,p}^{3\,s}\,
{\a_p^\dag}^{(s)\,i} \,\d^{jk}\\
&\qquad\qquad\qquad\qquad\qquad\qquad+ K_{-n_1}^{(3)} \,K_k \,\wt
N_{n_2\,p}^{3\,s}\, {\a_p^\dag}^{(s)\,j} \,\d^{il} +K_{-n_2}^{(3)}
\,K_k \,\wt N_{n_1\,p}^{3\,s}\, {\a_p^\dag}^{(s)\,i} \,\d^{jl}
\biggr)\,v^{kl}\\ -\biggl( & {\s^i}^{\dr_1 \r_1} \, K_{n_1}^{(3)} \,
\wt K^{\dr_2 \r_2} \, \wt N_{n_2\,p}^{3\,s}\, {\a_p^\dag}^{(s)\,j} +
{\s^j}^{\dr_1 \r_1} \, K_{n_2}^{(3)} \, \wt K^{\dr_2 \r_2} \, \wt
N_{n_1\,p}^{3\,s}\, {\a_p^\dag}^{(s)\,i} \biggr) \, s_{\r_1 \r_2}(Y)
\,s^*_{\dr_1 \dr_2}(Z)\\ -\biggl( & {\s^i}^{\dr_1 \r_1} \,
K_{-n_1}^{(3)} \, K^{\dr_2 \r_2} \, \wt N_{n_2\,p}^{3\,s}\,
{\a_p^\dag}^{(s)\,j} + {\s^j}^{\dr_1 \r_1} \, K_{-n_2}^{(3)} \,
K^{\dr_2 \r_2} \, \wt N_{n_1\,p}^{3\,s}\, {\a_p^\dag}^{(s)\,i} \biggr)
\, s^*_{\r_1 \r_2}(Y) \,s_{\dr_1 \dr_2}(Z) \Biggl] \la 3 | V\ra
\end{split}
\ee

% ************************************************************************** %
\section{More matrix elements}

Consider
\be
\begin{split}
\la 3 | \,\a_{n_1}^{(3)\,i}\,\a_{n_2}^{(3)\,j} \, \wt K^{\dg_1 \g_1}
|V\ra = ( K_{-n_1}^{(3)} \, {\s^i}^{\dg_1 \g_1} \wt
N^{3\,s}_{n_2\,p}\, \a_{p}^{\dag \,(s)\,j} + K_{-n_2}^{(3)} \,
{\s^j}^{\dg_1 \g_1} \wt N^{3\,s}_{n_1\,p}\, \a_{p}^{\dag \,(s)\,i}\\
+\wt K^{\dg_1 \g_1} \, \wt N^{3\,3}_{n_1\,n_2}\, \d^{i\,j}) \la 3 |
V\ra
\end{split}
\ee
and therefore
\be\label{NQ3elem}
\begin{split}
\la 3 &|\, \a_{n_1}^{(3)\,i}\,\a_{n_2}^{(3)\,j} \, |
Q_{3\,\b_{1}\db_{2}} \ra
=\frac{g_{2}\eta}{4\a_{3}^{3}}\sqrt{-\frac{\a'\kappa}{2}}\\
&\times\Bigl\{s_{\dg_{1}\db_{2}}(Z)t_{\b_{1}\g_{1}}(Y) \Bigl[
K_{-n_1}^{(3)} \, {\s^i}^{\dg_1 \g_1} \wt N^{3\,s}_{n_2\,p}\,
\a_{p}^{\dag \,(s)\,j} + K_{-n_2}^{(3)} \, {\s^j}^{\dg_1 \g_1} \wt
N^{3\,s}_{n_1\,p}\, \a_{p}^{\dag \,(s)\,i} \\
&~~\qquad\qquad\qquad\qquad+\wt K^{\dg_1 \g_1} \left( \wt
N^{3\,3}_{n_1\,n_2}\, \d^{i\,j} + \wt N^{3\,r}_{n_1\,q}\,
\a_{q}^{\dag\,(r)\,i}\, \wt N^{3\,s}_{n_2\,p}\, \a_{p}^{\dag\,(s)\,j}
\right) \Bigr]\\ &~~+i s_{\b_{1}\g_{2}}(Y)t_{\db_{2}\dg_{2}}^{*}(Z)
\wt K^{\dg_2 \g_2} \, \left( \wt N^{3\,3}_{n_1\,n_2}\, \d^{i\,j} + \wt
N^{3\,r}_{n_1\,q}\, \a_{q}^{\dag\,(r)\,i}\, \wt N^{3\,s}_{n_2\,p}\,
\a_{p}^{\dag\,(s)\,j} \right) \Bigr\}
\end{split}
\ee

\noindent We will need the following expressions:

\be
\begin{split}
w_{n\,\db_{1}\da_{2}}^{i\,j}&\equiv
[\b_{n\,\db_{1}\da_{2}},v^{i\,j}]=G_{|n|}^{(3)} \Biggl\{\delta^{i\,j}
\left( -\frac{1}{3}Z^{3}_{\db_{1}\da_{2}} -
\frac{1}{36}Y^{4}Z^{3}_{\db_{1}\da_{2}}\right) \\ &+\frac{i}{2}
\biggl[ \frac{1}{3}Y^{2\,i\,j}Z^{3}_{\db_{1}\da_{2}}
-\sigma^{i\,j}_{\dr_{1}\dg_{1}} \left(
\d^{\dr_{1}}_{\db_{1}}Z^{\dg_{1}}_{\da_{2}}
+\d^{\dg_{1}}_{\db_{1}}Z^{\dr_{1}}_{\da_{2}} \right) \left(
1+\frac{1}{12}Y^{4} \right) \biggr] -\frac{1}{2}Y^{2\,k\,(i}
\sigma^{j)\,k}_{\db_{1}\dg_{1}}Z^{\dg_{1}}_{\da_{2}} \Biggr\}
\end{split}
\ee
\be
\begin{split}
w_{n\,\db_{1}\da_{2}}^{i'\,j'}&\equiv
[\b_{n\,\db_{1}\da_{2}},v^{i'\,j'}]=G_{|n|}^{(3)}
\Biggl\{\delta^{i'\,j'} \left( \frac{1}{3}Z^{3}_{\db_{1}\da_{2}} -
\frac{1}{36}Y^{4}Z^{3}_{\db_{1}\da_{2}}\right) \\ &+\frac{i}{2}
\biggl[ -\frac{1}{3}Y^{2\,i'\,j'}Z^{3}_{\db_{1}\da_{2}}
-\sigma^{i'\,j'}_{\dr_{2}\dg_{2}} \left(
\d^{\dr_{2}}_{\da_{2}}Z^{\dg_{2}}_{\db_{1}}
+\d^{\dg_{2}}_{\da_{2}}Z^{\dr_{2}}_{\db_{1}} \right) \left(
1-\frac{1}{12}Y^{4} \right) \biggr] -\frac{1}{2}Y^{2\,k'\,(i'}
\sigma^{j')\,k'}_{\da_{2}\dg_{2}}Z^{\dg_{2}}_{\db_{1}} \Biggr\}
\end{split}
\ee
We'll also need the following
\be Q_{2\,\,\a_1\,\da_2}
\,\a_{n_1}^{\dag\,(3)\,k}\,\a_{n_2}^{\dag\,(3)\,l} \, | \a_3 \ra =
\frac{-\eta}{\sqrt{2\,|\a_3|}} \, \left( \,\O_{n_1} \,
{\s^k}_{\a_1}^{\dg_1} \, \a_{n_2}^{\dag\,l} \, \b^{\dag}_{n_1\,\,\dg_1
\da_2} + ~~ (k \leftrightarrow l, \,n_1 \leftrightarrow n_2 ) \right)
| \a_3 \ra \ee
or
\be |\l \ra = Q_2^{\b_1\,\db_2}
\,\a_{n_1}^{\dag\,(3)\,k}\,\a_{n_2}^{\dag\,(3)\,l} \, | \a_3 \ra =
\frac{-\eta}{\sqrt{2\,|\a_3|}} \, \left( \,\O_{n_1} \,
     {\s^k}^{\dg_1\,\b_1} \, \a_{n_2}^{\dag\,l} \,
     \b^{\dag\,\db_2}_{n_1\,\,\dg_1} + ~~ (k \leftrightarrow l, \,n_1
     \leftrightarrow n_2 ) \right) | \a_3 \ra \ee
allowing us to calculate,
\be\label{Q2onH3}
\begin{split}
\la \l &| H_3 \ra = \frac{g_2\,\a'}{8\,\a_3^3} \frac{\bar
\eta}{\sqrt{2|\a_3|}} {\s^k}_{\b_1\,\dg_1} \,\O_{n_1} \\
&\times\Biggl[ \left( -\b^{\dag\,\dg_1}_{q\,(r)\,\db_2} \, \wh
Q^{3\,r}_{n_1\,q} \right) \wt N^{3\,s}_{n_2\,p}\, \a_{p}^{\dag
\,(s)\,l} \biggl\{ \left[ K_i \wt K_j - \frac{\m\,\a}{\a'}\, \d_{ij}
\right]\,v^{ij} - \left[ K_{i'} \wt K_{j'} - \frac{\m\,\a}{\a'}\,
\d_{i'j'} \right]\,v^{i'j'}\\ &\qquad\qquad- K^{\dr_1 \r_1}\, \wt
K^{\dr_2 \r_2} \, s_{\r_1 \r_2}(Y) \,s^*_{\dr_1 \dr_2}(Z) - \wt
K^{\dr_1 \r_1}\, K^{\dr_2 \r_2} \, s^*_{\r_1 \r_2}(Y) \,s_{\dr_1
\dr_2}(Z) \biggr\}\\ &+\wt N^{3\,s}_{n_2\,p}\, \a_{p}^{\dag \,(s)\,l}
\biggl\{ \left[ K_i \wt K_j - \frac{\m\,\a}{\a'}\, \d_{ij}
\right]\,\left(w_{n_1}^{ij}\right)^{\dg_1}_{\db_2} - \left[ K_{i'} \wt
K_{j'} - \frac{\m\,\a}{\a'}\, \d_{i'j'}
\right]\,\left(w_{n_1}^{i'j'}\right)^{\dg_1}_{\db_2}\\ &\qquad\qquad +
G_{|n_1|}^{(3)}\,K^{\dr_1 \r_1}\, \wt K^{\dr_2 \r_2} \, s_{\r_1
\r_2}(Y) \biggl[ \d^{\dg_1}_{\dr_1} \, \e_{\db_2\,\dr_2} -\frac{i}{3}
\left( \d^{\dg_1}_{\dr_1} \, Z^2_{\db_2\,\dr_2} - \e_{\db_2\,\dr_2} \,
Z^{2\,\dg_1}_{\dr_1} + Z_{\dr_1\,\db_2} \, Z^{\dg_1}_{\dr_2} \right)
\biggr]\\ &\qquad\qquad + G_{|n_1|}^{(3)}\,\wt K^{\dr_1 \r_1}\,
K^{\dr_2 \r_2} \, s^*_{\r_1 \r_2}(Y) \biggl[ \d^{\dg_1}_{\dr_1} \,
\e_{\db_2\,\dr_2} +\frac{i}{3} \left( \d^{\dg_1}_{\dr_1} \,
Z^2_{\db_2\,\dr_2} - \e_{\db_2\,\dr_2} \, Z^{2\,\dg_1}_{\dr_1} +
Z_{\dr_1\,\db_2} \, Z^{\dg_1}_{\dr_2} \right) \biggr]\biggr\}\\
&+\left(K_{n_2}^{(3)} \, \wt K_j\, \d^{li} + K_{-n_2}^{(3)} \, K_i\,
\d^{lj} \right) \Bigl\{ \left( -\b^{\dag\,\dg_1}_{q\,(r)\,\db_2} \,
\wh Q^{3\,r}_{n_1\,q} \right) \, v^{ij} +
\left(w_{n_1}^{ij}\right)^{\dg_1}_{\db_2} \Bigr\}\\ &-\s^{l\,\dr_1 \,
\r_1} \, K_{n_2}^{(3)} \, \wt K^{\dr_2 \r_2} \, s_{\r_1 \r_2}(Y)
\biggl\{ s^*_{\dr_1 \dr_2}(Z)\,\left(
-\b^{\dag\,\dg_1}_{q\,(r)\,\db_2} \, \wh Q^{3\,r}_{n_1\,q} \right)\\
&\qquad\qquad\qquad\qquad\qquad\qquad\quad-G_{|n_1|}^{(3)}\,\left[
\d^{\dg_1}_{\dr_1} \, \e_{\db_2\,\dr_2} -\frac{i}{3} \left(
\d^{\dg_1}_{\dr_1} \, Z^2_{\db_2\,\dr_2} - \e_{\db_2\,\dr_2} \,
Z^{2\,\dg_1}_{\dr_1} + Z_{\dr_1\,\db_2} \, Z^{\dg_1}_{\dr_2} \right)
\right] \biggr\}\\ &-\s^{l\,\dr_1 \, \r_1} \, K_{-n_2}^{(3)} \,
K^{\dr_2 \r_2} \, s^{*}_{\r_1 \r_2}(Y) \biggl\{ s_{\dr_1
\dr_2}(Z)\,\left( -\b^{\dag\,\dg_1}_{q\,(r)\,\db_2} \, \wh
Q^{3\,r}_{n_1\,q} \right)\\
&\qquad\qquad\qquad\qquad\qquad\qquad\quad-G_{|n_1|}^{(3)}\,\left[
\d^{\dg_1}_{\dr_1} \, \e_{\db_2\,\dr_2} +\frac{i}{3} \left(
\d^{\dg_1}_{\dr_1} \, Z^2_{\db_2\,\dr_2} - \e_{\db_2\,\dr_2} \,
Z^{2\,\dg_1}_{\dr_1} + Z_{\dr_1\,\db_2} \, Z^{\dg_1}_{\dr_2} \right)
\right] \biggr\} \Biggr]\\ &+~~ (k \leftrightarrow l, \,n_1
\leftrightarrow n_2 )
\end{split}
\ee

% ************************************************************************** %
\section{Example calculation}
\label{appsec:example}

There is a more direct method of calculating amplitudes without
resorting to the intermediate state projectors introduced in section
\ref{sec:pankmassshift}. Consider a 2-string $\rightarrow$ 2-string
operator $M$; as an example $M$ could be $\la e | H_3\ra\la H_3 |
e\ra$, where $|e\ra$ is some external string state. Consider the
spacetime index structure of the matrix element

\be\label{matelem}
\la \a_{n_1}^i \a_{n_2}^j | M | \a^{k\,\dag}_{n_1} \a^{l\,\dag}_{n_2}
\ra = A \,\d^{ij} \d^{kl} + B \,\d^{ik} \d^{jl} + C \,\d^{il} \d^{jk}.
\ee

\ni Based on this structure, we now consider the $|[{\bf 9},{\bf
    1}]\ra$ state

\be
\left. |[{\bf 9}, {\bf
1}]\ra^{(ij)}\right. =
\frac{1}{\sqrt{2}}
\left(\a^{\dag\,i}_n\a^{\dag\,j}_{-n}+\a^{\dag\,j}_n\a^{\dag\,i}_{-n}
-\frac{1}{2}\d^{ij}\a^{\dag\,k}_n\a^{\dag\,k}_{-n}\right)|\a\ra.
\ee

\ni We find that

\be
{}^{(i,j)}\la [{\bf 9},{\bf
    1}]| M  |[{\bf 9},{\bf
    1}]\ra^{(i,j)} = (B+C) \left(1 + \frac{1}{2} \d^{ij}\right)
\ee

\ni where $1+\frac{1}{2}\d^{ij}$ is the normalization of the $|[{\bf 9},{\bf
    1}]\ra$ state, and is understood to be dropped in calculating an
    energy shift. Thus in calculating such a shift, we are instructed
    to simply calculate $B$ and $C$ in (\ref{matelem}).

The method is to take $|e\ra = \a^{\dag\,k}_{n_1}
\a^{\dag\,l}_{n_2} | \a_3 \ra$, and so for a general amplitude
involving the 3-string states $|A\ra$ and $|B\ra$, we expand the
2-string states $\la e|A\ra$ and $\la e| B\ra$ to the desired order in
intermediate oscillators and calculate

\be\label{howto}
 \la A | \a^{\dag\,k}_{n_1} \a^{\dag\,l}_{n_2} |   \a_3 \ra
\la  \a_3|\, \a^i_{n_1} \a^j_{n_2} | B \ra 
\ee

\noindent where for the $|[{\bf 9},{\bf 1}\ra$ state we sum only those
contributions proportional to either $\d^{ik}\d^{jl}$ or
$\d^{il}\d^{jk}$. For example if the contact term were being
calculated, $|A\ra = |B\ra = |Q_3\ra$, and we would use (\ref{NQ3elem})
expanded to quadratic order in oscillators in order to capture the
impurity conserving channel contribution. Of course the appropriate
level matching must be enforced, and we further have that

\be\label{vaccontr}
\la \text{3 string vacuum} |  \a_3 \ra \la  \a_3 |
\text{3 string vacuum} \ra = 2 r(1-r) 
\ee

\noindent where the factor of 2 counts the two ways of contracting the
internal string vacuua between right and left. Finally, the internal
momenta must be integrated over via $\int_0^1 d\,r$.

As an example we calculate the contribution from the double fermionic
intermediate state to the $H_3$ term of the mass shift for the trace
state of section \ref{sec:trace}. We take those pieces of
(\ref{NH3elem}) quadratic in fermionic oscillators (see the fourth
line of (\ref{NH3elem}))

\bsp
\la \a_3 |\, \a_{n_1}^{(3)\,i}\,\a_{n_2}^{(3)\,j} \, | H_3\ra =
f\frac{g_2\,\a'}{8\,\a_3^3} \biggl(
K_{n_1}^{(3)}\,K_{-n_2}^{(3)}\,\d^{ik}\,\d^{jl} +
K_{-n_1}^{(3)}\,K_{n_2}^{(3)}\,\d^{il}\,\d^{jk} \biggr) \d^{kl} \\
\times \left(
\frac{1}{2} \wh Q^{r_1\,r_2}_{q_1\,q_2}
\bigl(\b^{\a_1\a_2\,\dag}_{q_1(r_1)}\b^{\dag}_{q_2(r_2)\,\a_1\a_2}-
\b^{\da_1\da_2\,\dag}_{q_1(r_1)}\b^{\dag}_{q_2(r_2)\,\da_1\da_2}\bigr)\right) \\
\times \la \a_3 | \text{3 string vacuum}\ra
\end{split}
\ee

\ni where we have taken the leading delta function term of $v^{kl}$
(\ref{thisisv}). Another contribution stemming from the quadratic
pieces of $v^{kl}$ will be zero here because for the trace state we
take $i=j$ and sum. This kills the antisymmetric combination of
$\s$-matrices found in the quadratic terms of $v^{kl}$. Taking
$n_1=n=-n_2$ and acting $\frac{1}{2}\d^{ij}$ on the above element,
and then taking it's square modulus, we find

\bsp
\left| \frac{1}{2} \la \a_3 |\, \a_{n}^{(3)\,i}\,\a_{-n}^{(3)\,i} \, |
H_3\ra \right|^2 &= |f|^2
\left(\frac{g_2\a'}{4\a_3^3}\right)^2 \left[ K_n^2 + K_{-n}^2
  \right]^2 \left( \frac{1}{2}\wh Q^{r_1\,r_2}_{q_1\,q_2} \right)^* 
\frac{1}{2} \wh Q^{s_1\,s_2}_{p_1\,p_2}\\
 \la \text{3 string vacuum}& |  \a_3 \ra
 \bigl(\b^{\a_1\a_2}_{q_2(r_2)}
\b^{}_{q_1(r_1)\,\a_1\a_2}-
\b^{\da_1\da_2}_{q_2(r_2)}\b^{}_{q_1(r_1)\,\da_1\da_2}\bigr)\\
&\qquad\bigl(\b^{\dag\,\b_1\b_2}_{p_1(s_1)}\b^{\dag}_{p_2(s_2)\,\b_1\b_2}-
\b^{\dag\,\db_1\db_2}_{p_1(s_1)}\b^{\dag}_{p_2(s_2)\,\db_1\db_2}\bigr)
\la \a_3 | \text{3 string vacuum}\ra.
\end{split}
\ee

\ni Commuting the $\b$ oscillators though one another gives the
following factor (times two since dotted and undotted oscillators are
orthogonal)

\be\label{fermcontr}
4 \left( \d_{q_1 p_1} \d^{r_1 s_1} \d_{q_2 p_2} \d^{r_2 s_2} -
\d_{q_1 p_2} \d^{r_1 s_2} \d_{q_2 p_1} \d^{r_2 s_1} \right).
\ee

\ni Sums over $s_i$ and $p_i$ set these variables to their $r_i$ and
$q_i$ counterparts indicated by the delta functions. Level matching is
then imposed by adding the following factor before summing over $r_i$ and
$q_i$

\be
\left(\d^{r_1\,r_2} \d_{q_1+q_2} + (1 -
\d^{r_1\,r_2})\d_{q_1}\d_{q_2}\right)
\ee

\ni where the second term counts the contribution from the zero modes
where each intermediate string is excited by a single oscillator. For
the purpose of this example, we will ignore these as we are interested
in demonstrating the convergence of the sum over the remaining mode
number. We have

\be\label{delEexample}
\d E = \int_0^1 dr\, 2r(1-r) |f|^2
\left(\frac{g_2\a'}{4\a_3^3}\right)^2 
\left[ K_n^2 + K_{-n}^2 \right]^2 
\sum_q \sum_{s=1}^2 \frac{-\a_3}{2\left(\o_n - r^{-1}\o_q\right)} 
2\cdot 8\cdot \frac{1}{4} \left| \wh Q_{q\,-q}^{s\, s} \right|^2
\ee

\ni where we have included the energy denominator, used
(\ref{vaccontr}), and noted that the second term in (\ref{fermcontr})
gives the same result as the first on account of the antisymmetry of
$\wh Q^{r\,s}_{p\,q}$. The sum over the string label $s$ just gives a
factor of 2, as there is a 1$\lr$2 string symmetry running through all
equations. Also note that $f=r^{-1}(1-r)^{-1}$. The convergence of the
sum is evident from the form of $\wh Q^{1\,1}_{q\,-q}$ (see appendix \ref{app:relations})

\be
\left| \wh Q^{1\,1}_{q\,-q} \right|^2 = \left(\frac{1}{4\pi}\right)^2
\frac{q^2}{\o_q^4} \sim \frac{1}{q^2}
\ee

\ni and so the sum in (\ref{delEexample}) is convergent.

% ************************************************************************** %
% ************************************************************************** %
% ************************************************************************** %
\chapter{Plane-wave matrix model 2-loop effective action}
\label{app:pwmm2loop}

\section{The theta diagram}

The ``theta'' diagram is given by the combination of the two
three-vertices for the scalars of the first kind. It is the middle
diagram in figure \ref{fig:pwmm2loop}.

\noindent From the action we get the vertex as:
\begin{equation}
\frac{i\mu}{3}\Tr[ \epsilon_{\bar a \bar b \bar c}
 X^{\bar a} X^{\bar b} X^{\bar c}]
\end{equation}
so we can write the diagram as:
\begin{equation}
-\frac{1}{2}\frac{\mu^2}{3^2}\int_{-\frac{\beta}{2}}^{\frac{\beta}{2}}d\tau
\int_{-\frac{\beta}{2}}^{\frac{\beta}{2}} \epsilon_{\bar a \bar b \bar
  c}\epsilon_{\bar d \bar e \bar f} X^{\bar a}_{a b}(\tau)X^{\bar
  b}_{b c}(\tau)X^{\bar c}_{c a}(\tau)X^{\bar d}_{d e}(\tau')X^{\bar
  e}_{e f}(\tau')X^{\bar f}_{f d}(\tau')
\end{equation}
There is therefore three propagators between $X(\tau)$ and
$X(\tau')$'s the $\bar a,\bar b,\bar c$ and $\bar d,\bar e,\bar f$ all
range over 1,2,3 and the $\epsilon$ limits them to the totally
symmetric and totally antisymmetric combinations. That makes up for 6
on each side.  Furthermore the requirement that the diagram be planar
makes sure that if the $\bar a,\bar b,\bar c$ combination is symmetric
then $\bar d,\bar e,\bar f$ can not be anti-symmetric and vice
versa. The planar contractions also introduce a sign, due to their
mixed symmetry. There is therefore $6\times 3$ combinations of the
propagators with the summation over the indices:
\begin{equation}
\frac{\mu^2R^3 }{8\omega_1^3}\sum_{a b
c}\int_{-\frac{\beta}{2}}^{\frac{\beta}{2}}d\tau
\int_{-\frac{\beta}{2}}^{\frac{\beta}{2}} d\tau'
[g(\tau-\tau')+g^*(\tau'-\tau)]_{a b}[..]_{b c} [..]_{c a}
\end{equation}
where
\begin{equation}
\omega_1=\frac{\mu}{3}
\end{equation}
The diagram can then be written as
\begin{equation}
\frac{\mu^2R^3 }{8\omega_1^3}\sum_{a b
c}\int_{-\frac{\beta}{2}}^{\frac{\beta}{2}}d\tau
\int_{-\frac{\beta}{2}}^{\frac{\beta}{2}} d\tau'
\left[\frac{\phi_{ab}^{(\tau'-\tau-\beta)}}{1-\phi_{ab}^{-\beta}}+
\frac{\phi_{ab}^{*-(\tau'-\tau)}}{1-\phi_{ab}^{*-\beta}}\right]
\left[..\right]_{bc}\left[..\right]_{ca}\theta(\tau'-\tau)
\end{equation}
\begin{displaymath}
+\left[\frac{\phi_{ab}^{*-(\tau'-\tau-\beta)}}{1-\phi_{ab}^{*-\beta}}+
\frac{\phi_{ab}^{(\tau'-\tau)}}{1-\phi_{ab}^{-\beta}}\right]
[..]_{bc}[..]_{ca}\theta(\tau-\tau')
\end{displaymath}

\ni where $\phi_{ab}$ is defined in section \ref{sec:threeeff}.
The expression can also be written as:
\begin{equation}
\frac{\mu^2R^3}{8\omega_1^3}\sum_{a b
c}\int_{-\frac{\beta}{2}}^{\frac{\beta}{2}}d\tau
\int_{-\frac{\beta}{2}}^{\frac{\beta}{2}} d\tau' \left[\frac{{\cal A}_{a
b}\phi_{a b}^{(\tau'-\tau)}+{\cal B}_{a b}\phi_{a
b}^{*-(\tau'-\tau)}}{C_{a b}}\right][..]_{b c}[..]_{c
a}\theta(\tau'-\tau)
\end{equation}
\begin{displaymath}
+\left[\frac{{\cal A}^*_{a b}\phi_{a b}^{*-(\tau'-\tau)}+{\cal B}^*_{a b}\phi_{a
b}^{(\tau'-\tau)}}{C_{a b}}\right][..]_{b c}[..]_{c a}\theta(\tau-\tau')
\end{displaymath}

\noindent where again, quantities are defined in section
\ref{sec:threeeff}. It is possible to interchange $\tau$ and $\tau'$
in the second line of the integral in order to get everything
multiplied by the same Heaviside function. The second line is then
just a complex conjugate of the first because $C_{a b}$ is a real
quantity. It is then possible to write the diagram as:
\begin{equation}
\frac{\mu^2R^3 }{4\omega_1^3}\sum_{a b c}\frac{1}{C_{a b}C_{b c}C_{c
a}}\int_{-\frac{\beta}{2}}^{\frac{\beta}{2}}d\tau
\int_{\tau}^{\frac{\beta}{2}} d\tau' \text{Re}[({\cal A}_{a b}{\cal
A}_{b c}{\cal A}_{c a})(\phi_{a b}\phi_{b c}\phi_{c a})^{(\tau'-\tau)}
\end{equation}
\begin{displaymath}
+({\cal A}_{a b}{\cal A}_{b c}{\cal B}_{c a})(\phi_{a b}\phi_{b
c}\phi^{*-1}_{c a})^{(\tau'-\tau)}+...+ ({\cal B}_{a b}{\cal B}_{b
c}{\cal B}_{c a})(\phi^{*-1}_{a b}\phi^{*-1}_{b c}\phi^{*-1}_{c
a})^{(\tau'-\tau)}]
\end{displaymath}
This can be simplified by noticing that:
\begin{equation}
\phi_{a b}\phi_{b c}\phi_{c a}=e^{3\omega_1},\phi_{a b}\phi_{b
c}\phi^{*-1}_{c a}=e^{\omega_1},...
\end{equation}
Then diagram is given by:
\begin{equation}
\frac{\mu^2R^3 }{4\omega_1^3}\sum_{ab
c}\frac{1}{C_{a b}C_{b c}C_{c a}}\int_{-\frac{\beta}{2}}^{\frac{\beta}{2}}d\tau
\int_{\tau}^{\frac{\beta}{2}} d\tau'
\text{Re}({\cal A}_{a b}{\cal A}_{b c}{\cal A}_{c a})e^{3\omega_1(\tau'-\tau)}+
\end{equation}
\begin{displaymath}
\text{Re}({\cal A}_{a b}{\cal A}_{b c}{\cal B}_{c
a}+...)e^{\omega_1(\tau'-\tau)}+\text{Re}({\cal A}_{a b}{\cal B}_{b
c}{\cal B}_{c a}+...)e^{-\omega_1(\tau'-\tau)} +\text{Re}({\cal B}_{a
b}{\cal B}_{b c}{\cal B}_{c a})e^{-3\omega_1(\tau'-\tau)}
\end{displaymath}
Performing the integration:
\begin{equation}
\frac{\mu^2R^3 }{4\omega_1^3}\sum_{ab c}\frac{1}{C_{a b}C_{b c}C_{c
a}}\Biggl[ \text{Re}({\cal A}_{a b}{\cal A}_{b c}{\cal A}_{c
a})\frac{1+3\beta\omega_1-e^{3\beta\omega_1}}{9\omega_1^2}+
\end{equation}
\begin{displaymath}
\text{Re}({\cal A}_{a b}{\cal A}_{b c}{\cal B}_{c
a}+...)\frac{1+\beta\omega_1-e^{\beta\omega_1}}{\omega_1^2}+
\text{Re}({\cal A}_{a b}{\cal B}_{b c}{\cal B}_{c
a}+...)\frac{1-\beta\omega_1-e^{-\beta\omega_1}}{\omega_1^2}
\end{displaymath}
\begin{displaymath}
+\text{Re}({\cal B}_{a b}{\cal B}_{b c}{\cal B}_{c
a})\frac{1-3\beta\omega_1-e^{-3\beta\omega_1}}{9\omega_1^2}\Biggr]
\end{displaymath}
Using the given definitions of ${\cal A}$ and ${\cal B}$ from (\ref{ABC}) it is possible to
simplify the above

\begin{equation}
\boxed{
\begin{split}
\frac{27\beta R^3}{4\mu^2}&\sum_{abc}
\frac{1}{C^{\omega_1}_{ab} C^{\omega_1}_{bc} C^{\omega_1}_{ca} }
\Biggl[ 1 + e^{-2\beta\mu} - 9\,e^{-4\beta\mu/3}
+ 16\,e^{-\beta\mu} -9 \,e^{-2\beta\mu/3}\\
&+\left[\cos(\beta A_{a b})+\cos(\beta  A_{b c})+\cos(\beta A_{ca})\right]\\
&\qquad\times\left[ 2e^{-5\beta\mu/3} +  2e^{-\beta\mu/3} -8e^{-4\beta\mu/3}
-8e^{-2\beta\mu/3} + 12e^{-\beta\mu}\right] \Biggr]
\end{split}}
\end{equation}

\section{The figure-eight diagram}

This section is dedicated to the calculation of the first diagram in
figure \ref{fig:pwmm2loop}.

\noindent This comes from expanding the action to first order, $\exp(-S) \simeq 1 - S$,
and so we pick up a sign:

\begin{equation}
\begin{split}
\frac{R}{4}&\int_{-\beta/2}^{\beta/2}\left<\Tr[X^{i},X^{j}]^{2}\right>d\tau=
\frac{R}{2}\int_{-\beta/2}^{\beta/2}
\sum_{i<j}\left<\Tr[X^{i},X^{j}]^{2}\right>d\tau\\
=&\frac{R}{2}\int_{-\beta/2}^{\beta/2} \sum_{a<b}\left<\Tr[X^{{\bar
a}},X^{{\bar b}}]^{2}\right>d\tau+\frac{R}{2}\int_{-\beta/2}^{\beta/2}
\sum_{i<j}\left<\Tr[X^{i},X^{j}]^{2}\right>d\tau \\
&\qquad\qquad\qquad\qquad\qquad\qquad\qquad
+\frac{R}{2}\int_{-\beta/2}^{\beta/2} \sum_{{\bar
a}}\sum_{i}\left<\Tr[X^{{\bar a}}, X^{i}]^{2}\right> d\tau
\end{split}
\label{fig8}
\end{equation}

\noindent Let us consider one of the above terms

\begin{equation}
\begin{split}
\Tr([X^{{\bar a}},X^{{\bar b}}][X^{{\bar a}},X^{{\bar
b}}])&=\Tr((X^{{\bar a}}X^{{\bar b}}-X^{{\bar b}}X^{{\bar
a}})(X^{{\bar a}}X^{{\bar b}}-X^{{\bar b}}X^{{\bar a}}))\\
&=2\Tr(X^{{\bar a}}X^{{\bar b}}X^{{\bar a}}X^{{\bar b}}-X^{{\bar
a}}X^{{\bar a}}X^{{\bar b}}X^{{\bar b}})
\end{split}
\end{equation}

\noindent While the first term has a non-planar contribution, hence the whole
expression for the $\bar a$-flavor bosonic field would be

\begin{equation}
\begin{split}
R\sum_{\bar a< \bar b}\sum_{abcd}
&\int_{-\beta/2}^{\beta/2}(-\langle X_{ab}^{\bar a}X_{bc}^{\bar a}\rangle
\langle X_{cd}^{\bar b}X_{da}^{\bar b}\rangle)\,d\tau\\
&=-R\sum_{\bar a< \bar b}\sum_{abcd}
\int_{-\beta/2}^{\beta/2}\left(\frac{R}{2\omega_{1}}\right)^{2}\delta^{aa}\delta^{bb}
\delta_{ac}\delta_{bb}
\delta_{ac}\delta_{dd}
P_{ab}(\omega_{1})P_{cd}(\omega_{1})
\end{split}
\end{equation}

\noindent where

\begin{equation}
P_{ab}(\omega)\equiv(g+g^*_{-})_{ab}^{\omega}
\end{equation}

\noindent as per (\ref{ferm_thingy}) and (\ref{scalar_prop}). Therefore what we get is

\begin{equation}
-\left(\frac{R^3}{4\omega_{1}^{2}}\right)\frac{3\cdot2}{2}
\sum_{abd}\int_{-\beta/2}^{\beta/2}
P_{ab}(\omega_{1})P_{ad}(\omega_{1})\,d\tau
\end{equation}

\noindent The only difference for the $i$-flavor would be the value
of the $\Sigma_{i<j}$, hence putting everything together
we find the following result for (\ref{fig8}):

\begin{equation}
\begin{split}
-\frac{R^3}{4}\sum_{abd}\int_{-\beta/2}^{\beta/2}
\Biggl[\frac{3\cdot2}{2\omega_{1}^{2}}
P_{ab}(\omega_{1})&P_{ad}(\omega_{1})\\
&+\frac{6\cdot5}{2\omega_{2}^{2}}P_{ab}(\omega_{2})P_{ad}(\omega_{2})
+\frac{6\cdot3}{\omega_{1}\omega_{2}}P_{ab}(\omega_{1})P_{ad}(\omega_{2})
\Biggr]d\tau
\end{split}
\end{equation}

\noindent Considering that $\omega_{1}=\mu/3$ and $\omega_{2}=\mu/6$, then
the final expression would be

\begin{equation}
-\frac{ R^3}{4\mu^{2}}\sum_{abd}\int_{-\beta/2}^{\beta/2}\biggl[27P_{ab}(\omega_{1})
P_{ad}(\omega_{1})+540P_{ab}(\omega_{2})P_{ad}(\omega_{2})
+324P_{ab}(\omega_{1})P_{ad}(\omega_{2})\biggr]d\tau
\end{equation}

\noindent Now, consider the product of two $P$'s:

\begin{equation}
P_{ab}P_{ad} = (H_1 \theta + G_1 \bar\theta)(H_2 \theta + G_2 \bar\theta)
= H_1 H_2 \theta + G_1 G_2 \bar \theta
\end{equation}

\noindent It so happens that, when $\tau$ is set equal to $\tau'$

\begin{equation}
H_1 = H_2 = G_1 = G_2 = \frac{1}{C}(1-e^{-2\beta\omega})
\end{equation}

\noindent Using the fact that $[\theta(s) + \theta(-s)]_{s=0} \equiv 1$,
we arrive at the final form:

\begin{equation}
\boxed{
-\frac{27\beta R^{3}}{4\mu^{2}}\sum_{abd}\Biggl[ \frac{[1-e^{-2\beta\mu/3}]^{2}}{
C_{ab}^{\omega_{1}}C_{ad}^{\omega_{1}}}
+20\frac{[1-e^{-\beta\mu/3}]^{2}}{C_{ab}^{\omega_{2}}C_{ad}^{\omega_{2}}}
+12\frac{[1-e^{-2\beta\mu/3}][1-e^{-\beta\mu/3}]}{
C_{ab}^{\omega_{1}}C_{ad}^{\omega_{2}}}\Biggr]}
\end{equation}

% ************************************************************************** %
% ************************************************************************** %
% ************************************************************************** %
\chapter{$1/4$ BPS Wilson loop - chiral primary correlator}
\label{app:wilson}

% ************************************************************************** %
\section{Metric fluctuations}
\label{app1}

Given (\ref{fluct}) and (\ref{metric}), we must construct the traceless
symmetric double covariant derivative,

\be
D_{(\m} D_{\n)} \equiv \frac{1}{2} \left( D_\m D_\n + D_\n D_\m \right) -
\frac{1}{5} g_{\m \n} \, g^{\r \s} D_{\r \s}.
\ee

\noindent The action of $D_\m D_\n$ on a scalar field $\phi$
is,

\be
D_\m D_\n \phi = \del_\m \del_\n \phi - \G^\l_{\m \n} \del_{\l} \phi.
\ee

\noindent The Christoffel symbols for the $AdS$ geometry (\ref{metric}) are,

\bea
\G^{r_i}_{\phi_i \phi_i} = -r_i \qquad \G^y_{\phi_i \phi_i} =
\frac{r_i^2}{y} \qquad &&
\G^{\phi_i}_{\phi_i r_i} = \frac{1}{r_i} \qquad \G^{\phi_i}_{\phi_i y}
= -\frac{1}{y} \cr
\G^{y}_{r_i r_i} = \frac{1}{y} \qquad \G^{r_i}_{y r_i} &=&
-\frac{1}{y} \qquad \G^{y}_{y y} = -\frac{1}{y}
\eea

\noindent where $i=1,2$. The trace of $D_\m D_\n \,\phi$ is given by,

\be
g^{\m \n} D_\m D_\n = \sum_{i=1}^2 \left( y^2 \del_y^2 + y^2 \del_{r_i}^2 +
\frac{y^2}{r_i^2} \del_{\phi_i}^2 -3y \del_y + \frac{y^2}{r_i} \del_{r_i}
\right)\,\phi
\ee

\noindent Because of (\ref{prop}), we only keep those terms of $D_{(\m} D_{\n)}$
which contain derivatives in $y$. These are,

\be
D_{(y} D_{y)} = \frac{4}{5} \del_y^2  + \frac{8}{5 y} \del_y, \qquad
D_{(r_1} D_{r_1)} = \frac{1}{r_1^2} D_{(\phi_1} D_{\phi_1)} =  -\frac{1}{5} \del_y^2
-\frac{2}{5y} \del_y.
\ee

\noindent We now note that since the derivatives will be acting on
$y^J$ from the propagator (\ref{prop}), we may replace $\del_y^2 \rightarrow
J(J-1)/y^2$ and $y^{-1}\del_y \rightarrow J/y^2$.  Therefore the
metric fluctuations may be expressed as follows,

\bea
\d g_{y y} &=&
\left[ -\frac{6J}{5} + \frac{4}{J+1} \left( \frac{4}{5} J(J-1) +
  \frac{8}{5}J\right) \right] \frac{L^2}{y^2} = 2J \frac{L^2}{y^2}\cr
\d g_{r_1 r_1} &=& \frac{1}{r_1^2} \d g_{\phi_1 \phi_1} =
\left[ -\frac{6J}{5} -\frac{4}{J+1} \left( \frac{1}{5} J(J-1) +
  \frac{2}{5}J\right) \right] \frac{L^2}{y^2} = -2J \frac{L^2}{y^2}.
\eea

% ************************************************************************** %
\section{Spherical harmonics}
\label{app2}

The five-sphere is embedded in $\bR^6$ in the following manner,

 \bea &&x^1 = \sin \theta \cos \phi \qquad \qquad x^2
= \sin \theta \sin \phi \cr &&x^3 = \cos \theta \sin \r \cos \hat
\phi \qquad x^4 = \cos \theta \sin \r \sin \hat \phi \cr &&x^5 =
\cos \theta \cos \r \cos \tilde \phi \qquad x^6 = \cos \theta \cos
\r \sin \tilde \phi, \eea and has the metric

\be ds^2_{S^5} = d\theta^2 + \sin^2 \theta \,d\phi^2 + \cos^2 \theta
\, \left( d\r^2 + \sin^2\r\, d\hat\phi^2 + \cos^2 \r\, d \tilde
\phi^2 \right). \ee The embedding (\ref{embed}) takes $\r = \pi/2,
\hat \phi =0$, or $x^4 =x^5=x^6=0$. Note that $\r \in [0,\pi/2]$ while
$\theta \in [0,\pi]$. A general chiral primary normalized as in
(\ref{chiralprim}) may be written as,

\be
\frac{2^{J/2}}{\sqrt{\l^J J}} C^{I_1 \ldots I_J} \Tr 
\Phi_{I_1} \ldots \Phi_{I_J}
\ee

\noindent where $C^{I_1 \ldots I_J}$ is traceless symmetric and
$C^{I_1 \ldots I_J}{C^*}^{I_1 \ldots I_J} = 1$. The corresponding
spherical harmonic
is given by $Y_J(\theta,\phi) = C^{I_1 \ldots I_J} x^{I_1} \ldots
x^{I_J}$. A properly normalized (i.e. (\ref{chiralprim})) operator built on 
$\Tr(u \cdot \Phi)^J$ will then correspond to

\be
Y_J(\theta,\phi) ={\cal N}_J(u) \biggl[
u_1 \sin \theta \,\cos\phi +
u_2 \sin \theta \,\sin\phi +
u_3 \cos \theta \biggr]^J
\ee

\noindent for some normalization ${\cal N}_J(u)$. If we choose
$u_1=u_2=0$ and $u_3 = \pm i u_4 = 1$, i.e. the operator
$\Tr(\Phi_3 \pm i \Phi_4)^J/\sqrt{\l^J J}$, then ${\cal N}_J(u)
=2^{-J/2}$.

% ************************************************************************** %
\section{R-symmetry }\label{app3}

Let $ {\cal O}_J = \frac{1}{\sqrt{J\,\l^J}}\,\Tr
\left(\Phi_1 + i \Phi_2\right)^J $,
   Let $U$ be a rotation in
the $x^1$-$x^2$ plane. Then

\be \la {\cal O}_J(x) \, W[C_{\tiny 1/4}] \ra  = \la U\, {\cal
O}_J(x) \, W[C_{\tiny 1/4}]  \, U^\dag \ra = \la {\cal O}_J(U\,x) \,
U \, W[C_{\tiny 1/4}]  \, U^\dag \ra \ee  Examining $C_{\tiny 1/4}$
in (\ref{quartcircle}), we see that the spatial rotation acting on
$W[C_{\tiny 1/4}] $ may be realized by a shift in the contour
parameter $\t$, which can in turn by compensated by an R-symmetry
rotation $R$ in the $\theta^1$-$\theta^2$ plane, $ U\, W[C_{\tiny 1/4}]  \,
U^\dag = R\, W[C_{\tiny 1/4}]  \, R^\dag $.  Then,

\be \la {\cal O}_J(x) \, W[C_{\tiny 1/4}]  \ra = \la R\, {\cal
O}_J(Ux)\, R^\dag\, W[C_{\tiny 1/4}] \ra.  \label{proof}\ee The
operator expansion coefficient depends on the leading asymptotic in
large $x$ which is a function of only the length of $C_{\tiny 1/4}$
and $x^2$,

\be \la {\cal O}_J(x) \, W[C_{\tiny 1/4}]  \ra \simeq
\left(\frac{2\pi R}{4\pi^2 x^{2}}\right)^J\xi_J+\ldots
\label{proof1}\ee Performing the $\theta^1$-$\theta^2$ plane R-symmetry
transformation on ${\cal O}_J$ multiplies it by a phase $\exp(i J
\phi)$ so that,

\be \la R\, {\cal O}_J(Ux)\, R^\dag\, W[C_{\tiny 1/4}]  \ra \simeq
e^{iJ\phi}\left(\frac{2\pi R}{4\pi^2 (Ux)^{2}}\right)^J\xi_J+\ldots
= e^{iJ\phi}\left(\frac{2\pi R}{4\pi^2 x^{2}}\right)^J\xi_J+\ldots
\ee Using (\ref{proof}) and (\ref{proof1}), we have
$e^{iJ\phi}\,\xi_J = \xi_J$, i.e. $\xi_J=0$.

\bibliographystyle{plain}
\bibliography{phdthesis}

\end{document}